\documentclass[12pt,a4paper]{book-quix2}
\usepackage{makeidx}
\usepackage[T1]{fontenc}
\usepackage[english]{babel}
\usepackage{graphicx}
\usepackage{epsfig} 
\usepackage{wrapfig}
\usepackage{amsmath}
\usepackage{amssymb}
\usepackage{amsthm} 
\DeclareMathOperator{\arcsh}{arcsh}
\usepackage{dcolumn} 
\usepackage{bm} 
\usepackage{fancyhdr}
\usepackage{syntonly}
\usepackage{sidecap}
\renewcommand{\chaptermark}[1]{\markboth{#1}{}}
\renewcommand{\sectionmark}[1]{\markright{\thesection\ #1}}
\fancypagestyle{plain}{%
   \fancyhead{} 
}

\makeindex

\begin{document}
\def\·{$\cdot$}

\renewcommand{\tilde}{\widetilde}
\def\ie{{\em i.\hspace{0.2em}e.}}
\def\eg{{\em e.\hspace{0.2em}g.}}
\def\ea{{\em et~al.}}

\def\ef{\varepsilon_F}
\def\eps{\epsilon}
\def\uk{u_{\kv}}
\def\vk{v_{\kv}}
\def\xik{\xi_{\kv}}
\def\Ek{E_{\kv}}
\def\ek{\epsilon_{\kv}}
\def\ekp{\epsilon_{\kvp}}
\def\hm{\frac{\hbar^2}{2m}}
\def\hM{\frac{\hbar^2}{2M}}
\def\Pm{-\frac{\hbar^2}{2m}\nabla^2}

\def\De{\Delta}
\def\Dek{\Delta_{\kv}}
\def\mDe{\overline{\Delta}}
\def\Dekx{\Delta_{\kv}}
\def\de{\delta}
\def\Om{\Omega}
\def\om{\omega}
\def\dmu{\delta\mu}
\def\al{\alpha}
\def\alm{\alpha_{\rm max}}
\def\G{\Gamma}
\def\R{\rho}
\def\ru{\rho_\ua}
\def\rd{\rho_\da}
\def\vareps{\varepsilon}

\def\Raw{\Rightarrow}
\def\raw{\rightarrow}
\def\rp{\right)}
\def\lp{\left(}
\def\ra{\rangle}
\def\la{\langle}
\def\rra{\right\rangle}
\def\lla{\left\langle}

\def\psiod{\hat{\psi}^{\dagger}}
\def\psio{\hat{\psi}}
\def\Psio{\hat{\Psi}}
\def\cd{\hat{c}^{\dagger}}
\def\cop{\hat{c}}
\def\ua{\uparrow}
\def\da{\downarrow}
\def\muu{\mu_{\uparrow}}
\def\mud{\mu_{\downarrow}}
\def\ku{k_{\ua}}
\def\kd{k_{\da}}
\def\kut{\widetilde{k}_{\ua}}
\def\kdt{\widetilde{k}_{\da}}

\def\bcs{\textsc{bcs}}
\def\BCS{\textsc{bcs}}
\def\loff{\textsc{loff}}
\def\dfs{\textsc{dfs}}

\def\qmc{\textsc{qmc}}
\def\dmc{\textsc{dmc}}
\def\DMC{\textsc{dmc}}
\def\vmc{\textsc{vmc}}
\def\VMC{\textsc{vmc}}
\def\df{\textsc{df}}
\def\DF{\textsc{df}}

\def\intr{\int \mathrm{d}r\,}
\def\intR{\int \mathrm{d}\Rv\,}
\def\intx{\int \mathrm{d}^3x\,}
\def\inty{\int \mathrm{d}^3y\,}
\def\intyV{\int \frac{\mathrm{d}^3y}{\cal V}}
\def\intk{\int \frac{\mathrm{d}^3k}{(2\pi)^3}\,}
\def\intkp{\int \frac{\mathrm{d}^3k'}{(2\pi)^3}\,}
\def\sumn{\inv{\beta\hbar}\sum_{n}}
\def\sumnp{\inv{\beta\hbar}\sum_{n'}}
\newcommand{\parct}[1]{\frac{\partial #1}{\partial t}}
\newcommand{\parcta}[1]{\frac{\partial #1}{\partial\tau}}
\newcommand{\inv}[1]{\frac{1}{#1}}
\def\invV{\inv{\cal V}}

\def\xt{\xv\tau}
\def\yt{\yv\tau}
\def\xtp{\xvp\tau'}


\def\bea{\begin{align}}
\def\eea{\end{align}}
\def\be{\begin{equation}}
\def\ee{\end{equation}}
\def\bD{\bm{D}} 
\def\bI{\bm{I}} 
\def\bG{\bm{G}} 
\def\bS{\bm{\Sigma}} 
\def\bP{\bm{P}} 
\def\xv{\bm{x}}
\def\yv{\bm{y}}
\def\xvp{\bm{x'}}
\def\rv{\bm{r}}
\def\rvp{\bm{r'}}
\def\Rv{\bm{R}}
\def\kvp{\bm{k'}}
\def\kv{\bm{k}}
\def\kon{\kv,\om_n}
\def\konp{\kvp,\om_{n'}}
\def\qv{\bm{q}}
\def\pv{\bm{P}}
\def\hv{\bm{h}}
\def\zv{\bm{0}}
\def\ann{a_{nn}}
\def\deps{\delta\epsilon}

\def\fv{\bm{f}}
\def\Fv{\bm{F}}
\def\P{{\cal P}}


\def\nablab{\bm{\nabla}}
\def\Vt{V_{\rm trap}}
\def\Vext{V_{\rm ext}}
\def\Eloc{E_{\rm loc}}
\def\opt{{\rm opt}}
\def\F{{\rm F}}
\def\B{{\rm B}}
\def\t{\textsf{T}}
\def\T{\t}
\def\mbf{m_{BF}}
\def\mb{m_B}
\def\mf{m_F}
\def\abf{a_{BF}}
\def\abb{a_{BB}}
\def\tbb{\t_{\rm BB}}
\def\tbf{\t_{\rm BF}}
\def\lz{l_z} 
\def\Lix{$^{6}$Li}
\def\Lin{$^{7}$Li}
\def\Rb{$^{87}$Rb}
\def\K{$^{40}$K}
\def\het{$^{3}$He}
\def\heq{$^{4}$He}

\def\bdg{\textsc{b}d\textsc{g}}

\def\lens{\textsc{lens}}
\def\LENS{\textsc{lens}}
\def\jila{\textsc{jila}}
\def\JILA{\textsc{jila}}
\def\mit{\textsc{mit}}
\def\MIT{\textsc{mit}}


\def\ao{\hat{a}}
\def\bo{\hat{b}}
\def\tr{{\rm Tr}}
\def\z{{\cal Z}}

\def\D{\frac{\hbar^2}{2m}}
\def\Dd{\frac{\hbar^2}{4m}}
\def\DD{\frac{\hbar^2}{8m}}
\def\g{\gamma}
\def\R{\rho}


\def\n{\sqrt{n}}
\def\vxi{\vec{\xi}}
\def\mig{\inv{2}}
\newcommand{\dt}[1]{\frac{\partial #1}{\partial t}}
\newcommand{\fm}[1]{\frac{\delta #1}{\delta\psi^+_m}}

\title{Many-body studies\\ on atomic quantum systems}
\author{Jordi Mur Petit}

%
%
%
%
%
%
%
\ifx \lpt\printer \oddsidemargin= 1.2cm
\else             \oddsidemargin= 0.7cm \fi
%
%
  \newfont{\fonta}{cmss12 scaled\magstep5} 
  \newfont{\fontb}{cmss8 scaled\magstep5} 
  \newfont{\fontc}{cmssi8 scaled\magstep3} 
%


%

%





%
%

\setcounter{page}{1}
\thispagestyle{empty}
\ifx \lpt\printer \oddsidemargin=  0.42cm
\else             \oddsidemargin= -0.08cm \fi
\vspace*{1.5cm}
\section*{\protect\hspace*{1cm} Many-body studies \\ 
         \protect\hspace*{1cm} on atomic quantum systems}
\vspace*{2.5cm}

\begin{tabular*}{\textwidth}{l@{\extracolsep\fill}l}
 \hspace*{1cm} Mem\`oria presentada per 
   & { Programa de doctorat} \\
 \hspace*{1cm} {\bf Jordi Mur Petit}   
   & { \bf F\'{\i}sica Avan\c cada}\\
 \hspace*{1cm} per optar al t\'{\i}tol de 
   & { del Departament d'Estructura}\\
 \hspace*{1cm} Doctor en Ci\`encies F\'{\i}siques. 
   & { i Constituents de la Mat\`eria,} \\ 
 \hspace*{1cm} Barcelona, novembre 2005. 
   & { bienni 2000-2002,}\hspace*{2.2ex} \\
   & { Universitat de Barcelona.} \\
\end{tabular*} 

\vspace*{4cm} 
\begin{tabular}{ll}
  \hspace*{0.45cm} Director de la tesi: & Prof. Artur Polls Mart\'{\i}\\
  & Catedr\`atic de F\'{\i}sica At\`omica, Molecular i Nuclear\\
  & Universitat de Barcelona
 \end{tabular}

\vspace*{1.5cm} 
\begin{tabular}{ll}
  \hspace*{0.45cm} Tribunal:  
  & Dr. Mart\'{\i} Pi, president del tribunal\\
  & Dra. Muntsa Guilleumas, secret\`aria del tribunal\\
  & Prof. Maciej Lewenstein\\
  & Prof. Giancarlo C. Strinati\\
  & Dr. Jordi Boronat
 \end{tabular}
%

%
%
\newpage \thispagestyle{empty} \phantom{a}
%
\clearpage \setcounter{page}{1} \thispagestyle{empty} \vspace*{2cm}
\begin{flushright}
  {\em A la meva fam\'{\i}lia\\
       i en especial a la meva mare}
\end{flushright}
%
\newpage \thispagestyle{empty} \phantom{a}
\clearpage \setcounter{page}{1} \thispagestyle{empty}
%
%

\cleardoublepage

\setlength{\evensidemargin}{17cm}
\addtolength{\evensidemargin}{-\textwidth}
\addtolength{\evensidemargin}{-1in}
\setlength{\oddsidemargin}{2.0cm}
\setlength{\oddsidemargin}{4cm}
\addtolength{\oddsidemargin}{-1in}

\frontmatter

\phantomsection
\tableofcontents

\phantomsection
\listoffigures

\phantomsection
\listoftables

\fancyhead[LO]{\bfseries\leftmark}
\chapter{Agraïments}

Buf! Vet-ho aquí el primer que em ve al cap en pensar que
ja he acabat d'escriure el que hauria de ser el resum de la meva tasca els
darrers cinc anys i escaig. Unes cent seixanta pàgines de ciència <<pura i
dura>>. Quina pallisa, no? Però aquest resum estaria excessivament esbiaixat
cap a la feina feta si no dediqués ni que fossin quatre ratlles per mirar
enrere també en el pla personal. Ben mirat, difícilment hauria pogut fer
res del que conté aquesta tesi sense la interacció amb una petita multitud de
gent 
amb qui sens dubte he evolucionat també com a persona. Probablement més que no
pas com a científic; al capdavall, tot just ara puc començar a anomenar-me
així.

En primer lloc, haig d'agrair a l'Artur haver-me acollit com a doctorand i
haver-se <<atrevit>> a iniciar amb mi la recerca en un camp nou, el dels àtoms
freds, que l'any 2000 ---que llunyà sona, sembla que parli del segle
passat!--- estava en plena explosió... i encara no ha parat. Els mal-de-caps
que hem tingut perseguint la literatura per posar-nos al dia!
La teva voluntat pel diàleg
de ben segur ha estat ingredient imprescindible per
l'èxit d'aquesta empresa. 
Vull agrair també a la Muntsa, la nostra experta local en BEC,
estar sempre disposada a escoltar-me i xerrar una estona, ja sigui de
física o del que convingui.

No puc oblidar-me tampoc dels altres membres del grup, passats i presents, que
han fet de les moltes hores passades al departament una experiència d'allò més
enriquidora. 
L'Àngels, que em va acollir iniciament amb l'Artur, si bé després
els nostres camins científics s'han separat lleugerament. 
L'Isaac i, especialment, la Laura amb qui vaig compartir no només l'assoleiat
despatx 652 sinó també moltes bones estones.
I l'Assum, baixllobregatina com un servidor i amb qui vaig connectar
ràpidament (potser per allò de compartir el telèfon durant tant de temps?).
Dels primers temps a ECM també haig de recordar el Hans, il tedesco piu'
specialle che ho mai trovato, amante della montagna, di Catania e della
<<Cerveceria Catalana>>, con cui ho imparato una sacco su {\em pairing}.

Més recentment, ha estat especialment engrescadora i estimulant la relació amb 
els <<encontradors>>, l'Arnau i l'Oliver, amb qui hem demostrat que disposar
d'un pressupost zero no és obstacle per organitzar un exitós cicle de
xerrades. Les reunions d'organització, sovint sessions de deliri mental
conjunt, ens han dut a reinterpretar Einstein i fins i tot sortir a la pàgina
web de la UB i a <<el País>>. Què més podem demanar?
Amb l'Arnau també vam tenir el plaer d'estrenar el nou despatx <<cara al
sol>>, juntament amb el Chumi, que amb els seus apunts {\em tutti colori}
i cartells teatrals hi ha aportat un toc ben especial. 

Per acabar amb l'entorn <<laboral>>, vull mencionar 
{\em il professore Brualla}, de capacitat verbal inacabable, sigui en català,
anglès o italià i que sempre té una anècdota a la butxaca
per il$\cdot$lustrar una conversa.
El Jesús, <<català del sud>> que em va acollir durant un parell de visites a
València i pel qual vaig anar a Finlàndia ---afortunadament, a l'agost--- per
aprendre {\em density functional}. 
I, com no, la Mariona, que no es va espantar en fer
la beca de col$\cdot$laboració i s'ha incorporat definitivament al grup. 
Et desitjo molta sort en aquesta aventura que ara emprens.
Per començar, la cosa pinta bé: t'acaben d'atorgar una beca de la 
famosa(?)\ Fundació Agustí Pedro i Pons. La mateixa beca a mi em va permetre
de sobreviure durant el primer any de doctorat, cosa que vull agrair també
aquí.

Dit això potser hauria d'agrair també la Gene, no? 
Deixant de banda les lliçons de <<burocràcia creativa>> que les renovacions
anuals comporten, vull agrair sobretot les borses de viatge, que m'han permès
agafar el <<gustet>> a això de viatjar. He pogut anar per mitja Europa i un
bon tros dels E.U.A., i he fet una colla d'amics.
Primer vaig anar a Pisa, a treballar amb Adelchi Fabrocini: grazie per
accogliermi e per aiutarmi a imparare un po' la lingua di Dante. Anche altri
membri del gruppo (sopratutto Ignazio), del dipartimento (Marco e i suoi
collegi sperimentali, insieme a Eleonora) e dell'Università (ricordo
specialmente i miei coinquillini a via San Frediano: Manu, Gigi, Max 
e anche Maria Luisa) hanno fatto un buon lavoro in questo senso.

Posteriorment vaig anar a treballar amb el grup d'Anna Sanpera i Maciej
Lewenstein a Hannover. Els agraeixo, juntament amb els altres membres del grup
(Verònica, Kai, Florian, Helge, Alem) la seva amistat i les estimulants
xerrades mantingudes.

Finalment, fa pocs mesos he pogut <<fer les Amèriques>> visitant el grup de
Murray Holland al JILA de Boulder. My stay turned out to be too short to 
finish a project, but long enough to enjoy the hospitality and friendship 
of you and the people in your group (Meret, Stefano, Rajiv, Brian, Jochen, Lincoln 
and Marilú).
I also want to thank Svet, Michele and Cindy for friendship and illustrative
discussions. Y, como no, la peña de `exiliados hispanohablantes' (Manolo de
Madrid, Fernando, María, Elena y Bárbara, Vitelia, Manolo de Sevilla)
que hicieron más llevadera la distancia espacial y temporal con mi `casita'.

Parlant d'amics, vull mencionar també l'Hèctor, amic literalment <<de tota la vida>> 
i amb qui el bàsquet m'ha unit de nou per viure grans moments al Palau. 
I l'Ivan, amic amb qui em trobo menys del que voldria, gràcies al qual vaig
contactar amb l'Esther que, en una recerca digna de Sherlock Holmes, va ser
capaç de trobar una agulla en un paller, o el que és el mateix, una cita dins
tota l'obra de Shakespeare. 

I això em porta a recordar els de casa, 
indispensables en la meva vida i
sense els quals no seria res.
Durant aquests darrers anys hem passat per moltes coses; 
hi ha qui ha marxat, i qui ha arribat, 
i mai m'ha faltat el vostre suport. 
Sobretot vull destacar la meva mare, Teresa,
que és qui m'ha hagut de suportar més hores
---i això de vegades és ben difícil!---:
perquè la teva capacitat d'esforç i 
superació en l'adversitat són un model a seguir.
El meu pare, Miquel,
que no es va voler rendir mai en la dificultat:
cada cop que viatjo enyoro més els teus consells 
---que abans no sempre vaig saber apreciar.
Els meus germans, Rosa i Ferran, així com el Pep i la Viole:
no puc oblidar els vostres ànims en els meus primers viatges a fora,
quan no estava segur de com aniria tot plegat i rebre 
una trucada o un {\em mail} era sempre una alegria.
I els <<reis de la casa>>, l'Enric i la Núria,
font inesgotable d'anècdotes per compartir i recordar,
i amb qui és ben fàcil d'aprendre una cosa nova cada dia.


\vskip 1cm
\begin{flushright}

Sant Boi de Llobregat, 22 de novembre de 2005
\end{flushright}


\fancyhead[LO]{\bfseries\leftmark}
\chapter{Plan of the thesis}

During the last five years, 
I have been learning some of
the theoretical tools of Many-Body physics, and I have tried to apply them to
the fascinating world of quantum systems.
Thus, the studies presented in this thesis constitute an attempt to show the
usefulness of these techniques for the understanding of a variety of systems
ranging from ultracold gases (both fermionic and bosonic) to the more familiar
helium.
The range of techniques that we have used is comparable to the number of
systems under study: 
from the simple, but always illustrative mean-field approach, 
to the powerful Green's functions' formalism,
and from Monte Carlo methods to the Density Functional theory.

I have made an effort to present the various works in an understandable way.
I would like to thank Prof. Artur Polls, my advisor, and Dr. Llorenç Brualla, 
Dr. Muntsa Guilleumas, Prof. Jesús Navarro and Dr. Armen Sedrakian for carefully 
reading various parts of the manuscript, raising interesting questions and 
suggesting improvements.
In any case, of course, any error or misunderstanding should be attributed
only to me. I'll be glad to receive questions or comments on any point related
to this work.\footnote{You can contact me at \texttt{jordimp2003@yahoo.es}}

The plan of the thesis is as follows. In a first part, we study ultracold
gases of alkali atoms. The first three chapters are devoted to the evaluation
of the possibilities of having and detecting a superfluid system of
fermionic atoms. 
The first chapter introduces the idea of pairing in a Green's functions'
formulation of the theory of superconductivity by Bardeen, Cooper and
Schrieffer. 
Chapter~\ref{ch:asym} presents the application of this theory to the case of a
two-component system where the two species may have different densities, a
fact that has important consequences for the prospects of superfluidity.
This work was performed in collaboration with Dr. Hans-Josef Schulze who, at
that time, was a post-doc in Barcelona. Many of the results expounded here
have appeared in the article~\cite{pla01}.

Chapter~\ref{ch:dfs} generalizes the structure of the \BCS\ ground state to the
cases of finite-momentum Cooper pairs (the so-called \loff\ or \textsc{fflo}
phase) and deformed Fermi surfaces (\dfs).
This last state was first proposed by Prof. Herbert Müther and Dr. Armen
Sedrakian from Universität Tübingen, with whom we have worked together and
who I would like to thank for their warm hospitality during my visits to
Tübingen. As a result of this work, we have recently published the
paper~\cite{sedrakian-pra}.

Chapter~\ref{ch:2dbf} introduces for the first time bosons into this
thesis. They are used to cool a one-component system of fermions, and their
capacity to help the fermions to pair is studied, with emphasis in the case of a
two-dimensional configuration.
This research, which appeared in~\cite{pra04} and in~\cite{jpb04}, was
performed again with Dr. Schulze, then in Catania, and with Prof. Marcello
Baldo. I want to thank both of them, and the rest of the \textsc{infn} group 
for their friendship and hospitality during my visit in 2003. 

In chapter~\ref{ch:spin} we enter boldly into the world of
bosonic systems and analyze the dynamical evolution of a set of bosons
whose spin degree of freedom is free to evolve inside an optical trap, only
restricted by the conservation of magnetization. An attempt to incorporate
temperature effects is also performed.
Thanks to this project I could enjoy for a couple of months the stimulating
atmosphere in the group of Prof. Maciej Lewenstein and Prof. Anna Sanpera at
Universität Hannover, just before they moved to Catalonia. 
This work, developed in close collaboration with the experimental group of
Prof. K. Sengstock and Dr. K. Bongs from Universität Hamburg, has been
submitted for publication to Physical Review A~\cite{spinors-pra}.

At the conceptual half of the thesis (p.~\pageref{part:heli}), we leave
ultracold gases and start the study of helium-4 in two dimensions, which
constitutes the second part of the work.
First, in chapter~\ref{ch:heli-dmc}, we present a Monte Carlo study of the
ground-state structure and energetics of $^4$He puddles, with an estimation
of the line tension.
The results obtained here are incorporated in chapter~\ref{ch:heli-df} to
build a Density Functional appropiate to study two-dimensional helium
systems. This functional is then used to analyze a variety of such systems,
and a comparison with previous studies in two and three dimensions is
presented.
Both these works have been developed in collaboration with Prof. Jesús
Navarro, from CSIC-Universitat de València, and Dr. Antonio Sarsa, who is now
at Universidad de Córdoba. Some of the results presented here have appeared
in~\cite{prb03} and \cite{prb05}.

The conclusions of the thesis are finally summarized.

\fancyhead[LO]{\bfseries\rightmark}
\mainmatter

\pagenumbering{arabic}
\part{Ultracold gases\label{part:gasos}}

\chapter{The pairing solution}
\label{ch:pair-intro}

\textsf{
 \begin{quote}
 ---Bien parece ---respondió don Quijote--- que no estás cursado en esto
 de las aventuras: ellos son gigantes; y si tienes miedo, quítate de ahí,
 y ponte en oración en el espacio que yo voy a entrar con ellos en
 fiera y desigual batalla. 
 (...) 
 !`Non fuyades, cobardes y viles criaturas, que un solo caballero
 es el que os acomete!
 \end{quote}
 %
 \begin{flushright}
   {Miguel de Cervantes,}
   {\em Don Quijote de la Mancha} (I, 8)
 %
 \end{flushright}
}

\section{Historical background: before \bcs\ theory}
\label{sec:pair-hist}

The \bcs\  theory of superconductivity~\cite{bcs,schrieffer} has been
one of the most successful contributions to physics in the 20$^\mathrm{th}$
century, because it was the first microscopic theory of a truly
macroscopic quantum phenomenon. It explains the mechanisms behind
dissipation-free electric transport in a number of materials. Indeed, since
its formulation, this theory has been applied to such different systems as
solid metals and alloys~\cite{schrieffer,degennes}, 
atomic nuclei and neutron stars~\cite{bohr,morten03}, 
elementary particle physics (see, \eg,~\cite{casal}) and superfluid
$^3$He~\cite{osheroff72a,
leggett-rmp}.

The theory developed by Bardeen, Cooper and Schrieffer appeared as
a final step in the theoretical understanding 
of a long series of experimental discoveries. In 1911 Heike Kamerling
Onnes~\cite{onnes} observed that mercury below 4.2 K looses its electrical
resistance and enters `a new state, which, owing to its particular
electrical properties, can be called the state of
superconductivity'~\cite{onnes-nobel}. In 1933 Meissner and
Ochsenfeld~\cite{meissner} observed that a superconductor has also notable
magnetic properties, namely it is a perfect diamagnet: a (small) applied
magnetic field vanishes in the interior of a bulk superconductor. 
Another important contribution was the discovery in 1950 of the isotope effect
(\ie, the dependence of the critical temperature on the mass of the
ions of a superconducting solid) by Maxwell~\cite{maxw} and
Reynolds {\em et al.}~\cite{reyn}. Finally, let us mention the experimental
determination of the quantization of the magnetic flux traversing a
multiply-connected superconductor by Deaver and Fairbank~\cite{deav} and Doll
and Näbauer~\cite{doll} following an initial prediction of the London model. 

There were many theoretical attemps to explain the experimental results prior
to the theory of Bardeen, Cooper and Schrieffer. In a first phenomenological
theory, Gorter and 
Casimir~\cite{gorter} formulated a two-fluid model (similar to that of
Tisza~\cite{tisza1,tisza2} and Landau~\cite{landau-twofluid} for $^4$He) 
where electrons can be either in a superconducting state or in the 
normal state. At $T=0$ all electrons are in the superfluid, while the fraction
of normal electrons grows with temperature and finally equals one at the
\index{critical temperature!for pairing transition}critical temperatre $T_c$. 
The fraction of normal electrons was calculated by
minimazing a free energy interpolated between the limits of a 
superfluid system at $T=0$ and a normal one at $T=T_c$. This model,
however, had a number of artificial points (such as the way to construct the
free energy), just intended to reproduce the experimental results known at the
moment, but without a microscopic physical motivation.

A more successful theory was that of F. London and
H. London~\cite{londons,london} based on the electromagnetic phenomenology of
supercondutors. It explained the Meissner-Ochsenfeld effect~\cite{meissner},
introduced a `penetration length' $\lambda$ of magnetic fields into 
superconductors and also predicted the quantization of magnetic flux in a
multiply-connected superfluid. F. London's work also speculated on the
possible existence of a gap in the 
\index{spectrum!in superfluid phases}excitation spectrum of a superfluid, 
a crucial ingredient in \bcs\ theory.

In 1950 Ginzburg and Landau~\cite{ginz} formulated a non-local theory of
superconductivity by generalizing the ideas of the London brothers: they
introduced a spatially-varying `order parameter' closely related to Londons'
superfluid density. This theory was valid only near $T_c$, but nevertheless it
became of great interest with the discovery of type II superconductors. It is
worth noticing that Ginzburg and Landau's theory can be also derived from the
\bcs\ microscopic theory~\cite{gork59}.

\index{Cooper problem}
In 1956, Cooper~\cite{cooper} studied the problem of an interacting pair of
fermions above a frozen Fermi sea and showed that, for an attractive
interaction (no matter how weak it was), the pair would be bound, with an
exponentially small binding energy in the limit of small coupling. These bound
pairs are called `Cooper pairs', and are a central ingredient of the \bcs\
theory of superconductivity: Assuming that {\em all} the electrons
in a metal are forming such pairs one can show that the total energy of the
system is lowered with respect to that of the normal system.

\section[The Bardeen-Cooper-Schrieffer theory]
        {The Bardeen-Cooper-Schrieffer theory%
}
\label{sec:bcs}

In this section, we introduce the microscopic theory of
superconductivity of Bardeen, Cooper and Schrieffer (know as `\bcs\ theory')
in a way that will later allow us to apply it to a variety of atomic gases as
we shall do in the following chapters. We will start in Sect.~\ref{sub:cooper}
by introducing the idea of Cooper pairs and show how the presence of a Fermi
sea allows an interacting pair of fermions to bind itself at arbitrary
small attraction. This many-body effect (in the sense that the pair would not
be bound in the absence of the Fermi sea unless the attraction was greater
than a minimum value) is the essence of \bcs\ theory, which we present in a
formal way in Sect.~\ref{sub:formal}. Finally, we present some basic results
of the theory in Sect.~\ref{sub:symmetric}.

\subsection{Premier: Cooper pairs}
\label{sub:cooper}

\begin{figure}[t!]
  \caption[The Cooper problem]
	  {\label{fig:cooper}
	    The Cooper problem: two fermions in states $|\kv\ra$ and
	    $|-\kv\ra$ interact above a frozen Fermi sea (filled sphere). 
            The fermions form a pair (dashed line) with center-of-mass
	    momentum $\qv=\zv$ and can scatter into states $|\kvp\ra$ and
	    $|-\kvp\ra$ for $|\kvp|>k_F$.
	  }
  \begin{center}
    \includegraphics
        [width=0.5\textwidth]
	{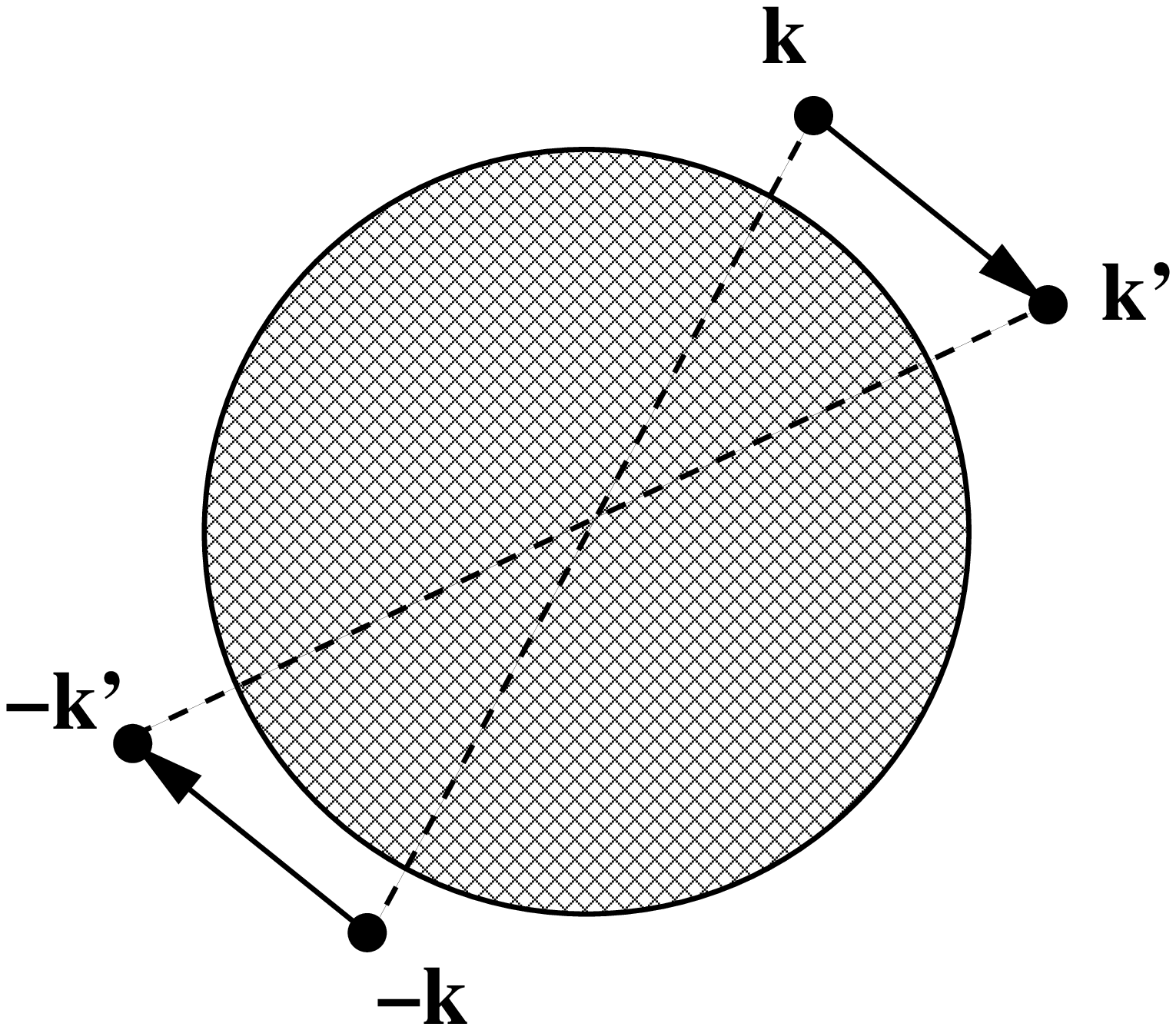}
  \end{center}
\end{figure}

Consider a pair of fermions of mass $m$ in homogeneous space above a 
`frozen' Fermi sea of non-interacting particles with Fermi energy 
$\ef=\hbar^2k_F^2/2m$. Let us
assume that these two particles are distinguishable (for example, they can
be electrons with different spins, atoms of the same chemical element with
different hyperfine spins, neutrons and protons, or quarks of different
colors), and that they interact through a two-body spin-independent 
potential. The presence of the Fermi sea forbids the particles from occupying
the energy levels below $\ef$ (see Fig.~\ref{fig:cooper}). Therefore, we can
take $\ef$ as their zero-energy level. This problem was originally solved by
Cooper~\cite{cooper}, so it is usually called the `Cooper problem' and the
bound pairs are known as `Cooper pairs'. The solution we describe is based
on~\cite{schrieffer}.

The wave function of a pair with momentum $\qv$ 
in the center-of-mass system (c.m.s.) can be written
\begin{equation*}
  \psi(\xv_1,\xv_2) = \phi_{\qv}(\rv)\exp\{i\qv\cdot\Rv\} \;,
\end{equation*}
where we have defined the centre-of-mass coordinate $\Rv=(\xv_1+\xv_2)/2$ and
the relative coordinate $\rv=\xv_1-\xv_2$. Taking for simplicity $\qv=\zv$,
we have
\begin{equation}
  \psi=\phi(\rv)=\sum_{k>k_F} a_{\kv} e^{i\kv\cdot \rv} =
   \sum_{k>k_F} a_{\kv} e^{i\kv\cdot \xv_1} e^{-i\kv\cdot \xv_2} \;.
\end{equation}
Here we have applied Pauli's exclusion principle to restrict the summations
over momenta to the region outside the filled Fermi sea. This wave
function can be interpreted as a superposition  of configurations with
single-particle wave functions with momenta 
$|\kv\ra$ and $|-\kv\ra$.

The corresponding Schrödinger equation can be written
\begin{align}
  (E-H_0)\psi &= V\psi \quad\Raw\quad 
  (E-2\ek)a_{\kv} = \sum_{k'>k_F} V_{\kv\kvp}a_{\kvp} \;,\\ 
  V_{\kv\kvp} &:= \lla \kv,-\kv | V | \kv', -\kv'\rra \;, \notag
\end{align}
where we have introduced the 
\index{spectrum!single-particle}single-particle spectrum $\ek$ 
corresponding to $H_0=-\hbar^2\nabla^2/(2m)$.

If we take the simple case of a constant, attractive interaction constrained
to a shell of width $\Omega$ above the Fermi surface, that is,
\begin{equation*}
  V_{\kv\kvp} = \left\{
    \begin{array}{cl}
      -|\lambda| & \ef<\ek,\ekp<\ef+\Omega \\
      0          & \mathrm{otherwise}
    \end{array}
    \right. \;,
\end{equation*}
then the lowest eigenenergy is
\begin{equation}
  E=2\ef -\frac{2\Omega}{\exp\left[\frac{2}{\nu(k_F)|\lambda|}\right]-1}
   \simeq 2\ef -2\Omega e^{-\frac{2}{\nu(k_F)|\lambda|}} \;,
\end{equation}
where $\nu(k_F)=mk_F/(2\pi^2\hbar^2)$ is the density of single-particle states
{\em for a single spin projection} at the Fermi surface. The last result is
valid in the \index{weak-coupling limit}weak-coupling approximation, 
$\nu(k_F)|\lambda|\ll 1$. 

Therefore, the pair will be bound ($E<2\ef$) no matter how small
$|\lambda|$ is. 
Note that this is a truly many-body effect as, in the absence of the
underlying, filled Fermi sea, the pair would not bind for too weak an
interaction.
Furthermore, the non-analytic behavior in the 
\index{weak-coupling limit}weak-coupling limit shows that
this result could not be found by perturbation theory, as it corresponds to a
deep modification of the ground state as compared to the ideal Fermi gas. 


\subsection{Formalism for the theory}
\label{sub:formal}

Let us study now a more realistic problem in which we have an 
{\em interacting} system of fermions with an internal degree of freedom,
that we indicate by a (pseudo)spin variable $s=\ua,\da$. This problem can be
modeled by the following second-quantized Hamiltonian
\begin{multline}
  \label{eq:bcs-ham}
  \hat H=
  \sum_s\intx \psiod_s(\xv)\left(-\hm\nabla^2\right)\psio_s(\xv) \\
  + \frac{1}{2} \intx\intx\!'\, V(\xv-\xvp)
  \psiod_{\da}(\xv)\psiod_{\ua}(\xvp)
  \psio_{\ua}(\xvp)\psio_{\da}(\xv)
\end{multline}
Here, $\psio_s(\xv)~(\psiod_s(\xv))$ is the operator that destroys (creates) a
particle of (pseudo)spin~$s$ at position $\xv$, and 
$V(\xv-\xvp)$ is the interaction potential between the fermions. 
We will be interested in low-density, low-temperature systems where the most
important contribution to interactions is the $s$-wave one.

We look for a solution to this problem using the tools of quantum field
theory, in particular, the Green's functions' formalism. 
We start linearizing the interaction and transforming it into a 
{\em  one-body} 
operator by means of a Hartree-Fock-like approximation,
\begin{multline}
  \hat V_{\rm HF}=\sum_s \intx\!\!\intx\!'\; V(\xv-\xvp)
        \Big\{
	  \lla \psiod_{s}(\xv)\psio_{s}(\xv) \rra 
	  \psiod_{-s}(\xvp)\psio_{-s}(\xvp) \\+
	  \lla \psiod_{s}(\xv)\psio_{-s}(\xvp) \rra 
	  \psiod_{-s}(\xvp)\psio_{s}(\xv) 
	\Big\} \:.
  \label{eq:HF}
\end{multline}
The first term inside the curly brackets is the direct (or Hartree) term,
while the second one is the exchange (or Fock) one.
In the next step, we consider a new term in the linearization process which
accounts for the possibility, outlined in the previous section, that two
fermions with opposite spins form a bound pair (cf. Sec. 51 in~\cite{fetter}
or Sec. 7-2 in~\cite{schrieffer}).
Even though $\hat H$ commutes with the number operator $\hat{N}$,
this pairing possibility is more easily implemented if we work in the
grand-canonical ensemble. Physically, we can understand this in the following
way: for a large number of particles $N$ in the system, the ground state
energies of the system with $N$ and $N\pm1$ particles are nearly degenerate if
we substract from them the chemical potential. Then, it is reasonable that a
wave function in Fock space, with only the average value $N=\la\hat N\ra$
fixed, is more flexible from a variational point of view than one with a fixed
number of particles. In this way, the function can be closer to the true
ground state  of the system immersed in the `bath' of condensed pairs. 
In fact, it is customary to assume that the Hartree-Fock potential above is
the same in the normal and superfluid states, so that we can disregard it from
now on as its inclusion would not affect the comparison between both states.
Therefore, we will work with the following effective (grand-canonical)
\index{BCS Hamiltonian}Hamiltonian 
\begin{multline}
  \hat K = 
  \sum_s\intx \psiod_s(\xv)\left(-\hm\nabla^2-\mu_s\right)\psio_s(\xv)
  +\int\!\mathrm{d}^3x \int\!\mathrm{d}^3x'\,V(\xv-\xvp) \times \\ \times
    \left[ \lla \psiod_{\da}(\xv)\psiod_{\ua}(\xvp) \rra 
           \psio_{\ua}(\xvp)\psio_{\da}(\xv) + 
	   \psiod_{\da}(\xv)\psiod_{\ua}(\xvp) 
	   \lla \psio_{\ua}(\xvp)\psio_{\da}(\xv) \rra \right].
  \label{eq:bcs-gc-ham}
\end{multline}
The angular brackets $\la \cdots \ra$ denote a thermal average with $\hat K$:
\begin{equation*}
  \la \cdots \ra := \frac
      {\mathrm{Tr}\left[e^{-\beta K}\cdots\right]}
      {\mathrm{Tr}\left[e^{-\beta K}\right]} \;,
      \qquad \beta=\frac{1}{k_B T} \;,
\end{equation*}
where $k_B$ stands for Boltzmann's constant.

Introducing now the imaginary-time variable
$\tau\in[0,\hbar\beta]\subset\mathbb{R}$ (see Appendix~\ref{app:suma}) and the
field operators in the Heisenberg picture,
\begin{subequations}
\begin{align}
  \psio_s(\xt) &:= e^{\hat K\tau/\hbar} \psio_s(\xv) e^{-\hat K\tau/\hbar} \\
  \psiod_s(\xt)&:= e^{\hat K\tau/\hbar} \psiod_s(\xv) e^{-\hat K\tau/\hbar} \:,
\end{align}
\end{subequations}
one can show that they satisfy these equations of motion:
\begin{subequations}
\begin{align}
  \hbar\parcta{\psio_{\ua}(\xt)} =
    &-\left(\Pm-\muu\right)\psio_{\ua}(\xt) \notag\\
    &+\inty V(\yv-\xv) \lla \psio_{\ua}(\xv)\psio_{\da}(\yv) \rra 
      \psiod_{\da}(\yt) \:,
      \displaybreak[0]\\
  \hbar\parcta{\psiod_{\da}(\xt)} =
    &+\left(\Pm-\mud\right)\psiod_{\da}(\xt) \notag\\
    &-\inty V(\xv-\yv) \lla \psiod_{\da}(\xv)\psiod_{\ua}(\yv) \rra 
    \psio_{\ua}(\yt) \:.
\end{align}
\end{subequations}

The solution of the problem is more naturally found by introducing the
temperature \index{Green's function}Green's function 
(or \index{propagator!normal}normal propagator) $G$ corresponding to
each species $s=\ua,\da$,
\begin{equation}
  G_{s}(\xt;\xtp) := 
     \lla -T_{\tau}\left[\psio_{s}(\xt)\psiod_{s}(\xtp)\right]\rra \;,
\end{equation}
where $T_{\tau}$ is the imaginary-time ordering operator~\cite{fetter}.
The equation of motion for $G_{\ua}$ is easily seen to be
\begin{multline}
  \label{eq:evolG_1}
  \hbar\parcta{G_{\ua}(\xt;\xtp)} = 
    -\hbar\delta(\tau-\tau')\delta(x-x')
    -\left(\Pm-\muu\right)G_{\ua}(\xt;\xtp) \\
    +\inty V(\yv-\xv) \lla \psio_{\ua}(\xv)\psio_{\da}(\yv) \rra
     \lla -T_{\tau}\left[\psiod_{\da}(\yt)\psiod_{\ua}(\xtp)\right]\rra \:.
\end{multline}
Defining the \index{propagator!anomalous}anomalous propagators 
$F$ and $F^{\dagger}$ by
\begin{subequations}
\begin{align}
  F(\xt;\xtp) &:=
     \lla -T_{\tau}\left[\psio_{\ua}(\xt)\psio_{\da}(\xtp)\right]\rra \;,\\
  F^{\dagger}(\xt;\xtp) &:=
     \lla -T_{\tau}\left[\psiod_{\da}(\xt)\psiod_{\ua}(\xtp)\right]\rra \;,
\end{align}
\end{subequations}
we can rewrite Eq.~(\ref{eq:evolG_1}) as
\begin{multline}
  \label{eq:evolG}
  \left[-\hbar\parcta{}-\left(\Pm-\muu\right)\right]
     G_{\ua}(\xt;\xtp) \\
     +\inv{\cal V}\inty \De(\xv,\yv) F^{\dagger}(\yt;\xtp) =
    \hbar\delta(\tau-\tau')\delta(\xv-\xvp) \,,
\end{multline}
where we defined the two-point gap function
\begin{equation}
  \label{eq:2p-gap}
  \De(\xv,\yv):=V(\yv-\xv) \lla\psio_{\ua}(\xv)\psio_{\da}(\yv)\rra {\cal V}
  \equiv -V(\yv-\xv)F(\xt^+,\yt) {\cal V}\;.
\end{equation}
Here we wrote explicitly the volume $\cal V$ of the system to make clear that
$\De(\xv,\yv)$ has dimensions of energy. Note that all 
\index{propagator}propagators in space
coordinates have dimensions of {\em density}. Moreover, note that $F$ and
$F^\dagger$ are {\em not} operators, neither is one the Hermitian adjoint of
the other (as the $\dagger$ symbol might seem to indicate): they are just
$c$-functions, as can be seen from their definitions as expectations values.

In a similar way, we can find the equations of motion for $G_{\da}$,
$F^\dagger$ and $F$:
\begin{align}
  \label{eq:evolGd}
  \left[\hbar\parcta{}-\left(\Pm-\mud\right)\right]
     G_{\da}(\xtp;\xt) && \nonumber\\
     + \inv{\cal V}\inty\De^*(\yv,\xv) F(\yt;\xtp) &=
    \hbar\delta(\tau-\tau')\delta(\xv-\xvp) \,, \\
  \label{eq:evolFd}
  \left[\hbar\parcta{}-\left(\Pm-\mud\right)\right]
     F^\dagger(\xt;\xtp)
     &= \intyV\De^*(\yv,\xv) G_{\ua}(\yt;\xtp) \,,\\
  \label{eq:evolF}
  \left[-\hbar\parcta{}-\left(\Pm-\muu\right)\right]
     F(\xt;\xtp) 
     &= \intyV\De(\xv,\yv) G_{\da}(\xtp;\yt) \,,
\end{align}


Then, in the absence of external fields, translational symmetry implies that a
Fourier transformation to momentum space can be carried out.
We do also a Fourier transformation of imaginary time to
\index{Matsubara frequencies}
(fermionic) Matsubara frequencies $\om_n=(2n+1)\pi/(\beta\hbar)$ of the normal
\index{propagators!normal!in momentum space}
\index{propagators!normal!in imaginary time}
propagators (see Appendix~\ref{app:suma})
\begin{align}
  G_s(\xt;\xtp) &\equiv G_s(\xv-\xvp;\tau-\tau') \nonumber\\
  &=\frac{1}{\beta\hbar}\sum_n \intk 
    e^{-i\om_n(\tau-\tau')} e^{i\kv\cdot(\xv-\xvp)} G_s(\kv,\om_n) \;.
\end{align}
with analogous definitions being valid for $F(\kv,\om_n)$ and
$F^\dagger(\kv,\om_n)$. 
With these definitions, the propagators in Fourier space have dimensions of
{\em time}.

Introducing now $\xi_{\kv s}=\ek-\mu_s$ to measure all excitation energies
from the corresponding chemical potentials, the dynamical equations in Fourier
space turn out to be
\begin{subequations}
\label{eq:kw}
\begin{align}
  \label{eq:Gukw}
  \lp i\hbar\om_n-\xi_{\kv\ua}\rp G_\ua(\kon) 
    + \De_{\kv}(\xv) F^\dagger(\kon) &= \hbar \;,\\
  \label{eq:Fdkw}
  \lp-i\hbar\om_n-\xi_{\kv\da}\rp F^\dagger(\kon)
    - \De^*_{\kv}(\xv) G_\ua(\kon) &= 0       \;,\\
  \label{eq:Fkw}
  \lp i\hbar\om_n-\xi_{\kv\ua}\rp F(\kon)
    - \De_{\kv}(\xv) G_\da(-\kv,-\om_n) &= 0       \;,\\
  \label{eq:Gdkw}
  \lp-i\hbar\om_n-\xi_{\kv\da}\rp G_\da(-\kv,-\om_n) 
    + \De^*_{\kv}(\xv) F(\kon) &= \hbar       \;,
\end{align}
\end{subequations}
where we defined
\begin{equation}
  \De_{\kv}(\xv) = \inv{\cal V} \inty \De(\xv,\yv)e^{i\kv\cdot(\yv-\xv)} \;.
  \label{eq:gap-fun}
\end{equation}
We will consider in this work only translationally invariant
systems. Therefore, the $\xv$-dependence can be dropped:
$\De_{\kv}(\xv)\equiv\De_{\kv}$.

Equations (\ref{eq:kw}) form a system of four coupled, algebraic
equations. To solve them, let us introduce the symmetric and antisymmetric
(with respect to the exchange of $\ua$ and $\da$ labels) 
\index{spectrum!in superfluid phases}quasiparticle spectra
\begin{subequations}
\label{eq:Esa}
\begin{align}
  E_S &:= \frac{\xi_{\kv\ua}+\xi_{\kv\da}}{2}
        =  \ek-\frac{\muu+\mud}{2}
        \equiv  \ek-\mu  \equiv \xi_{\kv} \;, \\
  E_A &:= \frac{\xi_{\kv\ua}-\xi_{\kv\da}}{2}
        = -\frac{\muu-\mud}{2}
        \equiv -\dmu  \;.
\end{align}
\end{subequations}
Here we have also defined the mean chemical potential $\mu=\lp\muu+\mud\rp/2$
and half the difference in chemical potentials $\dmu=\lp\muu-\mud\rp/2$. With
all these definitions, it is easy to see that
Eqs.~(\ref{eq:Fkw},\ref{eq:Gdkw}) are just the same as
Eqs.~(\ref{eq:Fdkw},\ref{eq:Gukw}) with the replacements 
$G_\ua(\kon)\raw-G_\da(-\kv,-\om_n)$, 
$\De\raw\De^*$,
$F^\dagger(\kon)\raw F(\kon)$
and $E_S\raw-E_S$. The solution is
\begin{subequations}
\begin{align}
  G_\ua(\kon) &=
    \hbar\frac{i\hbar\om_n-E_A+E_S}
              {\lp i\hbar\om_n-E_A\rp^2 - E_S^2 - |\Dekx|^2} \nonumber\\
   &=
    \hbar\frac{i\hbar\om_n+\ek-(\mu-\dmu)}
              {\lp i\hbar\om_n+\dmu\rp^2 - (\ek-\mu)^2 - |\Dekx|^2} \\
  \label{eq:Fkw-sol}
  F(\kon) &=
    -\hbar\frac{\Dekx}
              {\lp i\hbar\om_n-E_A\rp^2 - E_S^2 - |\Dekx|^2} \;.
	       \nonumber\\
   &=
    -\hbar\frac{\Dekx}
               {\lp i\hbar\om_n+\dmu\rp^2 - (\ek-\mu)^2 - |\Dekx|^2} \;.
\end{align}
\end{subequations}
The solution for $G_\da(\kon)$ is identical to $G_\ua(\kon)$ with the
substitution $E_A \raw -E_A$, while
$F^\dagger(\kon)=\left[F(\kon)\right]^*$. As expected, in the symmetric case
(\ie, $\dmu=0$) one can check that $G_\ua(\kon)=G_\da(\kon)$.

\index{excitation spectra!in \bcs\ theory}
The quasiparticle excitation spectra of the system are just the poles of the
\index{propagators!at zero temperature}
\index{propagators!at zero temperature!poles of}
propagators at zero temperature. These propagators can be found from those at
finite temperature by analytic continuation of the imaginary frequencies onto
the real axis $i\om_n\raw \omega\in\mathbb{R}$, and we have
\begin{equation}
  \label{eq:espectre}
  \Ek^{\pm}
  =-\dmu\pm\sqrt{\lp\eps_k-\mu\rp^2+|\Dekx|^2}
  =E_A\pm\sqrt{E_S^2+|\Dekx|^2}
  \;,
\end{equation}
which gives the typical \bcs\ dispersion relation
$\Ek=\sqrt{E_S^2+|\Dekx|^2}$ in the 
symmetric case. 
In this case there is a minimum energy $|\De_{k_F}|$ 
required to excite the system, from where the name `(energy) gap' derives.

Substitution of the solution~(\ref{eq:Fkw-sol}) into the definition of the gap
function~(\ref{eq:gap-fun}) yields the 
\index{gap equation} gap equation of a density-asymmetric system at finite
temperature:
\begin{align}
  \De_{\kvp} 
  &= \inty V(\yv) 
           \lla \psio_{\ua}(\zv)\psio_{\da}(\yv) \rra
		  e^{i\kvp\cdot\yv} \notag\\
  &= \inty V(\yv) \left[-F(\zv\tau^+,\yv\tau)\right] 
     e^{i\kvp\cdot\yv} \notag\\ 
  &= \inty V(\yv)e^{i\kvp\cdot\yv}
     \sumn \intk e^{-i\kv\cdot\yv} \left[-F(\kon)\right] \notag\\
  &= \inv{\beta} \inty\frac{\mathrm{d}^3k}{(2\pi)^3}
     V(\yv)e^{i\kvp\cdot\yv} e^{-i\kv\cdot\yv} 
     \sum_n \frac{\Dekx}{(i\hbar\om_n-E_A)^2-\Ek^2} \notag\\
  &= \intk \underline{\underline{
	 \mathrm{d}^3y
	 V(\yv) e^{i(\kvp-\kv)\cdot\yv} }}
     \frac{\Dekx}{2\Ek}\left[f_F\lp\Ek^+\rp - f_F\lp\Ek^-\rp \right] 
     \notag
     \\
     &= \intk V_{\kvp\kv}\, \frac{\Dek}{2\Ek}
     \left[f_F\lp\Ek^+\rp - f_F\lp\Ek^-\rp \right]  \,.
   \label{eq:gap-eq}
\end{align}
where we have identified the doubly-underlined term as the matrix element of
the potential in $\kv$-space, and used the techniques of
Appendix~\ref{app:suma} to evaluate the summation over Matsubara frequencies.


\subsection{An easy example: the symmetric case}
\label{sub:symmetric}
\begin{figure}[!bh]
  \begin{minipage}{5.5cm}
  \caption[Momentum distribution and excitation energies in \bcs]
	  {\label{fig:uv_k}
	    \bcs\ results for $\De/\mu=0.03$: (Top) Quasiparticle
	    distributions $u_{\kv}^2$ (black) and $v_{\kv}^2$ (green) and
	    (bottom) quasiparticle energy $\Ek$ (black) compared to the ideal 
	    dispersion relation $\ek-\mu$ (green) as a function of
	    momentum. The inset is a magnification of the region around the
	    Fermi surface, which shows the existance of a minimum energy value
	    $\De$ required to excite the system.
	  }
  \end{minipage}%
  \hfill
  \begin{minipage}{8.cm}
  \begin{center}
    \includegraphics
        [width=0.8\textwidth,clip=true]
	{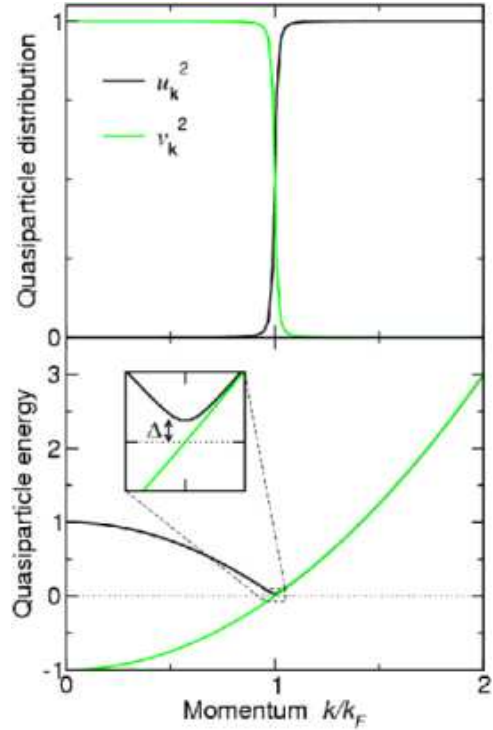}
  \end{center}
  \end{minipage}
\end{figure}

When both fermionic species have equal chemical potentials (which in the
homogeneous system corresponds to equal densities), the 
\index{propagator!normal}normal propagators for
up- and down-particles are the same and equal to
\begin{equation}
  G(\kon) = -\hbar\frac{i\hbar\om_n+\xik}
                          {\hbar^2\om_n^2 + \xik^2 + |\De|^2}
	     = \frac{u_{\kv}^2}{i\om_n-\Ek/\hbar} +
	       \frac{v_{\kv}^2}{i\om_n+\Ek/\hbar} \;,
\end{equation}
with
\begin{align*}
  u_{\kv}^2 = \frac{1}{2}\lp 1+ \frac{\xik}{\Ek} \rp \;, \quad
  v_{\kv}^2 = 1-u_{\kv}^2 \;, \quad
  \Ek^2 = \xik^2 + |\De|^2
\end{align*}
which are the typical results of \bcs\ theory for quasiparticles with
excitation spectra $\Ek$. $u_{\kv}$ can be interpreted as the 
probability amplitude of the paired state $|\kv\ua,-\kv\da\ra$ being
unoccupied, and $v_{\kv}$ the probability amplitude of it being occupied, that
is, there being simultaneously an up-fermion with momentum $\kv$ and another
down-fermion with momentum $-\kv$, thus forming a Cooper pair. Then $\Ek$ is
the excitation energy of the $(N+1)$-particle system when one particle is
added to the ground state of the $N$-particle system, or the excitation energy
of the $(N-1)$-particle system when one particle is removed from the ground
state of the $N$-particle system~\cite{schrieffer}. We show in
Fig.~\ref{fig:uv_k} characteristic curves for $u_{\kv}^2$ and $v_{\kv}^2$,
together with $\Ek$. From this plot the motivation for the name `gap' is
quite apparent: $\De$ is the energy distance between the ground state of the
paired system and its lowest excitations:
\begin{equation*}
  \left.\Ek\right|_{\mathrm{min}}=\De=\De_{k_F} \;.
\end{equation*}
In the general asymmetric case we have two different branches, one
of which can be gapless
; we will comment further on this point in chapter~\ref{ch:dfs}.

The value of the gap in a symmetric system is also readily calculated starting
from Eq.~(\ref{eq:gap-eq}). As now $E_A=0$, we have $\Ek^+=\Ek=-\Ek^-$, so
that $f_F(\Ek^+)-f_F(\Ek^-)=-\tanh(\beta\Ek/2)$. Therefore,
\begin{equation}
  \label{eq:gap-eq-sym}
  \De_{\kvp} = 
  -\intk V_{\kvp\kv}\, \frac{\Dek}{2\Ek} \tanh \lp\beta\frac{\Ek}{2}\rp \;.
\end{equation}
This equation is to be solved together with the number equation 
[compare with Eq.~(\ref{eq:sr})],
\begin{equation}
  \label{eq:num}
  \rho =
  \intk \left[1-\frac{\ek-\mu}{\sqrt{(\ek-\mu)^2+|\De|^2}}
                \tanh\lp\beta\frac{\Ek}{2}\rp \right] \;,
\end{equation}
which accounts for the relationship between the chemical potential and the
density of the system. However, in the 
\index{weak-coupling limit}weak-coupling limit in which we shall
be working, one can safely take $\mu=\ef$. Then Eqs.~(\ref{eq:gap-eq-sym}) and
(\ref{eq:num}) decouple and we only need to worry about the first one.


\vskip1em
There are two limiting cases of special interest:
\subsubsection{Zero temperature gap \mbox{$\De_{\rm sym}$}}
\begin{figure}[t]
  \caption[Diagrammatic representation of the $\T$-matrix]
	  {\label{fig:ladder}
	    Diagrammatic representation of the 
            \index{T-matrix}$\T$-matrix: 
            the solid lines
	    represent free \index{propagator}propagators, the dashed lines
	    stand for the bare 
	    potential $V$ between the two particles, and the wiggle is the
	    $\t$-matrix. 
	  }
  \begin{center}
    \includegraphics
        [width=0.9\textwidth]
	{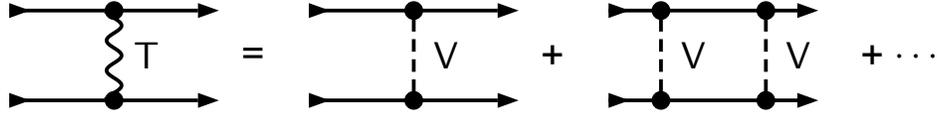}
  \end{center}
\end{figure}

When we set $T=1/(k_B\beta)=0$, the 
\index{gap equation!for symmetric system}gap equation reads
\begin{equation*}
  \De_{\kvp} = -\intk V_{\kvp\kv}\frac{\De_{\kv}}{2\Ek} \;.
\end{equation*}
For a contact interaction of the form $V(\xv-\xvp)=g\delta(\xv-\xvp)$, we have
$V_{\kvp\kv}=g=\mathrm{constant}$, and the integral is ultraviolet
divergent. In fact, this formulation of the gap equation has 
problems for potentials that do not have a well-defined transformation into
Fourier space. It is better to resort to the 
\index{T-matrix}$\T$-matrix, that is the
solution to the Lippmann-Schwinger equation and represents the repeated
action (summation of ladder diagrams, see Fig.~\ref{fig:ladder}) of the
potential $V$ in a propagating two-body system~\cite{messiah,pascual}
\begin{equation}
  \label{eq:t-matrix}
  \T = V + V\inv{E-H+i\varepsilon}V = V + V\inv{E-H_0+i\varepsilon}\T \;,
\end{equation}
and is generally well-defined for any interaction.
Then, the \index{gap equation!regularized}gap equation reads
\begin{equation}
  \label{eq:gap-eq-t}
  \De_{\kvp} = -\intk \T_{\kvp\kv} \lp\inv{2\Ek}-\inv{2\ek}\rp \De_{\kv} \;,
\end{equation}
which is more convenient for numerial treatment (see Section~\ref{sec:num}).

For dilute systems such as the atomic gases under current experimental
research, one has $k_F|a|\ll 1$. Therefore, as the integrand in the gap
equation is sharply peaked around the Fermi momentum $k_F$ in the
\index{weak-coupling limit}weak-coupling regime ($\De/\mu \ll 1$), 
we can substitute the momentum-dependent 
\index{T-matrix!at low momenta!in $s$-wave}
\index{T-matrix!and scattering length}
$\T$-matrix by its value at low momenta, 
which is a constant proportional to the scattering lenght $a$,
\begin{equation}
  \T_0 := \lim_{k',k\raw 0} \T_{\kvp\kv} = \frac{4\pi\hbar^2a}{m} \;.
\end{equation}
As $\T_0$ is a constant, it can be pulled out of the integral, and the gap
function becomes independent of momentum. With this simplification, and making
a change of variables to $x=\ek/\mu=\hbar^2\kv^2/(2m\mu)$ and $d=\De/\mu$, the
gap equation~(\ref{eq:gap-eq-t}) can be rewritten as
\begin{align}
  1&=
  -\frac{\T_0}{4\pi^2}\int_0^{\infty} \mathrm{d}k\,k^2
  \left[ \inv{\sqrt{\lp \ek-\mu \rp^2+\De^2}}-\inv{\ek}  \right] \notag\\
  &=
  -\frac{k_Fa}{\pi}\int_0^{\infty} \mathrm{d}x\,
  \left[ \sqrt{\frac{x}{\lp x-1 \rp^2+d^2}}-\sqrt{x}  \right]
  \label{eq:gapp}
\end{align}
where we have already integrated the angular variables and used the
approximation $\mu\approx\ef=\hbar^2k_F^2/(2m)$ valid in the 
\index{weak-coupling limit}weak-coupling
limit~\cite{fetter,schrieffer,bertsch}.
From this result, we see that the gap in this limit will be a function only of
the adimensional parameter $k_Fa$, irrespective of the specific form of the
interparticle potential. Moreover, we shall have a non-vanishing gap only for
an attractive interaction, $a<0$, as in the case of the Cooper problem
(Section~\ref{sub:cooper}). 

A quick, though approximate, evaluation of this integral can be done noting
that, for $d\ll1$, the integrand is sharply peaked around the Fermi surface
($x\approx1$), which allows us to simplify the first term inside the square
brackets taking $x=1$ in its numerator. The resulting integral is analytic, and
we obtain
\begin{equation*}
  \frac{\De_{\rm sym}}{\mu} = 2 \exp\lp\frac{\pi}{2k_Fa}\rp \qquad (a<0) \;.
\end{equation*}

A more careful evaluation 
of integral~(\ref{eq:gapp}) gives the well-known result first obtained by
Gor'kov and Melik-Barkhudarov \cite{gmb,bertsch}
\begin{equation}
  \label{eq:gap_sym}
  \frac{\De_{\rm sym}}{\mu} = \frac{8}{e^2}\exp\lp\frac{\pi}{2k_Fa}\rp  
  \approx 1.083 \exp\lp\frac{\pi}{2k_Fa}\rp
  \qquad (a<0) \;.
\end{equation}
Note that the functional dependency on $k_Fa$ is the same as in the previous
expression and only the pre-exponential factor has changed.

\subsubsection{Critical temperature \protect{\mbox{$T_c$}}}
\index{critical temperature!for pairing transition}
The critical temperature is defined as the lowest temperature where the gap
vanishes. Assuming again a low density, so that the gap is independent of
momentum and $\De_{\kvp}$ cancels with $\Dek$ in the numerator of the gap
equation, we can find $T_c$ from (\ref{eq:gap-eq}) while setting $\De=0$ in
$\Ek$:
\begin{equation*}
  1 =
  -g\intk\inv{2\xik}\tanh\lp\beta_c\frac{\xik}{2}\rp
  \qquad \lp\beta_c=\inv{k_BT_c}\rp \;.
\end{equation*}
This equation is solved in a way similar to that for the zero-temperature
gap~\cite{gmb}. The final result is
\begin{equation}
  \label{eq:T_c}
  \frac{k_BT_c}{\mu} = \frac{8e^{\gamma-2}}{\pi}\exp\lp\frac{\pi}{2k_Fa}\rp
  \;,
\end{equation}
where $\gamma=0.5772\cdots$ is Euler's constant. Note that $T_c$ has
the same functional dependence on $k_Fa$ as the gap. Therefore, in
the \index{weak-coupling limit}weak-coupling limit, 
the \bcs\ theory leads to a simple proportionality
relationship between the zero-temperature gap and the 
\index{critical temperature!for pairing transition}critical temperature for
the superfluid transition, independent of the physical system under study:
\begin{equation}
  \label{eq:Tc}
  \frac{k_BT_c}{\De_{\rm sym}} = \frac{e^{\gamma}}{\pi} \simeq 0.567 \;.
\end{equation}
This relationship ---which is reasonably well fulfilled by many superconducting
metals and alloys~\cite{fetter,kittel}, but not by high-temperature
superconductors~\cite{levin-physrep,htsc}--- will allow us to calculate only
the zero-temperature gaps. The transition temperatures into the superfluid
state for dilute, atomic gases can be obtained then from Eq.~(\ref{eq:Tc}).

\section{Summary}
In this chapter we have introduced the pairing solution for the ground state
of a system of interacting fermions. First, we have presented the concept of
`Cooper pairs'. Then, we have formulated the \bcs\ theory as a solution
in terms of Greens' functions at finite temperature of the interacting
Hamiltonian in a self-consistent Hartree-Fock-like approximation, in the
general case of a two-component system with density asymmetry, \ie, where the
two components may have different densities. 

Finally, we have obtained some analytic results for the gap at zero
temperature and the critical temperature for the case where both components
have equal densities. In chapter~\ref{ch:asym} we will study how the
energy gap, chemical potential and other physical parameters of the sytem are
affected when this last condition is not fulfilled.

We remark at this point that an equivalent solution for the Green's functions
for the full interacting Hamiltonian, can be found solving the
corresponding Dyson's equation, which in ($\kv,\om$)-space reads 
\begin{equation*}
  G(\kv,\om)^{-1} = G(\kv,\om)_0^{-1} - \Sigma(\kv,\om) \;.
\end{equation*}
Here $G_0^{-1}=\om-\xik+i\,\eta\:{\rm sgn}(\om)$ is the 
non-interacting propagator and $\Sigma$ is the self-energy, which has to be
determined in order to solve the problem. A further analysis of this approach
is beyond the scope of this introductory chapter and can be found
in~\cite{sedrakian-pra}.

\chapter[Pairing in density-asymmetric fermionic systems]
	{Pairing in density-asymmetric fermionic systems}%
\label{ch:asym}

\textsf{
  \begin{quote}
     Politiker, Ideologen, Theologen und Philosophen versuchen 
     immer und immer wieder, restlose Lösungen zu bieten, 
     fix und fertig geklärte Probleme. Das ist ihre Pflicht --- und es ist unsere, 
     der Schriftsteller, - die wir wissen, dass wir nichts rest- und widerstandslos 
     klären können - in die Zwischenräume einzudringen.
  \end{quote}
  \begin{flushright}
     Heinrich Böll, {\em Nobel lecture} 
  \end{flushright}
}

\section{Introduction}
\label{sec:asym-intro}

In Chapter~\ref{ch:pair-intro} we have developed a general formalism to
study the \bcs\ pairing mechanism in a system composed of two fermionic
species with chemical potentials $\mu_\ua$ and $\mu_\da$, which are not 
required to be equal. 
This is a subject of interest in a number of fields of research, such
as nuclear physics~\cite{sedr-prc}, elementary particles~\cite{casal} and
condensed matter physics~\cite{yeh}.
We are mainly interested in the application to dilute, atomic systems where a
great deal of work has been devoted since the pioneering contribution of Stoof
and coworkers~\cite{stoof96,houbiers}, who predicted the possibility to detect
a phase transition to a superfluid state in a trapped system of~\Lix\ atoms.

\begin{figure}[htb]
\caption[Evaporative cooling procedure]
	{\label{fig:cool}
	  Evaporative cooling procedure: in a trapped system of bosons at
	  temperature $T$, the most energetic ones are let escape, thus
	  reducing the mean energy per particle of the remaining
	  ensemble. After rethermalization, the system is at a lower
	  temperature $T'<T$.
	}
  \begin{center}
    \includegraphics
                  [width=0.7\textwidth,clip=true]
		  {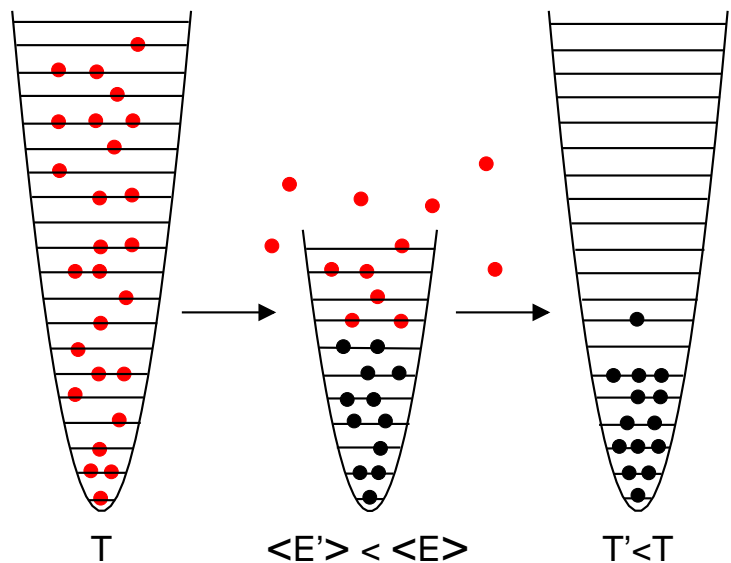}
  \end{center}
\end{figure}

After the achievement of Bose-Einstein condensation (BEC) in dilute gases 
of alkali atoms in 1995~\cite{jila95,mit95}, an important target in 
cooling atomic samples was to reach the degenerate regime in a gas 
of fermionic atoms. Indeed, the mechanism that allowed the production 
of BECs, known as {\em evaporative cooling}, relies intrinsically on the 
capacity of very cold atoms to quickly re-thermalize when some 
of them (those with higher energies) are let escape from the 
trap, see Fig~\ref{fig:cool}. 
At the very low temperatures of interest (of the order of $\mu$K or below),
the only remaining interaction in a dilute system are $s$-wave collisions. As
these are forbidden between indistinguishable fermions by Pauli's exclusion
principle~\cite{pascual,messiah}, this mechanism does not work to cool
one-component Fermi gases (\eg, a gas with spin-polarized atoms). 
For such a system, the loss of very energetic atoms implies, of course, a
decrease of the mean energy per particle, but in the absence of a
(re)thermalization mechanism, this just means that the system is not in
thermal equilibrium, but no redistribution of the remaining atoms in phase 
space occurs.

This limitation can be overcome if the trapped system is not composed just by
one kind of fermions. For example, one can trap atoms in two (or more) 
hyperfine states. In this case, there is no problem for the existence of
$s$-wave collisions between atoms belonging to different states and
this {\em simultaneous cooling} scheme works for such a multi-component
fermionic system in the same way as for a bosonic system, as was first shown
by DeMarco and Jin at \jila~\cite{jin99}.
Another possibility is to trap a mixture of bosons and fermions. 
In this case, the bosons are cooled in the usual way,
while the fermions cool down by thermal contact with the bosons, as $s$-wave
collisions between them are not forbidden. This mechanism 
is known as {\em sympathetic cooling} 
and it was first realized by the group
of Randy Hulet at Rice University, who reported experiments where a gas of
\Lix\ atoms (with fermionic character) was driven to quantum degeneracy 
by contact with a gas of \Lin\ (bosons)~\cite{tru01}.

Since these pioneering works, many groups have produced degenerate Fermi
gases around the world~[\textsc{jila, mit, lens}, Innsbruck, Paris,...]. 
One of the main goals of this research has been to create (and 
detect!%
)
a superfluid made up of atoms, in a sense analog to superfluid helium but
with the advantage (in principle) of a much weaker interaction due to the low
density of these systems (typically, in the absence of resonant phenomena such
as Feshbach resonances, $\rho|a|^3\ll 1$, while in helium $\rho a^3\sim1$),
which allows for an easier understanding of the physics behind the experiments. 
In particular, it is possible to use the mean-field \bcs\ theory of pairing in 
the \index{weak-coupling limit}weak-coupling approximation~\cite{stoof96,houbiers}.

In this Chapter, we will study which are the possibilities of having pairing
in a two-component fermionic system such as that in~\cite{jin99}. In
particular, 
we will focus on the relevant paper played by the difference in densities of
the two components. As we will show, this difference reduces dramatically
the size of the energy gap (and, therefore, the expected 
\index{critical temperature!for pairing transition}critical temperature 
for appearence of superfluidity) when considering
pairing between atoms belonging to different species. However, pairing between
atoms belonging to the same species can be {\em enhanced} by properly
adjusting the density of the other species.

\section{Pairing in \mbox{$s$}-wave}
\label{sec:swave}

Let us assume a fermionic system composed of two distinct species (which we
label $\ua$ and $\da$) 
of equal mass $m$ and with densities $\ru$ and $\rd$, 
or equivalently total density $\R=\ru+\rd$ and 
\index{density asymmetry!definition}
density asymmetry $\al=(\ru-\rd)/(\ru+\rd)$.  
For example, species $\ua$ and $\da$ can denote different hyperfine levels
of atomic gases, or `neutron' and `proton' in the context of nuclear
physics (where one should take into account both the spin and the isospin as
internal degrees of freedom of the `nucleon').
We also introduce the Fermi momenta
$k_s \equiv k_F^{(s)}=(6\pi^2\rho_s)^{1/3}$, and chemical potentials
$\mu_s=\hbar^2k_s^2 /(2m)$ ($s=\ua,\da$), 
together with $\mu=(\mu_\ua+\mu_\da)/2$ and $\delta\mu=(\mu_\ua-\mu_\da)/2$.

\begin{figure}
  \caption[The two possible lowest-order pairing interactions]
          {\label{fig:lowestorder}
	    The two possible lowest-order pairing interactions:
            (a) Direct $s$-wave interaction between different species.
            (b) Polarization-induced $p$-wave interaction between like
            species. 
            \textsf{V$_0$} and \textsf{T$_0$} are the $s$-wave ($L=0$) bare potential and
            $\T$-matrix between species $\ua$ and $\da$, respectively.
          }
  \begin{center}
    \includegraphics[width=0.7\textwidth]
                    {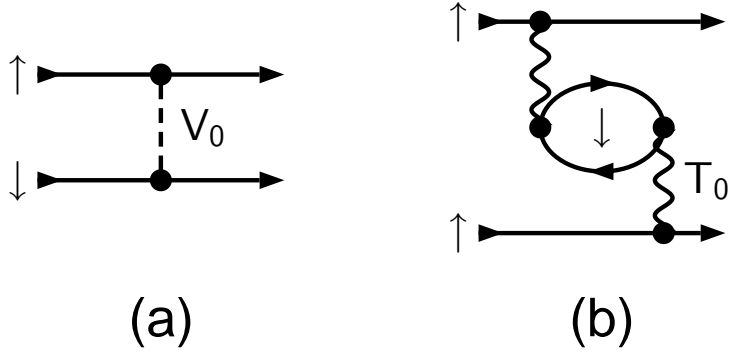}
  \end{center}
\end{figure}

For simplicity, we will for the moment consider an
idealized system 
where $\ua$ and $\da$ particles are interacting via a
potential $V$ with
$s$-wave scattering length $a$,
while direct interactions between like particles 
$\ua$-$\ua$ and $\da$-$\da$ are absent [cf. Eq.~(\ref{eq:bcs-ham}) and
Fig.~\ref{fig:lowestorder}(a)].
We are interested in the situation at very low density, $k_s|a| \ll 1$. We
have seen in Chapter~\ref{ch:pair-intro} that, in this limit, the pairing
properties of the symmetric system are completely determined by the scattering
\index{T-matrix!at low momenta}
length or, equivalently, the low-momentum $s$-wave $\T$-matrix, 
$\T_0 = 4\pi\hbar^2 a/m$. We will now study the general case where the two
species can have different densities, and analyze the corresponding pairing
gaps generated by this interaction $V$.
%

\begin{figure}[!bp]
  \begin{minipage}{6cm}
    \caption[\bcs\ momentum distributions in asymmetric matter]
	    {\label{fig:distrib}
	      \bcs\ momentum distributions of the species $\ua$ (black lines)
	      and $\da$ (green lines) in asymmetric matter. The continuous
	      lines stand for the normal matter ($\De=0$) distribution, the
	      dashed lines for superfluid matter with $\De=0.03\mu$, while the
	      dotted lines show the distributions corresponding to the more
	      coupled case $\De=0.07\mu$. %
              \index{density normalization!in pairing problem}[see Eqs.~(\ref{eq:rhos})]. 
              \index{forbidden region}\index{Pauli blocking}
	      The shaded area is the `forbidden region'
	      $[\mu-\deps,\mu+\deps]$ for the case $\De=0.07\mu$.
	    }
  \end{minipage}%
  \hfill
  \begin{minipage}{8cm}
  \begin{center}
  \includegraphics[width=0.9\textwidth,clip=true]
		  {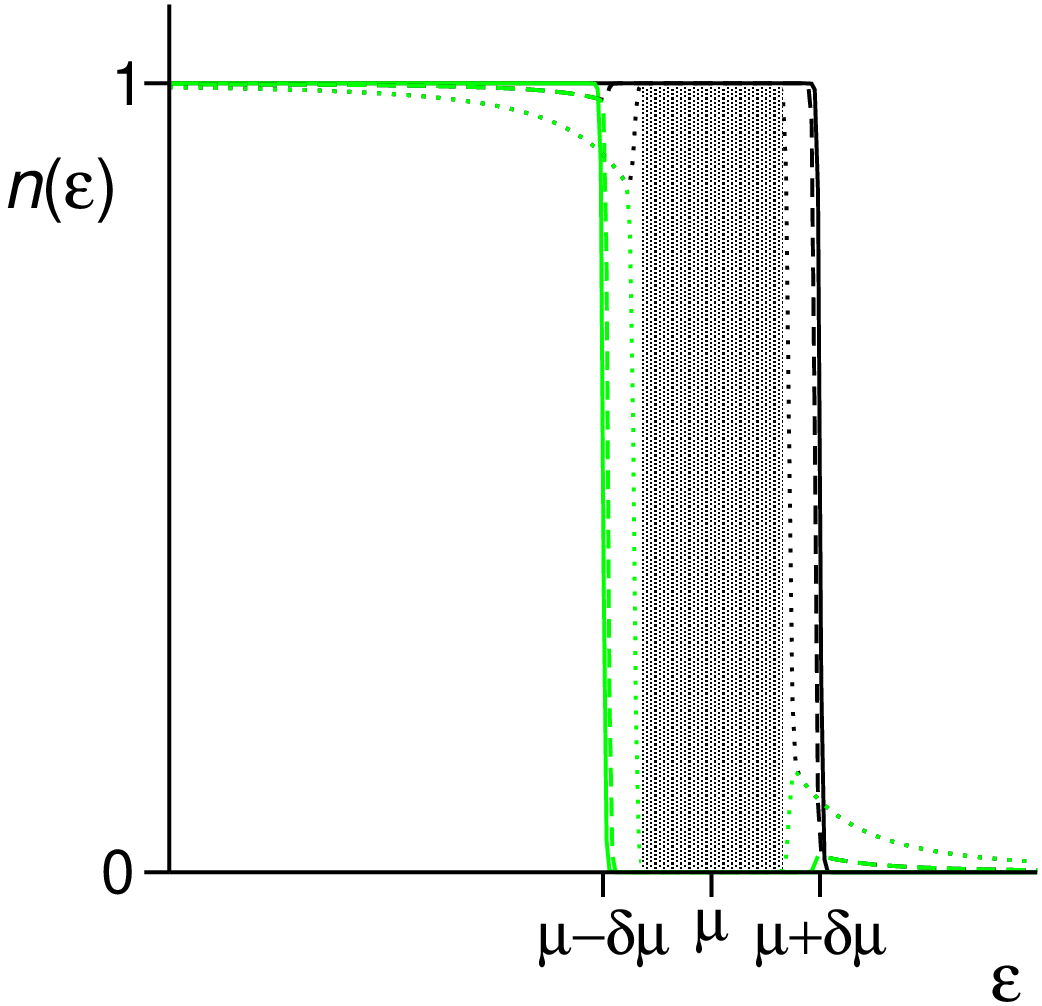}
  \end{center}
  \end{minipage}
\end{figure}

As we did for the symmetric system in Sect.~\ref{sub:symmetric}, we must solve
the \index{gap equation}gap equation~(\ref{eq:gap-eq}),
\begin{align*}
  \De_{\kvp} &= \intk V_{\kvp\kv}\, \frac{\Dek}{2\Ek}
     \left[f_F\lp\Ek^+\rp - f_F\lp\Ek^-\rp \right]  \:, \\
  \Ek^{\pm} &= E_A \pm \sqrt{E_S^2+|\De_{\kv}|^2} \;,
\end{align*}
together with the equations that fix the chemical potential of each species
(or, equivalently, $\mu$ and $\dmu$) from the values of their 
\index{density normalization!in pairing problem}densities.
These equations can be found again starting from the 
\index{propagator!normal}normal propagators and
summing over Matsubara frequencies (see Appendix~\ref{app:suma}):
\begin{subequations}
\begin{align}
  \ru(\kv) 
  &= \sumn -G_\ua(\kon) =
  \inv{2}\lp 1+\frac{\xik}{\Ek} \rp f(\Ek^+) +
  \inv{2}\lp 1-\frac{\xik}{\Ek} \rp f(\Ek^-) \notag \\
  &=
  \uk^2 f(\Ek^+) + \vk^2f(\Ek^-) 
  = 
  \uk^2\left[f(\Ek^+)-f(\Ek^-)\right] + f(\Ek^-)
  \:, \\
  \rd(\kv) 
  &= \sumn -G_\da(\kon) =
  1-f(\Ek^+) + 
  \uk^2\left[ f(\Ek^+) -f(\Ek^-)\right]
  \:.
\end{align}
\label{eq:rhos}
\end{subequations}
These distributions are shown in Fig.~\ref{fig:distrib} for $\dmu=0.1$ and
different values of the gap $\De$ to get some physical insight.
The sharp Fermi surfaces of the non-interacting system smear out for
increasing $\De$ reflecting the correlations introduced by the pairing
interaction. However, this `smearing' hardly penetrates into the region
between $\mud$ and $\muu$ as Pauli's principle forbids new $\ua$ particles to
enter there. Thus, the newly formed $\ua$-$\da$ pairs need to climb to states
$\ek>\muu$. 
This has a cost in kinetic energy, which is compensated by the pairing energy
gained. Clearly this mechanism cannot sustain arbitrarily large asymmetries,
as the kinetic energy investment grows rapidly with the width $\sim2\deps$ of
the forbidden region (see below). In fact, one can readily find the total
density $\R$ and density difference $\delta\rho$ to be
\begin{subequations}
\begin{align}
  \rho = \ru + \rd &= 
  \sum_{k} \Big[ 1 + \frac{\xik}{E_{\kv}} 
    \left[ f_F(E_{\kv}^+) - f_F(E_{\kv}^-) \right] \Big] \:,
  \label{eq:sr}
  \\
  \delta\rho = \ru - \rd &= \sum_{k} 
  \left[ f_F(E_{\kv}^+) + f_F(E_{\kv}^-) -1 \right] \:.
  \label{eq:dr}
\end{align}
\label{eq:rhos2}
\end{subequations}


At zero temperature, $f_F(E) = \Theta(-E)$ and, therefore,
\begin{align*} 
  f_F(E_{\kv}^+) - f_F(E_{\kv}^-) &= -\Theta(E_{\kv}^+) \:,
  \\
  f_F(E_{\kv}^+) + f_F(E_{\kv}^-) -1 &= \Theta(-E_{\kv}^+) \:,
\end{align*}
where we used the fact that $\Theta(x)=1-\Theta(-x)$.
From the second equation we see that the unpaired particles are to be found in
the energy interval $[\mu-\deps, \mu+\deps]$, with $\deps =
\sqrt{{\dmu}^2-{\De}^2}$ ($\approx\dmu$ for $\De<\mu$),
which does not contribute to the pairing interaction, as indicated by the
shaded area in Fig.~\ref{fig:distrib}.
This leads to a rapid decrease of the resulting gap when increasing the
asymmetry.%
\footnote{
  One could imagine also a promotion of all the $\ua$ particles responsible for
  the blocking effect to higher kinetic energy and having a continuous
  distribution of pairs at low momenta, but this would imply a larger kinetic
  energy investment.
}

In the \index{weak-coupling limit}weak-coupling case, $\De \ll \deps \ll \mu$, 
which is adequate in the low-density limit, the momentum distributions of the 
two species are anyhow very sharp and one obtains from Eqs.~(\ref{eq:rhos2})
\begin{align*}
  \rho &= \intk 
    \left[1+\frac{\xik}{E_{\kv}}\lp1-\Theta\lp-E_{\kv}^+\rp\rp\right]\\
  &=\inv{4\pi^2}\lp\frac{2m\mu}{\hbar^2}\rp^{3/2} 
    \left\{
    \lp\int_0^{1+\delta x}+\int_{1+\delta x}^\infty\rp\mathrm{d}x\,\sqrt{x}
    \right.\\
  &\qquad\left.
    -\int_0^{1-\delta x}\mathrm{d}x\,\sqrt{x}\frac{x-1}{\sqrt{(x-1)^2+d^2}}
    -\int_{1+\de x}^\infty\mathrm{d}x\,\sqrt{x}\frac{x-1}{\sqrt{(x-1)^2+d^2}}
    \right\}\\
  &\approx \inv{2\pi^2}\lp\frac{2m}{\hbar^2}\rp^{3/2}\frac{2}{3}\mu^{3/2}
\end{align*}
and
\begin{align*}
  \delta\rho &= \intk \Theta\lp\dmu-\sqrt{(\ek-\mu)^2-|\De|^2}\rp \\
  &= \inv{4\pi^2}\lp\frac{2m\mu}{\hbar^2}\rp^{3/2} 
     \int_0^\infty \mathrm{d}x\, \sqrt{x} \,
     \Theta\lp\delta x-\sqrt{(x-1)^2+d^2}\rp\\
  &\approx \inv{2\pi^2}\lp\frac{2m}{\hbar^2}\rp^{3/2}\mu^{1/2}\deps \;,
\end{align*}
where we used the shortening $\delta x=\deps/\mu$ and $d=\De/\mu$, and the
fact that $\dmu\ll\mu$. Therefore,
\be 
  \alpha =
  \frac{\delta\rho}{\rho} \approx \frac{3}{2}\frac{\deps}{\mu} \ll 1 \:,
\ee 
\ie, the width of the forbidden region $\deps$ is directly proportional to the
\index{density asymmetry! and forbidden region}
\index{forbidden region}\index{Pauli blocking}
density asymmetry $\alpha$. 
Analyzing in a similar way the gap equation, one can obtain these
relationships involving the parameter $\deps$
\cite{stoof96,sed97,umb2}, 
\begin{equation*}
  \dmu + \deps = {\rm const.} = \De_{\rm sym} \quad
  \Leftrightarrow \quad \De^2 = \De_{\rm sym}^2 - 2\De_{\rm sym}\deps
  \:, 
\end{equation*}
where $\De_{\rm sym}$ is the gap in symmetric matter of the same 
density $\R$ [Eq.~(\ref{eq:gap_sym})].
Altogether, we can write the gap as a function of the asymmetry thus:
\be 
  \frac{\De(\al)}{\De_{\rm sym}} = 
  \sqrt{ 1 - \frac{4\mu}{3\De_{\rm sym}}\alpha } \;.
  \label{eq:sgap}
\ee 
\index{density asymmetry!critical}
It vanishes at $\alpha_{\rm max}^{BCS} = 3\De_{\rm sym}/(4\mu)$, which is an
exponentially small number in the 
\index{weak-coupling limit}weak-coupling limit $k_F|a|\ll 1$.
Therefore, for very small asymmetries already, pairing generated by
the direct interaction between different species becomes impossible according
to the \bcs\ theory. 
However, this limitation can be partially overcome, as we will see in 
chapter~\ref{ch:dfs}.

\section{Numerical results}
\label{sec:num}

\subsection{The algorithm}
\label{sub:algor}
Let us now present our numerical procedure to solve the 
\index{gap equation!numerical solution}gap equation and check
the results with the previous analytical calculations.
We start rescaling all energies in units of the mean chemical potential,
\begin{align*}
  x   &= \ek/\mu   &   d &= \De/\mu &  \de x&= \deps/\mu \\
  E_x &= \Ek/\mu 
      & E_x^\pm &= -\frac{\dmu}{\mu}\pm E_x
      & x^\pm &=\Ek^\pm/\mu \:,
\end{align*}
Assuming a constant value for $V_{\kvp\kv}=g$, so that the gap
function will not depend on momentum, we rewrite the 
\index{gap equation}gap equation as
\begin{align*}
  \label{eq:gap-adim}
  1 &= g\frac{\lp2m\rp^{3/2}}{8\pi^2\hbar^3}\sqrt{\mu}
  \int_0^{\infty}\mathrm{d}x\, \sqrt{x}
  \inv{E_x}
    \left[ f_F\lp\mu x^+ \rp - f_F\lp\mu x^- \rp \right] 
  \:.
\end{align*}
To avoid a possible divergence in the 
\index{gap equation!regularized}gap equation due to the use of a contact
potential, we resort to the 
\index{T-matrix!in gap equation}
$\t$-matrix as in Sect.~\ref{sub:symmetric} (see also \cite{Strinati-epjb}). 
We get for the general low-density, asymmetric case 
\begin{equation}
  1 = \frac{a}{\pi}\sqrt{\frac{2m\mu}{\hbar^2}}
  \int_0^{\infty}\mathrm{d}x\, \sqrt{x}
  \left\{ \inv{E_x}
    \left[ f_F\lp\mu x^+ \rp - f_F\lp\mu x^- \rp \right] +\inv{x} 
  \right\} \:,
\label{eq:gap-adim}
\end{equation}
where we used $\T_0=4\pi\hbar^2a/m$ for a low-density system.
Regarding the densities, they are given by integrating Eqs.~(\ref{eq:rhos}).
We present below our numerical results for the solution of
\index{density normalization!in pairing problem}equations (\ref{eq:rhos}) 
and (\ref{eq:gap-adim}) obtained in the following way:
\begin{enumerate}
\item Introduce an `effective Fermi momentum' $k_\mu$ through the 
  chemical potential, $\mu\equiv\hbar^2k_\mu^2/(2m)$.
\item Define the functions
  \begin{align*}
    f &= \frac{\pi}{k_\mu a} -
         \int_0^{\infty}\mathrm{d}x\, \sqrt{x}
         \left\{ \inv{E_x}
         \left[ f_F\lp\mu x^+ \rp - f_F\lp\mu x^- \rp \right] + \inv{x}
         \right\} 
\\
    g &= \frac{4\pi^2\ru}{k_\mu^3} -
         \int_0^\infty \mathrm{d}x\, \sqrt{x} \left\{ 
	 u_x^2\left[f_F(\mu x^+)-f_F(\mu x^-)\right] + f_F(\mu x^-)
	 \right\}
\\
    h &= \frac{4\pi^2\rd}{k_\mu^3} -
         \int_0^\infty \mathrm{d}x\, \sqrt{x} \left\{
	 1-f_F(\mu x^+)+u_x^2\left[f_F(\mu x^+)-f_F(\mu x^-)\right]
	 \right\}
  \end{align*}
  with $u_x^2=[1+(x-1)/E_x]/2$.
  These are functions of the gap $d$ (through $E_x$) and the chemical
  potentials $\muu$ and $\mud$ (through $\mu$, $k_\mu$ and $x^\pm$).
  They must be zero at the solution of 
  \index{density normalization!in pairing problem}Eqs. (\ref{eq:rhos}) and
  (\ref{eq:gap-adim}). 
\item Assume some initial values $d=d^{(0)}$, 
 $\mu_{\ua\da}=\mu_{\ua\da}^{(0)}$. As we work at fixed total density $\rho$
\index{density asymmetry!definition}
 and density asymmetry $\de\rho=\rho\al$, we have used these guesses:
  \begin{align*}
    d^{(0)} &=\frac{\De(\al)}{\mu^{(0)}} 
              \quad\text{[cf. Eq.~(\ref{eq:sgap})],} &
    \mu_{\ua\da}^{(0)} &=\ef\lp 1 \pm \al\rp^{2/3} \:.
  \end{align*}
  with $\ef=\hbar^2\lp3\pi^2\rho\rp^{2/3}/(2m)$.
\item Insert these values into $f$ and find (\eg, by a Newton-Raphson method)
  a zero of $f$ as a function solely of $d$. Call this zero $d^{(1)}$.
\item Insert $d^{(1)}$ and $\mu_{\ua\da}^{(0)}$ into $g$, and find a zero
  of $g$ as a function of solely $\muu$. Call this zero $\muu^{(1)}$.
\item Insert $d^{(1)},~\muu^{(1)}$ and $\mud^{(0)}$ into $h$, and find a zero
  of $h$ as a function of solely $\mud$. Call this zero $\mud^{(1)}$.
\item If the new values for $d$ and $\mu_{\ua\da}$ make $|f|,~|g|$ and $|h|$
  smaller than a certain tolerance $\varepsilon$ (in our calculations,
  typically $\varepsilon\sim10^{-4}$), take them as the solution.
  Otherwise, go back to step 4 and try again until convergence is reached.
\end{enumerate}

\subsection{Symmetric case}
\label{sub:num-sym}

\begin{figure}[!bp]
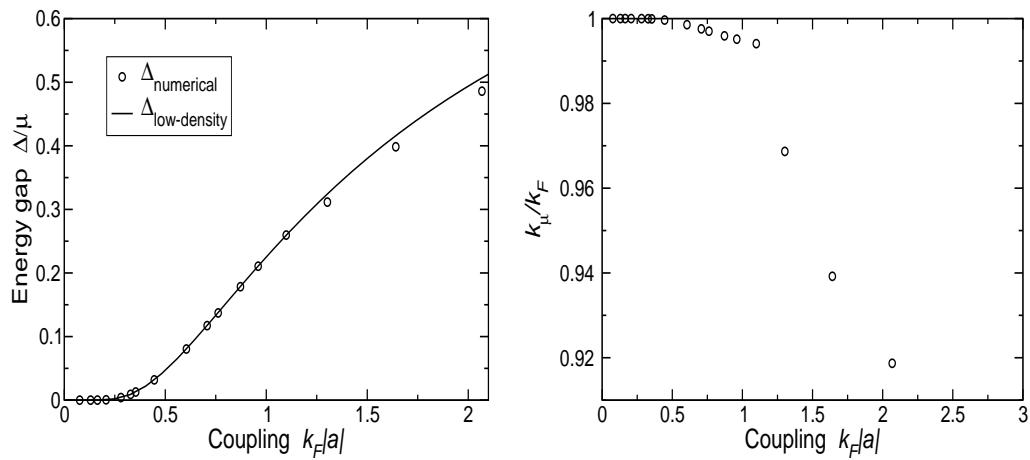

  \vspace{-0.8cm}\phantom{.}
  \caption[Numerical solution for the symmetric case varying $k_F|a|$]
	  {\label{fig:num_sym}
	    Solution of the \BCS\ equations for a temperature $T=0.1T_c$ and
	    varying coupling: 
	    (Left) Numerical solution for the gap $\De$ as a function of the
	    coupling and prediction of \BCS\ at low density,
	    Eq.~(\ref{eq:gap_sym}).
	    (Right) Effective Fermi momentum $k_\mu$ as a function of the 
	    coupling.
	    \vspace{10pt}\phantom{.}
	  }
  \begin{minipage}{0.49\textwidth}
    \begin{center}
      \includegraphics[width=0.95\textwidth,height=6cm,clip=true]
		      {pairing/prog/gap/explo1-gap_kfa}
    \end{center}
  \end{minipage}%
  \hfill
  \begin{minipage}{0.49\textwidth}
    \begin{center}
      \includegraphics[width=\textwidth,height=6cm,clip=true]
		      {pairing/prog/gap/explo1-kmu_kfa}
    \end{center}
  \end{minipage}
\end{figure}

First of all, in order to test our code, we analyze the dependence of the gap
and the chemical potential for the symmetric case, $\ku=\kd\equiv k_F$, at
low temperatures. We fix $T=0.1T_c$, with $T_c$ the \BCS\ weak-coupling
prediction for the 
\index{critical temperature!for pairing transition}critical temperature 
as given by Eq.~(\ref{eq:Tc}). 
The results that we obtain are plotted in Fig.~\ref{fig:num_sym}. 
In the left
panel we show with circles the calculated variation of the energy gap $\De$
with the coupling $k_F|a|$. For comparison, the solid curve is the \BCS\
prediction at low-density, Eq.~(\ref{eq:gap_sym}). We see that the agreement
is very good for all $k_F|a|<1$, and only deviates slightly for 
$k_F|a|\gtrsim 1.3$. 
The `effective' Fermi momentum $k_\mu=\sqrt{2m\mu}/\hbar$ (normalized to the
non-interacting Fermi momentum $k_F=\sqrt{2m\ef}/\hbar$) is plotted in the
right panel also as a function of the coupling. In this case, we find 
$k_\mu\approx k_F$ for $k_F|a|\leq 1$, with a smooth decrease that becomes
more pronounced above this value of the coupling. This is a signal of the
increasing modification of the system with respect to the non-interacting one
for such large couplings. In fact, for these values of $k_F|a|$ the mean-field
\BCS\ theory is not adequate, and more sophisticated methods need to be
used~\cite{holland01,griffin,bulgac04,falco2,strinati,levin-physrep}.
Therefore, we shall not work in the following in this strong-coupling
regime.

\begin{figure}[!bp]
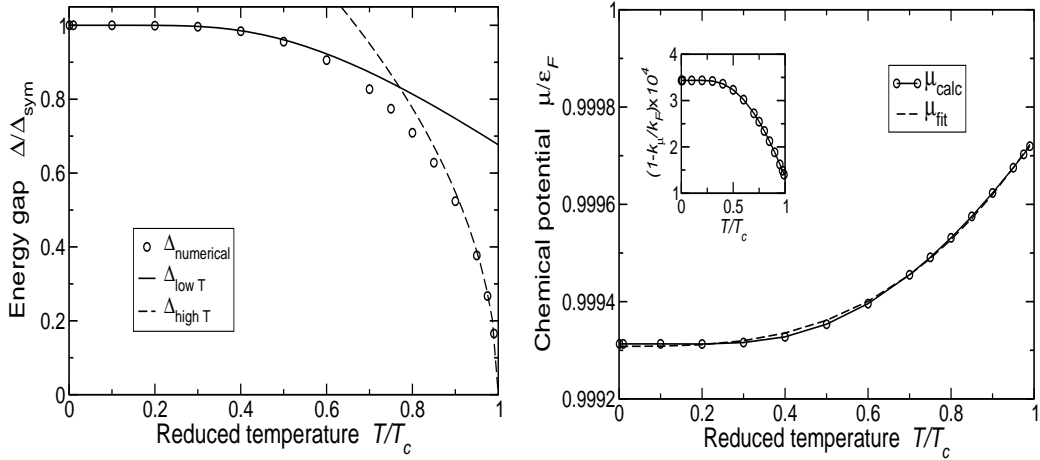

  \vspace{-0.9cm}\phantom{.}
  \caption[Numerical solution for the symmetric case varying $T$]
	  {\label{fig:num_sym_temp}
	    Solution of the \BCS\ equations at finite temperature 
	    for a fixed density $\rho=2\times10^{-12}$ cm$^{-3}$: 
	    energy gap (left) and chemical potential (right)
	    {\em vs.} reduced temperature $T/T_c$. 
	    See the text for details on the notation.
	    \vspace{10pt}\phantom{.}
	  }
  \begin{minipage}{0.48\textwidth}
    \begin{center}
      \includegraphics[width=\textwidth,height=6cm,clip=true]
		      {pairing/prog/gap/explo2-gap_T}
    \end{center}
  \end{minipage}%
  \hfill
  \begin{minipage}{0.49\textwidth}
    \begin{center}
      \includegraphics[width=\textwidth,height=6.12cm,clip=true]
		      {pairing/prog/gap/explo2-mu_T}
    \end{center}
  \end{minipage}
\end{figure}

Next we analyze the temperature dependence of the same quantities above. To
satisfy the \index{weak-coupling limit}weak-coupling condition just discussed, 
we fix the density to $\rho=2\times 10^{-12}$ cm$^{-3}$, 
so that $k_F|a|=0.445524797<1$. 
The results are shown in Fig.~\ref{fig:num_sym_temp}.
In the left panel, the energy gap [divided by the zero-temperature
value~(\ref{eq:gap_sym})] is plotted as a function of the reduced temperature
$T/T_c$, with $T_c$ given again by Eq.~(\ref{eq:Tc}). Our numerical results
are shown with empty circles, while the two lines are the analytical curves of
the \BCS\ model (see Ref. \cite{fetter}, p. 449) at low and `high'
temperatures: 
\begin{align*}
  \De_{\rm low~T} &= \De_{\rm sym}\left[
    1-\sqrt{2\pi\frac{k_BT}{\De_{\rm sym}}}e^{-\De_{\rm sym}/k_BT}
    \right] \:,& T/T_c \ll 1 \,, \\ 
  \De_{\rm high~T} &= 
    \sqrt{\frac{8\pi^2}{7\zeta(3)}}k_BT_c\lp1-\frac{T}{T_c}\rp^{1/2} 
    \:, & (T_c-T)/T_c \ll 1 \,.
\end{align*}
Both predictions agree well with the data in the corresponding ranges of
validity. More specifically, the low-temperature curve is very close to the
calculated values for $T/T_c\lesssim 0.5$, while the `high'-temperature one
is valid for $T/T_c\gtrsim 0.85$.

Regarding the chemical potential (shown by the circles and the solid curve,
which is a guide to the eye), it is always very similar to the non-interacting
one, \ie, $\ef$. However, the weak attraction between the fermions results in
a slight reduction of $\mu$, that diminishes when approaching the 
\index{critical temperature!for pairing transition}critical temperature, 
where the gap is expected to vanish (but the system is still
interacting, hence the difference between $\mu$ and $\ef$). The dependence of
$\mu$ on temperature can be well fitted by
\begin{equation*}
  \mu = \mu(T=0)
     \left[1-\gamma\lp{T}/{T_c}\rp^3\right] \:,
\end{equation*}
as shown by the dashed curve in the figure. Here $\mu(T=0)$ is the chemical
potential at zero temperature, and $\gamma$ an adjustable parameter, for which
we found the best value $\gamma=4.28\times10^{-4}$ 
($\chi^2_{\rm fit}=3.06\times10^{-10}$).
In an equivalent plot, the inset to the right figure shows the variation of
$1-k_\mu/k_F$ as a function of temperature.

\subsection{Asymmetric case}
\label{num-asym}
\begin{figure}[!bp]
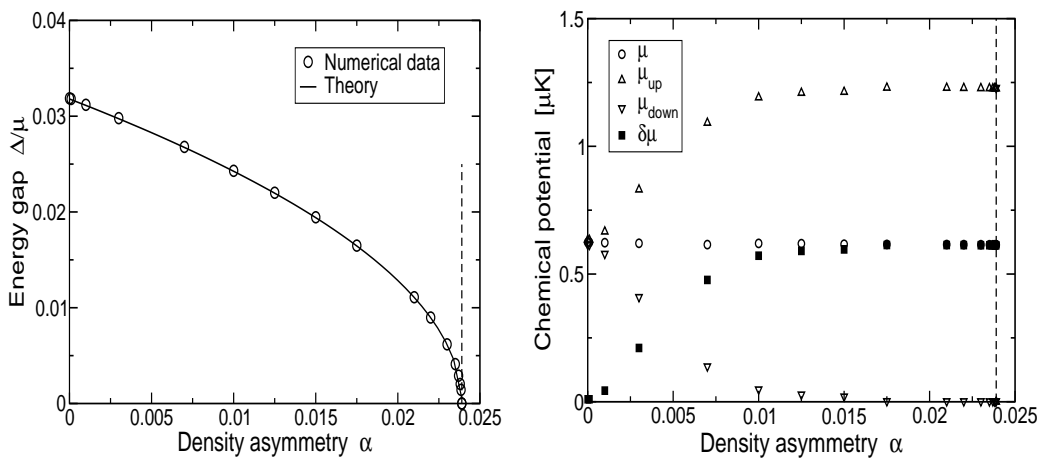

  \vspace{-0.9cm}\phantom{.}
  \caption[Numerical solution for the asymmetric case varying $\al$]
	  {\label{fig:num_asym}
           \index{gap!as a function of density asymmetry}
	    Solution of the \BCS\ equations for a fixed density
	    $\rho=2\times10^{-12}$ cm$^{-3}$ at zero temperature as a function
	    of the density asymmetry $\al$. (Left) Energy gap; (Right)
	    Chemical potentials: total $\mu$ (circles), $\muu$ (triangles up),
	    $\mud$ (triangles down) and difference $\dmu$ (filled squares).
	    \vspace{10pt}\phantom{.}
	  }
  \begin{minipage}{0.48\textwidth}
    \begin{center}
      \includegraphics[width=\textwidth,height=5.95cm,clip=true]
		      {pairing/prog/gap/explo3-gap_alfa}
    \end{center}
  \end{minipage}%
  \hfill
  \begin{minipage}{0.49\textwidth}
    \begin{center}
      \includegraphics[width=\textwidth,height=6cm,clip=true]
		      {pairing/prog/gap/explo3-mu_alfa}
    \end{center}
  \end{minipage}
\end{figure}

After this short overview of the symmetric case, we will study the dependence
of the gap on the density asymmetry $\al=\de\rho/\rho$ in order to check
numerically Eq.~(\ref{eq:sgap}). We consider $T=0$ and again the density
$\rho=2\times10^{12}$ cm$^{-3}$ ($k_F|a|\sim0.4$) as we have seen in the
previous paragraph that it is well into the \BCS\ validity regime. 
Our results are summarized in Fig.~\ref{fig:num_asym}.
On the left panel, we show how the energy gap $\De$ varies with increasing
asymmetry $\al$. The circles are the calculated data, while the line
corresponds to Eq.~(\ref{eq:sgap}), with the $\De_{\rm sym}$ given by
Eq.~(\ref{eq:gap_sym}); the dashed vertical line indicates the expected
\index{density asymmetry!critical}
asymmetry $\al_{\rm max}^{BCS}=2.4\%$ where the gap will vanish. The agreement
between the calculation and the expected behavior is excellent for all
asymmetries, even very close to this limiting asymmetry.

On the right panel, we show, for comparison with the symmetric case, the
dependence of the mean chemical potential $\mu$ together with the chemical
potentials $\mu_{\ua\da}$ and $\dmu$. We see that $\mu$ depends only very
weakly on $\al$, as compared to what happens with $\muu$ and $\mud$. These
start being equal in the symmetric case, and separate in an approximately
linear way for increasing $\al$ until $\al\approx0.01$, where they
`saturate' to the values that they will have for 
$\al=\al_{\rm max}^{BCS}$. As one would expect, an analogous behavior is
followed by $\dmu$.

As a conclusion, we have checked numerically the expected behavior of the gap
for the $s$-wave solution to the coupled equations determining the gap
and the densities. In particular, we have seen that the mean chemical
potential is essentially equal to the non-interacting value $\ef$ and that the
gap vanishes for very small asymmetries. 
At finite temperatures, as one would expect, we have checked also that the gap
decreases, and so the maximum asymmetry that would allow pairing is even
smaller.

\section{Pairing in \mbox{$p$}-wave}
\label{sec:pwave}

Therefore, for larger asymmetries only pairing between identical fermions can
take place. In this case, Pauli's exclusion principle demands the paired state
to be antisymmetric under exchange of the particles, which for spin-polarized
atoms means that they must be in a state of odd angular momentum $L$. The
leading contribution to the interaction will be the $p$-wave one.

We discuss in the following the pairing of $\ua$-particles mediated by the
polarization interaction due to particles pertaining to species $\da$ as shown
in Fig.~\ref{fig:lowestorder}(b).
We will check below that this contribution is dominant over the direct
$p$-wave $\ua$-$\ua$ interaction at low density.
Quantitatively the relevant interaction kernel reads 
\cite{gmb,pet,spp,kag88,bara} 
\be 
  \lla \kvp
  \left| \Gamma_\ua \right| \kv \rra = \frac{\Pi_\da(|\kv'-\kv|)}{2} \T_0^2
  \:.
  \label{eq:wtot}
\ee 
The factor 1/2 corrects for the fact that
conventionally the Lindhard function (pertaining to species $\da$),
\be 
  \Pi_\da(q) = -\frac{m k_\da}{\pi^2} 
  \left[\inv{2} + \frac{1-x^2}{4x}\ln\left|\frac{1+x}{1-x}\right| \right] 
  \ ,\quad x = \frac{q}{2\kd} \:,
  \label{eq:pi}
\ee 
contains a factor two for the spin orientations~\cite{fetter}, which is not
present in our case as we are dealing with fermions without internal degrees
of freedom.

One can see that $\Pi_{\da}<0$. Therefore, the effective 
interaction is {\em always attractive}, irrespective of the sign of
$a$~\cite{fay}. This is due to the absence of exchange diagrams, which makes 
$\Pi_\da$ attractive in contrast to the case of one species, where the
polarization effects reduce the $s$-wave \bcs\ gap by a factor 
$(4e)^{1/3} \approx 2.2$ even in the low-density limit, where one would
naively expect them to be negligible.
\cite{gmb,pet,spp}. 
One can therefore assert that, even in the limit
$k_F\raw0$ \bcs\ is not such a good aproximation, as it disregards these
higher-order interactions.

\begin{figure}[!b]
  \caption[Variation of the gap with density asymmetry]
	  {\label{fig:gap}
           \index{gap!as a function of density asymmetry}
	    (a,b) The functions $g$ and $h$, appearing in Eqs.~(\ref{eq:g})
	    and (\ref{eq:f}), respectively.
	    (c) The variation of the gap with asymmetry, Eq.~(\ref{eq:asy}),
	    for a density parameter $u=100$.
	    \vspace{10pt}\phantom{.}
	  }
  \centerline{\includegraphics[width=0.95\textwidth,clip=true]
		  {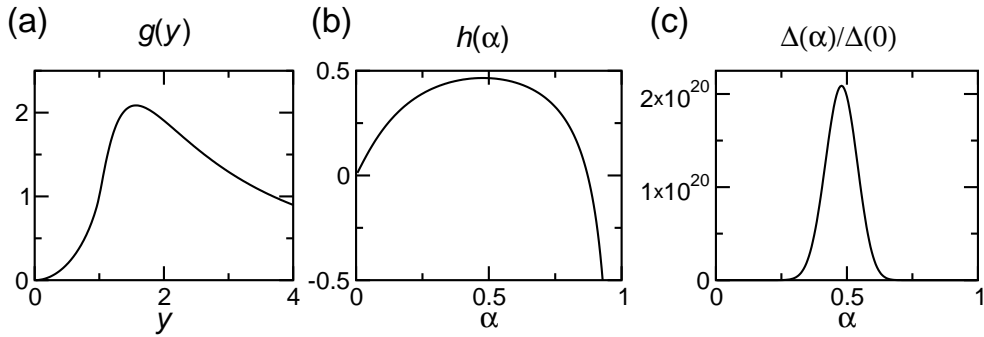} 
             }
\end{figure}

Now we project out the $p$-wave (\ie, $L=1$) contribution of the interaction
with the Legendre polynomial $P_1(z)=z$, with $z=\bm{\widehat{k}}' \cdot
\bm{\widehat{k}}$ the cosine of the angle between the in- and out-going
relative momenta, which in the range of the 
\index{weak-coupling limit}weak-coupling approximation we can
take on the Fermi surface~\cite{kag89,efre}
\begin{equation*} 
  \Gamma^{(L=1)}_\ua(k_\ua,k_\ua) =
  \frac{\T_0^2}{2}\, \inv{2}\!\int_{-1}^{+1} \!dz\, z\,
  \Pi_\da(\sqrt{2(1-z)}k_\ua) \:.
\end{equation*}
The integral of the Lindhard function can be evaluated making use of the
result~\cite{spp}
\begin{equation*}
  \int_0^{Z} \mathrm{d}x\,x^3\Pi(x) = 
  \inv{12}\left[
    2\ln|1-Z^2|+(5-3Z^2)Z^3\ln\left|\frac{1+Z}{1-Z}\right|+2Z^2+6Z^4
    \right]
\end{equation*}
resulting in
\begin{equation*}
  \Gamma^{(1)}_\ua(k_\ua,k_\ua)  
  = -\frac{8 a^2 \kd}{m} \frac{2\ln 2 -1}{5} g\Big(\frac{\ku}{\kd}\Big)\:,
  \label{eq:linkk}
\end{equation*}
where we introduced
\begin{multline}
  g(y) = -\inv{6(2\ln 2 -1) y^4} 
  \Big[ (4-10y^2) \ln\left| 1-y^2 \right| 
    \\
    - (5+y^2)y^3 \ln\left|\frac{1+y}{1-y}\right|
    + 4y^2+2y^4 \Big] \:.
  \label{eq:g}
\end{multline}
This function is normalized in symmetric matter, $g(1)=1$, and
it is plotted in Fig.~\ref{fig:gap}(a). 
It is customary to replace the numerical factor $(2\ln 2 -1)/5$
in Eq.~(\ref{eq:linkk}) by its approximate value 1/13.  

We use now the general result for the (angle-averaged) $L$-wave
pairing gap~\cite{fay,ander,kohn}, 
\be 
  \De_L(\ku) \raw c_L\, \muu
  \exp\!\left[ \frac{2\pi^2\hbar^2}{m \ku \T_L(\ku,\ku)} \right] \:, 
  \label{eq:aver-gap}
\ee 
with $\T_L$ the $L$-wave projection of the relevant interaction and $c_L$ a
constant of order unity.
Considering that for a pure polarization interaction $\Gamma_L$ the
leading order in density is $\T_L = \Gamma_L$, we get
\be
  \De_1(\ku) = c_1 \frac{\ku^2}{2m} 
  \exp\!\left[ -\frac{13(\pi/2)^2}{a^2 \ku \kd g(\ku/\kd)} \right] \:.
  \label{eq:dtwo}
\ee 
Taking into account the dependence of this expression on the two
Fermi momenta $k_{\ua\da}=[3\pi^2\rho(1\pm\al)]^{1/3}$, the final result for
the variation of the pairing gap with asymmetry $\al$ and total density
$\rho$ can be cast in the form
\be 
   \frac{\De_1(\rho,\al)}{\De_1(\rho,0)} = (1+\al)^{2/3}
   \exp\left[ u(\rho) h(\al) \right]
   \label{eq:asy}
\ee 
with 
\be 
  h(\al) = 1 -\inv{(1-\al^2)^{1/3} g\left[((1+\al)/(1-\al))^{1/3}\right] }
  \label{eq:f}
\ee 
and the density parameter 
\begin{equation*} 
  u(\rho) = 13 \left(\frac{\pi}{2{k_F} a} \right)^2 , 
    \quad {k}_F \equiv (3\pi^2\rho)^{1/3} \:. 
\end{equation*}
The function $h(\al)$ is displayed in Fig.~\ref{fig:gap}(b), where $\al=1$
corresponds to pure $\ua$-matter.
One notes a maximum at $\al_\mathrm{opt} \approx 0.478$
with an expansion $h(\al) \approx 0.465-1.343(\al-0.478)^2$.

At the optimal asymmetry, the gap is enhanced with
respect to that of the symmetric case [$h(0)=0$] by a factor 
$e^{0.465u} \approx 10^{0.20u}$.
In the low-density limit $u\raw\infty$ this represents an enormous
amplification at finite asymmetry.  
Around this peak, the variation of the gap with asymmetry is well described by
a Gaussian with width $\sigma=1/(1.64\sqrt{u})$.  
As an illustration of this effect, Fig.~\ref{fig:gap}(c) shows the ratio
(\ref{eq:asy}) for a value of the density parameter $u = 100$ (corresponding
to $k_F|a|\approx0.57$). 
As expected from the previous discussion, a peak of the order of $10^{20}$ is
observed, that becomes rapidly more pronounced and narrower with decreasing
density $\rho$ (increasing $u$), although of course at the same time the
absolute magnitude of the gap decreases strongly with decreasing density:
$\De_1(\R,0)\propto\exp(-u)$ [Eq.~(\ref{eq:dtwo})].

\begin{figure}
  \caption[Third order diagrams in a one-component system]
	  {\label{fig:higherorder}
	    Third-order diagrams in a one-component system: 
	    (a) Direct $p$-wave interaction, 
	    (b) Polarization contribution.
	  }
  \begin{center}
    \includegraphics[width=0.6\textwidth]
		    {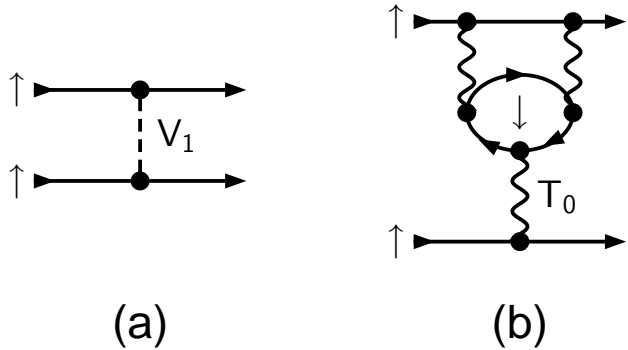}
  \end{center}
\end{figure}

Let us now briefly discuss higher-order effects, namely
contributions of order $(k_F a)^3$ in the denominator of
the exponent of Eq.~(\ref{eq:dtwo}). 
There are two principal sources of such effects, which are shown
diagramatically in Fig.~\ref{fig:higherorder}.
The first one, Fig.~\ref{fig:higherorder}(a), is the direct $p$-wave
interaction~\cite{you} between like species that we have neglected before.
\index{T-matrix!at low momenta!in $p$-wave}
\index{T-matrix!and scattering volume}
Parametrizing the low-density $p$-wave $\T$-matrix in the standard form
$\T_1(k,k') \approx 4\pi a_1^3 k k'/ m$, where $a_1^3$ is the $p$-wave
scattering volume, leads to a \bcs\ gap \cite{you} 
\be 
  \frac{\De_1}{\muu} =
  \frac{8}{e^{8/3}} \exp\!\left[ \frac{\pi}{2(\ku a_1)^3}\right] \:.
\ee
where we have explicitly determined the prefactor of the exponential.
This gap is much smaller than the $\da$-mediated one for
$\{|a|,|a_1|\}\ll\ku^{-1}$, which justifies the fact that we neglect this
effect in the previous calculation.
However, around a $p$-wave Feshbach resonance, where $|a_1|\gg|a|$, this
effect might be important, even though this kind of pairing has not been
observed in the experiments that explored the $p$-wave resonance of
\K\ in the hyperfine state $|F=9/2,m_F=-/2\ra$~\cite{cindy03} or \Lix\ in
$|F=1/2\ra$~\cite{salomon-p} (see also \cite{bohn} for theoretical
calculations of the position of the resonances, and \cite{chin00,weber}
for experiments on bosonic cesium).

The second type of third-order contributions are polarization effects
involving the $s$-wave scattering length $a$.  In contrast to the case
of a one-component system, where there are several relevant diagrams,
in the present two-com\-po\-nent system only one diagram exists,
Fig.~\ref{fig:higherorder}(b). Unfortunately it can only be computed
numerically, which we have not attempted. 

Finally, to fourth order, there is a large number of diagrams
contributing to the interaction kernel.
Moreover, at this order it is also necessary to take into
account retardation effects, \ie, the energy dependencies of gap
equation, interaction kernel, and self-energy need to be
considered~\cite{alex}.
All this can only be done numerically and was performed in
Ref.~\cite{efre} for the case of a one-component system.

In any case, the existence of higher-order corrections will not alter
the main conclusions drawn so far, namely the presence of a strongly
peaked Gaussian variation of the gap with asymmetry. They may, however, shift
this peak to a different density-dependent location, and also modify the
absolute size of the gap. Furthermore, we must note again that the
perturbative approach that we have followed is clearly limited to the
low-density range $k_F |a| < 1$.  For larger couplings, different theoretical
methods are required, which is an interesting field of current investigation
\cite{bb,ainsworth,holland01,griffin,bulgac04,strinati,falco,levin-physrep}.

\section{Summary}
In this chapter, we have studied the possibility of pairing in
a system composed of two distinct fermionic species, assuming that the direct
$p$-wave interaction between like species can be neglected.
First, we have studied analytically the gap produced by a direct,
attractive, $s$-wave interaction between different species. We have shown
that it produces a gap $\sim \exp\!\left[{\pi/ 2k_F a}\right]$
[Eq.~(\ref{eq:gap_sym})] only for very small asymmetries between the densities
of the two species, $\al\lesssim\De_{\rm sym}/\mu$, Eq.~(\ref{eq:sgap}).
We have numerically checked this behavior, and we have seen that, in fact,
the maximum asymmetry decreases with temperature.

However, a $p$-wave attraction between two fermions of the same
species produced by a polarization of the medium of the other species, 
can give rise to pairing over the whole range of asymmetry.
In practice a sharp Gaussian maximum around $\al\approx0.478$
($\ru/\rd \approx 2.83$, $\ku/\kd\approx$ 1.41) appears.
Unfortunately, the absolute magnitude of this gap is rather small, as it
depends strongly on the density of the system, 
$\sim \exp\!\left[-13({\pi/ 2k_F a})^2\right]$, 
which we assumed low in order to be in the range of validity of the \BCS\
mean-field approach. 

We argue that higher-order corrections may modify
quantitatively but not qualitatively these general features.


The experimental observation of both types of pairing in dilute, atomic vapors
is expected to be difficult.
For the case of $p$-wave pairing due to the extremely small size of the
expected gap, which translates into a very low transition temperature
[remember Eq.~(\ref{eq:Tc})], which is difficult to reach experimentally (see
Sect.~\ref{sec:asym-intro}).
For the $s$-wave case, the nearly perfect symmetry that is
required poses a difficult experimental challenge. However, the present
possibility to transform a two-component Fermi gas into a molecular
gas by crossing a Feshbach resonance~\cite{cindy04,grimm04} opens the door to
create a perfectly symmetric system by removing the atoms not converted into
molecules and, then, going back across the resonance, forming again a
two-component fermionic system, now with well-balanced populations.
Up to now, the experiments have focused on the strongly-interacting regime
$k_F|a|\gtrsim 1$ where superfluidity might arise at higher temperatures. Once
this has been succesfully achieved~\cite{kett-nat-vort}, one could imagine to
go further to the \index{weak-coupling limit}weak-coupling regime and 
transfer atoms in a controled way from one hyperfine state to the other 
by RF pulses, which would allow a precise experimental check of 
Eq.~(\ref{eq:sgap}) in case low enough temperatures were achieved.

\chapter
        {Pairing with broken space symmetries: \loff\ {\em vs.} \dfs}
\label{ch:dfs}

\textsf{
  \includegraphics[totalheight=4cm]{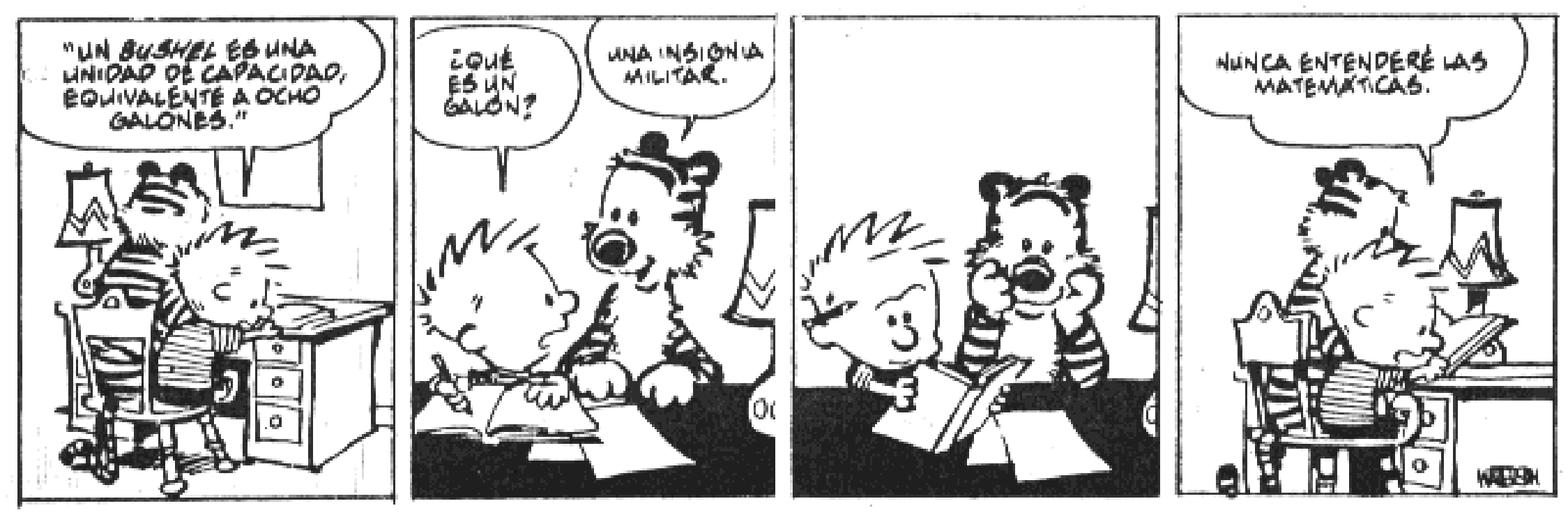}
  \begin{flushright}
    Bill Waterson, {\em Calvin and Hobbes}
  \end{flushright}
}

\section{Introduction}
\label{sec:dfs-intro}

In chapter~\ref{ch:pair-intro} we introduced the \bcs\ theory for pairing in a
fermionic system. 
We showed that this theory predicts a phase transition into a superfluid state
for temperatures of the order of $\De_{\rm sym}/k_B$, where in the
\index{weak-coupling limit}weak-coupling regime $k_F|a|\ll 1$ 
[see Eq.~(\ref{eq:gap_sym})]
\begin{equation}
  \frac{\De_{\rm sym}}{\mu} = \frac{8}{e^2}\exp\lp\frac{\pi}{2k_Fa}\rp \:.
\end{equation}
Stoof and coworkers were the first to point out that this result, when applied
to dilute systems of \Lix, predicts relatively large gaps for moderate
densities (\eg, $\rho\sim10^{12}$ cm$^{-3}$) due to the large (negative) value
of the $s$-wave scattering length of this species, $a=-2160a_B$
($a_B\approx0.53$ \AA\ is Bohr's radius). 
These large
gaps would translate in  transition temperatures of the order of nanokelvin
and, therefore, possibly within experimental reach in a short time after their
proposal in 1996~\cite{stoof96}. Regarding the problem on the density
asymmetry between the two species to pair, they also pointed out `that the
most favorable condition for the formation of Cooper pairs is that both
densities are equal'.
Then, they solved the gap equation in this most favorable situation, for a
trapped system of \Lix\ atoms using local density approximation
(\textsc{lda})~\cite{houbiers}.

In chapter~\ref{ch:asym} we have carefully analyzed why a two-component
symmetric system will always have $s$-wave gaps larger than asymmetric
systems. Besides, we have also shown that these $s$-wave gaps are
non-vanishing, within standard \bcs\ theory, for quite a narrow window of
density asymmetries $\al$. 
For $\al>\al_{\rm max}^{BCS}=3\De_{\rm sym}/(4\mu)$ [cf. Eq.~(\ref{eq:sgap})],
only $p$-wave pairing between atoms in the same hyperfine state seems
possible. However, we have seen that the size of this gap is 
rather small, and in fact it is difficult to expect this phase to be detected
experimentally. Now we will explore another way of overcoming the
density-asymmetry problem. To this end, we solve the gap equation on a wider
space of functions, allowing for more complex structures than the typical
rotationally-symmetric order parameter found in chapters \ref{ch:pair-intro}
and \ref{ch:asym}.

In references~\cite{sedr-prl,sedr-prc} it was shown that for
nuclear matter at saturation density 
---which is a strongly coupled system ($\De_0/\mu \sim 0.3$)---
a superconducting state featuring elliptically 
deformed Fermi surfaces (\dfs)
was preferable to the spherically symmetric \bcs\ state.
Following these ideas, we propose here to explore the weak-coupling regime of
this theory in connection with current experiments with ultracold, dilute
systems.

It has been known for a long time that the homogeneous 
\bcs\ phase can evolve into the Larkin-Ovchinnikov-Fulde-Ferrell 
(\loff) phase~\cite{lo,ff}, which can sustain asymmetries 
$\al > \al_{\rm max}^{BCS}$ by allowing the Cooper pairs to carry a
non-vanishing center-of-mass momentum. 
Note that, even though both the \loff\ and the \dfs\ phases break
the global space symmetries, they do it in fundamentally different ways: the
\loff\ phase breaks both rotational and translational symmetries due to 
the finite momentum of the pairs' condensate.
On the contrary, the \dfs\ phase breaks 
only the rotational symmetry [from O(3) down to O(2)].
We remark also that, after so many years since its proposal, only very
recently some experimental evidences of the detection of the \loff\ phase
have been reported in the heavy-fermion compound CeCoIn$_5$
(see~\cite{radovan03,bianchi03,watanabe04,kaku05,martin05}).
In this sense, it is interesting to note that atomic systems offer a novel
setting for studying the \loff\ phase under conditions that are more
favorable than those in solids (absence of lattice deffects, access to the
momentum distribution in the system through time-of-flight experiments) as
explored in~\cite{COMBESCOT,COMBESCOT_MORA1,COMBESCOT_MORA2,mizushima,yang05}.
These systems also offer the possibility of novel realizations of 
the \loff\ phase which for example invoke $p$-wave anisotropic 
interactions~\cite{COMBESCOT}.
There has been also much interest in the \loff\ phase in connection with
hadronic systems under extreme conditions where the interactions are mediated
by the strong force, see~\cite{alford,sedr01,bowers,casal}. In this context no
experimental detection of this phase has been reported yet.

We note that another possible configuration for a density-asymmetric system 
is the phase separation of the superconducting and normal phases in real space, 
such that the superconducting phase contains particles with the same chemical 
potentials (\ie, it is symmetric), while the normal phase remains asymmetric, 
see~\cite{Bedaque,BCR,CALDAS}. 
The description of such an heterogeneous phase 
requires knowledge of the poorly known surface tension between superconducting
and normal phases, and will not be attempted here.

We shall compare below the realizations of the \dfs\ and \loff\ phases in an
ultracold gas of $^6$Li atoms where the hyperfine 
states $|\!\!\ua\rangle=|F=3/2,m_F=3/2\rangle$ and 
$|\!\!\da\rangle=|3/2,1/2\rangle$ are simultaneously trapped. 
Here, $F$ denotes the total angular momentum of the atoms in units of
$\hbar$, 
while $m_F$ is its projection onto the $z$ axis of some reference frame.
Fermionic systems where two hyperfine levels are populated have been created
and studied experimentally with $^6$Li and $^{40}$K atoms 
(see \cite{jin99,mod02,thomas02,streck03,kett03,salomon03}, to quote a few
examples).
The mixture of different hyperfine components allows one to overcome the
problem of cooling fermions set by Pauli's exclusion principle, as indicated
in Sect.~\ref{sec:asym-intro}.

These systems are characterized by a hierarchy of length scales. The largest
scale is usually set by the harmonic trapping potential. As it is much larger
than any other scale in the system, we will neglect the finite-size effects 
for the moment, and assume an homogeneous system for our analysis. 
A step further would be to perform an \textsc{lda} calculation, 
in a way similar to that used in~\cite{houbiers} for the standard 
\bcs\ treatment or~\cite{chiof02} for pairing in a resonantly interacting 
Fermi gas.

The typical range of the interatomic, van der Waals forces is 
$R\le 10^{-6}$~cm while the de Broglie wavenumber of particles 
at the top of the Fermi sea is $k_F \sim 10^3-10^4$ cm$^{-1}$. 
Therefore $k_FR\ll 1$, and the interaction can be approximated 
by a zero-range force characterized by the $s$-wave scattering length $a$.
For the particular case of 
collisions between \Lix\ atoms in the above-mentioned states, 
$a=-2160a_B$, and we obtain $k_F|a|\simeq 0.04$. Therefore the system is in
the \index{weak-coupling limit}weak-coupling regime, 
since $\nu(k_F) g = 2 k_F|a|/\pi \ll 1$, where 
$\nu(k_F) = mk_F/(2\pi^2\hbar^2)$ is the density of states at the
Fermi surface of the non-interacting system, and $g = 4\pi\hbar^2 a/m$ is a
measure of the strength of the contact interaction. For larger values 
$k_F|a| \gtrsim 1$, 
the bound states need to be incorporated in the theory 
on the same footing that the pair correlations, 
and the formal treatment becomes more delicate.

To study the \loff\ phase, we will assume that its order parameter has a simple
plane-wave form. Even though more complex structures for the \loff\ order
parameter can be studied, we believe this would not change qualitatively our
results.
Furthermore, we will show that, allowing the Fermi surfaces of the species to
deform into ellipsoids, the range of asymmetries over which pairing is
possible is enlarged with respect to the predictions of the standard \bcs\ or
\loff\ theories. Also, the gap in asymmetric systems where pairing was still
possible within those frameworks, will be shown to be larger in the \dfs\
phase.

Finally, at the end of the chapter, we shall describe an experimental
signature of the \dfs\ phase that can be established in time-of-flight
experiments and that would allow one to distinguish the \dfs\ phase from the
competing phases.


\section{Breaking the symmetry: \loff\ and \dfs}
\label{sec:finiteP}

\subsection{Description of the \loff\ state}
While the \bcs\ ground state assumes that the fermions bound in a Cooper pair
have equal and opposite momenta (and spins),  
for fermionic systems with unequal numbers
of spin up and down particles this is not always true. In this situation, 
Larkin and Ovchinnikov~\cite{lo} and independently Fulde and Ferrell~\cite{ff} 
noted that the pairing is possible amongst pairs which have finite total 
momentum with respect to some fixed reference frame. 
The finite momentum $\bP$ changes the quasiparticle spectrum of the paired state. 
To see this, we can write down the 
\index{propagator!normal}normal propagator in that reference frame:
\begin{align}
\label{NORMAL}
  G^{N}_{\ua\da}(\kv,\bP;\om_n)
  =\left[ (\om_n +i\eta)
              -\hm\left(\frac{\bP}{2}\pm \kv\right)^2 \right]^{-1} \:.
\end{align}
Now, using Eq.~(\ref{eq:Esa}), the symmetric and anti-symmetric
parts of the 
\index{spectrum!in \loff\ phase}quasiparticle spectrum read
\begin{subequations}
  \begin{align} 
    E_S &= \hbar^2 \frac{P^2+4k^2}{8m} - \frac{\muu+\mud}{2} ,\\
    E_A &= \hbar^2 \frac{\bP\cdot\kv}{2m} - \frac{\muu-\mud}{2} .
  \end{align}
\label{eq:espectre-dfs}
\end{subequations}
Fortunately, the results of the previous chapters remain valid with the above 
redefinitions of $E_S$ and $E_A$. Note that the quantities of interest, in
particular the gap, now depend parametrically on the total
momentum. 
Interestingly, 
$E_A$ in (\ref{eq:espectre-dfs}) does not vanish in the limit of 
equal number of
spin-up and down particles (\ie, when $\mu_{\ua}=\mu_{\da}$). In this case,
the \loff\ state (a condensate of pairs all with momentum $\bP$) lowers the
energy of the system with respect to the normal (unpaired) state. 
Nevertheless, it is not the real ground state of the symmetric system, as it is
unstable with respect to the ordinary \bcs\ ground state. In fact, it is well
know that, for the symmetric system, the most favorable configuration for
pairing is for $\bP=0$.

\subsection{Description of the \dfs\ state}
We now turn to the deformations of the Fermi surfaces.
The two Fermi surfaces for spin-up and -down particles are defined in momentum
space for the non-interacting system by the fact that the energy of a
quasiparticle vanishes on them:
$$\xi_{\kv\ua/\da} = \epsilon_{\bP/2\pm\kv} -\mu_{\ua/\da} = 0 \:.$$
When the
states are filled isotropically within a sphere, the chemical potentials 
are related to the Fermi momenta $k_{s}$ ($s$=$\ua,\da$) as 
$\mu_{s} = \hbar^2k_{s}^2/(2m)$
(for simplicity we assume here that the temperature is zero).
To describe the deformations of the Fermi surfaces from their spherical 
shape we expand the 
\index{spectrum!expansion in spherical harmonics}
quasiparticle spectra in spherical harmonics
$\xi_{\kv s} = \sum_l \xi_{s}^{(l)} P_l(\chi)$, where $\chi$
is the cosine of the angle formed by the quasiparticle momentum 
and a fixed symmetry-breaking axis; 
$P_l(\chi)$ are the Legendre polynomials. 
The $l = 1$ term breaks translational symmetry by shifting the Fermi
surfaces without deforming them; this term corresponds to the \loff\ phase
and is already included by using $\eps_{\bP/2\pm\kv}$. 
Truncating the expansion at the second order ($l = 2$), we rewrite the
\index{spectrum!in \dfs\ phase}spectrum in the form
\cite{sedr-prl,sedr-prc}
\begin{equation}
  \xi_{\kv\ua/\da} = \epsilon_{\bP/2\pm\kv}-
  \mu_{\ua/\da}\left(1+\eta_{\ua/\da} \, \chi^2\right) , \notag
\end{equation}
where the parameters $\eta_{s}=\xi_{s}^{(2)}/\mu_s$ describe a
quadrupolar deformation of the Fermi surfaces. It is convenient to work with
the symmetrized $\Xi = (\eta_{\ua}+\eta_{\da})/2$ and
anti-symmetrized $\de\epsilon  = (\eta_{\ua}-\eta_{\da})/2$ 
combinations of $\eta_{\ua,\da}$. 
For simplicity, below we shall assume $\Xi = 0$,
and consider only two limiting  cases:  
$\de\epsilon\neq 0$ and  $\bP= 0$ (the \dfs\ phase) and
$\de\epsilon=0$  and $\bP\neq  0$ (the plane-wave \loff\ phase) 
[see Table~\ref{table:dfs} for a summary of the nomenclature used].
\begin{table}[!b]
  \caption{\label{table:dfs}
           The various candidates for a superfluid ground state studied.} 
  \begin{center}
    \begin{tabular}{|l|ccc|}
    \hline 
    Candidate phase & $\al$   &  $P$    & $\deps$ \\ \hline \hline
    Symmetric  \bcs &   0     &   0     &   0     \\
    Asymmetric \bcs & $\neq0$ &   0     &   0     \\
    \loff           & $\neq0$ & $\neq0$ &   0     \\
    \dfs            & $\neq0$ &   0     & $\neq0$ \\ \hline 
    \end{tabular}
  \end{center}
\end{table}
Thus, in the \dfs\ phase we have
\begin{equation} 
  \xi_{\kv\ua/\da} = \ek-\mu_{\ua/\da}\lp1\pm\deps\,\chi^2\rp \;.
  \label{eq:new-eps}
\end{equation}
Clearly, this expression vanishes at $k=k_s$ for $\chi=0$,
\ie, in the $xy$ plane in $k$-space. On the other hand, assuming $\deps>0$,
the $\ua$-Fermi sphere becomes elongated along the $z$ axis ($\xi_{\kv\ua}$
vanishes at $k>\ku$), while the $\da$-sphere is squeezed. 
The conservation of the densities $\ru$ and $\rd$ requires a recalculation of
the chemical potentials, which is done by integrating again the corresponding
momentum distributions.
The net effect is that the surfaces approach each other on the $xy$ plane%
, see Sect.~\ref{ssec:dfs-estimate} and Fig.~\ref{fig:dfs5}.

\section{The gap in the \bcs, \loff\ and \dfs\ phases}
\label{sec:dfs-gap}

Consider a trap loaded with $^6$Li atoms and assume that the net number 
of atoms in the trap is fixed while the system is maintained at constant 
temperature. Assume further that the number of atoms corresponds to a 
Fermi energy $\ef/k_B \equiv T_F=942$ nK, which in the uniforme and symmetric
case at $T=0$ would translate into a 
density of the system $\rho=3.8\times10^{12}$ cm$^{-3}$
and a Fermi momentum $k_F\approx  4.83\times10^4$~cm$^{-1}$. 
All the results below have been calculated for a {\em homogeneous} system at
this density and at a constant temperature $T=10$ nK $\ll T_F$, so the system
is in the strongly degenerate regime.
In the conditions of~\cite{hadzi}, this Fermi energy corresponds to about
$4\times10^5$ atoms in a single hyperfine component of \Lix\ [see
also~\cite{hadzi-tesi}, especially chapter 5].
Present experiments can control the partial densities in the two
different hyperfine states $|\!\!\ua\rangle=|3/2, 3/2\rangle$ and 
$|\!\!\da\rangle=|3/2,~1/2\rangle$ of \Lix\ by trans\-fer\-ring atoms
from one to the other using $\sim$76 MHz RF pulses~\cite{hadzi-tesi}.
Since the free-space triplet scattering length between $^6$Li atoms 
in these hyperfine states is $a =-2160a_B$, the
system is in the weakly coupled regime  $k_F|a| \ll 1$. 

\begin{figure}[!t]
    \caption[Dependence of \loff\ and \dfs\ gaps on asymmetry]
            {\label{fig:dfs1}
	    Dependence of the pairing gaps in the \loff\ phase (upper
	    panel) and the \dfs\ phase (lower panel) on the density-asymmetry
	    parameter for the values of the the pair momentum $P/k_F$ 
	    and deformation parameter $\de\epsilon$ indicated in
	    the legends. The Fermi momentum is $k_F=4.83\times10^{4}$~cm$^{-1}$
	    and the scattering length is $a=-2160a_B$, so that $k_F|a|=0.55$.
	    \vspace{10pt}\phantom{.}
	    }
    \includegraphics[width=0.9\textwidth,clip=true]
                    {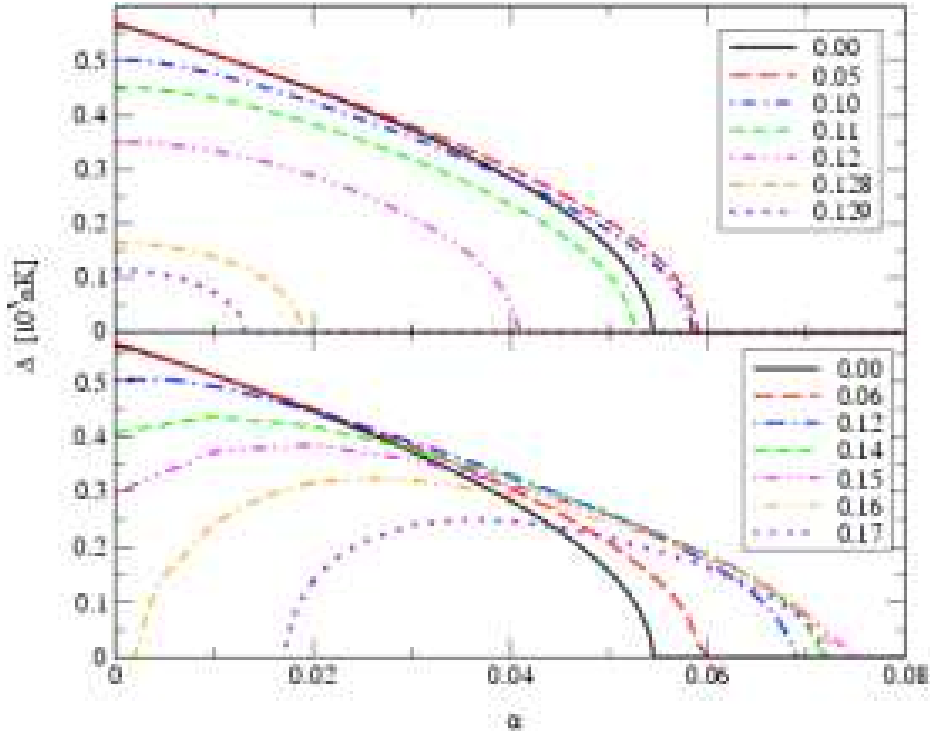}
\end{figure}

Without loss of generality, we assume a non-negative density asymmetry, \ie,
$\ru\geq \rd$. 
We show only results for $\de\epsilon \ge 0$ since we have checked that
it is the one that gives the lowest free energy. 
We remind that this $\deps>0$ corresponds to a cigar-like deformation of the
$\ua$ Fermi surface and a pancake-like deformation of the $\da$ Fermi surface,
as we will see explicitely in Fig.~\ref{fig:dfs5}.
The pairing gaps of the \loff\ and \dfs\ phases computed
from the (coupled) gap and number equations [(\ref{eq:gap-eq}) and
(\ref{eq:rhos})] are shown in Fig.~\ref{fig:dfs1} as a function of the
density-asymmetry parameter 
$\al = (\rho_{\ua}-\rho_{\da})/(\rho_{\ua}+\rho_{\da})$  for 
different values of the corresponding parameter signaling the symmetry
breaking: total momentum $P$ for the \loff\ phase (top panel) and deformation
$\deps$ for the \dfs\ phase (lower panel).
As expected, for vanishing asymmetries $\al\raw0$, the maximal
gap is attained by the standard \bcs\ ground state, which is indicated by the
solid, black line in both panels. The symmetry-breaking states that are very
different from it (\eg, a \loff\ state with $P/k_F\gtrsim0.1$ or a \dfs\
state with $\al\gtrsim0.1$) have notably smaller gaps.

The situation changes for increasing asymmetry. For example, the
\loff\ state with $P/k_F=0.05$ presents a larger gap than the asymmetric \bcs\
state for $\al>0.03$, a density asymmetry for which also the \dfs\ state with
$\deps=0.12$ has a gap larger than the \bcs\ one. Finally, for
$\al\geq\alm^{BCS}\approx0.055$, the \bcs\ state no longer presents a finite
gap, while both the \loff\ and \dfs\ phases `survive' up to higher
asymmetries: 
$\alm^{LOFF}\approx0.06$ (for $P/k_F=0.05-0.10$) and 
$\alm^{DFS}\approx0.075$ (for $\deps\approx0.15$), respectively.
It is also remarkable the presence of a `reentrance' phenomenon in the \dfs\
phase. For instance, consider the case $\deps=0.17$ (blue, dotted curve).
Such a large deformation shows a finite gap only for asymmetric
systems, and a lower $\al_{\rm cr\,1}\approx0.02$ and higher 
$\al_{\rm cr\,2}\approx0.07$ critical asymmetries can be identified. As the
quantity that determines the true ground state of the system is the free
energy and not the size of the gap, we cannot tell from Fig.~\ref{fig:dfs1}
what will be the structure of the ground state. However, before discussing
this point in detail, we shall study some aspects of the excitation spectrum 
of the system.

\section{Excitation spectra in the superfluid phases}
\label{sec:dfs-spectra}
\index{spectrum!in superfluid phases}

To elucidate the dominance of the phases with broken space symmetries
over the asymmetric \bcs\ state, it is useful to consider the modifications
implied by these phases to the quasiparticle spectra
\begin{align} 
  \Ek^{\pm} 
            &= E_A  \pm \sqrt{E_S^2 + |\De|^2} \:.
  \label{eq:qp-spectra}
\end{align}
These energies correspond to the poles of the 
\index{propagator!normal}propagators $G_{\ua\da}$
of chapter~\ref{ch:pair-intro}. Physically, $\Ek^+$ is the excitation energy
of the system when we move it from its ground state to a state with momentum
$\kv$ and (pseudo)spin $\ua$. 
For a non-interacting system for which the number of particles is not
considered fixed, this can be accomplished in two ways:
\begin{itemize}
\item adding an $\ua$-particle with momentum $\kv$, or
\item removing a $\da$-particle with momentum $-\kv$.
\end{itemize}
Conversely, $-\Ek^-$ is the excitation energy when the system is forced to
have $\kv$ momentum and $\da$ (pseudo)spin. 
Which is the process that ultimately will be required to actually perform
these excitations is essentially determined by Pauli's principle and the
interactions in the medium. For example, a system composed only of
$\ua$-particles ($\mu=\dmu=\muu/2$) at $T=0$, can only be excited to states
$|\kv,\ua\ra$ for momenta $k>\ku$, as for $k<\ku$ all states are already
occupied and Pauli's principle forbids a new $\ua$-particle to be added to the
system below its Fermi momentum. The resulting system will have excitation energy
$\hbar^2k^2/(2m)-\ef \equiv \Ek^+$. For $k<\ku$, the reachable states are
$|\kv,\da\ra$ with excitation energy $\ef-\hbar^2k^2/(2m) \equiv -\Ek^- >0$,
corresponding to the creation of a hole in the $\ua$-Fermi sea at $k$.
Similar ideas hold for the interacting system, and for other values of the
densities $\ru$ and $\rd$.

\begin{figure}[!h]
  \caption[\bcs\ and \dfs\ quasiparticle spectra]
          {\label{fig:dfs2}\index{spectrum!in superfluid phases}
	   Dependence of the quasiparticle spectra $\Ek^\ua$ (black lines) and
	   $\Ek^\da$ (green) on the momentum for the symmetric \bcs\ case
	   (left); an asymmetric \bcs\ case with $\al = 0.04$ (center); 
	   and the optimal $\deps=0.08$ \dfs\ result for the same 
	   asymmetry (right).
	   The $z$ axis is the symmetry-breaking axis of the \dfs\ phase.
	   In the top plots, the solid lines correspond to the spectra 
	   $\Ek^\ua$ (black) and $\Ek^\da$ (green) along the $k_z$ axis.
	   The dashed lines represent the results
	   on the $xy$ plane ($k_z=0$) with the same color coding.
	   The dotted line is at $\Ek=0$.
	   In the lower plots, $\Ek^\ua$ is given only for $k_z>0$ and
	   $\Ek^\da$ for $k_z<0$.
	   }
  \begin{center}
     \includegraphics[width=12cm,clip=true]
                     {pairing/matlab/espectres_separats}
     \includegraphics[width=11.5cm,clip=true]
                     {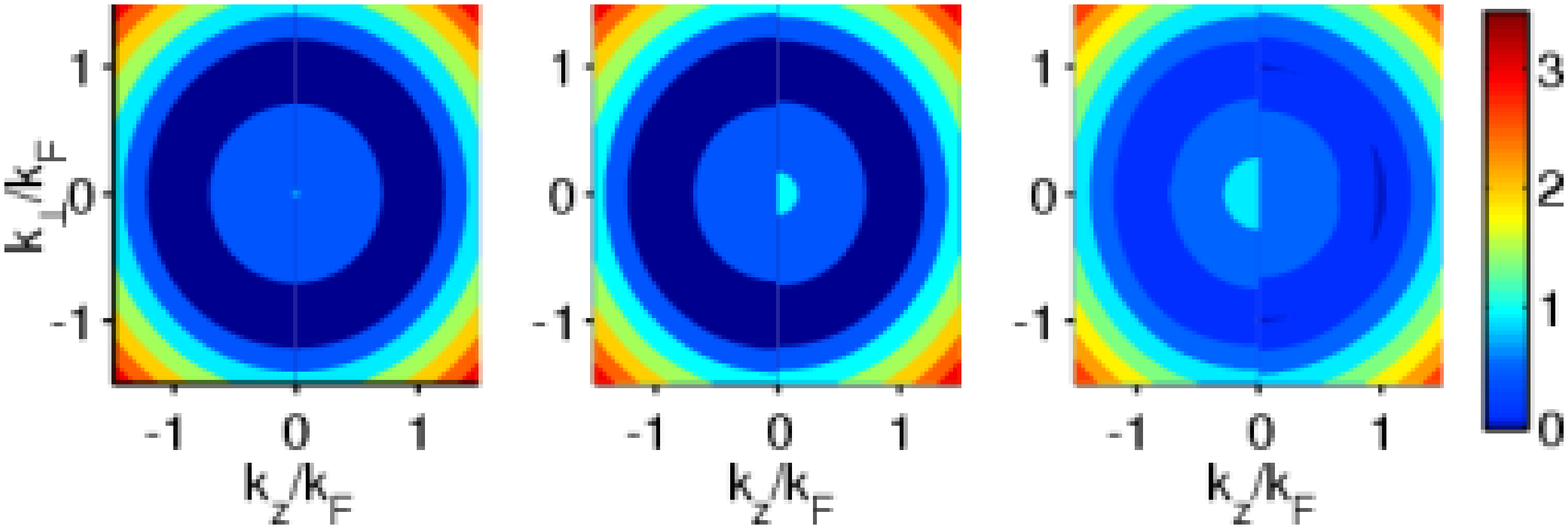}
  \end{center}%
\end{figure}

We call the reader's attention to the minus sign in front of $\Ek^-$: this is
due to the conventional way of assigning energies to hole excitations (see,
\eg, \cite{fetter}, pp.~74--75 ): the more negative they are, the more excited
state of the system we get. This is easily understood again in the pure
$\ua$-system, where removing an $\ua$-particle from $k=\ku$ leaves the system
in its ground state, and therefore the excitation energy is
$\eps_\mathrm{exc}(k=\ku)=0 \equiv -E^-_{\ku}$. On the other hand, removing a
particle at $k=0$ will leave the system in a highly excited state,
corresponding to having promoted [in the ($N-1$)-particle system] the
$\ua$-particle at $k=0$ to $k=\ku$; the corresponding excitation energy is 
$\eps_\mathrm{exc}(k=0)=\muu \equiv -E^-_0$.
Therefore, the excitation energies of the system
displayed in Fig.~\ref{fig:dfs2} are
\begin{itemize}
\item $\Ek^\ua:=\Ek^+=\sqrt{E_S^2+|\De|^2}+E_A$: excitation energy of the
      system with momentum $\kv$ and (pseudo)spin $\ua$;
\item $\Ek^\da:=-\Ek^-=\sqrt{E_S^2+|\De|^2}-E_A$: excitation energy of the
      system with momentum $\kv$ and (pseudo)spin $\da$.
\end{itemize}

Let us turn now to the effects of the density asymmetry on the solution 
of the gap equation. We have seen in the previous chapter that, in the 
asymmetric \bcs\ state,
$E_A$ acts in the gap equation
(\ref{eq:gap-eq}) to reduce the phase-space coherence between the
quasiparticles that pair. In other words, $E_A \neq 0$ introduces a
\index{forbidden region}
`forbidden region' for the momentum integration in the gap equation
[cf. Sect.~\ref{sec:swave}, especially Eq.~(\ref{eq:dr}) 
and Fig.~\ref{fig:distrib}]. 
The \bcs\ limit is recovered for $E_A=0$, with equal occupations for both
particles and perfectly matching $\ua$ and $\da$ Fermi surfaces.
This blocking effect is responsible for the reduction  
of the gap with increasing asymmetry and its disappearance above 
$\al=\alm^{BCS} \simeq 0.055$.

When the pairs move with a finite total momentum or the Fermi surfaces are
deformed (and taking the symmetry-breaking axis as $z$ axis),
the anti-symmetric part of the spectrum $E_A$ is modulated with the cosine of
the polar angle [cf. Eqs.~(\ref{eq:espectre-dfs}--\ref{eq:new-eps})].
In the plane-wave \loff\ phase $E_A\propto \chi=\cos\theta$, while in the
\dfs\ phase, which is the object of our primary interest, we have
\index{spectrum!in superfluid phases!angular dependence of}
\begin{subequations}
\begin{align}
  E_S &= \ek - \mu -\dmu\,\deps\, \chi^2 \\
  E_A &= - \dmu +\mu\,\deps\, \chi^2 \;.
\end{align}
\label{eq:sim-asim}
\end{subequations}
This angular variation acts to restore the phase-space coherence for some
values of $\chi$ at the cost of even lesser (than in the \bcs\ phase)
coherence for the other directions. 
\index{forbidden region}
That is, the width of the forbidden region in Fig.~\ref{fig:distrib}
now depends on the direction: it is reduced in the $xy$ plane and increased
on the $z$ axis.
This effect can be explicitely seen in Figure~\ref{fig:dfs2} which compares
the quasiparticle excitation spectra in the \bcs\ and \dfs\ phases. 
Let us comment carefully this figure, as it contain much information.
The first column shows the results for the usual symmetric \bcs\ case; the
second column has the results for the asymmetric \bcs\ case with a moderate
density asymmetry $\al=0.04$; finally, the third column contains the results
for the \dfs\ phase with the same asymmetry and the optimal deformation
$\deps=0.08$.%
\footnote{By `optimal' we mean `with lowest free energy', see below.}~%
\index{spectrum!in superfluid phases}
In all three columns, the top plot displays the excitation spectra $\Ek^\ua$
(black lines) and $\Ek^\da$ (green lines) close to the Fermi momentum $k_F$.
Solid lines correspond to the results along the symmetry-breaking $z$ axis,
while the dashed lines in the last column stand for the spectra in the $xy$
plane in $k$-space. 
The figure in the top-left corner is equivalent to the lower panel 
in Fig.~\ref{fig:uv_k}.
The spectra for the asymmetric \bcs\ case are shifted with respect to each
other due to the fact that $E_A\neq0$, cf. Eq.~\ref{eq:qp-spectra}, but keep
the rotational invariance.
\index{spectrum!in superfluid phases!angular dependence of}
Finally, the plot for the \dfs\ phase shows the effect of the
$\chi$-dependence of the spectra: the solid curves along the symmetry-breaking
axis (\ie, $\chi=1$ or $k_\perp=0$) 
have gone further apart, while the
dashed lines corresponding to $\chi=0$ ($k_z=0$) 
have approached one another.

To better evaluate the rotational properties of each phase, the lower plots
show $\Ek^{\ua\da}$ as a function of $k_z/k_F$ and $k_\perp/k_F$. 
Keeping $k=\sqrt{k_z^2+k_\perp^2}$ constant amounts to moving along circles 
on the $k_z-k_\perp$ plane, therefore exploring the angular dependence of 
the functions.
For the symmetric \bcs\ case, both $\da$ and $\ua$ excitation spectra are
equal, have a minimum at $k=k_F$ and are rotationally symmetric.
For an asymmetric system (second column) the $\ua$ and $\da$ spectra are no
longer equally deep, but $\Ek^\da$ is deeper than $\Ek^\ua$. Therefore, the
quasiparticles of one species, that are defined near the corresponding Fermi
surface, are far (in phase space) from those of the other species.
The third column shows the excitation spectra in the \dfs\ phase, as
given by Eqs.~(\ref{eq:espectre-dfs}) and (\ref{eq:qp-spectra}). Rotational
symmetry is now broken, as the spectra along $k_z$ (solid lines) are different
from the spectra perpendicular to $k_z$ (dashed lines).

\begin{figure} 
  \caption[\bcs, \loff\ and \dfs\ critical asymmetries]
          {\label{fig:dfs3}
	    Dependence of the critical asymmetry of the transition from 
	    the superfluid to the normal state on the pair momentum in the
	    \loff\ phase (solid line) and the deformation parameter in the
	    \dfs\ phase (dashed and dot-dashed lines).
	  }
  \begin{center}
    \includegraphics[width=8cm,clip=true]{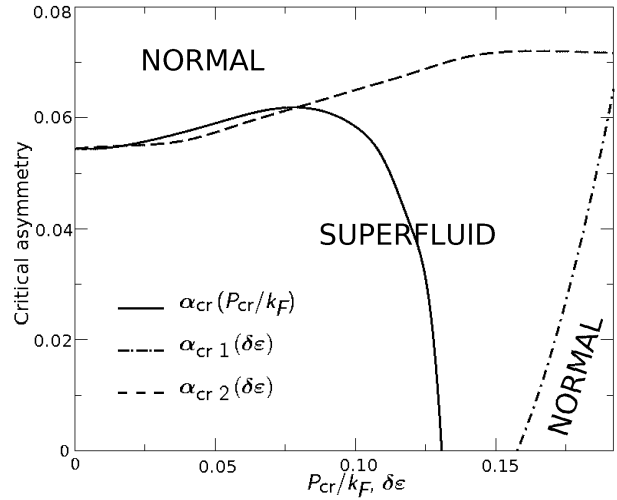}
  \end{center}
\end{figure}

\index{spectrum!in superfluid phases} 
A most remarkable feature in the \dfs\ is that the energy separation between the
quasiparticle spectra along the symmetry-breaking axis is considerably larger 
than in the asymmetric \bcs\ state; in the orthogonal directions the opposite
holds. 
Compared to the asymmetric \bcs\ state, the phase-space overlap
between pairs is accordingly decreased 
in the first region and increased 
in the second. 
The net result, displayed in Figure~\ref{fig:dfs3}, is the increase in
the value of the critical asymmetry $\alm$ at which superfluidity
vanishes. As noted above, at large asymmetries the \dfs\ phase exhibits the
re-entrance effect: pairing exists only for the deformed state between the
lower and upper critical deformations ($\al_{\rm cr\,1}\neq0$).
We note that to obtain this effect the recalculation of the chemical potentials
through the 
\index{density normalization!in pairing problem}normalization condition 
on the densities is essential, 
as it affects dramatically the value of $\dmu$, that enters $E_A$.

Notice also that in the asymmetric systems, the $\da$ spectrum is  gapless
in a region of momentum space defined by
\begin{equation}
  \Ek^\da \equiv E^\da(k_z,k_\perp) \leq 0 \;.
\end{equation}
The possibility of exciting the system without energy cost has important
consequences for the dynamical properties of the paired states,
such as the transport and the collective modes, and leads to a
number of  peculiarities in their
thermodynamics~\cite{GL1,sed97,umb2,GL3,GL41,GL42,GL5}.
That is, the macro-physical manifestations of the \loff\ and \dfs\ phases 
such as the response to density perturbations or electromagnetic
probes and the thermodynamic functions (heat capacity, etc) 
will differ from the ordinary \bcs\  phase due to the nodes 
{\em and anisotropy} of their spectra. We will show in Sect.~\ref{sec:dfs-exp} 
how such an anisotropy can be used to discriminate phases with broken space
symmetries in time-of-flight experiments. 
We remark finally that the phases with broken space symmetries (\loff\ and
\dfs) present a larger number of excitation modes because of the breaking of
global space symmetries; these additional modes are usually called
Goldstone modes~\cite{goldstone,peskin}.

\section{Determining the ground state}
\label{sec:dfs-gs}

\subsection{Calculation of the free energy}
The phase that must be identified as the equilibrium state at a given density
asymmetry and temperature is the one that has the lowest free energy. 
We have calculated the free energy for each candidate phase (\bcs, \loff, \dfs)
in the following way.

We have defined the free energy as usual by
\be
{\cal F} = E_{\rm kin}+E_{\rm pot} - T{\cal S},
\ee
where the first two terms comprise the internal energy which is the
statistical average of the Hamiltonian (\ref{eq:bcs-ham}), $T$ is the
temperature and  ${\cal S}$ the entropy, given for a gas of fermions 
by the well-known expression~\cite{fetter},
\begin{align*}
  {\cal S} &= -k_B \intk \sum_{s=\ua,\da} 
  \big\{ 
    \lp 1-n_s(\kv)\rp \ln\left[ 1-n_s(\kv) \right]
        +n_s(\kv) \ln\left[ n_s(\kv) \right]
  \big\} \:.
\end{align*}
For a contact potential of strenght $g$ , the potential energy is easily
evaluated, and we have 
\begin{align}
  E_{\rm kin}+E_{\rm pot} &= \intk \ek
    \left[n_{\ua}(\kv)+n_{\da}(\kv)\right] -\frac{\Delta^2}{g} \:, \\
  n_s(\kv) &= u^2(\kv)f(\Ek^s) + v^2(\kv)\lp1-f(\Ek^{-s})\rp \:.
  \label{eq:mom-distr}
\end{align}
The free energy of the undeformed normal state follows by setting in
the  above expressions $\Delta = 0 =\delta\epsilon$, while that of the \BCS\
phase is given by $\Delta\neq 0 = \deps$, and so on for the \loff\ and \dfs\ 
phases (see Table~\ref{table:dfs}).

\index{gap equation!regularization of}
Because of the contact form that we use for the interaction, the gap
equation and the superfluid kinetic energy need a regularization. There are
several ways to regularize the gap equation. We can write them formally
together in the form [cf. Eq.~(\ref{eq:gap-eq})]
\be
\label{GAP2}
1 = \frac{g}{2}
    \int_0^{\Lambda}\!\! \frac{\mathrm{d}k k^2}{(2\pi)^2}
    \int_{-1}^{1}\!\! \mathrm{d}\chi 
    \left( \frac{f_F(\Ek^+) - f_F(\Ek^-)}{\sqrt{E_S^2+\Delta^2}} + 
           \frac{\gamma}{\ek} \right) ,
\ee
where $f_F$ is the usual Fermi distribution function. The case $\gamma =1$ and
$\Lambda \to \infty$ corresponds to the common practice of
regularization~\cite{stoof96}, which combines the gap equation with the
\index{T-matrix}$\T$-matrix equation in free space, 
and is the one we have used in chapter 2. 
Choosing $\gamma =0$ and a finite $\Lambda$ corresponds to the cut-off
regularization of the original gap equation. The appropriate cutt-off in this
second scheme is found by requiring both schemes to give the same value for
the gap. Then, this $\Lambda$ is used to evaluate the kinetic energy
contribution to the free energy.
We note once more that Eq.~(\ref{GAP2}) must be solved together with the
\index{density normalization!in pairing problem}normalization constraints 
on the densities
\begin{align}
  \rho_s &= \intk n_s(\kv) \qquad (s=\ua,\da) \:.
  \label{eq:norms}
\end{align}

\subsection{Analytical estimation of the optimal deformation}
\label{ssec:dfs-estimate}
Before comenting the numerical results,
it is instructive to perform an analytical estimation of the deformation
$\deps$ of the Fermi surfaces that one expects will maximize the pairing
energy.

The physical idea behind the deformation of the Fermi surfaces of the pairing
species is to approach them in some regions of momentum space, so that
the quasiparticles' phase space has a sizeable overlap. If this
deformation had no kinetic energy cost, we can imagine that the optimal
deformation would be the one that effectively makes the two Fermi surfaces
be as close to each other as possible, given the constraint of the
conservation of the different densities. Thus, we will calculate which
deformation $\deps$ brings the outermost part of the Fermi surface of the
least populated species and the innermost part of the Fermi surface of 
the majority to touch. 
The optimal deformation is somewhat different that the one predicted by
this estimation because of the investment in kinetic energy that this
deformation requires. In any case, we can get a good estimate of the 
optimal $\deps$ in this simple way.

We assume for simplicity $T=0$ and start from Eq.~(\ref{eq:new-eps}). Here, we
must understand $\mu_s=\hbar^2\widetilde{k}_s^2/(2m)$, where $\widetilde{k}_s$
is such that the density of $s$-particles is conserved. 
For 
the case of $\ua$-particles:
\begin{align}
  \ru 
  &= 
  \intk \Theta\left[\kut^2\lp 1+\deps\cos^2\theta\rp-k^2\right] 
  \notag\\
  &=
  \inv{(2\pi)^2}\frac{\kut^3}{24}
  \left[ 2\sqrt{1+\deps}(5+2\deps)+6\frac{\arcsh{\sqrt{\deps}}}{\sqrt{\deps}}
  \right]
  \label{eq:rhoup}
\end{align}
The expression for $\rd(\kdt)$ is analogous with
$\deps\raw-\deps$. 
As the `true' Fermi momentum of $\ua$-particles is defined by
$\ru=\rho(1+\al)/2=\ku^3/(6\pi^2)$, we have%
\footnote{Here it is clear that we are working at fixed {\em density}
asymmetry, and not chemical potential difference, as was the original
treatment of Larkin and Ovchinnikov~\cite{lo} and Fulde and Ferrell~\cite{ff}
for the \loff\ phase, see below. In our approach, the chemical potentials come
through the 
\index{density normalization!in pairing problem}normalization of the densities, 
Eq.~(\ref{eq:norms}), which depends on the shape and location of the Fermi surfaces.}
\begin{align*}
  \frac{\ku^3}{\kd^3} &\equiv \frac{1+\al}{1-\al} \:, \notag \\
  &=\frac{\kut^3}{\kdt^3} \frac
    { 2\sqrt{1+\deps}(5+2\deps)+6\frac{\arcsh{\sqrt{\deps}}}{\sqrt{\deps}} }
    { 2\sqrt{1-\deps}(5-2\deps)+6\frac{\arcsh{\sqrt{\deps}}}{\sqrt{\deps}} }
  \approx \frac{\kut^3}{\kdt^3}\lp 1+ \frac{9}{8}\deps \rp
\end{align*}
where the last result follows for $\deps\ll 1$. 
In the plane perpendicular to the symmetry-breaking axis ($\cos\theta=0$),
each momentum distribution vanishes at the corresponding
$\widetilde{k}_s$. Therefore, we will have an exact matching of the two Fermi
surfaces in this plane for $\kut=\kdt$. Then, for a given $\al$ we can obtain
our estimation for the optimal deformation from the last equation:
\begin{equation} 
  \deps_{\rm opt} = \frac{16}{9}\frac{\al}{1-\al} \:.
\label{eq:opt-deps}
\end{equation}
Substitution of this value into (\ref{eq:rhoup}) gives the new value for $\kut$, 
and therefore the new chemical potential $\muu$, and analogously for $\mud$.
For $\al\ll1$ we expect a deformation 
$\deps_{\rm opt} \approx (16/9)\,\al \approx 2\al$. We will see that this
approximation agrees reasonably well with the numerical calculations
below, even though the overlap is not perfect in the numerical solution
because in this analysis we have disregarded the investment in kinetic energy.

\subsection{Numerical results}
\label{ssec:dfs-results}
\begin{figure}[!th]
  \caption[\loff\ and \dfs\ free energies vs. density asymmetry]
          {\label{fig:dfs4}
	   Difference of the free-energy density of the
	   plane-wave \loff\ (upper panel) and \dfs\ (lower panel) phases
	   with respect to the normal phase, as a function of
	   the asymmetry parameter $\al$, for the values of the pair
	   momentum $P/k_F$ and the deformation parameter $\de\epsilon$ 
	   indicated in the legends.
	  }
  \begin{center}
    \includegraphics[width=0.9\textwidth,clip=true]
                    {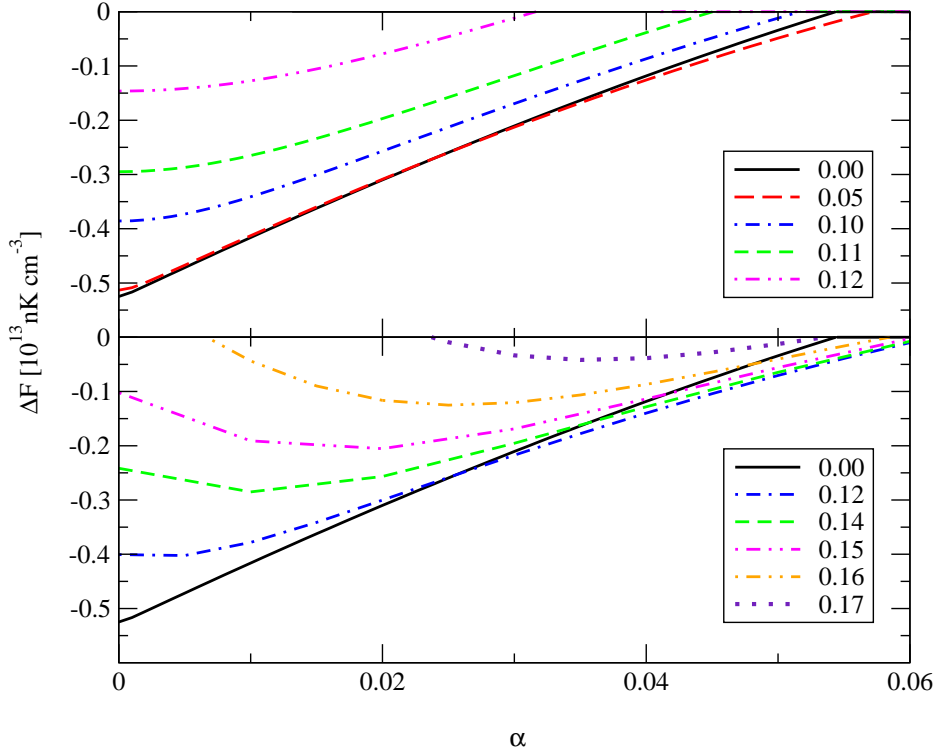}
  \end{center}
  \vspace{-1cm}\phantom{.}
\end{figure}

We plot in Fig.~\ref{fig:dfs4} the difference 
$\De {\cal F} = {\cal F}_S - {\cal F}_N$  between the
free energy ${\cal F}_S$ of the superfluid phase (either \bcs, \loff\ or \dfs) 
and the free energy ${\cal F}_N$ of the normal state ($\De,\deps,P=0$).
The \bcs\ result is plotted in both panels as the black solid line, while the
results for the \loff\ phase are 
summarized in the top panel for various values of $P/k_F$. The lower panel
contains the corresponding results for the \dfs\ phase for a variety of
$\deps$'s.
First of all, a simple comparison of this figure with \ref{fig:dfs1} shows
that $\De {\cal F}$ closely follows the behavior of $-\De^2$. However, the
contribution from the kinetic energy to ${\cal F}$ introduces some important
differences as, for example, the delay in the \dfs\ being preferable to the
normal phase, \eg\ for $\deps=0.17$ from $\al\approx0.018$ (where a non-vanishing
gap appears) to $\al\approx0.025$ (where $\De {\cal F}<0$).

The \loff\ phase is preferred to the normal and the
\bcs\ phases in a narrow window of asymmetries $0.04 
\le \al\le 0.057$ and for a total momentum of the pairs 
$P/k_F\sim 0.05$, as shown by the red dashed line lying below the solid
one. 
This is consistent with the results obtained in the scheme where the density
asymmetry is described by fixing the difference in chemical potentials of the
pairing species $\de\mu$.
In this approach, the critical value for the \bcs\ phase is $\de\mu_c^{BCS} =
0.707\De(0)$, while for the \loff\ phase  $\de\mu_c^{LOFF} = 0.755\De(0)$, 
where $\De(0) \equiv \De(\de\mu = 0)$~\cite{lo,ff}.
Note that while there is a non-trivial solution to the gap equation for 
$P/k_F\ge 0.1$ the gain in \index{pairing energy}pairing energy is less 
than the investment required
in kinetic energy due to the motion of the condensate, and the net free energy
of the \loff\ phase is greater than that of the asymmetric \bcs\ phase for
these momenta. However the \index{pairing energy}pairing energy of the 
\loff\ phase could be improved
by choosing a more complex form of the order parameter, \eg\ by keeping a
larger number of terms in its expansion in the Fourier series.

The \dfs\ phase is the ground state of the system in a wider range of
asymmetries $0.03\le \al\lesssim 0.06$  for the deformation parameter in the
range $0.12\le \de\epsilon\le 0.14$. For even larger deformations the gain in
\index{pairing energy}pairing energy does not compensate the investment 
in kinetic energy due to the
ellipsoidal stretching of the Fermi surfaces: the curve for $\deps=0.17$ lies
always above the \bcs\ one. We note finally that the free energy is
also affected by the re-entrance effect (\ie, restoration of pairing
correlations as the asymmetry is increased), but in fact we do not find that
this effect is responsible for the ground state at these weak couplings: the
free energy for the case $\deps=0.16$ lies in our calculations always above
those for $\deps\leq0.15$, which do not present re-entrance, in contrast to
what happens in more strongly-coupled systems such as neutron-proton pairing
at saturation density~\cite{sedr-prl} or `two color superconductivity'
between up and down quarks in dense hadronic matter~\cite{sedr-2sc}.

To summarize, the coherence is restored and the strength of
pair-cor\-re\-la\-tions is increased in the \loff\ phase due to the finite
momentum of the Cooper pairs. In the \dfs\ phase, the same is achieved by
stretching the spherical Fermi surfaces into ellipsoids. The fundamental
difference between these phases is that translational symmetry remains
intact for the \dfs\ phase, as it breaks only the rotational symmetry, while
the \loff\ phase breaks both symmetries. Quantitatively, the maximal value of
the gap and the absolute value of the ground-state free energy are larger in
the \dfs\ phase than in the \loff\ phase for asymmetries  $\al \gtrsim 0.04$. 
For these asymmetries both phases are favorable over the homogeneous \bcs\
phase. However, one should keep in mind that the \loff\ phase admits a variety
of lattice forms, and the plane-wave structure might not be the most favored
one~\cite{alford,sedr01,bowers,casal}.

\section{Detecting the \dfs\ phase in experiments}
\label{sec:dfs-exp}
Experimental evidence for the phases with broken space symmetries
can be obtained from the studies of their momentum distributions
which, unlike in the homogeneous phase, must be anisotropic in space.
Indeed, from Eqs.~(\ref{eq:qp-spectra}), (\ref{eq:sim-asim}) and
(\ref{eq:mom-distr}), we see that, for $\deps\neq0$, the probability of a
$s$-particle having momentum $\kv$, $n_s(\kv)$, will depend on $k_z^2$ and
$k_\perp^2$ separately or, equivalently, on $k=|\kv|$ and
$\chi=\cos\theta=k_z/k$, and not only on the scalar
combination $k^2=k_z^2+k_\perp^2$.

In this case, it is useful to analyze the angular anisotropy in momentum space
of the occupation numbers, that we define as
\begin{equation}
  \de n_{ s}(k) = | n_{ s}(k;\chi=1) - n_{ s}(k;\chi=0)| \;.
  \label{eq:aniso}
\end{equation}
This quantity vanishes for rotationally-invariant systems, and can be taken
as a measure of the degree of symmetry breaking of a given phase.
For example, $\de n_s(k)\equiv0~(\forall k)$ for the symmetric and asymmetric
\bcs\ solutions to the gap equation. 
For the \dfs\ phase, the momentum distributions for each species depend on $\chi$,
and we expect that $\de n_s$ will not vanish but present a maximum
around the corresponding Fermi momentum $k_s$.

\begin{figure}[!t]
  \caption[Occupation probabilites of the two hyperfine states]
          {\label{fig:dfs5}
	    Dependence of the occupation probabilities of two hyperfine 
	    states on the momentum. The Fermi momentum $k_F = 4.83\times10^4$
	    cm$^{-1}$ is indicated by the vertical line. The labeling of 
	    the lines is as follows: 
	    $\al = 0=\de\epsilon$ (black, solid line),  
	    $\al = 0.04$ and $\de\epsilon = 0$ (dashed lines);  
	    $\al = 0.04$, $\de\epsilon = 0.08$, 
	    $\chi = 0$ (dash-dotted) and $\chi = 1$ (dotted).
	    The blue lines are for $\ua$ particles (major population, see
            left), while the red ones are for $\da$ particles (minority).
	    The bell-shaped curves show the anisotropy as defined by
            Eq.~(\ref{eq:aniso})
	    for $\al = 0.04$, $\de\epsilon = 0.08$.
	    }
  \begin{center}
    \includegraphics[width=0.25\textwidth,height=0.44\textwidth,clip=true]
                    {pairing/dfs_mom_bn}%
    \hfill
    \includegraphics[width=0.7\textwidth,height=0.44\textwidth,clip=true]
		    {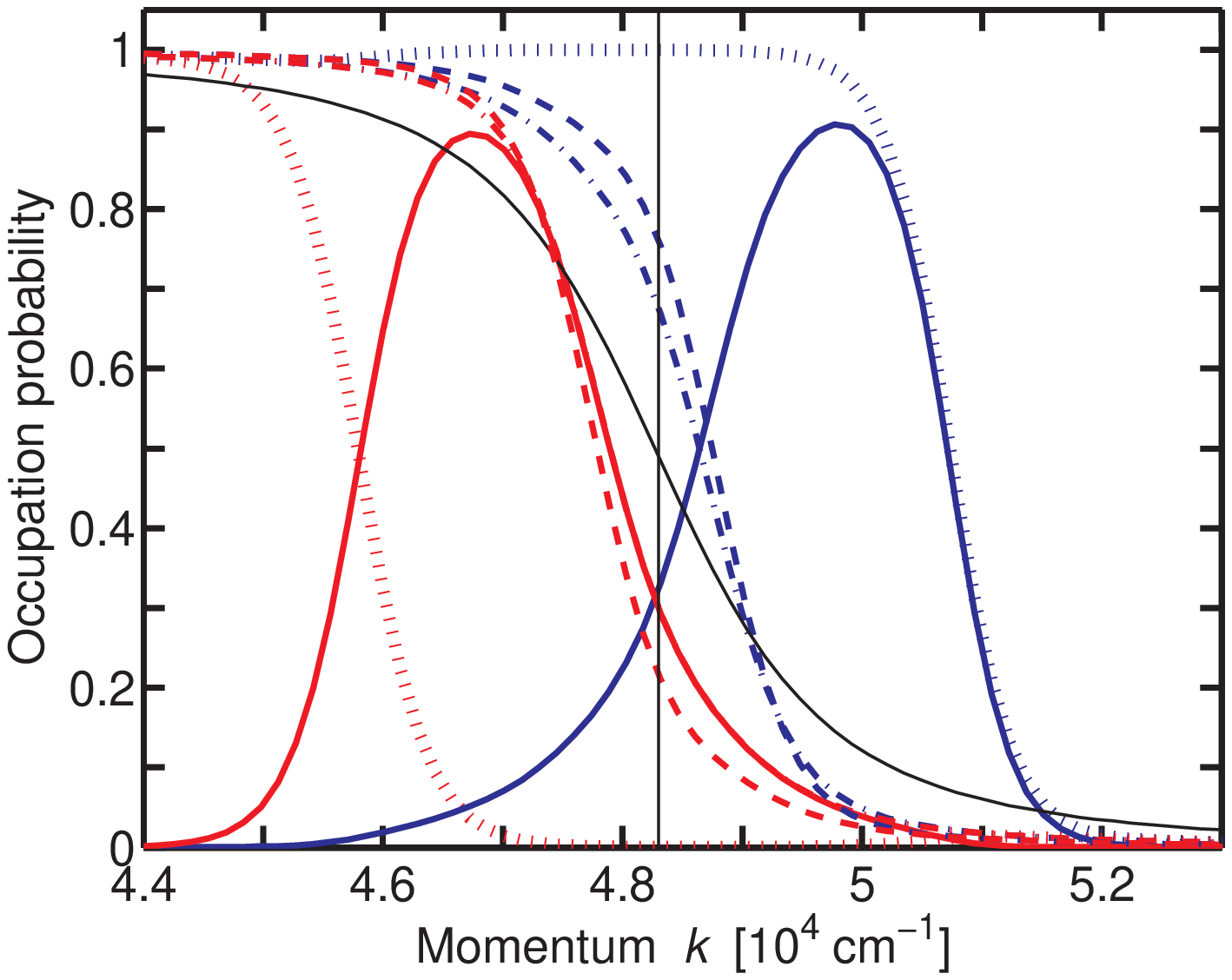}
  \end{center}
\end{figure}

We show in Fig.~\ref{fig:dfs5} the momentum distributions for the same system
as in Fig.~\ref{fig:dfs1}: $k_F=4.83\times10^4$ cm$^{-1}$ (indicated by the
vertical line) and $a=-2160a_B$. 
For clarity, we present only the data around $k_F$, where all the interesting
physics happens. 
There are four sets of curves corresponding to the symmetric \bcs\ phase
(thin solid line), asymmetric \bcs\ with $\al=0.04$ (dashed lines) and
\dfs\ with the same asymmetry and $\deps=0.08$ for $\chi=0$ (dash-dotted
lines) and $\chi=1$ (dotted lines), respectively.
For the $\al\neq0$ cases, blue lines correspond to $n_\ua(\kv)$ while red
lines are for $n_\da(\kv)$.
Finally, the thick, bell-shaped curves correspond to $\de n_s(k)$ with the
same color coding. 

The Fermi surfaces of the symmetric \bcs\ phase are spread around 
$k_F$ over a width of the order of $k_F\,\De/\ef$.
One should notice that some broadening of the surface 
is also due to the finite temperature assumed in the calculations.
The asymmetric \bcs\ distributions (dashed lines) show a similar shape, but 
with the characteristic decays centered around the corresponding Fermi momenta
$\kd\approx4.75\times10^4$ cm$^{-1}$, $\ku\approx4.91\times10^4$ cm$^{-1}$. At
the low temperature studied, the momentum distributions still fall off very
fast, and the two Fermi surfaces are separated $\sim0.16\times10^4$
\index{forbidden region}\index{Pauli blocking}
cm$^{-1}$, thus triggering the Pauli blocking effect commented in
Sect.~\ref{sec:swave}. 

The situation changes when looking at the \dfs\ results which show a clear
dependence on $\chi$.
First of all, we notice that the \dfs\ curves for $\chi=0$ (dash-dotted lines)
are almost indistinguishable, while the results
for $\chi=1$ (dotted lines) are further apart. This means that the
(ellipsoidal) Fermi surfaces of $\ua$ and $\da$ particles have approached each
other in the $xy$ plane in momentum space, while they have gone apart along
the $z$ direction (see Fig.~\ref{fig:dfs5}). 
Note also that the range of occupations covered by the \dfs\ curves includes the
asymmetric \bcs\ results. One could have expected the asymmetric \bcs\ results
to coincide with the \dfs\ ones for $\chi=0$, as in both cases the
symmetry-breaking term $\deps\, \chi^2$ vanishes [see
Eq.~(\ref{eq:sim-asim})]. However, the fully self-consistent solution of the 
gap equation together with the number equations involves a re-calculation of
the chemical potentials for the \dfs\ case, resulting in new effective Fermi
momenta $\tilde k_s$ (see Sect.~\ref{ssec:dfs-estimate}) which become very
similar to each other and to $k_F$. That is the reason why the \dfs\ results 
at $\chi=0$ are so close to each other, and relatively far from the asymmetric
\bcs\ ones. 
In addition, we obtain a {\em larger gap} (cf. Fig~\ref{fig:dfs1}), 
that translates into slightly softer
decays of the \dfs\ momentum distributions as compared with those for the
asymmetric \bcs\ case.

The anisotropy in the \dfs\ occupation probabilities along the
directions parallel and orthogonal to the symmetry-breaking axis reaches a
maximum of more than 90\% for both species centered at different momenta:
for $\de n_\da$, around 
$\tilde k_\da\sqrt{1-\deps}\approx4.63\times10^{4}$~cm$^{-1}$ 
[momentum for which $\de n_\da(\chi=0)\approx 1$ and 
$\de n_\da(\chi=1)\approx 0$]; 
for $\de n_\ua$, at 
$\tilde k_\ua\sqrt{1+\deps} \approx 5.07\times10^4$~cm$^{-1}$ 
[where $\de n_\ua(\chi=0)\approx 0$ and 
$\de n_\ua(\chi=1)\approx 1$].

In practice, this means that one expects to find more $\ua$ particles with
momenta $k>k_F$ in the $z$ direction ($\chi=1$) than in the radial direction
($\chi=0$), while the opposite holds for the $\da$ particles.
Thus, a direct way to detect the \dfs\ phase is a measurement of 
this {\em crossed anisotropy} in the momentum distributions of the trapped
atoms. 
Such a measurement can be realized by the time-of-flight technique
\cite{jila95,mit95,jin99}. 
This method uses the fact that after switching off the trap,
the atoms fly out freely and an image of their spatial distribution
taken after some time of flight
provides information on their momentum distribution when confined inside the
trap. Assuming that the system was in the deformed superfluid state
one would  detect a number of particles of type $\ua$ (majority) in the
direction of symmetry breaking by about 90\% larger than in any orthogonal
direction, and viceversa for the $\da$ particles (minority).
Therefore, the presence of this crossed anisotropy in the detected momentum
distributions would be an evidence for a `deformed superfluid state'
being the ground state of the system, as deformation alone
(\ie, without pairing) would not lower the energy so as to
produce a deformed non-superfluid ground state.
Note that this argument is equally valid for a homogeneous system or for an
atomic sample in a spherical trap, where no privileged direction is introduced
by the trapping potential. For a non-spherical trap, the normal-state momentum
distributions of both species are expected to be deformed {\em in the same
way}. Therefore, the detection of a crossed anisotropy in the momentum
distributions as discussed above would also be a strong case for the \dfs\
phase being the ground state of the system. However, a more specific
calculation (\eg, in local density approximation in the trap) would be
necessary to quantify this effect and study the influence of the trap
anisotropy on the momentum distributions in the different phases.

We remark finally that the direction of spontaneous symmetry breaking (in
$k$-space and, therefore, also in real space) is chosen by the system randomly
and needs to be located in an experiment to obtain maximum anisotropy.
Also, a clear distinction between the \dfs\ and the \loff\ phases can be
achieved in the time-of-flight experiments, since the latter predicts 
periodic momentum distributions. 

\section{Summary}

In this chapter we have studied the possibility to generalize the 
\bcs\ ground state by letting the Cooper pairs carry a finite center-of-mass
momentum (the so-called \loff\ phase) or by deforming the Fermi surfaces of
the pairing species (\dfs\ phase). For the density-symmetric case ($\al=0$),
the \bcs\ pair wave-function is the best among these three options, \ie, it
gives the lowest free energy for the system. However, for systems with
different populations of the two species ($\al\neq0$), this is only
true for very small asymmetries ($\al<0.03=3\%$). For larger asymmetries, the
\loff\ and/or \dfs\ phases become preferable. Our quantitative analysis has
shown that, at \index{weak-coupling limit}weak-coupling, 
the \dfs\ phase is prefered to the normal, \bcs\
and plane-wave \loff\ phases (though more general forms for the \loff\ phase
are possible and might lower further the free energy).
We note that similar conclusions have been earlier reported for
strongly-coupled systems such as neutron-proton pairing in asymmetric 
nuclear matter in the $^3S_1- ^3D_1$ channel~\cite{sedr-prl,sedr-prc} 
and pairing of up and down quarks of two colors in dense hadronic 
matter~\cite{sedr-2sc}. Therefore, we expect a
similar behavior for more strongly-coupled atomic systems, such as those
produced by means of magnetic Feshbach resonances to study the BCS-BEC
crossover~\cite{greiner03,kett03,streck03,grimm04,salomon04,thomas05}.

We have finally shown how the \dfs\ phase can be detected in an ultracold gas
of fermionic atoms in a spherical trap by studying the momentum
distribution of the released atoms in a time-of-flight experiment.


\chapter[Pairing in boson-fermion mixtures]
	{Pairing in boson-fermion mixtures}%
\label{ch:2dbf}

\textsf{
  \begin{quote}
    His house was perfect whether you liked food, or sleep, 
    or work, or story-telling, or singing, or just sitting 
    and thinking best, or a pleasant mixture of them all. 
  \end{quote}
  \begin{flushright}
    J. R. R. Tolkien, {\em The Hobbit} 
  \end{flushright}
}

\section{Introduction}

Mixtures of quantum fluids of different statistics 
have been an interesting field of research for a long time,
specially in the context of helium systems, where the fraction of $^3$He
(fermion) in a homogeneous $^4$He (boson) medium is limited to 
$x_{3\rm max}\approx6.5\%$. 
For $x_3>x_{3\rm max}$ 
the system becomes unstable and 
separates into a mixed phase with a \het\ concentration $x_{3\rm max}$ and
another phase with pure \het\ atoms.

Since the first achievement of Bose-Einstein condensation, the purpose 
of building an analogous ultracold Fermi system had to overcome the cooling
limitations established by Pauli's exclusion principle, that forbids
$s$-wave collisions between indistinguishable fermions. 
This problem can be solved in a boson-fermion mixture, as the bosonic
component can be `easily' cooled and, by thermal contact, drives the
fermionic component down to ultralow temperatures, as was first achieved at
Rice University when a gas of \Lix\ was driven to the degenerate regime by
sympathetically cooling it with \Lin~\cite{tru01}.

After this experiment, many others have shown the feasibility of studying
interesting quantum phenomena in ultracold fermionic systems. 
For example, the \lens\ group used \Rb-\K\ mixtures with varying numbers of
atoms of both species to study the collapse predicted by mean-field
theory~\cite{molmer,roth-feld02} and 
extracted a value for the interspecies scattering
length  $a\lp\mbox{\Rb-\K}\rp=-22$ nm~\cite{mod02}.%
\footnote{
  This value is currently under revision as determinations by similar, as well
  as other presumably more precise, methods have given a more moderate value
  $a=-15.0\pm0.8$ nm~\cite{michele04,bongs-priv}.
}

In this chapter, we address the problem of determining the fermion-fermion
interaction inside a dilute mixture of bosons and fermions. This problem is
important as medium effects can have dramatic consequences on the
behavior of the system. A classical example is that of superconductivity in
metals, where the phonon-mediated interactions between electrons give rise to
an attraction between the latter, thus triggering the Cooper instability of
the system and resulting in the superfluid behavior of the electronic
component~\cite{schrieffer}. First, we will review the three-dimensional
situation, already studied by Viverit \ea\cite{viv00,bijlsma}. Then, we will
study the two-dimensional case, where the reduction in the dimensionality of
the phase space makes one expect that correlations will play a more important
role as compared to the 3D situation. Indeed, today's available capabilities to
manipulate trapped, atomic gases allow for the production of effectively one-
and two-dimensional systems by the deformation of the traps or, more
efficiently, by shining on it an array of standing light-waves 
that produce a kind of periodic potential known as {\em optical lattice} (see,
\cite{guidoni,heck02} and references therein). We will end up applying the
developed model to the cases of \Rb-\K\ and \Lin-\Lix\ mixtures under current
experimental investigation.

\section{Three-dimensional mixtures}
\label{sec:pair-3d}

Let us start by considering the fermion-fermion ($FF$) interaction 
in three-dimensional space for a one-component fermionic system
in the presence of bosons. 
At the ultralow temperatures of interest $T\lesssim 10\mu$K and dilute
conditions of present experiments, only $s$-wave collisions are relevant
(unless they are absent). 
As for a system of indistinguishable fermions $s$-wave interactions are
forbidden by Pauli's principle, we will be left only with the boson-fermion
($BF$) and boson-boson ($BB$) interactions, while the direct $FF$ one can be
neglected. 
For a dilute and ultracold system, these collisions can be modeled in three
dimensions by a  contact interaction or, equivalently, by momentum-independent
\index{T-matrix!at low momenta}
$\t$-matrices,
\begin{align*}
  \tbf &= \frac{4\pi\hbar^2}{\mbf}\abf \\
  \tbb &= \frac{4\pi\hbar^2}{\mb}\abb \:,
\end{align*}
where $\mb$ ($\mf$) is the mass of a boson (fermion) and
$\mbf=\mb\mf/(\mb+\mf)$ is the reduced mass in a boson-fermion collision.

The boson-induced fermion-fermion interaction 
can be represented by the diagram in Fig.~\ref{fig:lowestorder}(b), where now
the bubble corresponds to a density fluctuation in the bosonic medium. 
Analogously, we can use a formula similar to Eq.~(\ref{eq:wtot}):
\begin{equation}
  \lla\kvp|\G_{F}|\kv\rra = \Pi_{B}^{\rm RPA}\lp|\kvp-\kv|\rp\tbf^2 \:.
\end{equation}
Here $\Pi_B^{\rm RPA}$ stands for the density-density response function of the
bosonic component in the random phase approximation (RPA).
Already from its diagrammatic representation [Fig.~\ref{fig:dia}(b)], one can
see that it is possible to evaluate $\Pi_{B}^{\rm RPA}$ as a series once
$\tbb$ and the bosonic Lindhard function $\Pi_{B}$ 
are known. For in- and out-going fermions on their Fermi surface
there will be no net energy transfer through the bosonic medium
[Fig.~\ref{fig:dia}(a)], and one can take the static limit $\om=0$ for the
bosonic response function~\cite{pines2,fetter}%
,
\begin{equation*}
  \Pi_{B}(q;\om=0) = -\frac{4\mb\R_B}{\hbar^2q^2} \:.
\end{equation*}
Altogether, we get
\begin{equation}
  \Pi_{B}^{\rm RPA}(q) = \frac{\Pi_B(q)}{1-\tbb\Pi_B(q)} \:.
  \label{eq:rpa}
\end{equation}
Note that, due to the low density of the systems we are interested in, we are
neglecting any fermionic influence on the bosonic component.

\begin{figure}
  \caption[Boson-induced interactions in two and three dimensions]
	  {\label{fig:h-3d}
	    The functions 
	    governing the boson-induced fermion-fermion interaction 
	    in three (lower curve) and two (upper curve) dimensions. 
	    The dashed lines indicate the position and value of the
	    corresponding maxima.
	    The inset shows $h_{\rm 2D}$ together with the quadratic
	    approximation around its maximum (thin line with crosses).
	  }
  \begin{center}
    \includegraphics[width=0.7\textwidth,clip=true]{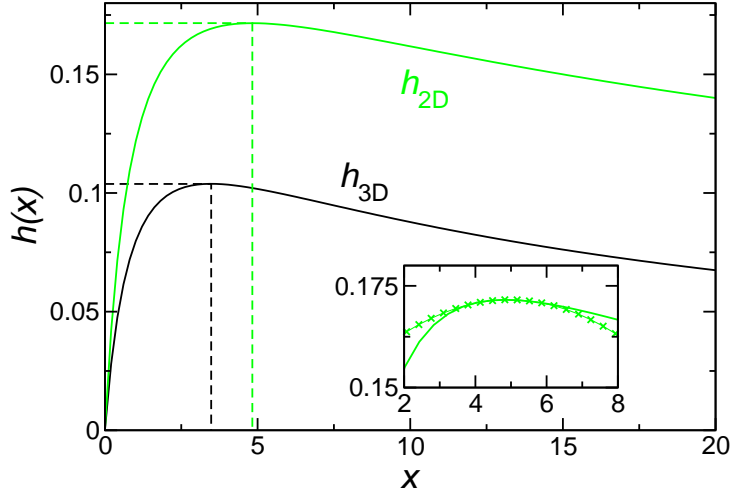}
  \end{center}
\end{figure}

As in Sect.~\ref{sec:pwave}, we must now project this polarization-induced
interaction onto its $p$-wave contribution with the corresponding Legendre
polynomial. Again, for a weakly-interacting system it is sufficient to
consider only the case where both in- and out-going fermions are on their
Fermi surface:
\begin{equation}
   \G_F^{(L=1)} (k_F, k_F) =
   {\tbf^2} \int_{-1}^{+1} \frac{dz}{2} z
   \Pi^{\rm RPA}_B(\sqrt{2(1-z)}k_F)
   = -\frac{\tbf^2}{\tbb} h_{\rm 3D}\!\lp x \rp \:,
   \label{eq:linkk-3d}
\end{equation}
with $x=\hbar^2k_F^2/(\mb\R_B\tbb)=2(\ef/\mu_B)(m_F/m_B)$ 
[for a dilute, homo\-ge\-neous boso\-nic system, $\mu_B=\tbb\R_B$], and the
function
\begin{equation*}
   h_{\rm 3D}(x) :=
   \frac{2}{x}\left[\lp\inv{x}+\inv{2}\rp\ln\lp1+x\rp -1 \right] \:,
\end{equation*}
which is plotted in Fig.~\ref{fig:h-3d}.

This function presents a broad maximum at
($x^{3D}_\opt\approx3.48,~h^{3D}_\opt\approx0.10$). 
This means that the induced interaction is optimized for a particular
boson/fermion ratio given by
\begin{equation*}
  \frac{\R_F^{2/3}}{\R_B} \approx 2.88\,\abb \:,
\end{equation*}
or, in terms of energies,
\begin{equation*}
  \frac{\ef}{\mu_B} \approx 1.74\,\frac{m_B}{m_F} \:.
\end{equation*}

For example, for the case of mixtures of \Rb\ (boson) and \K\ (fermion)
($\abb=5.2$ nm~\cite{kem02}),
this relationship gives 
$\R_F=1.8\times10^{14}$ cm$^{-3}$ for the case $\R_B=10^{16}$ cm$^{-3}$, 
or $\ef=138$ nK for $\mu_B=36$ nK.
\index{gap!optimized in three dimensions}
Estimating the resulting $p$-wave gap for the optimal boson concentration by
means of Eq.~(\ref{eq:aver-gap}), we get for these densities
$\De_1/\mu_F\sim 10^{-7}$, which corresponds to an unattainable 
\index{critical temperature!for pairing transition}critical temperature.

\section{Two-dimensional mixtures}
\label{sec:pair-2d}

Pairing in two dimensions has the peculiar feature that, for an attractive
$s$-wave interaction between two different fermionic species, a two-particle
bound state (of binding energy $E_b$) is always present, no matter how weak
the attraction is. 
Therefore the system enters the strong-coupling regime even at
low densities~\cite{schmitt,randl,rand,petrov03},
forming a Bose condensate of fermion pairs characterized by
\begin{align*}
 \mu_F &\raw \ef - E_b/2 \:, \\
 \De_{s} &\raw \sqrt{2 E_b e_F} \:,
\end{align*}
where $\De_{s}$ is the $s$-wave pairing gap in 2D.
This is in clear contrast to the three-dimensional case for two reasons: (a)
in 3D, a two-body (\ie, in a vacuum) bound state requires a minimum strength
of the (attractive) potential, and (b) a bound state in the many-body system
is only very weakly bound in the low density limit
(the gap vanishes exponentially with $k_F \rightarrow 0$). 

In three dimensions a small excess of one type of fermions implied an
important reduction of the gap, and eventually its disappearance even for very
small asymmetries (see Sect.~\ref{sec:swave}). 
The situation in the strong-coupling regime is very different. 
Here, the system will form all possible pairs, while the remaining
particles just stay in their original unpaired states, in close analogy with
the behavior of nuclear matter at densities low enough to allow for the
formation of deuterons (deeply bound neutron-proton pairs)~\cite{nozieres}.

We therefore exclude in the following this `trivial' case,
and focus on 
treating a system of identical (spin-polarized) fermions.
The first possibility of pairing concerns the $p$-wave pairing gap, 
$\De_1 \equiv \De_{L=1}(k_F)$, which in the low-density limit 
is given by the 
\index{weak-coupling limit!in two dimensions}weak-coupling result~\cite{rand}
\be
 \frac{\De_1}{\mu_F} = 
 c_1 \exp\!\left[-\frac{2\pi\hbar^2}{m_F \t_F}\right] \:,
\label{eq:wc}
\ee
where $c_1$ is a constant of order unity
and
\begin{align}
 \label{eq:tf}
 \t_F &= \t_F^{(L=1)} (k_F, k_F; 2\mu_F)
 = \int_{0}^{\pi} \frac{d\phi}{\pi} \cos{\phi} 
 \lla \kvp | \t(2\mu_F) | \kv \rra \;,\\
 &|\kv|=|\kvp|=k_F \:,
 \qquad\cos{\phi} = \bm{\widehat{k}}' \cdot \bm{\widehat{k}} \;,\notag
\end{align}
\index{T-matrix!in gap equation}
is the relevant $\t$-matrix element of the interaction, 
computed at collision energy $2\mu_F$.
Note that the two-dimensionality of the system is reflected already in
the integration to be performed over the angle $\phi$, in contrast to the 3D
case, where we could integrate directly over 
$z=\cos(\phi)$ [cf. Eq.~(\ref{eq:linkk-3d})]. 

\begin{figure}
  \caption[Diagrammatic contributions to the induced interaction]
    {\label{fig:dia}
      (a) Polarization interaction $\Gamma$ between two fermions (dashed
      lines) mediated by the presence of bosons (solid lines).
      The labels indicate the momentum and energy of each line.
      For condensate bosons and fermions
      on the Fermi surface, $\hv=\zv, \omega=0$. 
      (b) Diagrams contributing to the boson bubble in RPA; 
      the last one is an example of a backwardgoing diagram, negligible when
      $\mu_B\raw0$. 
      Here, thick solid lines are full \index{propagator}propagators, 
      thin solid lines are free
      propagators, and wiggles represent interactions.
    }
  \begin{center}
    \includegraphics[height=4.5cm]{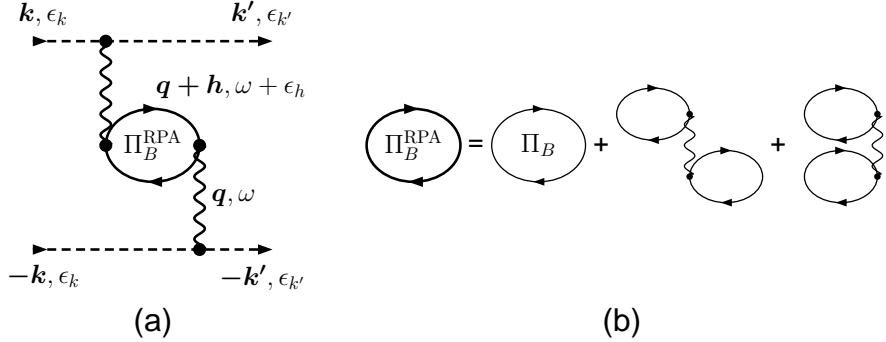} 
  \end{center}
\end{figure}

Let us consider a homogeneous mixture of bosons and fermions, and disregard
for the moment the possibility of a direct $p$-wave interaction between the
latter.
The most important contribution to the $FF$ interaction is then given by the
density fluctuations in the boson condensate sketched in 
Fig.~\ref{fig:dia}.

\subsection{Constant \mbox{$\t$}-matrices}
\index{T-matrix!in two dimensions!assumed constant}
As a first approach to the problem, we take $\tbf$ and $\tbb$ to be constants
as in three dimensions. Thus, the treatment is completely analogous to that of 
Sect.~\ref{sec:pair-3d}, and we just have to evaluate the integral in
Eq.~(\ref{eq:tf}). We obtain
\begin{align}
 \Gamma^{(L=1)}_{k_F k_F} 
 &= {\tbf^2} \int_{0}^{\pi} \frac{d\phi}{\pi} \cos{\phi} \; 
    \Pi^*_B(\sqrt{2(1-\cos{\phi})}k_F)
 = -\frac{\tbf^2}{\tbb} h_{\rm 2D}\left( x \right) \:,
 \label{eq:linkk-2d}
\end{align}
where $x =\hbar^2k_F^2/(m_B\tbb\rho_B)$ as before, and
\begin{equation}
 h_{\rm 2D}(x) := \frac{2+x}{x\sqrt{1+x}} - \frac{2}{x} \:.
 \label{eq:h2d}
\end{equation}
This function has a maximum located at
($x^{2D}_{\rm opt}=2/(\sqrt{2}-1)\approx 4.828,
~h^{2D}_{\rm opt}= 3-2\sqrt{2}\approx 0.172$), 
and can in its vicinity be approximated by a parabola,
as shown in the inset of Fig.~\ref{fig:h-3d}.
This translates into a sharp Gaussian peak for the gap
function, according to the 
\index{weak-coupling limit!in two dimensions}weak-coupling 
result~(\ref{eq:wc}).

Note that, even though $x^{3D}_{\rm opt}$ and $x^{2D}_{\rm opt}$ are both
dimensionless and similar in magnitude, the change in dimensionality
translates into different relationships of these parameters with the
densities. In fact, $x\propto\R_F^{2/3}/\R_B$ in three dimensions while
$x\propto\R_F/\R_B$ in 2D. Thus, the maximum boson-induced fermion-fermion
attraction would be reached in two dimensions for a density ratio
\begin{equation}
  \frac{\R_F}{\R_B} = \inv{2\pi}\frac{2}{\sqrt{2}-1}\frac{m_B\tbb}{\hbar^2}
  \approx 0.7685 \frac{m_B\tbb}{\hbar^2}
  \label{eq:xopt}
\end{equation}
if the collision $\t$-matrices were considered constant.

\subsection{Energy-dependent \mbox{$\t$}-matrices}
\index{T-matrix!in two dimensions!energy-dependent}

In two dimensions the $s$-wave scattering matrices $\tbf$ and $\tbb$
are not constant for a vanishing collision energy, but present a
logarithmic dependence on the c.m.s.~energy $E$ of the two-particle state
\cite{adhi,morgan}, \ie,
\be
 \lla \kvp | \T(\pv_{\rm CM}=\zv,E\raw 0) | \kv \rra \raw
 \frac{2\pi\hbar^2}{m_{\rm red}} \inv{\ln{(E_0/|E|)}} \:,
\label{eq:2dt}
\ee
where $m_{\rm red}$ is the reduced mass of the colliding particles
and $E_0\gg |E|$ is a parameter (with dimensions of energy) characterizing 
low-energy scattering. 
Therefore, we need to determine what are the possible collisions occurring in
the mixture and what is the scattering energy corresponding to each one.
As we consider elastic collisions ---\ie\ no change in the internal state of
the atoms is permitted---, the relevant energy for the collision will be given 
by the square root of the invariant $E_{\rm inv}^2=Q_\nu Q^\nu$ ($Q$ is the 
total 4-momentum of the collision), minus the rest energy.

\begin{figure}
  \caption[Possible collision events in a Bose-Fermi mixture]
	  {\label{fig:coll}
	    Possible collision events in the mixture, according to
	    Eq.~(\ref{eq:col}). Dashed lines denote fermions,
	    solid lines bosons, and wiggles represent interactions. 
	    The labels indicate the momentum and energy of each particle.
	  }
  \begin{center}
    \includegraphics[width=0.7\textwidth] 
		    {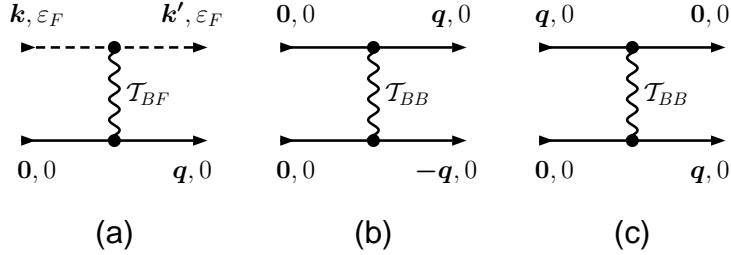}
  \end{center}
\end{figure}

As indicated above, we are initially neglecting direct $FF$ interactions, so
we need to consider only the following events (shown diagrammatically in
Fig.~\ref{fig:coll}):

\begin{subequations}
\label{eq:col}
\begin{itemize}
\item[(a)] a fermion on its Fermi surface (\ie, with 4-momentum 
  $(\hbar\kv,E_F=\sqrt{m_F^2+\hbar^2k_F^2})$ in the laboratory frame) scatters
  off a boson in the condensate (with 4-momentum $(\zv, m_B)$).

  The total momentum is $Q=(\hbar\kv,E_F+m_B)$, and so, assuming
  $\hbar k_F\ll m_F$, we obtain
  \begin{equation}
    E_{\rm inv}=m_B+m_F+\inv{2}\frac{m_B}{m_B+m_F}\frac{\hbar^2k_F^2}{m_F} \:
    \Rightarrow
    E_{\rm sc} =\frac{m_{BF}}{m_F}\ef \:,
    \label{eq:col1}
  \end{equation}
  with $m_{BF}=m_B m_F/(m_B+m_F)$ the reduced mass of the boson-fermion
  system.
%
\item[(b)] two condensate bosons $(\zv,m_B)$ scatter off each other. Here
  the calculation is simple:
  \begin{equation}
    E_{\rm sc} = 0 \:.
    \label{eq:col2}
  \end{equation}
  Because of this, the contribution of the backwardgoing diagrams like the
  last one in Fig.~\ref{fig:dia}(b) vanishes.
%
\item[(c)] a boson in the condensate scatters off a non-condensate boson with
  4-momentum $(\hbar\qv,E_B=\sqrt{m_B^2+\hbar^2q^2})$. After removing a term
  $2m_B$, we get
  \begin{equation}
    E_{\rm sc}=\frac{\hbar^2q^2}{4m_B} \qquad(\hbar q\ll m_B)\:.
    \label{eq:col3}
  \end{equation}
\end{itemize}
\end{subequations}

\index{T-matrix!in two dimensions!energy-dependent}
Thus, the transition matrices now depend explicitly on energy:
\begin{subequations}
\begin{align}
  \tbb(q) &= \frac{4\pi\hbar^2}{m_B} \inv{\ln{(4 m_B E_{BB}/\hbar^2q^2)}} \: ,
  \label{eq:tbb}\\
  \tbf(k_F) &= 
  \frac{2\pi\hbar^2}{m_{BF}} \inv{\ln{(2m_F^2E_{BF}/m_{BF}\hbar^2k_F^2)}} \: ,
  \label{eq:tbf}
\end{align}
\end{subequations}
where $E_{BB}$ and $E_{BF}$ are the parameters characterizing
low-energy $s$-wave $BB$ and $BF$ collisions, respectively.

As noted above, within the approximation $\mu_B=0$ (or more precisely
$\mu_B\ll \mu_F$), events of type (b) do not contribute [$\tbb(E=0)=0$]
and should therefore be removed from the series defining
$\Pi_B^{\rm RPA}$. In practice, this is performed by replacing 
$\tbb\Pi_B \raw \tbb\Pi_B/2$ in the denominator of Eq.~(\ref{eq:rpa}). 
Thus, the polarization-induced interaction reads
\begin{subequations}
\label{eq:pwave-2d}
\begin{align}
 \Gamma_F^{(L=1)}(k_F, k_F) &=
 -\frac{m_B \tbf^2(\eta)}{2\pi} h(x,\zeta) \;,
 \\
 h(x,\zeta) &=
 \int_{0}^{\pi} \frac{d\phi}{\pi} 
     \frac{\cos{\phi}}{x(1-\cos{\phi})-\inv{\ln{[\zeta(1 - \cos{\phi})]}} }
     \;,
\end{align}
\end{subequations}
where we introduced $x={\hbar^2 k_F^2}/(4\pi\R_B)={\R_F}/{\R_B}$, and 
the following dimensionless parameters
\begin{align*}
  \eta(k_F) &:= \frac{\hbar^2k_F^2}{2m_F}\frac{m_{BF}}{m_F}
  =\frac{m_{BF}}{m_F}\frac{\ef}{E_{BF}}  \\
  \zeta(k_F) &:= \frac{\hbar^2k_F^2}{2m_F}\inv{E_{BB}}
  =\frac{m_{F}}{m_B}\frac{\ef}{E_{BB}} \:.
\end{align*}
These give a measure of the characteristic energy of the fermions relative to
the scattering parameters $E_{BF}$ and $E_{BB}$. 
We remark that the condition $\zeta \ll 1$ must hold for the use of
Eq.~(\ref{eq:2dt}) to be valid.

\index{gap!optimized in two dimensions}
Given a fermion density (and, therefore, $\zeta$), the maximum
value of $h$ is reached for an optimal ratio $x_{\rm opt}(\zeta)$.
For $\zeta\raw0$, one obtains the following quasi-linear dependences on 
$\ln \zeta$: 
\begin{subequations}
\label{eq:opt}
\begin{align}
 x_{\rm opt}^{-1}(y) &\raw -0.58693\, \big[\ln\zeta -0.35197\big] \:,
\\
 h_{\rm opt}(y) &\raw \phantom{-}0.52022\, \big[\ln\zeta -4.3257\big] \:. 
\end{align}
\end{subequations}
\begin{figure}
  \caption[The optimal values $x_{\rm opt}$ and $h_{\rm opt}$ for the 
           induced interaction
          ]
	  {\label{fig:h}
	    The optimal values $x_{\rm opt}$ and $h_{\rm opt}$ for the pairing
            interaction, Eq.~(\ref{eq:pwave-2d}).
	    The dashed lines indicate the asymptotic behaviors,
           Eq.~(\ref{eq:opt}).
	  }
  \begin{center}
    \includegraphics[width=0.49\textwidth,height=0.38\textwidth]
		    {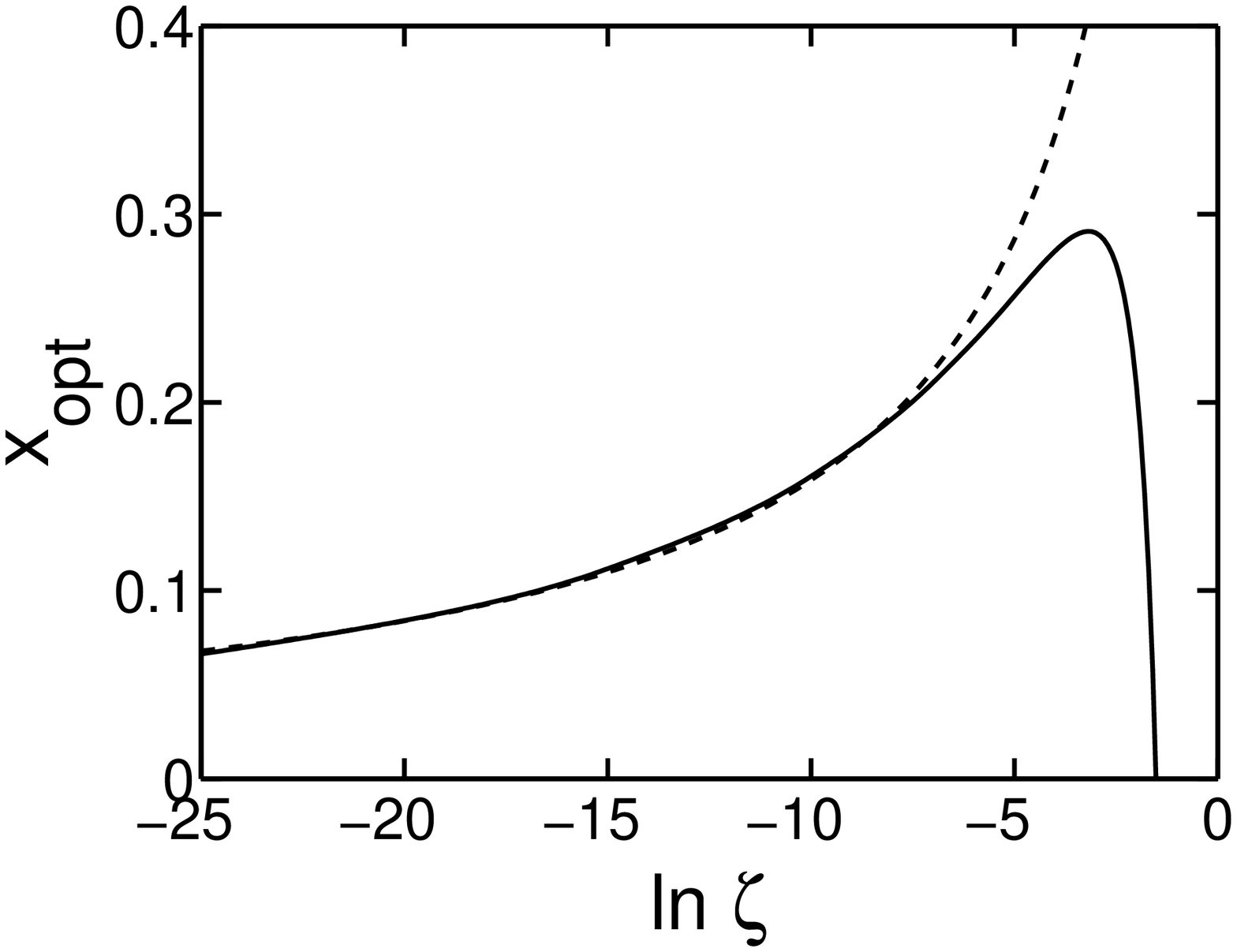}%
    \hfill
    \includegraphics[width=0.47\textwidth,height=0.38\textwidth]
		    {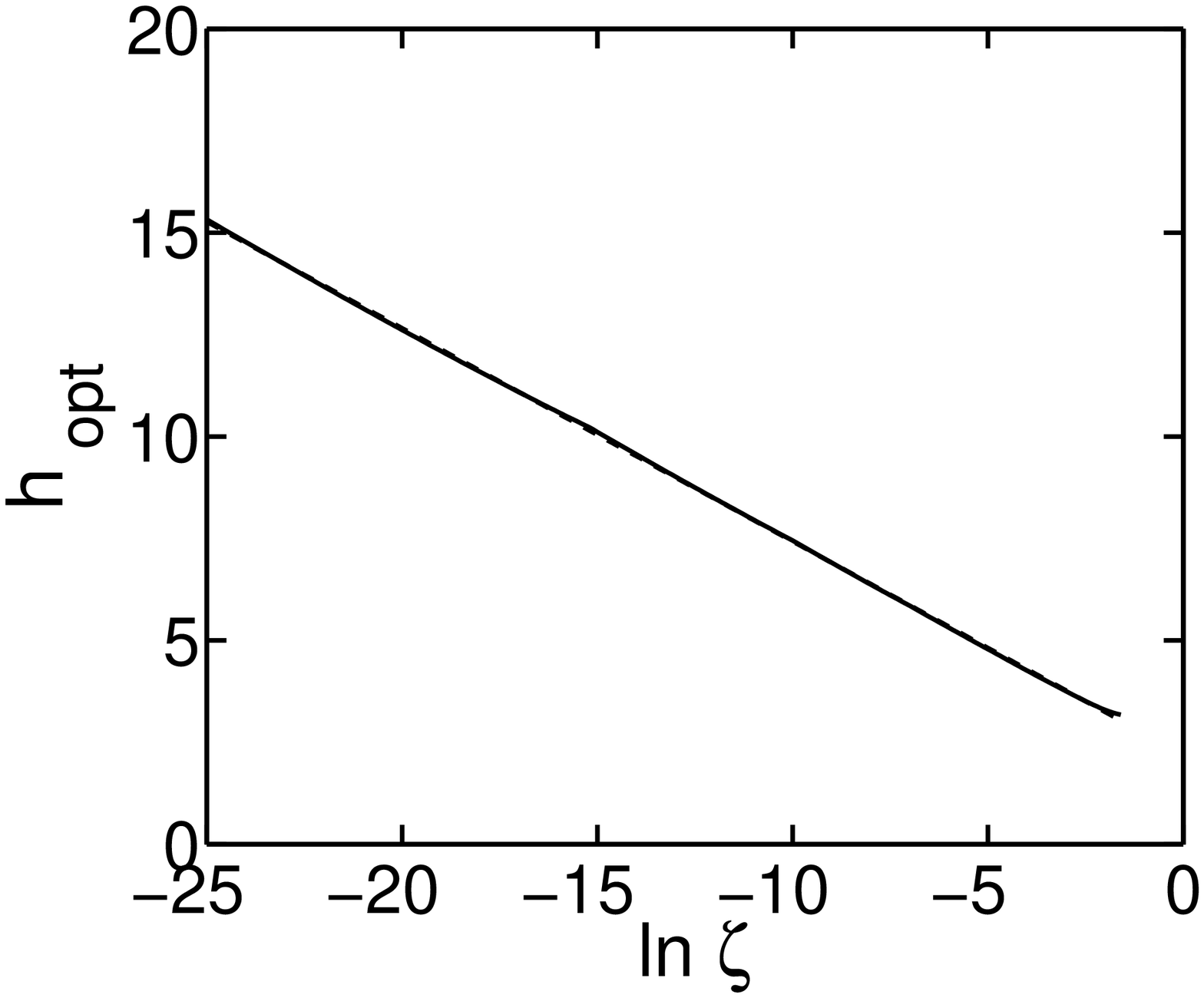}
  \end{center}
\end{figure}
\index{gap!optimized in two dimensions}
The accuracy of these approximations can be observed in Fig.~\ref{fig:h}
that shows the numerically computed optimal value $x_{\rm opt}(\zeta)$ 
[left panel] and the corresponding $h_{\rm opt}(\zeta)=h(x_{\rm opt}, \zeta)$
[right panel] as a function of $\ln\zeta$ (solid lines), compared with the
approximations above (dashed lines). One readily notices the very good
agreement between both calculations, especially for $h_{\rm opt}$, for which
already at $\zeta \approx \exp(-2) \approx 0.13$ the two curves become
indistinguishable at this scale.

Taking all these facts into account, 
the value of the pairing gap under optimal conditions becomes
\be
 \ln\frac{\De_1}{c_1\mu_F} \raw 
 - \frac{m_{BF}^2}{m_B m_F} \;
   \frac{\big[\ln{\eta}\big]^2}{0.52022\big[\ln{\zeta} - 4.3257\big]}
 \:.
\label{eq:optgap}
\ee
This means that the position and value of the maximum induced interaction
will depend logarithmically on the Fermi momentum. 

We compare now the importance of this induced interaction with a possible
direct fermion-fermion $p$-wave interaction, which at low density is given
by~\cite{rand},
\begin{equation*}
 T_F^{(L=1)}(k_F,k_F;2\mu_F) 
 \approx \frac{4\hbar^2}{m_F} \frac{\mu_F}{E_1} 
 \propto \rho_F \:,
\end{equation*}
where $E_1$ is the parameter characterizing 2D low-density $p$-wave scattering.
As the boson mediated attraction, Eqs.~(\ref{eq:pwave-2d},\ref{eq:opt}), at low
density depends only logarithmically on the fermion density, it will dominate
in this limit.
For the same reason, any fermionic polarization corrections can also
be neglected.

Finally, we analyze the assumption $\mu_B \ll \mu_F$ used above.
The boson chemical potential is deter\-mi\-ned
by~\cite{schick,fisher,stoof93,lee}
\begin{equation*}
  \mu_B = \rho_B \tbb(E=\mu_B) =
  \frac{4\pi\hbar^2\rho_B}{m_B} \inv{\ln{(E_{BB}/\mu_B)}} \:,
\end{equation*}
while for free fermions we have
\begin{equation*}
  \mu_F =  \frac{2\pi\hbar^2\rho_F}{m_F} \:.
\end{equation*}
Since the logarithm in the low-density domain is always large,
we have the sufficient condition 
\begin{equation*}
  \frac{2m_F}{m_B}\frac{\R_B}{\R_F} \lesssim 1 ~~\Rightarrow~~
  \frac{\rho_F}{\rho_B}\equiv x \gtrsim \frac{2m_F}{m_B} \:.
\end{equation*}
which will be well satisfied in atomic mixtures ($2m_F/m_B\sim 1$) in the
regime of validity of~(\ref{eq:opt}).

\subsection{Prospects of experimental detection}
\begin{figure}
  \caption[The pairing gap for optimal boson concentration]
	  {\label{fig:d}
	    The pairing gap for optimal boson concentration,
	    Eq.~(\ref{eq:optgap}), as a function of 
	    $\eta=m_{BF}\ef/(m_F E_{BF})$ and
	    $\zeta=m_{F}\ef/(m_B E_{BB})$:
	    (left) for a \Rb-\K\ mixture; (right) comparison with a 
	    \Lin-\Lix\ mixture for fixed $\zeta$.
	  }
  \begin{center}
    \includegraphics[width=0.49\textwidth]
		    {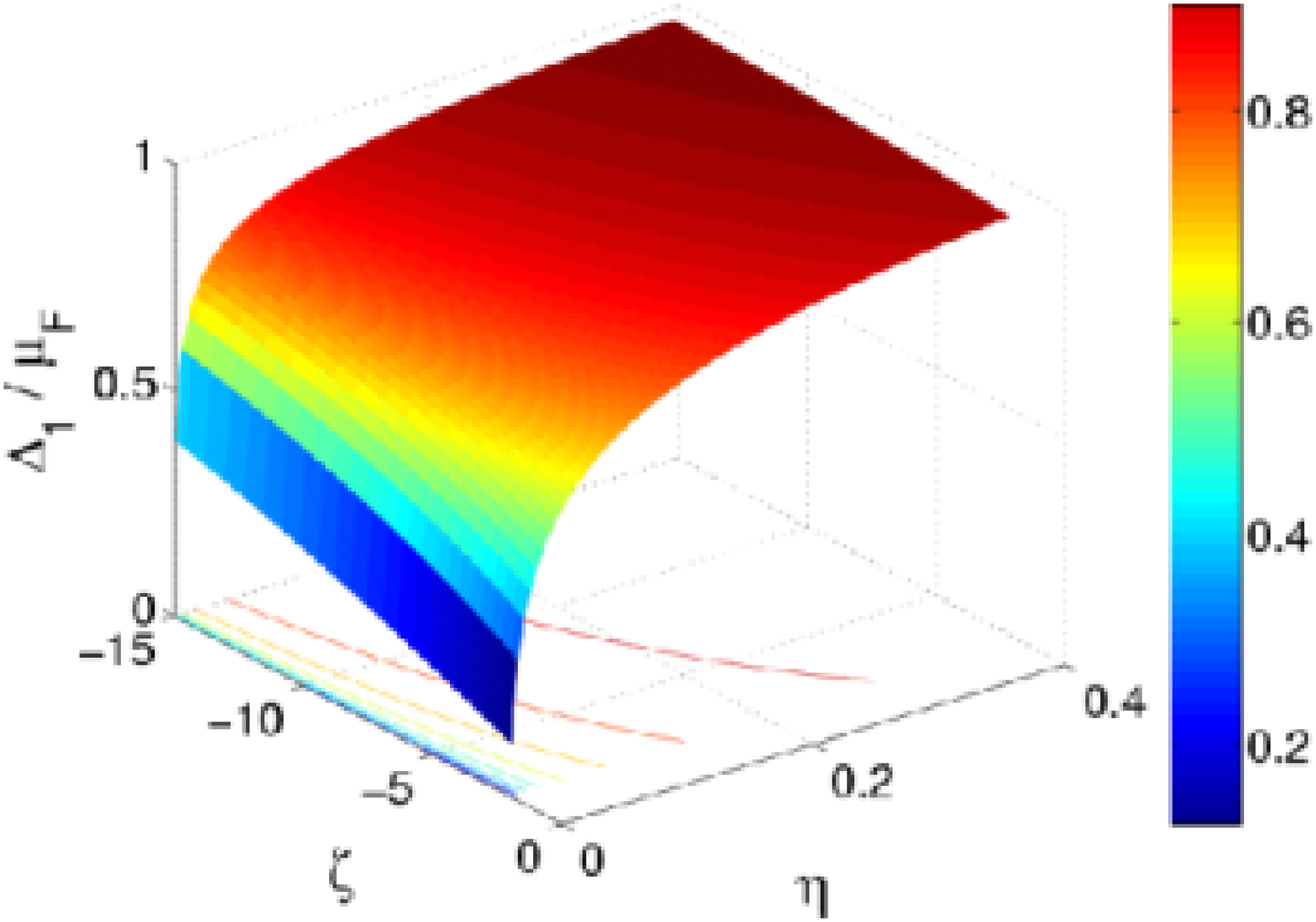}
    \hfill
    \includegraphics[width=0.49\textwidth]
		    {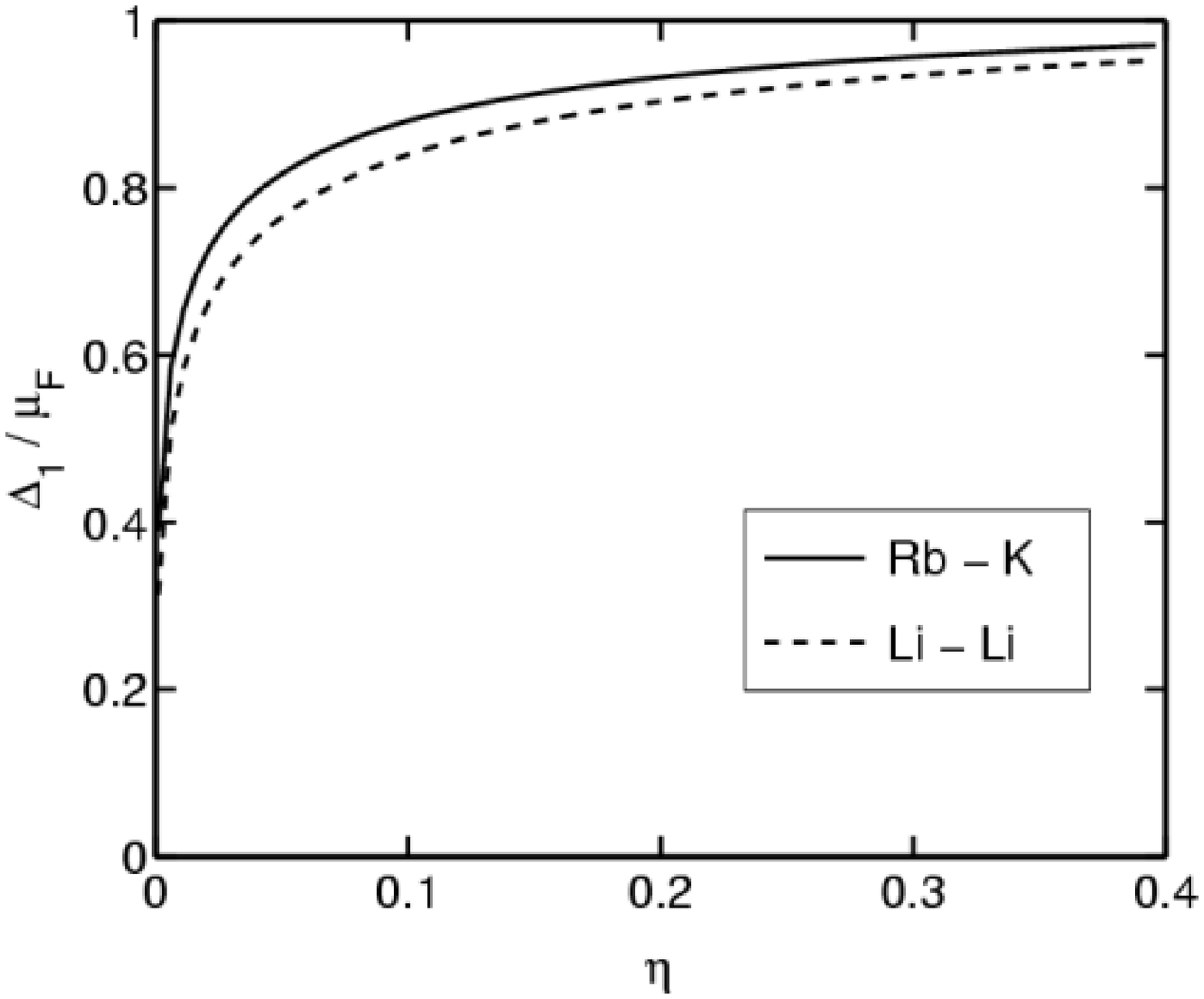}
  \end{center}
\end{figure}

In order to estimate typical sizes of the expected gap,
we plot in Fig.~\ref{fig:d} the gap $\De_1/\mu_F$, 
according to Eq.~(\ref{eq:optgap}),
as a function of the 
parameters $\eta$ and $\zeta$ defined above
(assuming for simplicity $c_1=1$). 
On the left we show the calculation for a \Rb-\K\ mixture.
We see that quite large gaps $\De_1\sim\mu_F$ are achievable, and that 
it is mainly $\eta\propto\ef/E_{BF}$ that determines the size of the gap. 
Indeed, the lines of constant value of the gap are almost parallel to
the $\zeta$ axis, as shown by the contour lines in the $\eta-\zeta$ plane,
except for $\eta\gtrsim 0.1$, when the gap is already large
($\De_1/\mu_F\gtrsim0.8$).
On the right part of the figure, we compare these results for the \Rb-\K\
mixture with the corresponding ones for a \Lin-\Lix\ mixture. The general
behavior is essentially the same, but the gap is a little bit smaller in the
last case, the difference being due to the different value of the masses'
coefficient in Eq.~(\ref{eq:optgap}), whose value is 0.22 for the first
mixture and 0.25 for the second one. Even though this difference is small, the
fact that appears in the exponential gives rise to the difference in the
figure.
In summary, we expect quite large gaps $\De_1 \lesssim \mu_F$ with
fermion chemical potentials $\mu_F \lesssim E_{BF}$, while $E_{BB}$ should
play a minor role.

In order to translate this condition into experimental quantities,
we use the results of Refs.~\cite{petrov00,petrov01,lee}, 
relating the 2D scattering parameter $E_0$ to the value of the 3D scattering
length $a_{3D}$, for a system confined in a strongly anisotropic trap 
characterized by frequencies $\om_\perp$ and $\om_z \gg \om_\perp$.
Assuming that these frequencies are the same for the two trapped species,
the treatment of~\cite{petrov01} is equally valid for a mixture if we
substitute the masses $m$ there by twice the reduced mass of the pair.
Thus, for the boson-fermion collision parameter, 
we obtain
\begin{equation}
 \frac{\mu_F}{E_{BF}} = 
 \frac{\pi}{B} \frac{\mu_F}{\hbar\om_z }
 \exp{\left(-\sqrt{2\pi}\frac{\lz}{a_{BF}}\right)} 
 \:,
 \label{eq:petri}
\end{equation}
where $B\approx0.915$ and $\lz=\hbar/\sqrt{2m_{BF}\om_z}$.
Since at the same time for a 2D situation the condition 
$\mu_F \ll \hbar\om_z$ must be fullfilled, 
one can only expect observable gaps if the exponential term is not too small.
We can distinguish two cases:
\begin{itemize}
\item[(i)] $a_{BF}>0$: in this case the ratio $l_z/a_{BF}$ should be minimized
  as much as possible. One possibility is to strongly compress the trap in the
  $z$ direction, or even better use a one-dimensional optical lattice to
  divide the trapped gas in a set of quasi-2D sub-systems. Another option is
  using a Feshbach resonance to enhance the repulsion; in this case, problems
  of stability might appear, as for strong interspecies repulsions the system
  may prefer to spatially separate the two species.
\item[(ii)] $a_{BF}<0$: in this case the exponential term is never small and
  pairing can be expected as long as $\mu_F/(\hbar\om_z)$ is not too
  small. Using the Thomas-Fermi approximation 
  $\mu_F=\sqrt{2N_F}\hbar\omega_\perp$ for the chemical potential of a
  non-interacting, strictly two-dimensional Fermi gas in a (in-plane) 
  harmonic trap of frequency $\omega_\perp$, this last condition can be 
  expressed as
  \begin{equation}
    \frac{\mu_F}{\hbar\om_z} = \frac{\om_\perp}{\om_z}\sqrt{2N_F} \:.
    \label{eq:ratio}
  \end{equation}
  Let us take typical values from~\cite{mod03}, where around
  $5\times10^4$ atoms of \K\ were sympathetically cooled with \Rb\ down to
  $T/T_F=0.3$ ($T_F=430$ nK) in a magnetic trap with frequencies
  $\om_z=2\pi\times 317$ Hz, $\om_a=2\pi\times 24$ Hz, and then submitted to
  an optical lattice with associated harmonic frequencies in the minima 
  $\om_l=2\pi\times 43$ kHz.
  Therefore $\om_\perp/\om_z \raw \om_z/\om_l \approx 1/136$, and we get a
  value $\sim 2$ for the ratio in Eq.~(\ref{eq:ratio}), which is close to the
  requirement for 2D behavior.
\end{itemize}

Let us note also that Eq.~(\ref{eq:petri}) applied to quasi-2D, bosonic \Rb\
trapped with the same $\om_z$ as above gives $\mu_F/E_{BB}\propto\zeta \ll 1$,
thus ensuring a broad range of $\mu_F/E_{BF}$ over which large gaps are
expected (see Fig.~\ref{fig:d}).


\section{Discussion}

Under favorable circumstances, $p$-wave pairing gaps 
of the order of the Fermi energy seem to be achievable, and
comparable to those predicted for $s$-wave pairing in quasi-2D
two-component Fermi gases~\cite{petrov03}.
However, more precise quantitative predictions cannot be made in 
this regime within the present approach since, when 
$\mu_F\approx \{E_{BF},E_{BB}\}$, 
also the asymptotic expression Eq.~(\ref{eq:2dt}) becomes invalid.
We stress that the same effect in three dimensions is less
efficient in increasing the size of the gap, and one expects
$\De_1/\mu_F\leq0.1$~\cite{pet}.

Also the stability of a Bose-Fermi mixture in two dimensions is as yet an
unexplored subject.
In three dimensions, this topic has been studied by different authors
\cite{molmer,viv00,miya,roth-feld02,pu02,search02,roth02,viv-gio02} that reach
similar conclusions.
Here we briefly comment on the implications from these
studies 
that might be applied to our case, but a more precise analysis of the
two-dimensional case would be of high interest.

According to Ref.~\cite{viv00}, in a three-dimensional boson-fermion mixture
one can expect to find one of the following situations:
\begin{itemize}
\item[(i)] a fermionic phase and a bosonic phase, separated from each other;
\item[(ii)] a fermionic phase and a boson-fermion mixture;
\item[(iii)] a single uniform mixture.
\end{itemize}
In case (i) there is no boson-fermion induced interaction nor sympathetic
cooling.
In case (ii) these problems are overcome, but only a fraction of the fermions
is efficiently cooled and can undergo the superfluid transition. 
Therefore, the interesting situation is that of case (iii). 
This can be obtained if there is attraction between bosons and fermions 
(to avoid their spatial separation), 
but in this case the system may collapse due to this same attraction as
predicted in \cite{roth-feld02} and observed in \cite{mod02}.
This will happen if, \eg, the number of
bosons exceeds some critical number $N_{\rm cr}$, which will depend on
$a_{BB}$ and $a_{BF}$. 
For a uniform system, we know that $a_{BB}>0$ is required in order to avoid
the collapse of the bosonic component. 
Roth and Feldmeier have shown that this condition also stabilizes
significantly the mixtures, even for $a_{BF}<0$, while
the case $a_{BF}>0$ rapidly gives rise to spatial separation
of the two components~\cite{roth-feld02}.

Applying these arguments to the mixtures used in typical atomic experiments, 
we see that the case \Lin-\Lix\ 
with $a_{BB}=-1.5\;\rm nm$ and $a_{BF}=2.2\;\rm nm$~\cite{tru01}
does not correspond to the optimal stability conditions.
However, the presence of the trapping stabilizes the system so that
experiments can be performed. 
On the other hand, for the \Rb-\K\ mixture,
where $a_{BB}=5.2$ nm~\cite{kem02} and $a_{BF}=-22$ nm~\cite{mod02},
the stability conditions for the homogeneous case are fully satisfied. 

In conclusion, the \Rb-\K\ mixture seems to be the best available candidate
within present-day ultracold boson-fermion mixtures to explore the outlined
possibility of a $p$-wave superfluid transition in a (quasi)-2D system, both
because of its demonstrated stability against phase separation and collapse,
and because the expected gaps $\De_1\sim\mu_F$ are larger that those for
\Lin-\Lix\ mixtures.


\chapter{Dynamics of spin-1 condensates at finite temperatures}
\label{ch:spin}

\index{spinor condensates}

\textsf{
 \begin{quote}
 Fu mio padre il primo ad accorgersi che qualcosa stava cambiando.
 Io ero appisolato e il suo grido mi svegli\`o:
 \\
 --- Attenzione! Qui si tocca!
 \\
 Sotto di noi la materia della nebula, da fluida che era sempre
 stata, cominciava a condensarsi.
 \end{quote}
 %
 \begin{flushright}
   {Italo Calvino,}
   {\em Sul far del giorno
   (Le Cosmicomiche) }
 \end{flushright}
}


\section{Introduction}

In the previous chapters we have studied the prospects for a superfluid
transition in a low density fermionic gas with two species of fermions
(chapters~\ref{ch:asym} and \ref{ch:dfs}) or in a mixture of a bosonic and a
fermionic species (chapter~\ref{ch:2dbf}). We will analyze now the behavior
of another multicomponent system, this one with only bosonic components,
namely a so-called {\em spinor condensate}.

We have seen in Section~\ref{sec:asym-intro} how alkali atoms can be trapped
by means of an inhomogeneous magnetic field. In this situation, the same
magnetic field that traps the atoms by coupling to their magnetic moment,
freezes their spin degree of freedom. Thus, the atoms behave in practice as
spinless bosons, and the structure and dynamics of low density gases can be
well described by the famous Gross-Pitaevskii equation for a scalar order
parameter $\psi(\rv,t)=\la\Psi(\rv,t)\ra$ in an external potential 
$V_{\rm ext}$,
\begin{equation*}
  i\hbar\parct{\psi(\rv,t)} = -\hM\nabla^2\psi(\rv,t) 
                              + V_{\rm ext}(\rv,t)\psi(\rv,t) 
			      + g|\psi(\rv,t)|^2\psi(\rv,t) \:,
\end{equation*}
with the coupling strength $g=4\pi\hbar^2a/M$ associated to a contact
\index{scattering length!in single-component condensate}
interaction with scattering length $a$ for particles of mass
$M$~\cite{gross,pita,dalfo-rmp}.

The situation is different when the atoms are trapped by means of an electric
field interacting with their electric dipole moment, $\bm{d}$. Even if
alkali atoms have ground states 
with vanishing $\bm{d}$, 
when they are
placed in an electric field, they acquire an
induced electric moment $\bm{d}\neq\bm{0}$. Its value is proportional to the
electric field, with the proportionality constant given by the polarizability
of the atom. In a first approximation, the polarizability is independent of 
the magnetic quantum number $m$ for atoms of the same hyperfine spin
$f$~\cite{Ho98}. Therefore, it is possible to store
atoms in all the states $|f;m=-f,\cdots,f\ra$ in an {\em optical} trap.

The experimental procedure is typically as follows: First, the atoms are
trapped with a magnetic field. Then, a set of lasers is switched on, defining
an optical trap superimposed to the magnetic one. Finally, the magnetic trap
is switched off, and the atoms remain in the optical potential. The first
realization of such a scheme was done by 
the group of W. Ketterle at \MIT\ in 1997~\cite{sk98}.
The remarkable degree of experimental control on lasers allows for a great
capacity to modify the trap in a controlled way. In particular, it permits the
production of very elongated traps, where quasi-one-dimensional atomic systems
are realized.

The ability to build such optical traps led the Ketterle group to the
production of the first spinor condensate with $^{23}$Na atoms~\cite{Sten98}.
The ground state of $^{23}$Na has total spin $f=1$ and the atoms can be in
three hyperfine states, $m=-1,0,1$. In the first stage of the
experiment, atoms in the $|f,m\ra=|1,-1\ra$ state ---which is low magnetic
field seeking--- are magnetically trapped. Then, they are transferred to the
optical  trap, where arbitrary populations of the three states are prepared
using radio-frequency transitions (Landau-Zener sweeps)~\cite{sk98}. Recently,
similar experiments have been accomplished with $^{87}$Rb atoms in
Hamburg~\cite{hamb2}, Georgia Tech~\cite{Chang04}, 
Gakushuin~\cite{Kuwamoto04} and Berkeley~\cite{Higbie05}.

It is interesting to note also that spinor condensates are closely related
to the pseudo-spin-1/2 systems realized 
in a purely magnetic trap by the group of Eric Cornell at \JILA\
since 1997~\cite{spin1o2-a,spin1o2-b,spin1o2-c}.
In these experiments, $^{87}$Rb atoms where trapped in the state
$|f=1,m=-1\ra$ and coupled via a two-photon transition to the state
$|2,1\ra$. Even though the states in the $f=2$ manifold are metastable, the
lifetime of the system (a few seconds) is long enough to perform
measurements. In such systems, remarkable results as spin waves at
$T>T_c$~\cite{McGuirk} and decoherence effects~\cite{Lewandowski} have been
observed, as well as interlaced vortex lattices with orthorombic
symmetry~\cite{schweik}. This surprising behavior, which contrasts with the
`usual' hexagonal Abrikosov lattices encountered in one-component atomic
BECs, had been foreseen among others by Mueller and Ho~\cite{mueller-ho} and
by Kasamatsu \ea~\cite{kasamatsu}.
Studies on vortex and spin textures for $f=1$ spinor condensates can be found
in~\cite{mueller} and~\cite{reij}.
Other properties of two-component Bose-Einstein condensates studied in the
literature are: collective modes~\cite{pu-bigelow,svid,kasa04}, 
spin structures in the ground and vortex 
states~\cite{ho-shenoy,fetter-svid,reij-duine}, 
solitons~\cite{ohberg-santos,busch-ang,kevre,ruos04,berloff},
formation of spin domains~\cite{kasa-tsubo}
and Josephson-like oscillations~\cite{ohberg99,lin00,sols}.
We finally quote also the experiments on cold gases with two different bosonic
atoms by the group of M. Inguscio and G. Modugno at
\LENS, who succesfully created a BEC of $^{41}$K atoms by
sympathetic cooling with $^{87}$Rb~\cite{modu-science};
other mixtures under study are
Rb-Cs~\cite{arimondo,marco-tesi},
and Li-Cs~\cite{mosk}.

The theoretical study of integer-$f$ systems in the context of ultracold gases
was initiated by T.-L. Ho~\cite{Ho98}, Ohmi and Machida~\cite{machida-jpsj}
and Law \ea~\cite{law98}.
The correct treatment of the spin degrees of freedom of the individual atoms 
requires to consider the vectorial character of the field operator in spin 
space. In the mean-field approximation, this feature is transferred to the 
order parameter,
\begin{equation}
  \la{\vec{\Psi}}\ra = \vec\psi 
  = \left( \begin{array}{c}
             \psi_f \\ \psi_{f-1} \\ \vdots \\ \psi_{-f}
           \end{array} \right)
  = \left( \begin{array}{c}
             |\psi_f|     e^{i\theta_f}     \\ 
	     |\psi_{f-1}| e^{i\theta_{f-1}} \\ 
	     \vdots \\ 
	     |\psi_{-f}|  e^{i\theta_{-f}}
  \end{array} \right) \:.
\label{eq:orderparam}
\end{equation}
Alternatively, this can be interpreted as having a different order 
parameter for each spin projection.

An additional peculiarity of these systems is that spin-exchanging
interactions allow for a transference of population between the different
components $m=-f,\cdots,f$, subject only to the conservation of the total
number of particles ${\cal N}=\sum_m N_m$ and of the magnetization of the
system, defined as ${\cal M}=\sum_m m\times N_m$, being $N_m$ the number of
atoms in the $m$ component.
For instance, two atoms in the state $|f=1,m=0\ra$ may
collide and become two atoms in the states $|1,+1\ra$ and $|1,-1\ra$. Thus,
the structure of the exact ground state, as well as the dynamics, will depend
on the spin-dependent part of the Hamiltonian.
In fact, it is the importance of the spin degree of freedom in these
multicomponent systems that has awarded them the name 
{\em spinor condensates}, even if the term `spinor' would not be rigorous
from a purely mathematical point of view.
\footnote{An easy introduction to the mathematical concept for `spinor' can
  be found in~\cite{wiki}. More formal texts are~\cite{cartan} for the
  mathematically-oriented reader, and~\cite{corson,rindler,morse-fesh} for the
  physics-oriented one.}

\section{Formalism}
As usual for ultracold gases, the low density and temperature of these systems
allow to substitute the interatomic potential by a contact pseudopotential,
characterized by its scattering length. To take into account the fact that a
collision between two spin-$f$ atoms can happen in any of the channels of
total spin $F=0,1,2,\cdots,2f$, this interaction is now written as
\begin{equation}
  V 
  =\delta(\rv_1-\rv_2)\, \sum_{F=0}^{2f} g_F {\cal P}_F \:,
  \label{eq:pot}
\end{equation}
where 
${\cal P}_F$ projects the two-particle state onto the subspace of total spin
$F$, in which the coupling $g_F$ is related to the scattering
\index{scattering length!in spinor condensate}
length $a_F$ by $g_F=4\pi\hbar^2a_F/M$. 
Symmetry constrains $F$ to take even values for identical bosons, and odd
values for identical fermions. 
Denoting by $\fv_i$ the spin of atom $i=1,2$ in a collision, and using the
identity $\bm{F}^2=(\fv_1+\fv_2)^2$, we have
\begin{equation*}
  \fv_1\cdot\fv_2 = \sum_{F=0}^{2f} \lambda_F\P_F  \:,
  \qquad
  \lambda_F=\frac{F(F+1)-2f(f+1)}{2}
  \:,
\end{equation*}
where the summation over $F$ takes into account the possibility of the
collisions occurring through different spin channels.
This expression will be useful to substitute (some of) the projectors in
Eq.~(\ref{eq:pot}) by products of atomic spin operators.

\begin{figure}[!b]
  \caption[The Euler angles $\{\al,\beta,\gamma\}$ that define a rotation]
	  {\label{fig:Euler}
	    The Euler angles $\{\al,\beta,\gamma\}$ that define a rotation
	    from the coordinate system $xyz$ to $x'y'z'$.
	  }
  \begin{center}
    \includegraphics[width=0.99\textwidth]{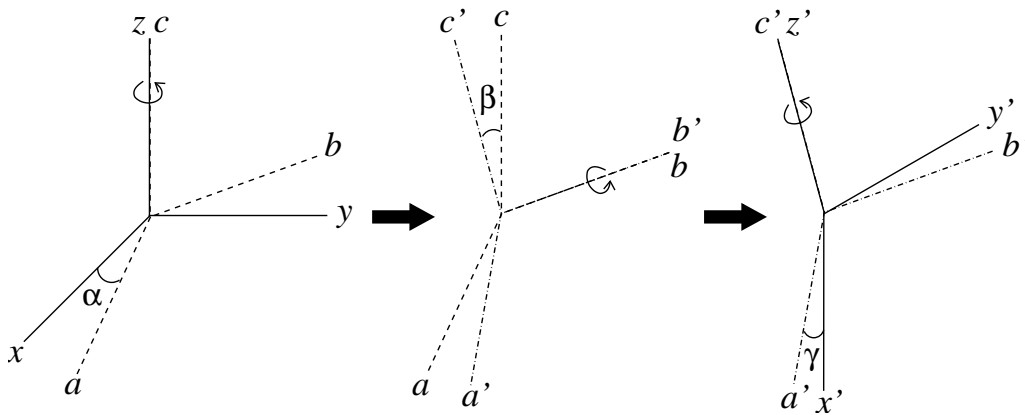}
  \end{center}
\end{figure}

We will for definiteness present the results for the case of bosons with spin
$f=1$ (for instance, $^{23}$Na and $^{87}$Rb in their electronic ground
state), while more detailed calculations for both $f=1$ and $f=2$ are given
in Appendix~\ref{app:spinor}.
In this case, the previous expressions together with the normalization of the
projectors, $\sum_F \P_F = \P_0 + \P_2 = 1$,
allow us to write the potential as
\begin{equation*}
  V   = c_0 + c_2\, \fv_1\cdot\fv_2  \:,
\end{equation*}
with the coupling strengths given by
\begin{subequations}
\label{eq:couplings}
\begin{align}
  c_0 &= \frac{g_0+2g_2}{3} = \frac{4\pi\hbar^2}{3M} \lp a_0+2a_2\rp \:, \\
  c_2 &= \frac{g_2-g_0}{3} = \frac{4\pi\hbar^2}{3M} \lp a_2-a_0\rp \:.
\end{align}
\end{subequations}
Thus, the interaction has a spin-independent part ($\propto c_0$) and a
contribution ($\propto c_2$) that couples atoms with different magnetic
quantum numbers $m$, keeping constant the total value $M=m_1+m_2$ for the
pair. 
We remark that this potential is invariant under rotations
$R(\al,\beta,\gamma)$ in spin space, where $\{\al,\beta,\gamma\}$ denote the
Euler angles defining the rotation (see Fig~\ref{fig:Euler}).

Once the potential is defined, we can write down the Hamiltonian in
second-quantized form:
\begin{multline}
  H = \int \mathrm{d}^3r \left\{
    \Psi^\dagger_m \left(-\frac{\hbar^2}{2M}{\bm \nabla}^2
                   + V_{\rm ext}\right) \Psi_m \right.   \\
\left.
    +\frac{c_0}{2} \Psi^\dagger_m \Psi^\dagger_{j} \Psi_{j} \Psi_m
    +\frac{c_2}{2}
      \Psi^\dagger_m\Psi^\dagger_{j} {\Fv}_{mk} \cdot 
                            {\Fv}_{jl} \Psi_{l}\Psi_k 
\right\} 
\:,
%
\label{eq:Hamiltonian}
\end{multline}
where $\Psi_m(\rv)$ $(\Psi_m^\dagger(\rv))$ is the field operator that
annihilates (creates) an atom in the hyperfine state $|f=1, m=1,0,-1 \rangle$
at point $\rv$, and summation over repeated indices is to be understood. 
The trapping potential $V_{\rm ext}(\rv)$ is assumed harmonic
and spin independent.
Finally, $\Fv$ denotes the vector of spin-1 Pauli matrices~\cite{Ho98,Zhang04}
\begin{equation*}
  F_x =
  \frac{1}{\sqrt2}\left(\begin{array}{ccc}
                          0 & 1 & 0 \\
                          1 & 0 & 1 \\
                          0 & 1 & 0 \\
                        \end{array}
                  \right)
  , ~~
  F_y =
  \frac{1}{i\sqrt2}\left(\begin{array}{ccc}
                           0 &  1 & 0 \\
                          -1 &  0 & 1 \\
                           0 & -1 & 0 \\
                         \end{array}
                  \right)
  , ~~
  F_z =
  \left(\begin{array}{ccc}
          1 & 0 &  0 \\
          0 & 0 &  0 \\
          0 & 0 & -1 \\
  \end{array}
  \right) .
\end{equation*}

\subsection{Ground state of the homogeneous system}
\label{sec:gs}
Let us first consider the case of a homogeneous system from a
mean-field point of view: we substitute the field operators $\Psi_m$ by
their expectation values $\psi_m$. Let us also define the spin components
$\xi_m$ of the mean-field order parameter through
\begin{equation}
  \vec\psi := \lp \begin{array}{ccc}
                   \psi_1 \\ \psi_0 \\ \psi_{-1}
		 \end{array} \rp
           \equiv \sqrt{n} 
	     \lp \begin{array}{ccc}
                   \xi_1 \\ \xi_0 \\ \xi_{-1}
	     \end{array} \rp
	     \:,
\end{equation}
where $n$ is the total density of the system and $\sum_m |\xi_m|^2=1$.
The general form to describe a state of a homogeneous spin-1 system with 
magnetization $\cal M$ is
\begin{equation*}
  \vec{\xi} = e^{i\theta} \lp 
    \begin{array}{c}
      e^{-i\phi}\sqrt{\frac{1-|\xi_0|^2+\cal M}{2}} \\
      \rule[-3ex]{0pt}{3em} |\xi_0| \\ 
      e^{i\phi}\sqrt{\frac{1-|\xi_0|^2-\cal M}{2}}
    \end{array} \rp \:,
\end{equation*}
where $\theta$ and $\phi$ are arbitrary phases, 
and $|\xi_0|\in\mathbb{R}$ is a free parameter.

The interaction part of the Hamiltonian reads
\begin{align*}
  H_{\rm int} &= \int\mathrm{d}^3r 
                 \frac{n^2}{2}\left[c_0 + c_2 \la\Fv\ra^2 \right] 
  \:, \\
  \la \Fv \ra &:= \sum_{kl} \xi_k^* \Fv_{kl} \xi_l
  \:.
\end{align*}
\index{spinor condensates!ground state structure}
Therefore, the spin structure of the ground state is determined by
the {\em sign} of $c_2\propto a_2-a_0$:
\begin{itemize}
\item \framebox{$c_2>0$:} the energy is minimized by setting $\la\Fv\ra=0$,
  and the system is usually called `antiferromagnetic'~\cite{Ho98}. 
  An analogy with condensed-matter antiferromagnetism is to be taken
  with care as there the interaction $\fv_1\cdot\fv_2$ is typically between
  spins in different positions of space, which implies a spatial ordering of
  them, while here the interaction is local. An example of atom with
  `antiferromagnetic' interactions is $^{23}$Na in the $f=1$ manifold.

  This structure for the ground state is achieved in the state
  $\vec\xi^{\,T}=(0, 1, 0)$ ($T$ denotes transpose) or more generally, due to
  rotational symmetry in spin space, in any of the states
  \begin{equation*}
    \vec{\xi}_{AF} = e^{i\theta} \lp \begin{array}{c}
                                      -\inv{\sqrt2}e^{-i\al}\sin\beta \\
				      \cos\beta \\
                                      \inv{\sqrt2}e^{i\al}\sin\beta
				     \end{array} \rp \:,
  \end{equation*}
  where $\theta$ is a global phase that does not affect 
  any expectation value, and $\{\al,\beta\}$ are the Euler angles that fix the
  quantization axis.
  This ground state presents a $U(1)\times S^2$ symmetry and is usually
  referred to as a {\em polar ground state}~\cite{Ho98}.
\item \framebox{$c_2<0$:} the energy is minimized by setting $|\la\Fv\ra|=1$,
  and the system is called `ferromagnetic'~\cite{Ho98}. The ground state
  configuration of the spin components reads
  \begin{equation*}
    \vec{\xi}_F = \lp \begin{array}{c}
                      1 \\ 0 \\0
		      \end{array} \rp
                \sim~
		  e^{i(\theta-\gamma)}
		  \lp \begin{array}{c}
		      e^{-i\al}\cos^2\frac{\beta}{2} \\
		      \sqrt{2}\cos\frac{\beta}{2}\sin\frac{\beta}{2} \\
		      e^{i\al}\sin^2\frac{\beta}{2} \\
		\end{array} \rp \:.
  \end{equation*}
  Note the coupling of the Euler angle $\gamma$ to the global phase $\theta$.
  The symmetry is now $SO(3)$~\cite{Ho98}.
\end{itemize}

A similar treatment follows for larger spin values, the main difference being
the greater number of scattering parameters that enter into the play and,
thus, the richer phase diagram that emerges. 
For instance, the case of $f=2$ has been
studied by Ciobanu \ea, who show that there are 3 possible phases for
the ground state depending on the values of the couplings 
$c_0\propto 4a_2+3a_4$, $c_1\propto 7a_0-10a_2+3a_4$
and $c_2\propto a_4-a_2$~\cite{ciobanu} (see also~\cite{koashi-ueda}).

It is also worth noticing that the presence of a homogeneous magnetic field
can drastically affect the structure of the ground state, as was analyzed by
Zhang \ea\ for the case of $f=1$~\cite{Zhang03} and for $f=2$ by Ciobanu
\ea~\cite{ciobanu} and more recently by Saito and Ueda~\cite{Saito05}.
Also the dynamics of the system is strongly influenced by external magnetic
fields as studied theoretically for the case of spin 2 by Saito and
Ueda~\cite{saito-2}, and shown experimentally by the
Hamburg~\cite{Schmaljohann04} and Gakushuin~\cite{Kuwamoto04} groups.
This notwithstanding, in order to reduce the number of parameters entering our
simulations and thus have a clearer picture of the effects of temperature, we
will restrict our analysis to the case of a vanishing magnetic field.

\subsection{Dynamical equations and transfer of population}
\index{spinor condensates!dynamical evolution of}
\index{transfer of population}
We focus now on the dynamical evolution of a {\em trapped} $f=1$ condensate
and, in particular, on the possible influence on it of thermal 
effects. 
In experiments, practically any initial configuration 
$\bm{N}(t=0)=(N_1,N_0,N_{-1})$ can be produced,%
\footnote{Here $\bm{N}$ is to be understood just as a shorthand for
 $(N_1,N_0,N_{-1})$, and not as a true vector in spin space.}
and 
it is interesting to analyze \eg\ whether the system is
expected to converge in the time scale of an experiment to its ground-state
configuration, 
or the spin structures that can appear in the dynamical process.

The equations of motion for the mean-field order parameters can be derived
from the energy expression in a mean-field formulation, 
\begin{multline}
  {\cal E}[\vec{\psi}] = \int\mathrm{d}^3r \,\bigg\{ 
    \psi^*_m \left(-\frac{\hbar^2}{2M}{\bm \nabla}^2
                   + V_{\rm ext}\right) \psi_m   \\
    +\frac{c_0}{2} \psi^*_m \psi^*_{j} \psi_{j} \psi_m
    +\frac{c_2}{2} 
      \psi^*_m\psi^*_{j} {\Fv}_{mk} \cdot {\Fv}_{jl} \psi_{l}\psi_k 
    \bigg\} \,,
\label{eq:functional}
\end{multline}
by functional differentiation according to 
$i \hbar \partial \psi_m/\partial t= \delta {\cal E}/\delta \psi_m^*$.
One readily obtains 
(see Appendix~\ref{app:spinor} for details)
\cite{Pu99,Zhang03} 

\begin{equation}
i \hbar \frac{\partial \psi_m}{\partial t} =
 \left[-\frac{\hbar^2}{2M}{\bm \nabla}^2 +V^{\rm eff}_{m}\right]\psi_{m}
           + c_2 T_{m}^*  \,,
\label{eq:dyneqs2}
\end{equation}
where we have defined the {\em transference terms}
\index{transfer of population}
\begin{equation}
  \left.\begin{array}{c}
    T^*_{\pm 1}=\psi_0^2 \psi_{\mp 1}^* \:, \\
    T^*_{0}=2 \psi_{1} \psi_0^* \psi_{-1}
  \end{array}\right.
  \label{eq:trans}
\end{equation}
which render these equations different from the usual number-conserving
Gross-Pitaevskii equation for one-component condensates.
In fact, the $c_2$-term in the Hamiltonian does not couple directly the
different $|f,m\ra$ states in which one atom can be found 
---as the radio-frequency field couples the $|1,-1\ra$ and $|2,1\ra$ states in
the \jila\ experiments---, but the two-atom states 
$|Z\ra:=|f_1,m_1\ra|f_2,m_2\ra=|1,0\ra|1,0\ra$ and 
$|U\ra:=|1,1\ra|1,-1\ra$ (properly symmetryzed).
\index{effective potentials}
The {\em effective potentials} that will determine the spatial dynamics of
each component read
\begin{equation}
  \left.\begin{array}{r@{}c@{}l}
    V^{\rm eff}_{\pm 1} &=& V_{\rm ext}+ c_0 n
                                     + c_2 \lp\pm n_{1}+n_0\mp n_{-1}\rp 
    \\
    V^{\rm eff}_{0}     &=& V_{\rm ext}+ c_0 n + c_2(n_{1}+n_{-1})
  \end{array} \right. 
  \,,
\label{eq:veff}
\end{equation}
with $n_m(\rv)=|\psi_m(\rv)|^2$ being the density of atoms in the $|1,m\ra$
state and $n(\rv)=\sum_m n_m(\rv)$ the total density, 
\index{density normalization!in spinor condensate}normalized 
to the total number of atoms ${\cal N}$.
The population of the hyperfine state $|1, m \rangle$
is $N_m=\int \mathrm{d}^3r\, |\psi_m|^2$, 
and $N_{1}+N_0+N_{-1}\equiv {\cal N}$ is a constant of the motion.

In analogy to what happens for the Gross-Pitaevskii equation of a
spin-polarized condensate~\cite{dalfo-rmp}, equations (\ref{eq:dyneqs2})
can be rewritten in the form of continuity equations, but now a balance term
\index{transfer of population}
accounting for the transfer of populations between the components will appear
on the right-hand side. Indeed, with the usual definition of the quantum
current applied to each component,
\begin{equation*}
  \bm{j}_m= \frac{\hbar}{2iM} (\psi_m^* {\bm \nabla} \psi_m 
                               -\psi_m {\bm \nabla} \psi_m^*) \:,
\end{equation*}
one readily finds
\begin{equation}
  \frac{\partial n_m}{\partial t} + {\bm \nabla} \cdot \bm{j}_m=
  \delta \dot n_m(\rv,t)
  \,, \\
\label{eq:continuity}
\end{equation}
where $\delta \dot{n}_m(\rv,t)=-(2 c_2/\hbar)\,{\rm Im}[T_m \psi_m]$
\index{transfer of population}
is the transfer of populations between spin components per unit time.

The dynamical equations for $m=\pm 1$ are 
invariant under time-reversal symmetry; 
therefore, 
$\delta \dot n_1=\delta \dot n_{-1}$, 
and the conservation 
of magnetization implies
$\delta \dot n_0 = -2\delta \dot n_{\pm 1} =
(2 c_2/i \hbar) \,(\psi_{1} \psi_0^{*2} \psi_{-1}-
\psi_{1}^* \psi_0^2 \psi_{-1}^*)$.

\section{Numerical procedure}

We have studied a system of $^{87}$Rb atoms in their electronic ground state
manifold $f=1$, trapped in a very elongated trap 
as in the Hamburg experiment~\cite{Schmaljohann04}.
In particular, we have considered 20$\,$000 atoms confined in a trap
characterized by frequencies $\om_\perp=891$ Hz and $\om_z=21$ Hz, which
results in an aspect ratio $\om_z/\om_\perp=0.024\ll1$. 
For this atom, $a_2=100.4a_B$ and $a_0=101.8a_B$~\cite{kem02}, which
results in $c_2<0$, and the expected behavior is `ferromagnetic'.
Moreover, we have simulated a purely one-dimensional system in a trap of
frequency $\om_z$, as it has been shown that elongated spin-1 systems can
be safely considered one-dimensional for most purposes~\cite{Zhang05}.

\index{effective potentials}
\index{transfer of population}
We have solved Eqs.~(\ref{eq:dyneqs2}) and analyzed them in terms of the
transference terms (\ref{eq:trans}) and the effective potentials
(\ref{eq:veff}). To this end, at each time step $dt$ we have split the
evolution in two parts: the kinetic energy part, and the rest of the
Hamiltonian. The evolution due to the kinetic energy has been solved by
transforming the wave function into momentum space by a Fast Fourier Transform
subroutine, and then the operator $\exp[-ip^2/(2m)\,dt]$ has been applied to
it. Finally, the wave function is transformed back into real space, where the
effect of the trapping potential and the coupling terms has been taken into
account by means of a fourth order Runge-Kutta algorithm. This splitting of
the full evolution in two parts introduces, in principle, higher order terms
when the separate parts do not commute with each other 
[see Eq.~(\ref{eq:bch})]. However, for short enough time steps $dt$, these
terms can be neglected. To check the accuracy of our procedure, we also solved
the evolution entirely with the Runge-Kutta algorithm, computing the gradient
terms numerically in $r$-space:
both methods give the same results for the test cases analyzed, but the split
method can make use of much larger time steps and, thus, calculations are faster.

We consider that initially a quasi-pure condensate in the $m=0$ spin component
is populated. Numerically, spin mixing requires at least a small seed of
atoms populating the other components, but we keep the total magnetization
equal to zero 
for the following reason:
It has been shown that, for the case of zero magnetization, 
not all the phases $\theta_m$ in Eq.~(\ref{eq:orderparam}) affect
the spin dynamics, but it
is governed by only one relative phase
$\Theta=2\theta_0-\theta_1-\theta_{-1}$~\cite{Pu99}%
. Therefore, we set
$\theta_1=\theta_{-1}=0$ and vary only $\theta_0$. This relative phase is a
very important parameter, 
as it can freeze completely the spin dynamics, or make it 
faster or slower~\cite{Pu99}.
In the experimental
scheme, the phases are not well controlled, and vary from shot to shot. Thus,
in order, to reproduce experimental results, we will randomly draw 20
values for $\Theta$ in the range $[0,2\pi)$, and make an average of the
corresponding results.

\section{Dynamical evolution at zero temperature}
\label{sec:spin-T0}

First of all, we perform several simulations at zero temperature, 
to be compared with the work of Yi \ea~\cite{Yi02}. These authors work in 
the so-called Single Mode Approximation (\textsc{sma}) which, based on the fact
that $|c_2|\ll c_0$, assumes $\xi_m(\rv)=\xi_m$~\cite{Pu99}. This
approximation allows 
for some analytical insight into the properties of condensates with spin
degree of freedom, and has therefore been extensively used.
However, its accuracy was analyzed by Pu \ea~\cite{Pu99}, who established
that it is only adequate for small particle numbers, when the contribution
from the $c_2$ term is negligible. As ${\cal N}$ increases, the contribution
from the spin-exchange term is more and more relevant
and the \textsc{sma} becomes invalid. 
In our system, ${\cal N}=2\times10^4$ and we have checked that indeed the
density profiles for the different components cannot be transformed into one
another by a simple rescaling, \ie\
$n_m(\rv)/ n_{m'}(\rv) \not= \mbox{ constant } \forall\rv~(m\neq m')$
(see Figs.~\ref{fig:spin2} and \ref{fig:spin5} below). 
Therefore, we will not use the \textsc{sma} in our work.

\begin{figure}[!bh]
  \caption[Population of the spin components {\em vs.} time at $T=0$]
	  {\label{fig:spin1}
	    Population of the spin components as a function of time for the
	    initial configuration 
	$(N_1/{\cal N},N_0/{\cal N},N_{-1}/{\cal N})=(0.5 \%,99 \%,0.5 \%)$
	    and $T=0$: the light, green lines stand for $N_0(t)$, while the
	    dark, black lines are $N_{\pm1}(t)$. 
	    The results obtained with one initial phase $\Theta=0$ are denoted
	    by dashed lines, while solid lines show numerical results averaged
	    over 20 random initial relative phases $\Theta$.
	  }
  \begin{center}
    \includegraphics[width=0.9\textwidth,clip=true]{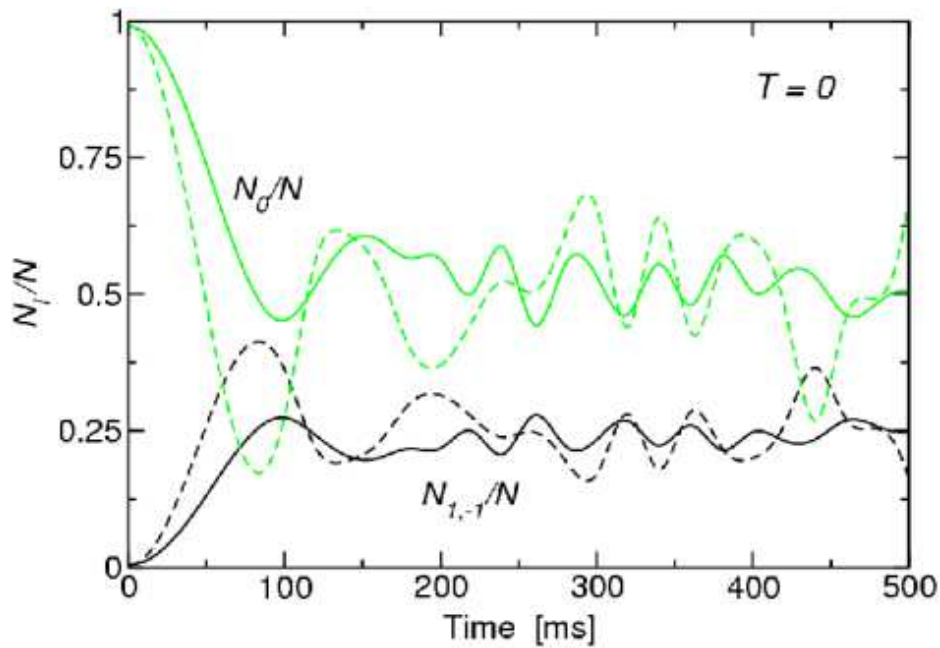}
  \end{center}
\end{figure}

The results of our simulations at zero temperature are summarized in
Figs.~\ref{fig:spin1}--\ref{fig:spin3}.
In figure~\ref{fig:spin1} we plot the population of each spin component
as a function of time, for the initial populations
$(N_1/{\cal N},N_0/{\cal N},N_{-1}/{\cal N})=(0.5 \%,99 \%,0.5 \%)$.
\index{spinor condensates!dynamical evolution of}
Dashed lines correspond to the dynamical evolution with an initial
relative phase $\Theta=0$, and solid lines stand for the average over $20$
random initial relative phases.
In absence of a magnetic field gradient the dynamical evolution of the
$m= \pm 1$ spin components is the same, and the curves $N_1(t)$ and $N_{-1}(t)$
coincide; besides 
\index{density normalization!in spinor condensate}$\sum N_m(t)={\cal N}$ 
is always fulfilled.

For the case of fixed $\Theta=0$ (dashed lines), we observe several
\index{population oscillations}
oscillations of the populations as a function of time, with a period
$\tau\sim 50$ ms. These oscillations are reminiscent of Josephson
oscillations between the two-atom states
$|1,0\ra|1,0\ra$ and $|1,1\ra|1,-1\ra$, coupled by the
$c_2$-term of the Hamiltonian, but appear to be damped. 
We have checked that these oscillations are coherent in
a homogeneous system, \ie, after a certain period 
$\tau_J\approx \hbar/(c_2\overline{n})\approx50$ ms 
($\overline{n}=10^{14}$ cm$^{-3}$) one has 
$\bm{N}(t=\tau_J)=\bm{N}(t=0)$, 
while at $t=\tau_J/2$ the populations are reversed, 
$N_{\pm1}(t=\tau_J/2)=N_0(0)/2$ and 
$N_0(t=\tau_J/2)=2N_{\pm1}(0)$.
In the case of the trapped gas, the inherent nonlinear character of the
interacting system, together with the discreteness of its 
\index{spectrum!discreteness induces collapse}spectrum, induce a
`collapse'~\cite{collapse1,collapse2} or
`dephasing'~\cite{villain,leggett-bec,oktel-levi} of the oscillations.
\index{population oscillations}
Moreover, as the period of oscillation depends strongly on the relative phase
$\Theta$, an average over various values induces a further
reduction of the amplitude of the oscillations, as shown by the solid lines in
the figure.
At much longer times, it has been shown that the full quantum solution 
presents revivals of the exchange
of populations~\cite{Pu99}. Nevertheless, this result cannot be retrieved in a
mean-field calculation such as ours~\cite{collapse2}.
%

In both our calculations (with or without average), 
the magnetization ${\cal M}=\sum_m m\times N_m$ 
is conserved during the time evolution, as it
has been experimentally observed \cite{Chang04}.
\index{population oscillations!are not coherent}
\index{population oscillations!and dynamical instability}
The oscillations between the populations of the $m=0$ and $m=\pm 1$ states
are not fully coherent but present a dynamical instability around 
$t \sim 100$ ms, when the large amplitude oscillations become small amplitude
oscillations~\cite{Saito05}.
\index{population oscillations}
At this point, the populations start to oscillate around the
ground state configuration of the system with zero magnetization which, in
absence of an applied magnetic field, is $(25\%,50\%,25\%)$~\cite{Zhang03}.
We note that these numerical results are in qualitative agreement with
the experimental measurements of Ref.~\cite{Chang04} obtained in a strongly
anisotropic disk-shaped trap, where the relaxation to the steady
state is also not monotonic but presents a few damped oscillations.
 

To further understand the spin dynamics, we now analyze
the time evolution of the density profiles of the different spin components in
the trap, in search for the possible existence of spin waves or the formation
\index{spin domains}
of spin domains. We plot the evolution corresponding to the run with fixed
$\Theta=0$ in Fig.~\ref{fig:spin2}.
\begin{figure}[!b]
  \caption[Density profiles of the spin components {\em vs.} time at
	  $T=0$]
	  {\label{fig:spin2}
	    Density profiles of the spin components at the times
	    indicated (in ms) and $T=0$, with the same color labelling as in
	    Fig.~\ref{fig:spin1}. The initial configuration corresponds to
	    $\bm{N}(t=0)/{\cal N}=(0.5 \%,99 \%,0.5 \%)$, 
	    and $\Theta=0$ (dashed lines in Fig.~\ref{fig:spin1}). 
	  }
  \begin{center}
    \includegraphics[width=0.84\columnwidth,clip=true]{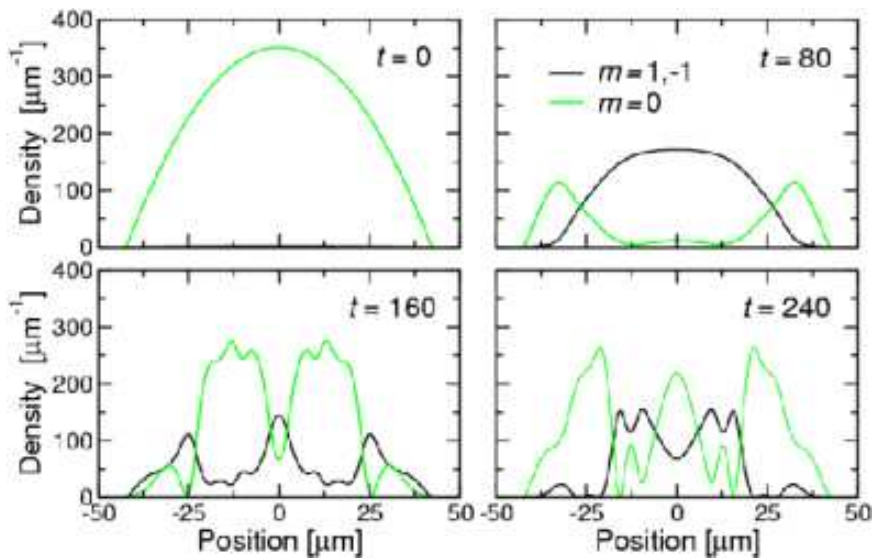}
  \end{center}
\end{figure}
At the initial stages of the evolution, the population of the $m=0$ 
component decreases due to the spin-exchange interaction, 
and the $\pm 1$ spin components 
start to be populated by equal amounts, keeping the
magnetization of the initial state. In fact, the magnetization is 
{\em locally} conserved along the time evolution, a fact not required by the
symmetries of the Hamiltonian.

For times $t < 100$ ms, the conversion of atoms from $m=0$ to $m=\pm1$ occurs
mainly in the central part of the condensate, where the density is higher and
thus the coupling among the different spin components is more effective, see
Eq.~(\ref{eq:veff}).
\index{population oscillations!and dynamical instability}
Later, at the time when the dynamical instability sets in, the $\pm1$ spin
components swing back to the $0$ component, 
\index{spin domains}
giving rise to the formation of a spin structure with small domains. 
The $m=0$ and the $m=\pm1$ domains are miscible, forming what
can be named as `mixed domains'. This fact is related to the
`ferromagnetic' character of the interaction: if in the $m=\pm1$ domains the
spin is locally oriented (anti)parallel to the quantization axis, and in the
$m=0$ domains it is perpendicular to it, in these `mixed domains' the spin is
locally oriented along a direction neither fully parallel nor perpendicular to
the quantization axis. Thus, the succession of domains should reflect a smooth
twist of the spin along the system, as also suggested by~\cite{Saito05}.

\index{spin domains}
The identical location of the $m=\pm1$ domains is due to the symmetry in the
corresponding initial profiles, together  with the fact that the dynamical
equations are symmetric under the relabelling $1\leftrightarrow-1$.
Note also that the number of small spin domains does not grow indefinitely,
\index{spin domains!size}
but it is limited by a characteristic size $l_{\rm dom}$ 
that depends on the internal coupling between different spin components.
We remark that the total density profile is constant during all the
simulation, as expected for a trapped condensate in the Thomas-Fermi regime. 
Indeed, in this regime the spin-independent interaction 
$\sim c_0 n \sim 100$ nK is dominant over the kinetic energy terms,
thus preventing the formation of total-density modulations~\cite{Dettmer02}.

\index{population oscillations!and dynamical instability}
\index{spin domains!and dynamical instability}
It has been argued that the appearance of the small domains 
is intimately related to a dynamical instability of the system. 
Therefore, the time scale of this process can be estimated from an
analysis of the 
\index{spectrum! and dynamical instability}spectrum of the system.
For two-component condensates, Kasamatsu and Tsubota~\cite{kasa-tsubo} have
\index{spin domains!time of appearance}
observed a very fast appearance of structure in the spin density distribution
of a $^{23}$Na system. They estimate the time for the growth of domains as 
$\tau_{\rm growth}=2\pi/|\Omega_-|$, where $\Omega_-$ is the excitation
frequency with largest imaginary part. In their case,
$\tau_{\rm growth}\approx 25$ ms. 

A similar analysis for $f=1$ condensates has been recently presented by Zhang
\ea~\cite{capullo}. For the case of `ferromagnetic' interactions, the
instability is expected to emerge at times $\sim 2\pi\hbar/(|c_2|n)$. 
For $^{87}$Rb at densities $n\sim n(z=0)\approx5\times10^{14}$ cm$^{-3}$, 
this estimation is around 80 ms, which is in fair agreement with our results,
cf. Fig.~\ref{fig:spin2}. They also estimate the typical size of a spin domain
\index{spin domains!size}
at long times as  $\lambda_{\rm dom}=2\pi/k_{\rm inst}$, where $k_{\rm inst}$
is the highest momentum for which an unstable mode exists 
[\ie, Im$\,\om(k_{\rm inst})\neq 0$]. With this formula, they get  
$\lambda_{\rm dom}\approx 13~\mu$m, a value consistent with our results. 
\index{population oscillations!and dynamical instability!in `antiferromagnetic' systems}
\index{spin domains!and dynamical instability!in `antiferromagnetic' systems}
For `antiferromagnetic' systems such as $f=1$ $^{23}$Na, they predict no
instability and, therefore, no domain formation, a point that we will address
in Sect.~\ref{ssec:sodi}.



\index{spinor condensates!dynamical evolution of}
\index{effective potentials}
It is possible to interpret the dynamical evolution of the various spin 
components in terms of the effective potential $V^{\rm eff}_m(z,t)$ felt
by each one 
and through the continuity equations~(\ref{eq:continuity}).
To this end, we plot in Fig.~\ref{fig:spin3}
the effective potentials $V^{\rm eff}_0$ and $V^{\rm eff}_1=V^{\rm eff}_{-1}$
[top panels], and the local transfer of population 
\index{transfer of population}
$\partial_t n_0(z,t)$ and $\partial_t n_1(z,t)=\partial_t n_{-1}(z,t)$ 
[bottom panels] at times $t=$0, 40 and 80 ms,
for the same initial conditions as in Fig.~\ref{fig:spin2}.
\begin{figure}[!b]
  \caption[Effective potentials and population transfers {\em vs.} time at
	  $T=0$]
	  {\label{fig:spin3} 
	   \index{transfer of population} 
	   \index{effective potentials}
	    Effective potentials $V^{\rm eff}_m(z,t)$ in kHz [top] 
	    and population transfers $\delta\dot n_m(z,t)$ in 
	    $\mu$m$^{-1}\cdot$ms$^{-1}$ [bottom] at $T=0$ 
	    for the initial configuration 
	    $\bm{N}(t=0)/{\cal N}=(0.5 \%,99 \%,0.5 \%)$ 
	    and $\Theta=0$ (dashed lines in Fig.~\ref{fig:spin1}), 
	    at times $t=0$, 40 and 80 ms.
	  }
  \begin{center}
    \includegraphics[width=0.84\columnwidth,clip=true]{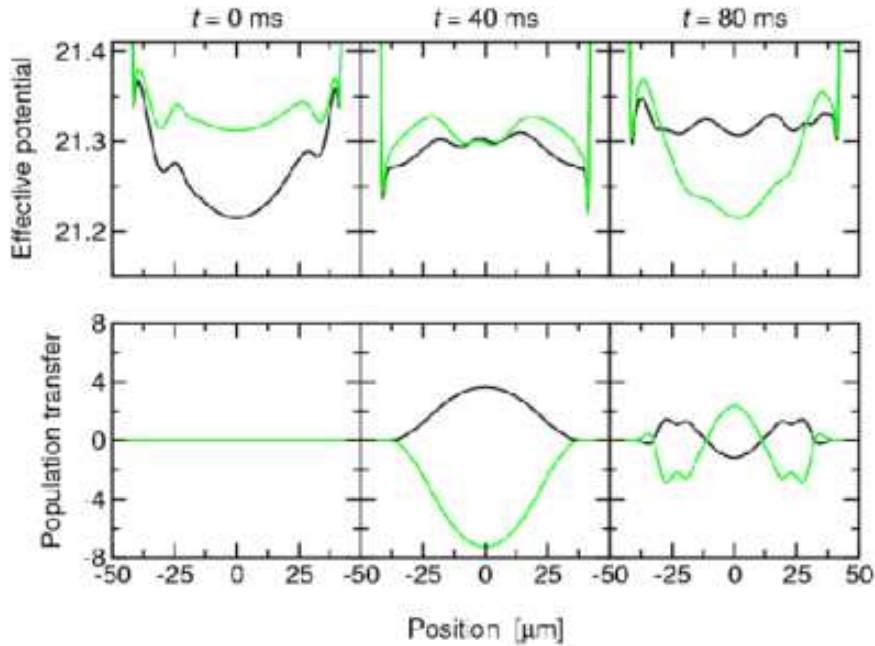}
  \end{center}
\end{figure}
As is clear in Eq.~(\ref{eq:continuity}), $\partial_t n_m$ has two 
contributions.
For a fixed time, $\delta \dot n_m(z,t)$ represents the local variation in
the occupation of the $|1,m\ra$ state due to spin-exchanging collisions,
while $-\nablab\cdot\bm{j}_m$ takes into account the change in $n_m$
due only to motion of $m$-atoms.
At the initial stages of the evolution, $t\lesssim40$ ms, 
the densities of the $m=\pm1$ components are very small and the dominant 
contribution is $\delta \dot n_m$ for all components.
For example, at $t=40$ ms $\delta \dot n_0(z,t)<0$ and consequently
$\delta \dot n_1(z,t)=\delta \dot n_{-1}(z,t)>0$: the 
$c_2$-term in the Hamiltonian is favoring the conversion of $|1,0\ra|1,0\ra$
pairs into $|1,1\ra|1,-1\ra$,
as seen in Fig.~\ref{fig:spin1} for the first stages of the evolution.
At $t=0$ this is also true, but it cannot be appreciated from the scale of the
figure.
Moreover, since the minimum and maximum of $\delta \dot n_0(z,t)$ and 
$\delta \dot n_{\pm 1}(z,t)$,  respectively are at the center of the
condensate, the spin-exchange mainly occurs at the central region, as we have
already observed in Fig.~\ref{fig:spin2}.

At $t=80$ ms $\delta \dot n_{\pm 1}$ is positive at the boundaries 
of the condensate, and negative in the central region, whereas 
$\delta \dot n_0$ has the opposite behavior. Therefore, the population with
$m = \pm 1$ is converting into $m=0$ atoms at the center while still 
increasing at the boundaries. In fact, at this time, the total tranfers
$\Delta \dot N_m(t)=\int\mathrm{d}z\,\delta\dot n_m(z,t)$ between $m=0$ and
$m=\pm1$ almost balance, which is reflected in the vanishing slope of the
curves $N_m(t)$ of Fig.~\ref{fig:spin1}.
At this stage of the evolution, the population of the $m=\pm1$ states
starts to be relevant, and the contribution of the currents is no longer
negligible. Typically, it has the same sign as the spin-exchange contribution,
but it is smaller in magnitude.

\index{effective potentials}
\index{transfer of population}
These features can also be understood in energy terms from the panels showing
the effective potentials. For example, at $t=0$ the higher energy for the
$m=0$ atoms induces them to convert into $m=\pm1$ pairs, especially in the
center of the cloud, where the difference $V^{\rm eff}_0-V^{\rm eff}_{\pm1}$
is larger.
At time $t=40$~ms, $V^{\rm eff}_0=V^{\rm eff}_{\pm1}$ at the center and, therefore,
the transfer of atoms will slow down there and reverse its sense shortly after,
while the population of the $\pm1$ components will continue to grow on the
sides of the cloud (cf. Fig.~\ref{fig:spin2}).
Later on, at $t=80$ ms, the energy of the $m=0$ component is still larger on
the boundaries than that of the $m=\pm1$ components, but it has become smaller
in the center of the trap; therefore, 0-atoms will continue to convert into
$\pm1$-atoms in the boundaries, but the reverse will happen at the center, as
confirms the lower panel. 


\section{Dynamical evolution at finite temperature}
\label{sec:spin-Tfinita}

\subsection{Introducing temperature fluctuations}
\index{thermal fluctuations}
We would like now to analyze if thermal effects may have some noticeable
effect on the dynamics of spinor condensates, making their expected evolution
differ from the results found in the previous section for $T=0$.

For single-component condensates, temperature effects on the condensate
fraction, dynamics, and damping of excitation modes have been studied
extensively. Usually, at low enough temperatures, thermal excitations 
can be accounted for within the Bogoliubov-de Gennes (\bdg)~\cite{degennes,pines2} 
or Hartree-Fock-Popov (\textsc{hfp})~\cite{ZNG99} frameworks.
Recently, this last scheme has been applied to study finite temperature effects
in the equilibrium density distribution of the condensed and non-condensed
components of $f=1$ atoms optically trapped~\cite{Zhang04}. In our work, we
will make use of the \bdg\ theory to describe the thermal clouds present in
multicomponent condensates.

For a highly elongated one-component system, it has been shown that at low
temperatures \index{thermal fluctuations}thermal fluctuations are relevant 
in the phases of the field
operator, while they can be disregarded on the density profile of the
ground state~\cite{kane-kada,reatto-ch}. 
To be precise, three regimes can be identified~\cite{Petrov00b}, according to
the value of the temperature relative to the {\em \mbox{degeneracy} temperature} 
$T_{\rm deg}={\cal N}\hbar\om$~\cite{kett-vandru} and a 
\index{critical temperature!for phase fluctuations}critical temperature
for phase fluctuations $T_\theta=T_{\rm deg}\hbar\om/\mu$: 
(a) a true condensate (density and phase fluctuations are small in the ground
state) at $T\ll T_\theta$; 
(b) a quasicondensate for $T_\theta \ll T \ll T_{\rm deg}$: the
density has the same profile as for the true condensate, but the phase
fluctuates on scales smaller than the cloud size, and the coherence
properties of the phase are drastically modified;
(c) finally, for $T\gtrsim T_{\rm deg}$, both phase and density fluctuate, and
the system is no longer quantum degenerate.

In the case of a one-dimensional, harmonically trapped single-component condensate, 
\index{thermal fluctuations!Bogoliubov-de Gennes description}
the Bogoliubov-de Gennes equations for the low-lying excitations,
\begin{align}
  \eps_j u_j &= \lp -\hm\nabla^2 + V_{\rm ext} -\mu +2gn \rp u_j + gn_0 v_j \\
 -\eps_j v_j &= \lp -\hm\nabla^2 + V_{\rm ext} -\mu +2gn \rp v_j + gn_0 u_j \:,
  \label{eq:BdG}
\end{align}
can be solved exactly~\cite{spectrum}, obtaining the corresponding
\index{spectrum! of one-dimensional trapped system}
energy eigenvalues $\eps_j=\sqrt{j(j+1)/2}\,\hbar\om$ and wave functions.
These can be written in terms of Legendre polynomials as%
\footnote{For quasi-1D condensates, the phase fluctuations at $T=0$ are
  described by Jacobi polynomials in contrast to exactly 1D condensates 
  (see~\cite{Petrov00b,spectrum}). Nevertheless, already at $T\simeq 0.2 T_c$
  quantitative difference between thermal effects in quasi-1D and strictly 1D
  cases becomes irrelevant (compare \cite{Dettmer02,Kreutzmann03}).}
\begin{equation*}
  f_j(z) = u_j(z) + v_j(z) = 
  \left\{
    \frac{2j+1}{R_{\rm TF}} 
    \frac{\mu}{\epsilon_j}
    \left[ 1-\lp\frac{z}{R_{\rm TF}}\rp^2 \right]
  \right\}^{1/2}
  \!P_j\!\lp\frac{z}{R_{\rm TF}}\rp \:,
\end{equation*}
with the chemical potential given by $\mu=\hbar\om(3{\cal N}\al/\sqrt{32})^{2/3}$,
where $\al$ gives a measure of the strength of the interactions as compared to
the trapping potential energy~\cite{Petrov00b}, and 
$R_{\rm TF}=\sqrt{2\mu/m}/\om$ 
is the Thomas-Fermi radius of the initial condensate.

\index{thermal fluctuations}
As the density fluctuations with respect to the
equilibrium profile $n_{\rm eq}(z)$ may be disregarded, one can write 
$\Psi(z,0)=\sqrt{n_{\rm eq}(z)}\exp[i\hat\theta(z)]$ for the particle
anihilation operator, with the phase operator given by~\cite{Petrov00b}
\begin{equation*}
  \hat\theta(z) =
  \inv{\sqrt{4n_{\rm eq}(z)}} \sum_j f_j(z) \hat{a}_j + {\rm H.c.} \:,
\end{equation*}
where $\hat{a}_j$ anihilates a quasiparticle in mode $j$.
Thus, in the temperature range $T_\theta \ll T \ll T_{\rm deg}$, 
where only phase fluctuations are relevant, one can simulate the 
\index{thermal fluctuations}thermal fluctuations in a
one-dimensional condensate by generating a thermal phase
$\theta^{\rm th}(z)$ according to
\begin{equation*}
\theta^{\rm th}(z) =
  \sum_{j=1}^\infty
  \left\{
    \frac{2j+1}{R_{\rm TF}} 
    \frac{\mu}{\epsilon_j}
    \left[ 1-\lp\frac{z}{R_{\rm TF}}\rp^2 \right]
  \right\}^{1/2}
  \frac{a_j+a_j^{*}}{\sqrt{4 n_{\rm eq}(z)}}
  \,P_j\!\lp\frac{z}{R_{\rm TF}}\rp
\end{equation*}
and adding it to the mean-field order parameter before the real-time
simulation starts~\cite{Dettmer01,Kreutzmann03}.
\index{thermal fluctuations}
An analogous treatment for the case of a multicomponent condensate shows
that, assuming that initially almost all the atoms are in a single component,
the \bdg\ equations for the non-condensate atoms depend mainly on the 
excited fraction in this most populated hyperfine state. Therefore, one
can simulate again temperature effect by introducing one
such fluctuating phase for each spin component,
\begin{equation}
\theta^{\rm th}_m(z) =
  \sum_{j=1}^\infty
  \left\{
    \frac{2j+1}{R_{\rm TF}} 
    \frac{\mu}{\epsilon_j}
    \left[ 1-\lp\frac{z}{R_{\rm TF}}\rp^2 \right]
  \right\}^{1/2}
  \frac{a_j^m+a_j^{m*}}{\sqrt{4 n_{\rm eq}(z)}}
  \,P_j\!\lp\frac{z}{R_{\rm TF}}\rp \:,
\label{eq:Tphase}
\end{equation}
that will be added to the corresponding $\theta_m$ of
Eq.~(\ref{eq:orderparam}).

\index{thermal fluctuations}
Here $a_j^m$ ($a_j^{m*}$) are complex amplitudes that replace the
quasi-particle annihilation (creation) operators in the mean-field
approach. In the numerical calculation, in order to reproduce the
quantum-statistical properties of the phase fluctuations,
$a_j^m$ and $a_j^{m*}$ are sampled as random numbers with zero mean and
$\langle |a_j^m|^2 \rangle=
[{\rm exp}(\epsilon_j/k_BT)-1]^{-1}$, the occupation number
for the quasi-particle mode $j$~\cite{Dettmer01}.
Let us note that this stochastic procedure will translate in different 
initial conditions for $m=1$ than $m=-1$, and will
therefore break the symmetry between them.

\subsection{Results for a `ferromagnetic' system: $^{87}$Rb }

\begin{figure}[t]
  \caption[Population of the spin components {\em vs.} time at $T=0.2T_c$]
	  {\label{fig:spin4}
	    Population of the spin components as a function of time for the
	    initial configuration 
	    $\bm{N}(t=0)/{\cal N}=(0.5 \%,99 \%,0.5 \%)$
	    and $T=0.2 T_c$.
	    The labelling of the lines is as in Fig.~\ref{fig:spin1}.
	  }
  \begin{center}
    \includegraphics[width=0.9\columnwidth,clip=true]{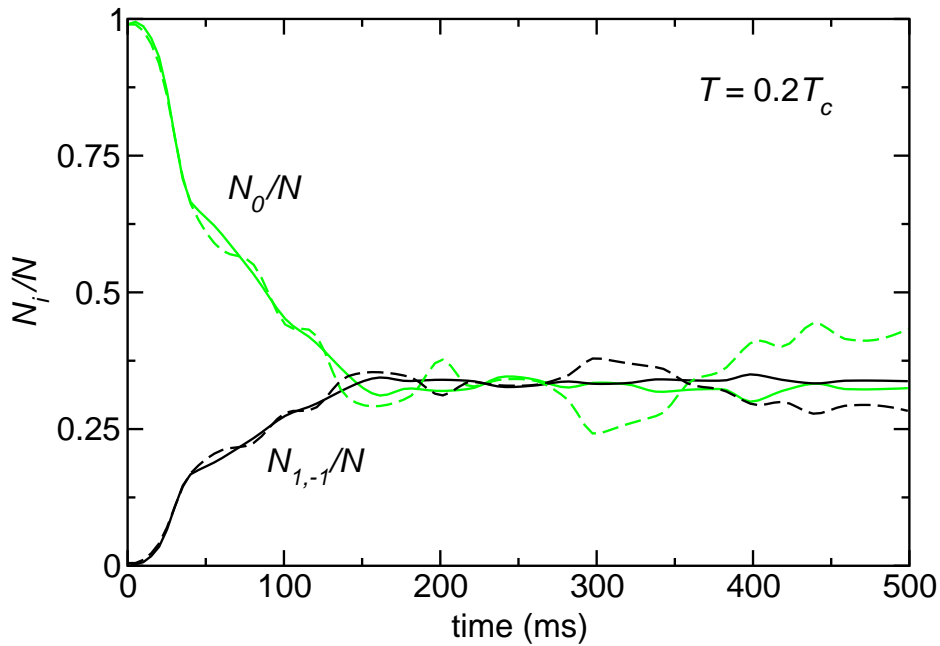}
  \end{center}
\end{figure}

We have performed a series of simulations following the scheme above
to introduce temperature fluctuations in the phases. 
The dynamical evolution
\index{spinor condensates!dynamical evolution of!at finite temperature}
of the population of the spin components that we obtain is plotted in
Fig.~\ref{fig:spin4} for the case of $T=0.2 T_c$ and the same initial
populations as in Fig.~\ref{fig:spin1}, \ie, 
$\bm{N}/{\cal N}=(0.5\%, 99\%, 0.5\%)$.
Here, $T_c={\cal N} \hbar \omega_z /\ln(2{\cal N})$ is the 
\index{critical temperature!for BEC in one dimension}critical
temperature of Bose-Einstein condensation for a single-component 1D Bose gas
in a harmonic trap of frequency $\omega_z$~\cite{kett-vandru}.
As in previous figures, solid lines correspond to the numerical
results averaged over 20 random values for the phase $\Theta$, and 
dashed lines to a single run with $\Theta=0$.
\index{population oscillations!at finite temperature}
One observes that the interaction of the condensate atoms with the thermal
clouds smears out the oscillations present at zero temperature
(cf. Fig.~\ref{fig:spin1}) and leads to an asymptotic configuration with all
components equally populated.
We note that this spin distribution has also been experimentally
obtained from some initial preparations with zero total
spin~\cite{Schmaljohann04}.
This effect can be understood in terms of the free energy of the system,
${\cal F = E}-T{\cal S}$: as the spin-related contribution to the energy, 
$\sim c_2 n\sim$ 1 nK, is relatively small compared to typical experimental 
temperatures $T_{\rm exp}\sim $ 100 nK~\cite{Sten98}, 
the entropy contribution to ${\cal F}$ dominates, and the equilibrium 
configuration will be the one that maximizes ${\cal S}$, \ie, with equipartition.

In a spin-polarized, elongated condensate, the occurrence of phase
\index{thermal fluctuations}
fluctuations due to thermal excitations induces modulations of the 
density during an expansion of the system 
in a time scale of the order of 10 ms~\cite{Kreutzmann03}.
These density fluctuations are not expected inside the trap as the mean
field prevents them, but in a time-of-flight experiment they can be used to
quantify the phase fluctuations of the trapped system and also as a
thermometry probe~\cite{Kreutzmann03}.
In a multicomponent condensate, one can follow the same argument to show that
modulations of the total density should not be expected in the trap.

\begin{figure}[tbh]
  \caption[Density profiles of the spin components {\em vs.} time at
            $T=0.2T_c$] 
	  {\label{fig:spin5}
	    Density profiles of the spin components at different
	    times (in ms) at $T=0.2 T_c$ for the same initial configuration
	    as before (dashed line in Fig.\ref{fig:spin1}).
	    Again, the total density keeps the same profile throughout the
            simulation.
	  }
  \begin{center}
    \includegraphics[width=0.85\columnwidth,clip=true]{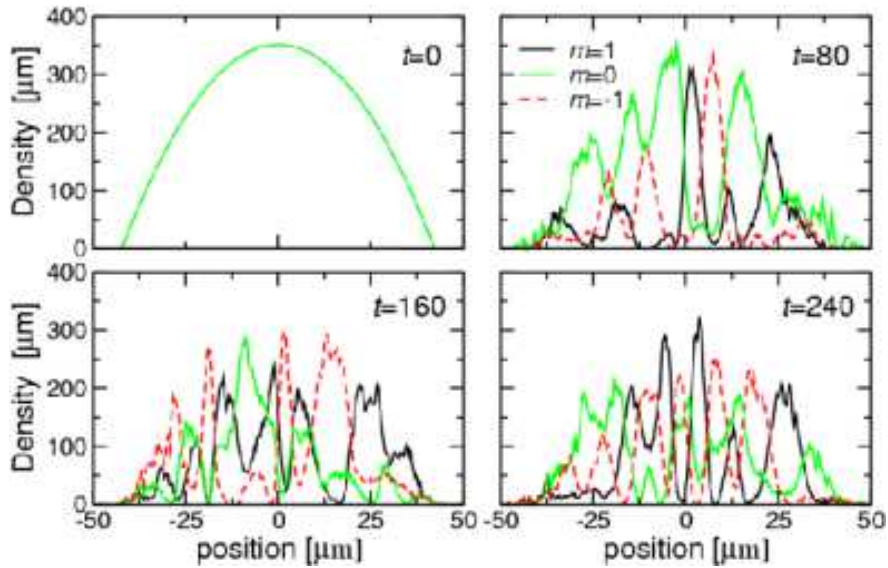}
  \end{center}
\end{figure}

\index{spin domains!as spin-density fluctuations}
However, the situation is different for the densities of the various spin
components. Indeed, the spin-exchange interaction ($\sim c_2n$) is much
weaker than the scalar mean field ($\sim c_0n$), and it is not strong enough
to prevent spin-density fluctuations, which lead to the formation of spin 
domains, cf. Fig.~\ref{fig:spin5}.
\index{spin domains!time of appearance}
This process is much faster at finite temperature than at $T=0$; 
for example, the number of spin domains at time $t=80$ ms for a temperature
$T=0.2T_c$ is larger than at time $t=240$ ms for $T=0$. 
Such a fast domain formation has also been observed in simulations of
pseudo-spin-1/2 systems~\cite{kasa-tsubo}.
This result contrasts with the relatively long times for the emergence of
domains predicted by the theory and simulations of Zhang \ea~\cite{capullo}
and obtained in the simulations by Saito and Ueda~\cite{Saito05}.
We think that this discrepancy may be due to the presence of the thermal
clouds.
Indeed, it has been argued that spin oscillations in the 
thermal cloud can considerably affect the spin dynamics in a
finite-temperature condensate~\cite{McGuirk} (see also~\cite{oktel-levi}).
With regards to this point, it is interesting to note that, 
\index{thermal fluctuations}
even though our way of introducing thermal phase fluctuations does not 
take into account possible spin correlations among the clouds, 
the fact that the initial system is almost completely in one component 
may prevent such correlations from being relevant for the later dynamics.

Moreover, the density profiles for the $m=\pm1$ components at finite
temperature are no longer equal as was the case at $T=0$. 
This is due to the existence of different thermal clouds for each
component~\cite{Lewandowski,hamb1,hamb2}, which is accounted for in the
simulation by having different random profiles for the corresponding thermal
phases, $\theta^{\rm th}_1(z) \neq \theta^{\rm th}_{-1}(z)$.
As a consequence, the local magnetization $n_1(z,t)-n_{-1}(z,t)$ is no longer
constant as at $T=0$. Nevertheless, as expected, the total magnetization is
still a conserved quantity along the time evolution.

\index{population oscillations!and dynamical instability}
\index{spin domains!size}
The typical size of the domains $l_{\rm dom}$ can be estimated from our data
to be $\sim$10 $\mu$m. Ueda~\cite{Ueda01} and Saito and Ueda~\cite{Saito05}
have studied the excitation modes of the system within a Bogoliubov-de Gennes
scheme, and determined that it should undergo a dynamical instability through
modes related to spin waves which carry angular momentum $\pm\hbar$. The most
unstable modes according to them are those with a momentum $k_{\rm inst}$ that
satisfies $2\pi/k_{\rm inst}\approx l_{\rm dom}$.
Their simulations (for 1D and 2D systems) also present this typical
domain size~\cite{Saito05}.
They also note that this size should depend on the presence
of a homogeneous magnetic field: for larger fields, the domain size would
grow, reaching $\sim 16$ $\mu$m for $B=1$ G.
This value is closer to the experimental results~\cite{capullo-nat} than
ours.
Also a recent work in an elongated 3D system by Zhang \ea~\cite{capullo} 
reports a value similar to the experimental one including in the simulations
a magnetic field $B=0.3$~G.
Nevertheless, it is also possible that other thermal effects might be
important. 
\index{thermal fluctuations}
For example, spin excitation modes should be taken into account in
our simulation on an equal footing to the phase fluctuations.
To this end, the wave functions corresponding to the 
\index{spectrum!of spin excitations}spectrum of spin
excitations found in~\cite{Ho98,Saito05,capullo} need to be calculated. Then,
they should be thermally populated at the beginning of the simulation in a
way similar to what we have done for the phase fluctuations. 

Finally, we note that the results obtained by Saito and Ueda are for a system
at zero temperature. The reason why they obtain a dynamical separation of the
$\pm1$ domains ---contrary to our results, see Fig.~\ref{fig:spin2}--- is that
their starting input profiles have a small magnetization, which triggers the
formation and development of spin waves along the system. Also, the dynamical
\index{population oscillations!and dynamical instability}
\index{spin domains!time of appearance}
instability appears in their calculations at later times $t\sim 300$~ms, even
though they have more particles in the system, ${\cal N}=4\times10^6$. These
differences are most probably due to the fact that the inhomogeneity in the
initial conditions that we impose through the phase fluctuations is more
important, and favors the appearance of domains.

\subsection{An `antiferromagnetic' system: 
  $^{87}\mbox{Rb}_\mathrm{AF}$}
\label{ssec:sodi}

\index{population oscillations!in `antiferromagnetic' systems}
After studying the system of $^{87}$Rb ($f=1$), that is `ferromagnetic',
we turn our attention to an `antiferromagnetic' case, to check the
prediction of Zhang \ea\ that for such systems no dynamical instabilities 
occur and, therefore, no domain formation is to be expected~\cite{capullo}. 
A system naturally `antiferromagnetic' is $^{23}$Na in the $f=1$ hyperfine 
manifold. However, in order to do a direct comparison with the results of 
the previous section, we prefer to work in a model system, 
which is formed by $^{87}$Rb
atoms with $f=1$ but where we have changed the sign of the coupling constant
$c_2$. In this way, only the spin dynamics is expected to change, while
nothing different should happen to the total density as compared to the
previous results. We will denote this artificial version of rubidium as
$^{87}\mbox{Rb}_\mathrm{AF}$.

\begin{figure}
  \caption[Populations {\em vs.} time for an `antiferromagnetic' system] 
	  {\label{fig:anti-pobls}
	    Population of the spin components as a function of time for the
            initial configuration $\bm{N}/{\cal N}=(0.5\%,99\%,0.5\%)$ 
	    and $\Theta=0$ for the `antiferromagnetic' system
            $^{87}\mbox{Rb}_\mathrm{AF}$ (see text). 
	    The dashed lines are the results at $T=0$ while the solid lines
            have been calculated at $T=0.2T_c$.
	  }
  \begin{center}
    \includegraphics[width=0.8\columnwidth,clip=true]
		    {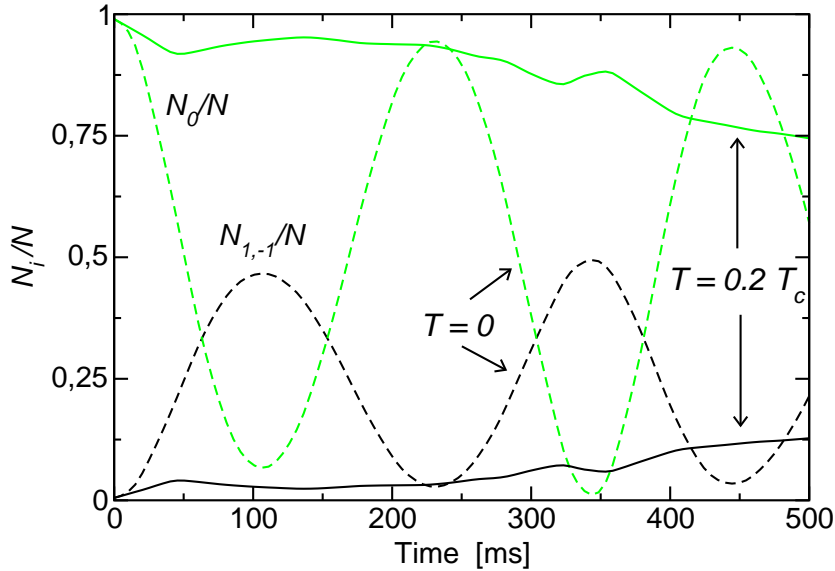}
  \end{center}
  \vskip-0.5cm
\end{figure}

\begin{figure} 
  \caption[Density profiles for an `antiferromagnetic' system at $T$=$0.2T_c$]
	  {\label{fig:anti-perfils}
	    Density profiles for the `antiferromagnetic' system
	    $^{87}\mbox{Rb}_\mathrm{AF}$ for times $t=0$, 80, 160, 240 and 500
	    ms.
	    The dashed lines are the results at $T=0$ while the solid lines
            have been calculated at $T=0.2T_c$.
            The initial configuration is $\bm{N}/{\cal N}=(0.5\%,99\%,0.5\%)$ 
	    with $\Theta=0$.
	    The $m=0$ profile is plotted in green and the $m=1$ profile in
	    black, while the red line in the last plot is for the $m=-1$
	    component. 
	  }
  \begin{center}
    \includegraphics[width=0.8\columnwidth,clip=true]
		    {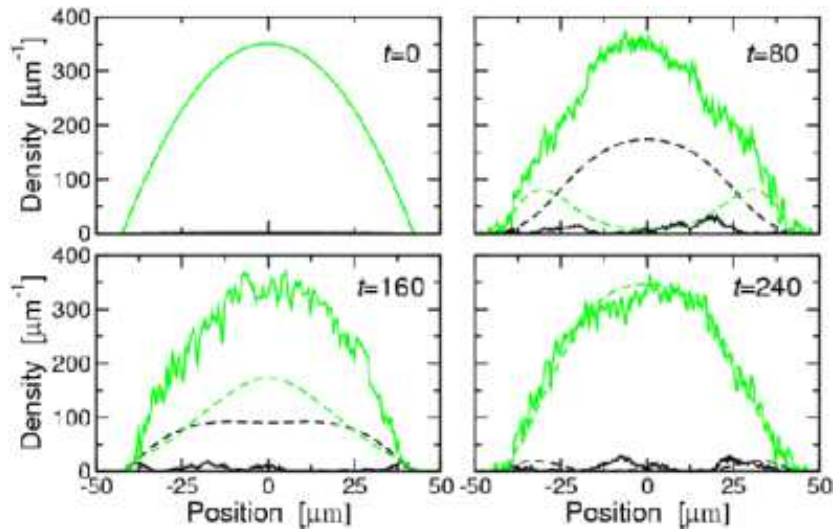}
  \end{center}
\end{figure}

As before, we have performed simulations at zero and finite temperature. The
results are presented in Figs.~\ref{fig:anti-pobls}--\ref{fig:anti-perfils}
and they are readily seen to be quite different from those corresponding to
the ferromagnetic case.
In figure~\ref{fig:anti-pobls} we present the evolution of the population of
the different spin components at zero and finite temperature. 
The zero-temperature results, displayed by dashed lines, 
consist in almost perfect
\index{population oscillations!in `antiferromagnetic' system}
oscillations between the initial population 
$\bm{N}/{\cal N}=(0.5\%, 99\%, 0.5\%)$ and the 
exact mean-field ground state $(50\%, 0\%, 50\%)$~\cite{Zhang03}. 
This behavior can be interpreted as Rabi-like oscillations of pairs of atoms
between the states $|U\ra$ and $|Z\ra$, defined above 
[see discussion after Eq.~(\ref{eq:trans})]. 
Indeed, the ground-state 
configuration is basically formed by ${\cal N}/2$ pairs in state $|U\ra$ and
the initial one 
has the same number of $|Z\ra$ pairs; moreover, both states
satisfy $|\la\Fv\ra|=0$ and are almost degenerate.
However, the exchange of populations is not complete,
and the overall data cannot be fitted by a simple sinusoidal function, but
the oscillations seem to `speed up' as time goes by: the two first maxima of
$N_0/{\cal N}$ are separated $\approx230$ ms, while the second and the third
are $\approx 215$ ms apart.
Again, this fact should be attributed to the nonlinearity of the sytem. 

The results for the simulation at $T=0.2T_c$ (solid lines) present a very
different behavior. For the case $\Theta=0$, there is
only a very slow decay of the population of the $m=0$ component, with the
corresponding increase of the number of atoms in $m=\pm1$.
This is due in part to the fact that the chosen starting configuration is
energetically very close to the ground state for the antiferromagnetic
system%
, which will result in a long thermalization time.
We also remark that in neither the zero-temperature or the finite-temperature
cases a steady-state is observed in the simulated time of 500 ms, contrary to
what happened for the `ferromagnetic' case (cf. Figs.~\ref{fig:spin1} and
\ref{fig:spin4}). One may expect, however, to converge to 
an equilibrium state with longer simulations.

\index{transfer of population}
The very slow evolution of the spin transfer in the finite-temperature case is
also reflected in the density profiles, that we present in
Fig.~\ref{fig:anti-perfils}. Again, solid lines stand for the $T=0.2T_c$ case,
while dashed lines are for the $T=0$ calculation, and the color coding is the
same as in previous figures.
At zero temperature, in the initial stages $t\lesssim80$ ms, we see an
evolution resembling that of the `ferromagnetic' $^{87}$Rb 
(compare Fig.~\ref{fig:spin2}). However, for $t=240$ ms the system practically
turns back to its initial distribution, a behavior very different from the
`ferromagnetic' one, where one could observe the formation of spin
domains. The differences are still more pronounced for the simulation at 
finite temperature, where there is essentially no evolution beyond some small
fluctuations, coming from the random thermal phases. Only at very
long times $t\sim500$ ms (not shown in the figure for clarity) the
$m=\pm1$ clouds start to be visible on the scale of the figure.

\index{spin domains!absence in `antiferromagnetic' systems}
Compared to the `ferromagnetic' case in Fig.~\ref{fig:spin5}, one remarks
the difficulty in forming separate spin domains, even though
the thermal phases are different for all components. 
This must be attributed to the `antiferromagnetic' character of the 
system under study, that tends to keep atoms in states $m=\pm1$ together 
to minimize the local spin, $|\la F\ra|(z)=0$. 
On the other hand, one expects a separation of $m=\pm1$ atoms from $m=0$ 
atoms, as observed experimentally for the case of $^{23}$Na~\cite{Sten98}, 
which also explains the difficulty in populating the
$m=\pm1$ components starting from an almost pure 0-condensate.

\section{Conclusions}

In this chapter, we have studied the spin dynamics of a spin-1 $^{87}$Rb 
condensate in a highly elongated trap, which we have modeled as a
one-dimensional system. 
We have solved the three coupled dynamical equations for the spin components 
within a mean-field framework without any further approximations. 
This is in fact necessary, since approximated treatments, such as \textsc{sma},
frequently mask some of the aspects of the dynamics.
We have also considered finite temperature effects. For one-dimensional
systems at very low temperatures, these effects manifest themselves basically
as phase fluctuations, which we have incorporated using the approach of
Ref.~\cite{Dettmer02} based on a thermo-statistical population of the
excitations of the system.

We have found that the spin dynamics towards the steady state is not monotonic 
but rather slowly damped, involving a coherent exchange of population
between the various spin components. 
\index{population oscillations!at finite temperature}
At finite temperature, the coherent oscillations of populations among the $m=0$
and $m=\pm1$ states are strongly damped. 
The internal coupling of the spin components together with the presence of
a different thermal cloud for each component lead to the formation of
\index{spin domains}
\index{spin domains!size}
numerous spin domains within the condensate. The size of these domains
defines a characteristic length scale in the system, 
$l_{\rm dom}\sim10~\mu$m, which does not decrease with time, and seems to be
intrinsically determined by a dynamical instability of the excitation modes of
the `ferromagnetic' system, without noticeable thermal effects. Indeed,
similar simulations for a system in which we have artificially 
changed the sign of the spin-exchange interaction present no formation of
domains even at finite temperatures, 
as expected from theoretical considerations~\cite{capullo}.

For a `ferromagnetic' condensate at zero temperature and with initially 
almost all atoms in the $m=0$ component and zero magnetization,
the populations of the spin components oscillate around the
configuration which corresponds to the ground state, $(25\%, 50\%, 25\%)$.
On the other hand, at finite temperature the interaction of the condensate
atoms with their thermal clouds leads to equipartition in
populations $(1/3,1/3,1/3)$.
However, recent simulations at zero temperature with a very small initial 
magnetization ${\cal M/N}=10^{-4}$ also drive the system to stationary states 
far from the ground state. 
Moreover, in these simulations formation of multiple spin
domains is also observed. Therefore, the true role of thermal effects on the
dynamics (beyond that of breaking the symmetry of the initial conditions of the
$m=\pm1$ components) is not clear yet, and deserves further consideration.

The results for an `antiferromagnetic' case are very different. 
In particular, the dynamics of populations and density profiles is very slow,
and no domains appear for times $t\lesssim500$ ms starting from an
almost-pure $m=0$ condensate. We have explained this result by the
`antiferromagnetic' character of the interactions in this system 
together with the quasi-degeneracy of the initial configuration with the
mean-field ground state.

Our results might also be relevant to the question of decoherence. 
Our simulations, and in particular the finding of very fast domain formation,
suggest that decoherence is enhanced as the number of components in the system
is enlarged.
Of course, there are many open questions connected to this, \eg, whether the
formation of domains goes along with a loss of phase relations, and gives rise
to some enhanced (generalized) phase fluctuations. It seems also interesting
\index{spin domains!time of appearance}
to study the mechanism underlying the speeding of domain formation by
temperature. To this end, one might consider incorporating spin-density
fluctuations already at the start of the simulation in a way similar to what
has been done for the phase fluctuations. 


\part{Two-dimensional helium systems\label{part:heli}}

\chapter*{Motivation: helium as `the' quantum liquid}
\addcontentsline{toc}{chapter}{Motivation}
\label{ch:motiv}

In this second part of the thesis, we aim at studying a variety of helium
systems, which are quite different from the ultracold, dilute gases.
Helium has for a long time been considered `the' quantum fluid, as many
of its physical properties can be directly related to the quantum laws:
from the fact that helium remains liquid down to zero temperature at
saturation vapor pressure, to the superfluidity present in both the bosonic
$^4$He and the fermionic $^3$He species.
In fact, since the pioneering experiments of H. Kamerling Onnes to cool down
helium and other substances~\cite{onnes,onnes-nobel}, until the more recent
claim for the observation of a supersolid phase in helium~\cite{supersolid},
the study of helium has received a lot of attention. This attention has
rewarded us with a more profound understanding not only of this atomic
species, but of the implications of quantum laws when applied to a many-body
system.

Among the most striking experiments, one can mention: the liquefaction of
helium by H. Kamerling Onnes (1908); the discovery of $^4$He superfluidity by
Pyotr Kapitza~\cite{kapi} and Allen and Misener~\cite{allen}%
; the discovery of $^3$He
superfluidity by Osheroff \ea~\cite{osheroff72a,osheroff72b}.
For later reference, we would also like to cite the production of helium
clusters by supersonic expansion of a helium gas by Becker \ea~\cite{becker}
--which finally led to the observation of the extremelly-weakly-bound helium
dimer 
by Luo \ea~\cite{luo}, and 
by Schöllkopf and Toennies~\cite{toennies}-- 
and the production of effectively low-dimensional helium systems by adsorbtion
on a graphite by Bretz \ea~\cite{bre73}.

A few theoretical works worth being cited as well are:
the study of the role of the excitation spectrum in determining the superfluid
behavior of $^4$He by Landau~\cite{landau-twofluid,landau47}
and the introduction of correlations to account for the strong repulsion at
short distances between two helium atoms by Jastrow~\cite{jastrow} and by
Feynman~\cite{feynman,feynman2}.
Finally, we would also like to emphasize the original application of density
functional techniques 
to the study of atomic quantum liquids by Stringari~\cite{str84,str85}.

Recently, a great deal of work has been devoted to study
quantum liquids in restricted geometries~\cite{kro02}.
One important feature of these systems is that their internal structure
becomes more easily observable than in bulk liquids due to the restricted
motion of the particles in the confining potential.  Among these systems the
study of quantum films has received particular attention. They consist
of liquid helium adsorbed to a more-or-less attractive flat surface. 
In 1973, M. Bretz \ea~\cite{bre73} observed for the first time 
the adsorption of $^4$He onto the basal plane of graphite. In the last
few years, adsorption properties of helium on different substrates
such as carbon, alkali and alkaline-earth flat surfaces, carbon
nanotubes and aerogels have become a fertile topic of research.

The structure and growth of thin films of $^4$He adsorbed to a substrate was 
studied by Clements \ea~\cite{cle93} employing the optimized
hypernetted-chain Euler-Lagrange theory with realistic atom-atom
interactions. It turns out that films with low surface coverages (where all
atoms cover the surface with a thickness corresponding to a single atom), can
be reasonably well described by a two-dimensional model. In connection with
these systems, an interesting question naturally arises as how physics depends
on the dimensionality of the space.

The homogeneous 2D liquid has been studied using different theoretical
methods, such as molecular dynamics~\cite{cam71} and quantum Monte
Carlo simulations, either Green's Function~\cite{whi88} or
diffusion~\cite{gio96} techniques. The inhomogeneous case was studied
by Krishnamachari and Chester who used a shadow variational wave function to
describe 2D puddles of liquid $^4$He~\cite{kri99}.

In the following two chapters we present our studies on $^4$He systems in two
dimensions (2D).
In Chapter~\ref{ch:heli-dmc} we report results of Variational (VMC) and
Diffusion Monte Carlo (\DMC) calculations. First, we introduce the Monte Carlo
techniques. Then, we show our results for 2D clusters with a finite number of
helium atoms. We study both the energetics and the density profiles of these
systems.

In Chapter~\ref{ch:heli-df} we present a similar study developed in the
framework of the Density Functional techniques, which allow us to study much
larger clusters of $^4$He.
We start the chapter giving a short introduction to theoretical background of
our work: the Hohenberg-Kohn theorem and Density Functional theory.
Then we present how to use the data obtained from the \DMC\ calculations to
set up a zero-range Density Functional. With this functional, we perform
calculations for large clusters, which would be computationally prohibitive
for a \DMC\ calculation.
Finally, we compare the results obtained with both methods.
%

\chapter[Quantum Monte Carlo study of $^4$He clusters]
    {Quantum Monte Carlo study of two--dimensional $^4$He clusters}
\label{ch:heli-dmc}

\textsf{
 \begin{quote}
 Your bait of falsehood takes this carp of truth:\\
 And thus do we of wisdom and of reach,\\
 With windlasses and with assays of bias,\\
 By indirections find directions out.
 \end{quote}
 %
 \begin{flushright}
   {William Shakespeare,}
   {\em Hamlet} (II, 1)
 \end{flushright}
}

\section{Short introduction to Quantum Monte Carlo methods}
\label{sec:qmc}

\index{QMC methods}

Quantum Monte Carlo (\qmc) methods are powerful numerical techniques to
solve the Schrödinger equation for interacting many-body systems.
There is a large variety of these methods: from the easy-to-program
Variational Monte Carlo (\vmc), to the more powerful Diffusion Monte Carlo
(\dmc) and Green's Function Monte Carlo (\textsc{gfmc}) and the
more sophisticated Path Integral Monte Carlo (\textsc{pimc})
that is able to cope with finite-temperature problems.
One can find in the literature a number of introductions to and comparisons
among the various methods, see \eg~\cite{ceperley-rmp,guardiola,hlr94} and
also~\cite{astrak-tesi,ajw}. Here we will just present a concise introduction
to the philosophy and algorithms actually used in our calculations (\vmc\ and
\dmc).

\subsection{Importance sampling and Metropolis algorithm}
\label{ssec:importance}

Let us denote by $\Phi$ a wave function corresponding to some
state of our interacting system. The postulates of Quantum Mechanics say that
\begin{equation}
  p(\Rv) = \frac{|\Phi(\Rv,t)|^2}{\intR|\Phi(\Rv,t)|^2} \mathrm{d}\Rv \;,
\end{equation}
with $\Rv=\{\rv_1,\rv_2,\cdots,\rv_N\}$, is the probability that the $N$
particles of the system (in our case, $^4$He atoms in a droplet) 
are located within a volum $\mathrm{d}\Rv$ around the
positions $\{\rv_1,\rv_2,\cdots,\rv_N\}$, at time
$t$~\cite{pascual,messiah}. Indeed, $\Phi$ contains all the information 
that could in principle be retrieved from the system. 

However, we are usually interested only on averages of operators weighed with
$p(\Rv)$,
\begin{equation*}
  \la \hat A\ra = \frac{\la \Phi | \hat A | \Phi \ra}
                       {\la\Phi|\Phi\ra}
		= \intR p(\Rv) A(\Rv) \:.
\end{equation*}
For example, the energy of a state is just the expectation value of the
Hamiltonian in such a state:
\begin{align}
  E[\Phi] &= \frac{\la \Phi | \hat H | \Phi \ra}{\la\Phi|\Phi\ra}
           = T+ V = \notag \\
  &\!\!\!\!= \frac{\sum_{i=1}^N \int\mathrm{d}\Rv\,\Phi^*(\Rv) 
           \frac{-\hbar^2\nabla_i^2}{2m}\Phi(\Rv)}
          {\intR |\Phi(\Rv)|^2}
   + \frac{\sum_{i<j}\int\mathrm{d}\Rv\, {\hat V}(\rv_i,\rv_j)|\Phi(\Rv)|^2}
          {\intR |\Phi(\Rv)|^2}
  \;,
  \label{eq:ener-phi}
\end{align}
where we assumed that only two-body interactions $\hat V$ are present.
In order to know $E[\Phi]$
not all the information contained in $\Phi$ is required,
but just some information on its spatial variations (to
evaluate the kinetic energy $T$) 
and the two-point distribution function obtained once the coordinates of $N-2$
particles have been integrated out.
Therefore, two questions must be addressed: (1) how to find a good
approximation to the wave function $\Phi$ of interest (usually, the
ground-state wave function); and (2) how to calculate these averages once the
wave function is given.

One could na\"{\i}vely think in a $\mathbb{R}^{Nd}$-generalization of the 
simple trapezoidal approximation with $M$ points to a one-dimensional 
integral,
\begin{equation*}
  \int_a^b \mathrm{d}x \, f(x) \approx \frac{b-a}{M}\sum_{k=1}^M f(x_k) \,,
  \qquad x_k=a+\frac{k}{M}(b-a) \,
\end{equation*}
the main difference being that now the integration is over $Nd$ dimensions.
However, the exponential increase in size of the (hyper)volume of
integration with the number of particles in the system, together with the fact
that most of this (hyper)space usually contributes poorly to the total value
of the integral, make such `traditional' integration rules
(trapezoidal, Simpson's, orthogonal polynomials, etc.) unapplicable, and one
has to resort to more sophisticated methods
such as Monte Carlo techniques. 

Here one evaluates averages such as those in~(\ref{eq:ener-phi})
{\em stochastically}: a number $N_W$ of points (called {\em walkers})
$\{\Rv_k\}~(k=1,\cdots,N_W)$ are randomly drawn in $\mathbb{R}^{Nd}$ 
according to the probability distribution $p(\Rv)$. 
Therefore, more points are generated where they have more weight, and very few
(if any) where the contribution to the integral is small. 
The fact that the points are sorted according to $|\Phi|^2$ is what makes
these quadrature techniques so powerful to perform multi-dimensional
integrations. One usually refers to the function used to weigh the 
configurations (in the present case, $\Phi$) as the `importance function', 
and the technique `importance sampling'.

Then, the values of the energy of the system (or whatever the observable to
estimate is) in such configurations $e_k=e(\Rv_k)$ are recorded. For a large
enough number of points $\Rv_k$, the Central Limit theorem guarantees that
\begin{equation}
  E[\Phi] = \intR e(\Rv) p(\Rv) = \inv{N_W}\sum_k e_k \:. 
  \label{eq:ener-import-sampling}
\end{equation}

\begin{figure}
  \caption[Flow chart of the Metropolis algorithm]
	  {\label{fig:metro}
	    Flow chart of the Metropolis algorithm: equilibration
	    process (left) and averaging process (right).}
  \begin{center}
    \begin{minipage}{0.495\textwidth}
      \includegraphics*[width=0.95\textwidth,totalheight=0.9\textheight]
		       {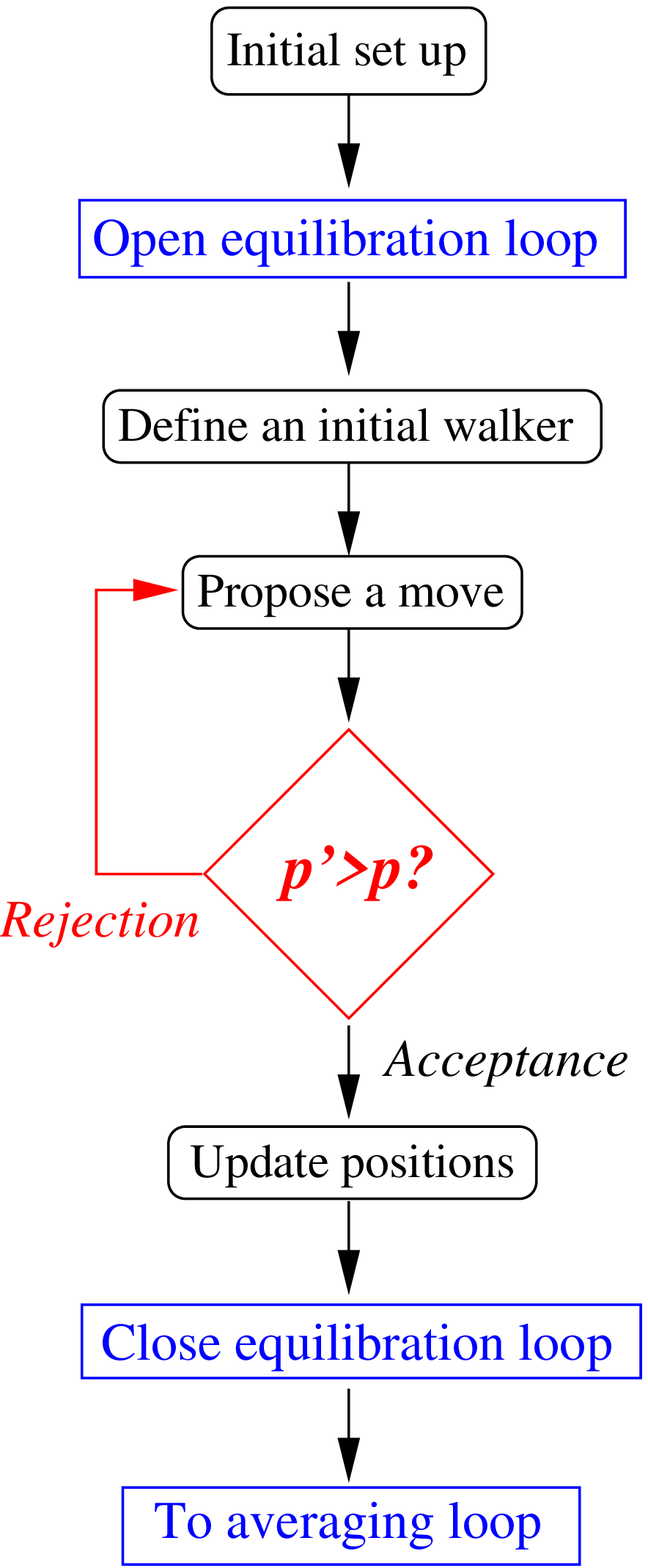}
    \end{minipage}
    \begin{minipage}{0.495\textwidth}
      \includegraphics*[width=0.95\textwidth,totalheight=0.9\textheight]
		       {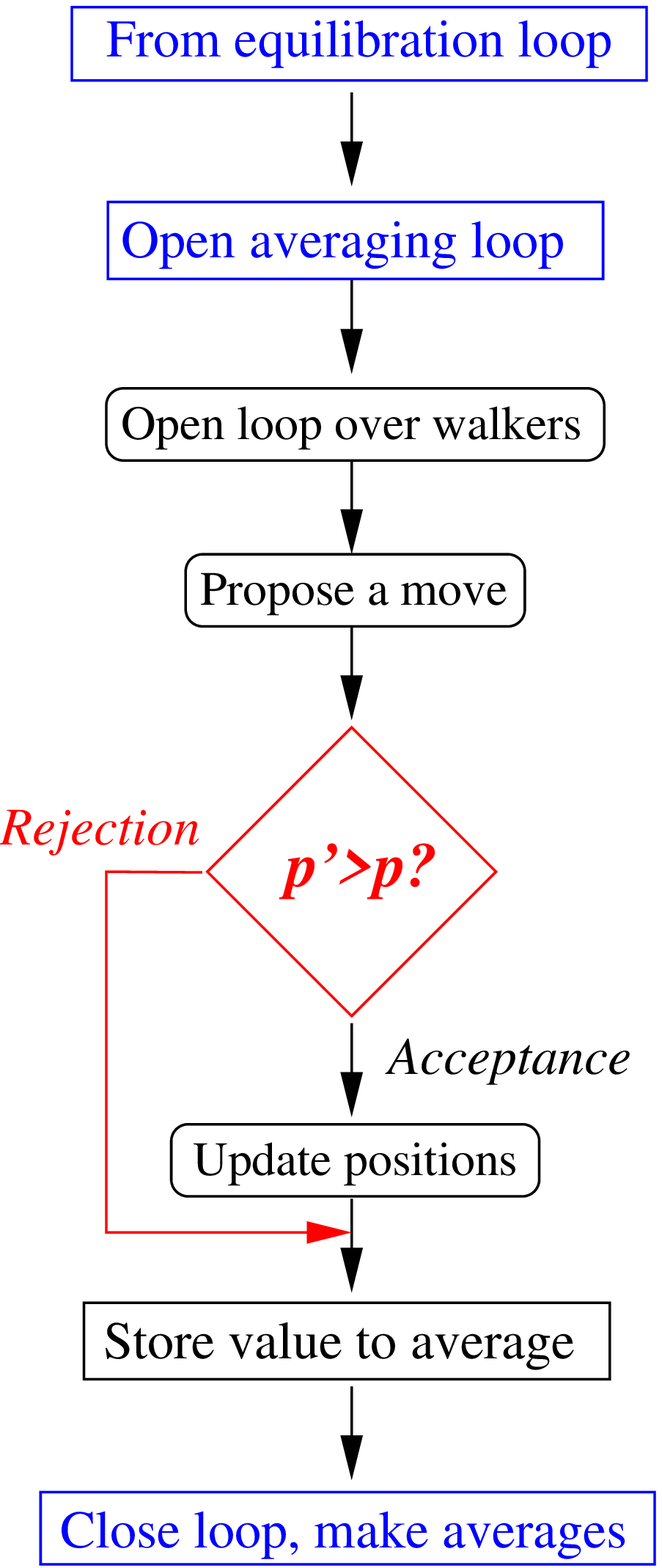}
    \end{minipage}
  \end{center}
\end{figure}

One way to obtain a set of points in $\mathbb{R}^{Nd}$ distributed 
according to $p(\Rv)$ is by means of the Metropolis algorithm~\cite{metro}:
\begin{enumerate}
\item Start by generating $N$ positions $\{\rv_i\}~(i=1,\cdots,N$) 
  distributed randomly within configuration space. 
  These points define an starting walker, $\Rv$.
\item For each position, perform a random move as
  \begin{equation*}
    \rv'_i = \rv_i + \ell \bm{\xi}_i \,,
  \end{equation*}
  where $\ell$ is the typical size of the random step, and $\bm{\xi}_i$ is a
  three-component vector whose elements are random numbers uniformly
  distributed in $[-\tfrac{1}{2},\tfrac{1}{2}]$. Thus, each particle 
  in the walker has moved from its original position to a new one 
  inside a box of side $\ell$ around it.
\item The probabilities of the system being in the configuration corresponding
  to $\Rv=\{\rv_1,\cdots,\rv_N\}$ and its proposed successor
  $\Rv'=\{\rv'_1,\cdots,\rv'_N\}$ are calculated. Then, the proposed movement
  $\Rv\rightarrow\Rv'$ is {\em accepted} with a probability:
  \begin{equation*}
    P(\Rv\raw\Rv')=
      \begin{cases}
	1              & p(\Rv')\geq p(\Rv) \\
	p(\Rv')/p(\Rv) & p(\Rv') < p(\Rv) \\
      \end{cases} \:,
  \end{equation*}
  If the movement is accepted, we set $\rv_i=\rv'_i~\forall i$. 
  Otherwise, we keep $\Rv$ as is; that is, the new walker is identical to its
  predecessor.
\item After a certain number of iterations of steps 2--3 (`equilibration
  loop'), the set of points in the last walker should be distributed 
  according to $p(\Rv)$. 
  Then, it is time to start saving data to perform the average of
  Eq.~(\ref{eq:ener-import-sampling}). To this end, one performs again a
  number of iterations of steps 2--3, but now the quantity to be averaged is
  stored with the value corresponding to $\Rv$ after the acceptance/rejection
  step (`averaging loop').
  That is, if the new walker is identical to its predecessor, 
  one must record the same value $e(\Rv)$ saved in the previous loop.
\end{enumerate}
A chart illustrating this process is in Fig.~\ref{fig:metro}.
The size of the step $\ell$ must be chosen carefully: too short a
step will result in a large acceptance of moves, but also in very correlated 
configurations and a long time to reach equilibrium. On the other hand, too
long a step will probably result in a large number of rejected moves, and also
a long time for equilibration. A good choice is the one that results in an
acceptance around $50\%-70\%$~\cite{guardiola}. 
In our calculations, $\ell\approx 6$ \AA.

It is also important to note that both the equilibration and
averaging loops have to be repeated a number of times, called {\em blocks}.
Indeed, each configuration produced in a step of the Metropolis algorithm is
strongly correlated with the previous one. Thus, a direct average as suggested
in (\ref{eq:ener-import-sampling}) would produce typically too small a
variance. In order to have uncorrelated data $e_k$ to average, one divides the
simulation in a set of blocks, each of which composed of a certain number of
simulation steps.
For each block, an average of the energies accumulated in it is performed, 
resulting in a `block estimate' for the the energy $e_{\rm block}$.
A careful analysis on the length of the blocks has to be performed 
in order to ensure that the configurations in the different blocks are
uncorrelated, so that the $e_{\rm block}$ values are effectively
independent (see~\cite{fly,night-nato}).
Once this has been done, one can finally give an estimate for the energy
and the associated uncertainty,
\begin{align*}
  \bar{e} &= \inv{N_{\rm blocks}}\sum_{\rm block=1}^{N_{\rm blocks}} e_{\rm block} \\
  \De e   &= \sqrt{ \inv{N_{\rm blocks}-1}\sum_{\rm block=1}^{N_{\rm blocks}} 
                    \lp e_{\rm block} - \bar{e}\rp^2  } \:,
\end{align*}
where $N_{\rm blocks}$ is the number of blocks in the averaging loop. 
In our simulations, typically $N_{\rm blocks} \approx 100$, and the number of 
Monte Carlo steps per block $N_{\rm steps} \approx 1000$.

\subsection{Variational Monte Carlo}

\index{QMC methods!Variational Monte Carlo (\vmc)}

The Variational Monte Carlo (\VMC) method relies on the `Variational
Theorem', which can be stated in this way: 
\newtheorem*{vtheo}{Variational theorem}
\begin{vtheo}
  The expectation value of the Hamiltonian over an arbitrary wave function
  $\Phi$ is larger than, or equal to, its expectation value on the exact
  ground state, $E_0$.
  The equality only holds if $\Phi$ coincides with the exact ground state wave
  function, $\Psi_0$.
\end{vtheo}
\begin{proof}
  This theorem is easily proved expanding $\Phi$ on the basis of eigenstates
  of the Hamiltonian $\{\Psi_n\}~(n=0,1,2,\cdots)$ (with corresponding
  eigenenergies $\{E_n\}$). Without loss of generality, we assume that $\Phi$
  is already normalized to unity, so that we can write
  \begin{align*}
    |\Phi\ra &= \sum_n c_n |\Psi_n\ra \,, \\
    c_n &:= \la \Psi_n | \Phi \ra \,, 
    \qquad |c_n|\in[0,1] \,, \quad \sum_n |c_n|^2 = 1 \:.
  \end{align*}
  Now, we insert this expression into the expectation value of the
  Hamiltonian:
  \begin{align*}
    \la \Phi | \hat H | \Phi \ra 
    &= \sum_{nm} c_n^* c_m \la \Psi_n^* | H | \Psi_m \ra 
    = \sum_{nm} c_n^* c_m \la \Psi_n^* | E_m | \Psi_m \ra =\\
    &=\sum_{nm} c_n^* c_m E_m \delta_{nm} = \sum_n |c_n|^2 E_n \geq E_0 \:,
  \end{align*}
  where we assumed that the eigenstates are properly orthonormalized and
  ordered in such a way that $E_0 \leq E_1 \leq \cdots$.
\end{proof}

In a \vmc\ calculation, one proposes a trial wave function $\Phi_T$ with one
or more (say $p$) variational parameters $\{\alpha_i\}~(i=1,\cdots,p)$. 
The functional form assumed should be based on all the available physical
information of the system, and not too many variational parameters should be
included in order to simplify the minimization process.
The scope is to find the values of the parameters $\{\overline\al_i\}$ that
minimize the expectation value
\begin{equation}
  E_T(\{\al_i\}) := E[\Phi_T] 
  = \frac{\la \Phi_T | \hat H | \Phi_T \ra}
         {\la \Phi_T | \Phi_T \ra}
 \:.
  \label{eq:ener-vmc}
\end{equation}
The corresponding wave function $\Phi(\{\overline\al_i\})$ will be the best
approximation to $\Psi_0$ with the functional form of $\Phi$.

One evaluates the expectation value $E[\Phi_T]$ stochastically as outlined
above. 
For the case of two-body interactions, 
and taking for simplicity a central interparticle potential,
$V(\rv_i,\rv_j)=V(|\rv_i-\rv_j|)$, we have
\begin{equation}
  e(\Rv)= -\sum_{i=1}^N \hm\frac{\nabla_i^2 \Phi_T(\Rv)}{\Phi_T(\Rv)}
  + \sum_{i<j} V(|\rv_i-\rv_j|) \:.
\end{equation}
Note that this expression is only meaningful for $\Phi_T(\Rv)\neq 0$. 
This is not a problem since configurations for which $\Phi_T(\Rv)=0$ 
are in fact discarded by the Metropolis algorithm.
There will be no problems in implementing this algorithm for Bose sytems,
whose ground states are known to have no zeroes~\cite{ajp02}.
The situation is more complex for fermionic systems, where Pauli's principle
forces $\Psi(\Rv)$ to vanish whenever $\Rv$ contains two indistinguishable
particles in the same position, \eg\ $\Phi(\rv_1,\cdots, \rv, \cdots, \rv,
\cdots, \rv_N)\equiv0$. This is known as the `sign problem' for fermionic
systems, and a number of methods have been developed to overcome it. However,
as we are interested in the study of $^4$He, which is bosonic and will not
suffer this problem, we will not discuss further this point, but refer the
reader to the literature~\cite{cep-web,hlr94,ajw,astrak-tesi}.

A good feature of \VMC\ calculations is the `control' on the wave function
that one has.
Indeed, the calculated value~(\ref{eq:ener-vmc}) will depend on the
functional form of $\Phi_T$ and the values of its variational parameters
$\al_i$. Therefore, one can get insight into the physics of the problem by
checking different forms for $\Phi_T$, which parameters affect more the value
of $E_T$ and, finally, what are the optimal values for these parameters.

Also, if one has been able to find the {\em exact} ground-state wave function,
$\Psi_0$, one can make averages as that in Eq.~(\ref{eq:ener-phi}) for any
operator in order to retrieve more information on the system, which might be
too difficult to extract analytically. 
For example, 
the exact ground state of a one-dimensional system of hard-core bosons is
known to be the absolute value of a Slater determinant of plane waves.
In this case, despite having the exact $\Psi_0$, analytical calculations
are quite difficult, and Monte Carlo methods have been useful to obtain
its momentum distribution, correlation functions, etc.~\cite{astrak-tesi}.

As a drawback, \VMC\ is unable to correct by itself any `bad' behavior
(symmetries, long- or short- range orders, etc.) that we may have introduced
in $\Phi$. That is, if we have assumed a wave function with a different 
symmetry than $\Psi_0$, the \VMC\ calculation will be equally valid, 
but the upper bound we obtain for the ground state energy may be far 
from its true value.
Any other expectation value that we calculate will be equally biased.
A way to overcome these problems is to perform a Diffusion Monte Carlo,
calculation which is able to 
provide the {\em exact} ground state energy for a Bose system.

\subsection{Diffusion Monte Carlo}
\label{ssec:dmc}

\index{QMC methods!Diffusion Monte Carlo (\dmc)}
The Diffusion Monte Carlo (\dmc) method consists of a stochastic solution of
the Schrödinger equation,
\begin{equation*}
  i\hbar\parct{} \Phi(\Rv,t) = \hat H \Phi(\Rv, t) \:.
\end{equation*}
We introduce an `imaginary time' variable%
\footnote{The resemblance between this variable and the time variable of
  finite-temperature 
  \index{Green's function}\index{propagator}Green's functions 
  in Sect.~\ref{sec:bcs} and Appendix~\ref{app:suma} should not lead 
  to confusion. The latter can be related to the temperature of the system, 
  while $\tau$ has no physical meaning in the Diffusion Monte Carlo formalism.}
$\tau=it$, so that
\begin{equation}
  -\hbar\parcta{}\Phi(\Rv,\tau) = (\hat H -E_T) \Phi(\Rv,\tau) \:,
\end{equation}
being $E_T$ an (arbitrary) energy shift whose utility will become clear below.
The formal time-dependent solution to this equation can be easily expressed in
terms of the eigenfunctions and eigenenergies of the Hamiltonian as
\begin{multline*}
  \Phi(\Rv,\tau) = \sum_{n=0}^\infty c_n \Psi_n(\Rv,\tau)
  = \sum_n c_n e^{-(\hat H-E_T)\tau/\hbar} \Psi_n(\Rv,0) \\
  = \sum_n c_n e^{-(E_n-E_T)\tau/\hbar} \Psi_n(\Rv,0)
  \stackrel{\tau\raw\infty}{\longrightarrow} 
  c_0 \Psi_0(\Rv,0)e^{-(E_0-E_T)\tau/\hbar} \:.
\end{multline*}
Therefore, a trial function propagated on imaginary time, will converge to
the ground state of the system, as long as it is not orthogonal to it
($c_0\neq0$). In general, the propagation will converge to the less
energetic state with a non-vanishing overlap with $\Phi$,
while the other components are exponentially suppressed.%
\footnote{This fact allows for a \qmc\ search also of {\em excited} states,
  see \eg~\cite{cep88,boro96}.} 
Moreover, if the energy shift $E_T$ equals the ground-state energy $E_0$,
$\Phi(\Rv,\tau)$ has a stationary behavior at large $\tau$, which coincides
with the ground state solution (up to a normalization constant).

Let us see now how to perform this propagation on imaginary time.
To this end, we consider a Hamiltonian with two-particle interactions $V$ and
a possible external potential $\Vext$,
\begin{equation*}
  \hat H = -\hm\sum_{i=1}^N \nabla_i^2 
           +\sum_{i=1}^N \Vext(\rv_i)
           +\sum_{i<j}V(r_{ij})
\end{equation*}
where $r_{ij}:=|\rv_i-\rv_j|$. The Schrödinger equation in imaginary time for
this Hamiltonian can be cast in the form
\begin{equation}
  -\hbar\parcta{}\Phi(\Rv,t) = -D\nabla_{\Rv}^2\Phi + V(\Rv)\Phi - E_T\Phi \:,
  \label{eq:schr}
\end{equation}
where we defined $D=\hbar^2/(2m)$ and 
$V(\Rv)=\sum_i \Vext(\rv_i) + \sum_{i<j} V(r_{ij})$, and
$\nabla_{\Rv}^2$ is a shorthand notation for $\sum_i \nabla_i^2$.
Next, we define a function $f$ as the product of the exact ground-state wave 
function, $\Psi$,%
\footnote{We drop the subscript `0' for notational simplicity.}
and a trial one
\begin{equation*}
  f(\Rv,\tau) := \Phi_T(\Rv)\Psi(\Rv,\tau) \:.
\end{equation*}
One can optimize the convergence of the \dmc\ procedure by starting from a
`reasonable' trial function, typically obtained in a previous \vmc\
calculation.
Substitution of this expression into (\ref{eq:schr}) gives an equation for $f$
that is fully equivalent to the Schrödinger equation,
\begin{equation}
  -\hbar\parcta{}f=- D\nabla_{\Rv}^2 f 
                  + D\nablab_{\Rv}\cdot\left[\bm{F}f\right]
		  + (\Eloc-E_T)f \:,
\label{eq:sch-f}
\end{equation}
with the definitions
\begin{subequations}
\begin{align}
  \bm{F} &:= \frac{2}{\Phi_T}\nablab_{\Rv}\Phi_T &\text{quantum drift force,}
  \\
  \Eloc   &:= \inv{\Phi_T} \hat H \Phi_T         & \text{local energy.}
  \label{eq:local}
\end{align}
\end{subequations}
The use of $f$ instead of $\Psi$ is recommended because it 
reduces the variance of the calculated observables.

The formal solution of Eq.~(\ref{eq:sch-f}) in $\Rv$-space can be written as
\begin{multline}
  f(\Rv,\tau)\equiv\la \Rv | f(\tau) \ra = 
  \intR' \la\Rv | e^{-(\hat H -E_T)\tau/\hbar} | \Rv'\ra \la\Rv'|f(0)\ra
  \equiv\\
  \equiv\intR' G(\Rv,\Rv',\tau) f(\Rv',0) \:,
\end{multline}
where we have introduced the 
\index{Green's function}\index{propagator}Green's function
\begin{equation*}
  G(\Rv,\Rv',\tau) := \la \Rv | e^{-(\hat H -E_T)\tau/\hbar} | \Rv'\ra \:.
\end{equation*}
Up to this point, no approximations have been made, which means that if we
were able to know the exact $G$, we could get $f(\Rv,\tau)$ for any $\tau$
and obtain from it the ground state wave function $\Psi$. 
However, usually one does not know $G$ exactly, but it can only be 
found for small time steps. 
Then, the value of $f(\Rv,\tau)$ for $\tau\raw\infty$ is found iteratively,
\begin{equation*}
  f(\Rv,\tau+\de\tau) = \intR' G(\Rv,\Rv',\de\tau)f(\Rv',\tau) \:.
\end{equation*}

To obtain a reasonable approximation for 
\index{Green's function}\index{propagator}$G$, let us split the operator
acting of $f$ in Eq.~(\ref{eq:sch-f}) into three parts as follows
\begin{subequations}
\begin{align}
  \hat H_1 &= -D\nabla_{\Rv}^2 
  \label{eq:diff} \\
  \hat H_2 &= D\left[\lp\nablab_{\Rv}\cdot\bm{F}\rp 
                     + \bm{F}\cdot\nablab_{\Rv}\right] 
  \label{eq:drift} \\
  \hat H_3 &= \Eloc(\Rv)-E_T \:.
  \label{eq:branch}
\end{align}
\end{subequations}
We define also the 
\index{Green's function}\index{propagator}Green's functions 
characteristic of these operators:
\begin{equation*}
  G_i(\Rv,\Rv',\tau) := \la \Rv | e^{-\hat H_i\tau/\hbar} | \Rv' \ra \:.
\end{equation*}
Even though the operators $\hat H_i$ do not commute with each other, we can
use the Baker-Campbell-Hausdorff formula~\cite{bch1,bch2,bch3,bch4,bch5},
\begin{equation}
  \exp{\hat X}\exp{\hat Y} = 
  \exp\lp \hat X+\hat Y+\tfrac{1}{2}[\hat X,\hat Y]+\cdots \rp \:,
  \label{eq:bch}
\end{equation}
to approximate the exponential to some fixed order for short time
lapses $\de\tau$; for example, to first order, we have
\begin{equation}
  e^{-(\hat H-E_T)\de\tau/\hbar} = e^{-\hat H_1\de\tau/\hbar} 
  e^{-\hat H_2\de\tau/\hbar} e^{-\hat H_3\de\tau/\hbar} + O(\de\tau^2)
\label{eq:order}
\end{equation}
Similar expressions can be built with a higher degree of accuracy (\ie,
valid up to a higher power of $\de\tau$). 
Then, the \index{Green's function}\index{propagator}full Green's function will read
\begin{equation*}
  G(\Rv,\Rv',\tau) = 
  \iint\mathrm{d}\Rv_1\mathrm{d}\Rv_2 \,
  G_1(\Rv,\Rv_1,\tau) G_2(\Rv_1,\Rv_2,\tau)G_3(\Rv_2,\Rv',\tau) \:.
\end{equation*}

The solutions for the separate 
\index{Green's function}\index{propagator}Green's functions are easy to find. For
example, $\hat H_1$ corresponds to a purely diffusive problem, and the
solution is well-know to be that of a random walk
\begin{equation*}
  G_1(\Rv,\Rv',\tau)=\lp4\pi D\tau\rp^{-3N/2}
                     \exp\left[-\frac{\lp\Rv-\Rv'\rp^2}{4D\tau}\right] \:.
\end{equation*}
The effect of $\hat H_2$ is the same as that of a classical drift 
force $\bm{F}$ pointing towards the regions where the trial wave 
function is maximal:
\begin{align*}
  G_2(\Rv,\Rv',\tau) &=\de\lp\Rv-\Rv(\tau)\rp \:,\\
  \Rv(\tau) &\text{ such that } 
            \left\{ \begin{array}{l}
	            \Rv(0)=\Rv' \\
		    \parcta{\Rv}=D\bm{F}\!\lp\Rv(\tau)\rp
		    \end{array}
	    \right. \:.
\end{align*}
Finally, the Green's function corresponding to $\hat H_3$ is
\begin{equation*}
    G_3(\Rv,\Rv',\tau) = \exp[(E_T-\Eloc(\Rv))\tau/\hbar]\de\lp\Rv-\Rv'\rp
    \:.\\
\end{equation*}
The effect of $G_1$ and $G_2$ is implemented by proposing random movements of
walkers of the type
\begin{equation*}
  \Rv_i \raw \Rv'_i = \Rv_i + \bm{\xi}_i + \frac{\hbar^2 t}{m}\bm{F} \:,
\end{equation*}
where $\bm{\xi}_i$ is a $3N$-component vector whose elements are random
numbers generated according to a $Normal(0,1)$ distribution.
The $G_3$ term is usually called the `branching term' as it is the only
contribution to $G$ that does {\em not} conserve the weights of the walkers as
provided by $\Phi_T$: walkers with lower $\Eloc$ adquire larger weights, while
those with larger $\Eloc$ have smaller weights. This is implemented in the
algorithm in the following way:
\begin{enumerate}
\item The factor $s=\exp[(E_T-\Eloc(\Rv))\tau/\hbar]$ is computed for each
  walker.
\item A number $\xi$ is randomly generated with uniform distribution in
  [0,1].
\item The integer part of $n_{\rm sons} = s+\xi$ is calculated.
\item The walker $\Rv$ is replicated $n_{\rm sons}$ times. (In practice, we
  set a maximum value $n_{\rm sons}=10$ in our calculations.)
\end{enumerate}
Thus, walkers with lower $\Eloc$ `reproduce', while those with larger
$\Eloc$ `die out'. This behavior will ultimately {\em correct} any
component of the assumed $\Phi_T$ orthogonal to  $\Psi$. 
Note that if $\Phi_T$ happens to be an eigenstate of $\hat H$,
$\Eloc(\Rv)\equiv\Eloc$ [cf. Eq.~(\ref{eq:local})]: all walkers have the 
same weight and no `reproduction effects' take place.
In the chart of Fig.~\ref{fig:metro}, 
this process takes place {\em after} updating the positions and {\em before}
storing the values for the averages.

\subsection{Building the trial function}

From the previous discussion, we see that a crucial ingredient in both \vmc\
and \dmc\ calculations is the definition of the trial wave function
$\Phi_T$. Indeed, the `velocity' to converge from a given $\Phi_T$ to the
exact ground state wave function will depend on the `distance' between these
two functions, and the accuracy of the average values obtained with the
converged $f$ will also depend on it. Therefore, we must construct $\Phi_T$
carefully.

For a Hamiltonian with central, two-body interactions, even if they are strong
as in the case of helium, experience has shown that a good guess has
the Jastrow form~\cite{jastrow,artur-ictp}
\begin{equation}
  \Phi_T = {\cal N} \left[ \prod_{i=1}^N f_1(\rv_i) \right]
                    \left[ \prod_{j<k}   f_2(r_{jk}) \right] \:.
\end{equation}
Here $\cal N$ is the normalization constant and $f_1$ is a one-body
function accounting for the effects of a possible external potential.
In our case of two-dimensional droplets, we simulate the self-trapping of the
systems by introducing single-particle factors of the forms
\begin{subequations}
\begin{align}
  f_1 (r) &= \exp\left[ -\frac{\al^2}{2}r^2\right]  &\text{harmonic;} \\
  f_1 (r) &= \exp\left[ -{\al} r\right]   &\text{exponential}.
\end{align}
\end{subequations}
As it will be shown later, however, only the translationally invariant part 
of $\prod f_1$ is to be used as there is no external trapping potential that
locates the droplet in space.

\begin{figure}[tb]
  \caption[Aziz HFD-B(HE) interaction potential for $^4$He atoms]
	  {\label{fig:aziz}
	    Aziz HFD-B(HE) interaction potential for $^4$He atoms.}
  \begin{center}
    \includegraphics*[width=10cm]{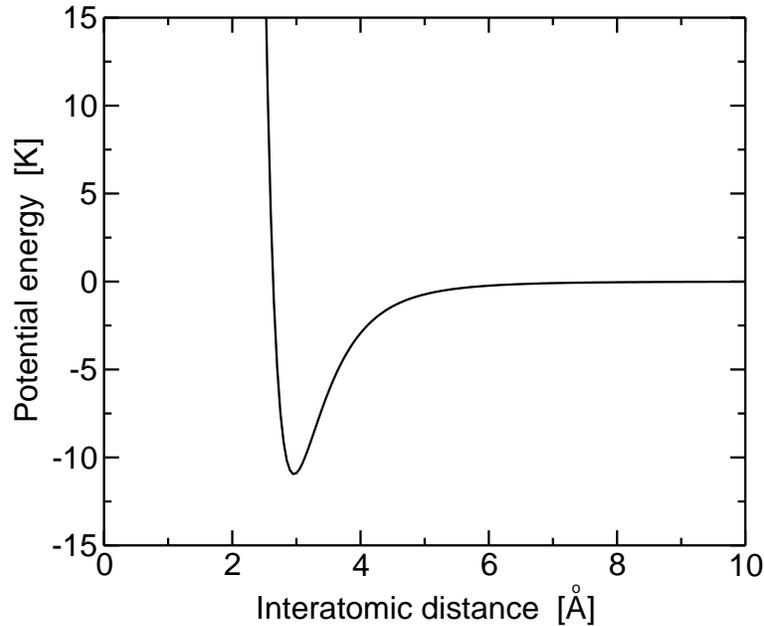}
  \end{center}
\end{figure}

On the other hand, $f_2$ contains two-body correlations and should go to unity
for large inter-particle distances. In particular, for studies on liquid
helium the following form introduced by McMillan~\cite{mcm65} has shown to
be useful:%
\footnote{A more general approach would be that of searching the {\em optimal}
  functions $f_1$ and $f_2$ by solving the Euler-Lagrange equations
  $\de\la\hat H\ra / \de f_i = 0$, see~\cite{artur-ictp}. 
  We will not attempt this here, as the functional forms for the $f_i$ that we
  use have already been proved to be a good starting point for \vmc\
  calculations with helium and, ultimately, the \dmc\ calculation in
  Sect.~\ref{sec:dmc} will `correct' any bad behavior of the trial wave
  function. For the same reason, and in order to ease the operations to be
  performed on it, we are not including three-body correlations into
  $\Phi_T$.}
\begin{equation}
  f_2(r) = \exp\left[ -\inv{2}\lp\frac{b}{r}\rp^\nu \right] \:.
\end{equation}
Here $b$ and $\nu$ act as variational parameters with respect to
which the expectation value $\la \Phi_T | \hat H | \Phi_T \ra$ will be
minimized. 
For helium in three dimensions, the values $b\approx3$ \AA\ and
$\nu\approx5$ provide a good {\em Ansatz} for the calculations,
as they account for the strong repulsion at short distances
\index{He-He interaction!Aziz potential}
$r\lesssim\sigma=2.556$ \AA\ characteristic of the helium-helium interaction,
which in our calculations is described by the Aziz HFD-B(HE) 
potential~\cite{azi87}, see Fig.~\ref{fig:aziz}. 
As these values are related basically to the two-body 
potential, we think they will also be a reasonable starting point for our
two-dimensional case. We note that these values give the correct behavior of
the correlation function at short distances but they do not produce the
correct behavior at long distances. However, we expect this not to be so
important for finite systems such as the small droplets we will be dealing
with~\cite{panda86}.

\section{\vmc\ ground-state energies}
\label{sec:vmc}
To study a system of $N$ $^4$He atoms in two dimensions 
we start from the following trial wave function
\begin{equation}
  \label{eq:wf}
  \Phi_T({\bf R}) = \prod_{i<j} \exp \left[ 
    - \frac{1}{2} \left(\frac{b}{r_{ij}}\right)^{\nu}
    - \frac{\alpha^2}{2N} r^2_{ij} \right] \,.
\end{equation}
The first term inside the brackets corresponds to the simple McMillan
two-body correlation%
, while the second term is the translationally invariant part
of a harmonic oscillator (\textsc{ho}) wave function with parameter $\alpha$,
extracted from the relation
\begin{equation*}
  \sum_{i=1}^N r_i^2 = \inv{N}\sum_{i<j}r_{ij}^2 + \inv{N}R_{\rm CM}^2 \:,
\end{equation*}
to roughly confine the system to a region of radius $\sim \alpha^{-1}$.

In our calculations, we employ the value $\hbar^2/m_4 = 12.1194$~K~\AA$^2$
for the atom mass. The parameters $b$ and $\nu$ have been fixed to the values
3.00~\AA\ and 5, respectively thus restricting the variational search to the
\textsc{ho} parameter $\alpha$. The optimal values found as a function of the
number of particles in the droplet are given in Table~\ref{tab:var_ene} 
together with the expectation value of the Hamiltonian and the separate
contributions of kinetic and potential energies. It can be seen that the total
energy results from an important cancellation between these contributions,
which is in fact larger than in the 3D case. Let us recall that in 
3D bulk, the energy per particle results from adding $\approx$  14~K of
kinetic energy to  $\approx -21$~K of potential energy. In 2D, both kinetic
and potential contributions are very similar, hence calculations are more 
delicate owing to the statistical uncertainties.
\begin{table}
  \begin{center}
    \caption[\VMC\ ground-state energy per particle of various $^4$He
             clusters] 
	    {\label{tab:var_ene}
	      Variational results for the ground-state energy per particle
              $E/N$ of 2D $^4$He puddles of various cluster sizes. The optimal
              confining \textsc{ho} parameter $\alpha$ is given in \AA$^{-1}$
              and all energies are in K. The expectation values of the kinetic
              and the potential energies are also displayed. The column
              labelled KC refers to the \VMC\ results of Ref.~\cite{kri99}.
	      The numbers in parenthesis stand for the statistical uncertainty 
	      on the last figure.
	    }
    \vspace{0.3cm}
    \newcolumntype{d}[1]{D{.}{.}{#1}} 
    \newcommand{\mul}[1]{\multicolumn{1}{c}{#1}} 
    \newcommand{\mulb}[1]{\multicolumn{1}{c|}{#1}} 
    \newcommand{\mulbb}[1]{\multicolumn{1}{|c|}{#1}} 
    \begin{tabular}{|d{0}|d{4}d{6}d{6}d{6}d{6}|}
      \hline
      \mulbb{$N$}& \mul{$\al$}& \mul{$E/N$}& \mul{$T/N$}& \mul{$V/N$}&
      \mulb{KC} \\
      \hline
      \hline
      8      & ~0.1565 & -0.2239(2) & 1.3003(6) & -1.5242(5) & \mulb{---} \\
      16     & ~0.129  & -0.3510(2) & 1.7354(6) & -2.0864(5) & -0.380(8)\\
      36     & ~0.094  & -0.4532(4) & 2.031(3)  & -2.484(3)  & -0.471(7)\\
      64     & ~0.073  & -0.4961(7) & 2.159(2)  & -2.655(2)  & -0.528(5)\\
      121    & ~0.054  & -0.5241(6) & 2.223(2)  & -2.747(2)  & -0.570(7)\\
      165    & ~0.047  & -0.5328(3) & 2.289(1)  & -2.822(1)  & -0.602(7)\\
      512    & ~0.0266 & -0.5493(5) & 2.282(3)  & -2.831(3)  & -0.621(2)\\
      \infty & ~0.0000 & -0.6904(8) & 4.312(2)  & -5.003(1)  & \mulb{---} \\
      \hline
    \end{tabular}
  \end{center}
\end{table}

The last column of Table~\ref{tab:var_ene} reports the \VMC\ results of 
Krishnamachari and Chester~\cite{kri99}. As compared with their 
results, our calculations provide smaller binding energies in spite of the
fact that the interaction used in~\cite{kri99} is an older version of the Aziz
potential, which tends to underbind the systems. This is due to their use of
shadow wave functions, which contain more elaborate correlations not present
in our simple trial wave function.
Finally, the \VMC\ energy for the bulk system corresponds to the saturation
density obtained in the \DMC\ calculation of Ref.~\cite{gio96}, 
$\rho_0= 0.04344$ \AA$^{-2}$.

We have also performed calculations using a different trial wave function, 
replacing the translationally invariant \textsc{ho} part by an exponential
one, \ie
\begin{equation}
  \label{eq:wf-expo}
  \Phi_T({\bf R}) = \prod_{i<j} \exp \left[ 
    - \frac{1}{2} \left(\frac{b}{r_{ij}}\right)^{\nu}
    - \frac{\alpha}{2} r_{ij} \right],
\end{equation}
to check whether a larger tail in the wave function results in more
binding. Actually, we do not find significant differences for small values of
$N$. For instance, in the case $N=8$, using the same values for $b$ and $\nu$
as before, we get $E/N=-0.2178(5)$ K, $T/N=1.266(2)$ K,
$V/N=-1.484(2)$ K for $\alpha=0.035$ \AA$^{-1}$. When the values of
$b$, $\nu$ and $\alpha$ are all optimized, we obtain a slightly larger
binding energy, $E/N=-0.2267(8)$ K for $b=3.04$ \AA, $\nu=5.0$ and
$\alpha=0.035$ \AA$^{-1}$. For greater values of $N$, the harmonic
{\em Ansatz} tends to provide more binding than the exponential. For
example, with the exponential {\em Ansatz}, for $N=16$ we get $E/N=-0.1816(7)$
K for \mbox{$\alpha=0.023$ \AA$^{-1}$,} and optimizing the different
parameters we get $E/N=-0.2514(6)$ K, with $b=3.04$ \AA. In conclusion,
the correlated \textsc{ho} wave function (\ref{eq:wf}) seems good enough
to be used as importance function in the \DMC\ calculations.

\section{\dmc\ ground-state energies}
\label{sec:dmc}

After finding appropriate trial wave functions $\Phi_T(\al)$ with the previous
\vmc\ calculation, we will now perform \dmc\ calculations to find the
exact ground state energy and density profiles of the $^4$He two-dimensional
clusters. We have used both first and second order~\cite{bor94} propagators
[see Eq.~(\ref{eq:order})] in the present work and both of them provide the
same extrapolated energy, within statistical uncertainties, using the optimal
trial function of the \vmc\ calculation as guiding function.

Our simulations have been carried out with a population of typically 400
walkers. As usual, some `equilibration' runs are first done to establish the
asymptotic region of the short time propagator. Then the average blocks are
performed. This procedure is done for several values of the time step
$\de\tau$. Finally a fit of the different obtained energies 
$E_{\rm DMC}(\de\tau)$, either linear or quadratic, has been carried out to
obtain the extrapolated energy, which is the one we report.
For example, for $N=16$ the time steps $\de\tau=$0.0001, 0.0002, 0.0003, and
0.0004 have been used to perform the extrapolation. In general, the
statistical uncertainty for each time step is of the order of the uncertainty
in the extrapolated value.

In Table~\ref{tab:dmc_ene} we present the results of our linear and quadratic
\DMC\ calculations of the total energy per particle for puddles containing $N$
atoms. The linear and quadratic \DMC\ results are compatible within their
error bars.
\begin{table}[t]
  \begin{center}
    \caption[\DMC\ ground-state energy per particle of various $^4$He
             clusters]
	    {\label{tab:dmc_ene}
	      Energy per particle (in K) for 2D $^4$He puddles for various
              cluster sizes obtained with the linear and quadratic \DMC\
              algorithms.
	      The figures in parenthesis stand for the statistical
              uncertainty.
	    }
    \vspace{0.3cm}
    \newcolumntype{d}[1]{D{.}{.}{#1}} 
    \newcommand{\mul}[1]{\multicolumn{1}{c}{#1}} 
    \newcommand{\mulb}[1]{\multicolumn{1}{c|}{#1}} 
    \newcommand{\mulB}[1]{\multicolumn{1}{|c}{#1}} 
    \begin{tabular}{|d{0}@{~~}d{6}d{6}|}
      \hline
      \mulB{$N$} & \mul{linear} & \mulb{quadratic} \\
      \hline
      \hline
      8      & -0.2613(4) & -0.2612(2)    \\
      16     & -0.4263(4) & -0.426(1)     \\
      36     & -0.578(2)  & -0.575(3)     \\
      64     & -0.658(4)  & -0.652(4)     \\
      121    & -0.710(2)  &  \mulb{---} \\
      \infty & -0.899(2)  & -0.8971(6)    \\
      \hline
    \end{tabular}
  \end{center}
\end{table}
We have reproduced the results of the binding energy per particle of
homogeneous 2D liquid $^4$He at the equilibrium density
$\rho_0^{\DMC}=0.04344(2)$ \AA$^{-2}$ obtained in Ref.~\cite{gio96},
where the same version of the Aziz potential was used. For this case, the
simulations have been carried out for a system of 64 atoms with periodic
boundary conditions, for which the errors due to finite size effects are
smaller than the statistical uncertainties~\cite{whi88}.
As expected, the \DMC\ results lower the corresponding energies 
obtained by \VMC\ either with our simple variational wave function or with a 
shadow wave function~\cite{kri99} up to $\sim 25 \%$ in the case of the 
bulk system.
It is worth stressing that the final \DMC\ result for the energy does not 
depend on the starting trial wave function for a boson system, a fact that in
the present case has been numerically checked for the Gaussian and the
exponential {\em Ansätze}, Eqs.~(\ref{eq:wf}) and~(\ref{eq:wf-expo}). Indeed,
for boson systems the \DMC\ method provides the exact ground-state energy,
within statistical uncertainties, given a long enough computing time.

\section{Energy and line tension: introducing the mass formula}

\index{mass formula}
\index{line tension}

For a saturating self-bound system, the ground-state energy per
particle can be expanded in a series of powers in the variable
$N^{-1/d}$, where $N$ is the number of constituents and $d$ is the
dimensionality of the space. This is the well-known mass
formula~\cite{weizs,bethe36}, which in the present case reads
\begin{equation}
  \label{eq:mass-f}
  E(N)/N = \varepsilon_{b} + \varepsilon_{l} z + \varepsilon_{c} z^2 + \cdots
\end{equation}
with $z=N^{-1/2}$. The two first coefficients of this expansion correspond to
the bulk energy $\varepsilon_{b}$ and the line energy $\varepsilon_{l}$ (the
2D equivalent of the surface energy for 3D systems), out of which the line
tension $\lambda$ is defined by $2 \pi r_0 \lambda = \varepsilon_{l}$. Here
$r_0$ is the unit radius, that is the radius of a disk whose surface is equal
to the inverse of the equilibrium density of the 2D bulk liquid, \ie\ 
$\rho_0 \pi r_0^2=1$.
Finally, $\varepsilon_c$ is the so-called curvature energy.

\begin{figure}[tb]
  \caption[Energies per particle of $N$-atom puddles {\em vs.} $N^{-1/2}$]
	  {\label{fig:energies}
	    Energies per particle (in K) of $N$-atom puddles as a function of
	    $N^{-1/2}$, obtained from our \VMC\ (empty squares) and linear
	    \DMC\ (filled circles) calculations. 
	    Solid lines stand for parabolic fits to the data. 
	    The dashed line is a straight line between the $N=8$ and bulk
	    \DMC\ values.}
  \begin{center}
    \includegraphics*[width=11cm]{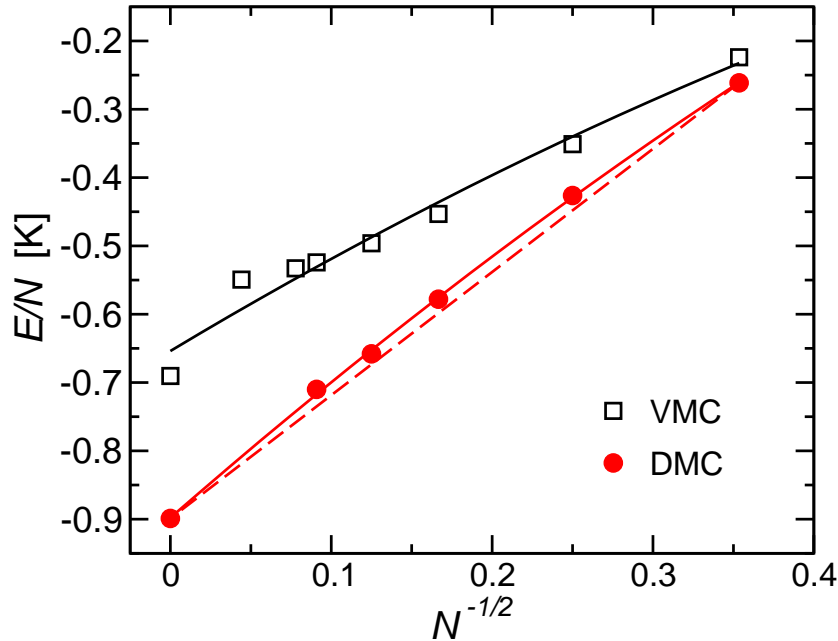}
  \end{center}
\end{figure}

Our calculated ground-state energies (Tables~\ref{tab:var_ene} and
~\ref{tab:dmc_ene}) are plotted in Fig.~\ref{fig:energies} as a function of 
$N^{-1/2}$. (Regarding the \dmc\ data, we just plot the results obtained with
the linear algorithm for the sake of clarity.)
One can see that the differences between our \VMC\ and \DMC\ energies
increase with the number of atoms in the puddle. This shows that the \dmc\
calculation proceeds as expected, getting rid of the `incorrect' features of
our simple guiding function, which becomes ever less accurate as we add more
particles (and, therefore, more correlations) to the system. This could be
improved by including, for example, three-body correlations in the guiding
function for a purely \vmc\ calculation, but nevertheless it is adequate for
the importance sampling in the present \DMC\ calculation.

\index{mass formula}
We have fitted these energies to a parabolic mass formula like 
Eq.~(\ref{eq:mass-f}). The coefficients of the fit are given in 
Table~\ref{tab:coefs}, together with the deduced line tension.
\begin{table}[tb]
  \begin{center}
    \caption[Coefficients of a fit of the mass formula and line tension]
	    {\label{tab:coefs} \index{mass formula} \index{line tension}
	      Coefficients (in K) of a parabolic fit of the mass formula, as
	      given in Eq.~(\ref{eq:mass-f}). The last column displays the
	      deduced line tension (in K~\AA$^{-1}$).
	    }
    \vspace{0.3cm}
    \newcolumntype{d}[1]{D{.}{.}{#1}} 
    \newcommand{\mul}[1]{\multicolumn{1}{c}{#1}} 
    \newcommand{\mulb}[1]{\multicolumn{1}{c|}{#1}} 
    \newcommand{\mulB}[1]{\multicolumn{1}{|c}{#1}} 
    \begin{tabular}{|c|d{5}d{4}d{4}|d{5}|}
      \hline
      Method & \mul{$\vareps_b$} & \mul{$\vareps_l$} & \mulb{$\vareps_c$} &
      \mulb{$\lambda$} \\
      \hline
      \hline
      \VMC & -0.654(1) & 1.41(1) & -0.62(2) & 0.083(1) \\
      \DMC & -0.898(2) & 2.05(2) & -0.71(3) & 0.121(1) \\
      \hline
    \end{tabular}
  \end{center}
\end{table}
Notice that the coefficient $\varepsilon_b$ of the \dmc\ calculation is
identical, within statistical uncertainties, to the bulk energy per particle 
of Table~\ref{tab:dmc_ene}; on the contrary, the same does not happen for the
\vmc\ values.
This gives us confidence on Eq.~(\ref{eq:mass-f}) being suitable for the
fit of the \dmc\ data. In fact, the $\chi^2$ of the \dmc\ fit is very small,
$\chi^2=5.7\times10^{-6}$.
\index{line tension}
Regarding the line tension, and despite using a different version of the Aziz
potential and a different trial function, we observe that our \VMC\ estimate is
rather close to the one reported in~\cite{kri99}, $\lambda_{\rm KC}=0.07$
K/\AA. However, both \VMC\ results are remarkably different from the \DMC\
line tension that we find, namely $\lambda_{\rm DMC}=(0.121\pm0.001)$ K/\AA. 
 
In all cases, the line energy coefficient is approximately minus twice 
the volume energy term, similarly to the 3D case~\cite{panda86}. Also the
curvature coefficient is not small and therefore one expects curvature effects
to be important.
To stress this fact, we have also plotted in the figure a straight dashed
line between the $N=8$ and bulk \DMC\ values. In fact, a linear fit of the
\DMC\ energies gives coefficients $\varepsilon_b=-0.885$~K and
$\varepsilon_l=1.80$~K, which are appreciably different from the previous 
ones. The  bulk energy extrapolated from this linear fit differs from the
directly calculated value, and the corresponding line energy is closer to the
variational one.

\begin{figure}[tb]
  \caption[Why do we loose binding energy in a finite system?]
	  {\label{fig:curvatura}
	    Why do we loose binding energy in a finite system? On the left
	    side we see a semiinfinite medium (shaded region) in contact with
	    a vacuum (white): the lack of particles on the right-hand-side 
	    (indicated by the arrows) of the interface (solid line) gives rise
	    to a reduction of their binding energy and, therefore, also of the
	    total binding energy of the system. On the right graph we see what
	    happens when the interface is curved (from the dashed line to the
	    solid one): even more binding is lost due to the further decrease
	    in the number of neighbors for the atoms close to the frontier.} 
  \begin{center}
    \includegraphics*[width=11cm]{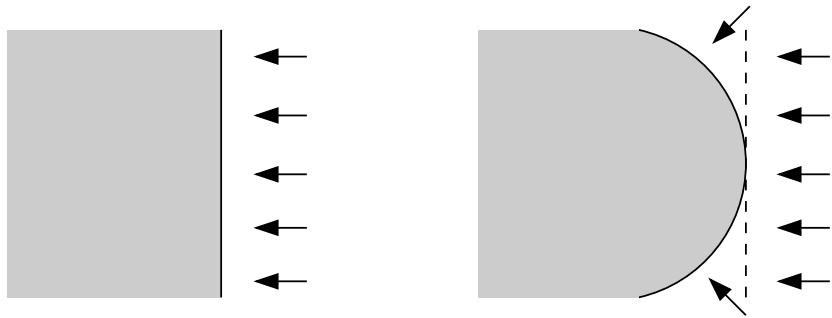}
  \end{center}
\end{figure}

We note that in both \VMC\ and \DMC\ cases the extracted $\varepsilon_c$ is
negative,  \ie\ the binding energy is a convex function of $z$ as it also
happens for the 3D clusters~\cite{panda86}. This is in contrast with the value
of $\varepsilon_c$ reported in Ref.~\cite{kri99} which was positive
but rather smaller in absolute value and with larger error bars.
Actually, as argued in Ref.~\cite{panda86} for 3D clusters, 
one would expect the curvature correction to the energy to be positive, as it
corresponds physically to a lack of binding of the surface atoms due to the
curvature of the surface (see Fig.~\ref{fig:curvatura}) and, so, to an
increase in the total energy of the system. Therefore, one should take the
extracted value for $\varepsilon_c$ with certain caution and not emphasize its
physical significance. Most probably it is due to the small size of the
clusters used to build the mass formula, which ultimately forces the $E/N$
curve to approach the abscisa axis.
In fact, we will see in Chapter~\ref{ch:heli-df} that $\varepsilon_c$ turns
out to be positive when the smaller droplets are discarded and much larger
ones 
are taken into account.

In any case, we have checked that, with the present \dmc\ data, the value and
sign of $\varepsilon_c$ are stable against different possible fits, \eg,
changing the number of data points used to build the fit, or using a cubic
mass formula.  
Also, the values of the two first coefficients $\varepsilon_b,\varepsilon_l$
are quite robust against all performed fits. As an illustration, 
if one takes out the bulk point, the predicted bulk energy per 
\index{line tension}
particle and line tension are equal to the reported ones within 1\% 
and 5\% respectively. 
Therefore, we believe that the extracted line tension should be reliable,
as it also happens for \vmc\ calculations of
three-dimensional clusters~\cite{panda86}, where the extracted surface tension
is in agreement with the experimental one even if $\varepsilon_c<0$ in
the fit.

\section{\VMC\ and \DMC\ density profiles}

\subsection{Calculating pure estimators}

\index{QMC methods!Diffusion Monte Carlo!calculation of pure estimators}

The calculation of observables given by operators that do not commute
with the Hamiltonian poses a new problem to the \DMC\ method. After
convergence, the walkers are distributed according to the so-called
mixed probability distribution given by the product of the exact and
the trial wave functions,
$f(\Rv,\tau\raw\infty)=\Psi(\Rv)\Phi_T(\Rv)$. Therefore averaging the local
values of the operator does not give the exact expectation value unless the
operator commutes with the Hamiltonian. The result obtained by straightforward
averaging is the so-called {\em mixed estimator},
\begin{equation*}
  \la A(\Rv) \ra_{\rm mixed} = 
  \frac{ \la \Psi | A | \Phi_T \ra } { \la \Psi | \Phi_T \ra } \:,
\end{equation*}
which is of first order error in the
trial wave function. 
A first option devised to overcome this problem is the use of 
{\em extrapolated estimators}~\cite{hlr94},
\begin{equation*}
  \la A(\Rv) \ra_{\rm extrapolated} = 
  2 \la A(\Rv) \ra_{\rm mixed} -
  \la A(\Rv) \ra_{\rm variational} \:,
\end{equation*}
where the variational estimate is calculated with $\Phi_T$ alone.
However, also extrapolated estimates are trial-function dependent and
biased~\cite{hlr94,cas95}.

Several alternatives have been presented in the literature in order to obtain
`unbiased' or `pure' (\ie, trial-function-independent and statistically
exact) values. 
We have used the {\em forward} or {\em future walking}
technique~\cite{hlr94} to calculate unbiased density profiles. 
The key ingredient to correct the mixed estimator is to 
include as a weight in the sampling the quotient 
$\Psi({\bf R})/\Phi_T({\bf R})$ for each walker, 
\begin{equation*}
  \frac{ \la \Psi | A | \Psi \ra } { \la \Psi | \Psi \ra } \equiv
  \frac{ \la \Psi | A \frac{\Psi}{\Phi_T}| \Phi_T \ra } 
       { \la \Psi | \frac{\Psi}{\Phi_T}  | \Phi_T \ra } =:
  \la A \ra_{\rm pure} \:.
\end{equation*}
The value of this weight for each walker is given by its asymptotic number of
descendants~\cite{liu74}:
\begin{equation}
  \frac{\Psi(\Rv)}{\Phi_T(\Rv)} = n_{\rm sons}(\Rv;\tau\raw\infty) \:.
  \label{eq:sons}
\end{equation}
Various algorithms have been proposed in order to compute this quantity. In
this work we use an algorithm by Casulleras and Boronat~\cite{cas95} that
constitutes a simple and efficient implementation of the future walking
method.

Let us assume that we are already in the asymptotic regime where the usual
Monte Carlo procedure would start to accumulate data to compute (mixed)
averages. We define a new variable $P$ that accumulates the value of the
quantity of interest, $A(\Rv)$, for some Monte Carlo steps (labeled here by
$\nu=1,\cdots,M$) within this asymptotic regime
\begin{equation*}
  P_i=\sum_{\nu=1}^{M} A(\Rv_i^\nu) \:.
\end{equation*}
Here the same value $A(\Rv_i)$ is transmitted to {\em all} descendants of
walker $\Rv_i$, while the death of a walker $\Rv_j$ at any time implies that
no more values will be added to the corresponding $P_j$ at later times. 
Thus, each value $A(\Rv_i)$ will appear in the set of 
$\{ P_i \}~(i=1,\cdots,N_f)$ 
at the end of the interval as many times as descendants of the 
walker $\Rv_i$ have appeared, \ie, with a weight $\propto n_{\rm sons}(\Rv_i)$,
which is the desired one [cf. Eq.~(\ref{eq:sons})].
Therefore, averaging $P_i$ at the end of this interval we have 
\begin{equation}
  \overline{P} :=
  \inv{M\times N_f}\sum_{i=1}^{N_f} P_i
  \equiv \la A \ra_{\rm pure} 
  \:.
  \label{eq:forwalk}
\end{equation}

This scheme is easily implemented in a usual \dmc\ calculation, as one can
determine a value for $\overline{P}$ at the end of each averaging block
setting $M=$ {\em number of steps in a block}. At the end of the simulation, 
from the set of $\overline{P}$ over the different blocks we can get an
estimate for $\la A \ra_{\rm pure}$ and its statistical uncertainty.
A study of the dependence of $\overline{P}$ on $M$, typically shows an initial
transient period during which the estimator `gets rid' of the properties
coming from the guiding function. After this period, and for a wide range of
block lengths $M$, the computed average gives an unbiased estimate of the
expectation value of the operator~\cite{cas95}. Therefore, studying this
dependence, one can adjust $M$ so as to obtain a pure estimate with a small
statistical uncertainty.

\subsection{Numerical results}

\begin{figure}[t]
  \caption[Convergence of the future walking method]
	  {\label{fig:pure}
	    Convergence of the future walking method: density profiles for a
	    puddle containing $N=36$ atoms as obtained by the future walking
	    method with block lenghts $M=100$ (grey line) and $M=1000$ (black
	    line).}
  \begin{center}
      \includegraphics*[width=0.6\textwidth]
		       {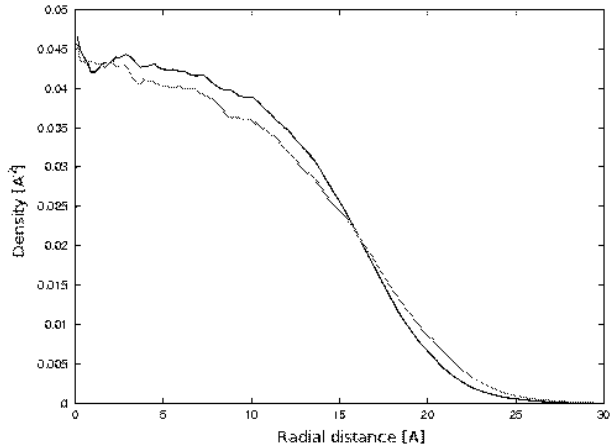}
  \end{center}
\end{figure}

To obtain pure \dmc\ estimates for the density profiles of $^4$He puddles, we
have performed a careful analysis of the behavior of the profiles with
varying block lenghts $M$ over which they were calculated. 
Typically, a range $M=100-1000$ has been explored, showing that
for $M=500$ the computed quantity had already converged.
For example, we show in Fig.~\ref{fig:pure} the estimated density profiles
for block lengths $M=100$ (grey line) and $M=1000$ (black line). The
improvement from one to the other is noticeable: the estimation obtained 
with a larger block length has a steeper surface, which indicates a 
\index{line tension} larger line tension.

\begin{figure}[t]
  \caption[Density profiles of $^4$He puddles with various numbers of atoms]
	  {\label{fig:profiles}
	    Density profiles of $^4$He puddles with various number of atoms,
	    $N=8$ (circles), 16 (squares), 36 (diamonds), 64 (triangles up)
	    and 121 (triangles down), obtained from our pure estimators for
	    the linear \DMC\ calculations. The solid horizontal line indicates
	    the saturation density of the homogeneous system. The dashed line
	    is the \VMC\ profile for $N=36$ with a Gaussian trial
	    function. The figure also contains the fits to the data provided
	    by a generalized Fermi function, as explained in the text.
	  }
  \begin{center}
    \includegraphics*[width=11.5cm]{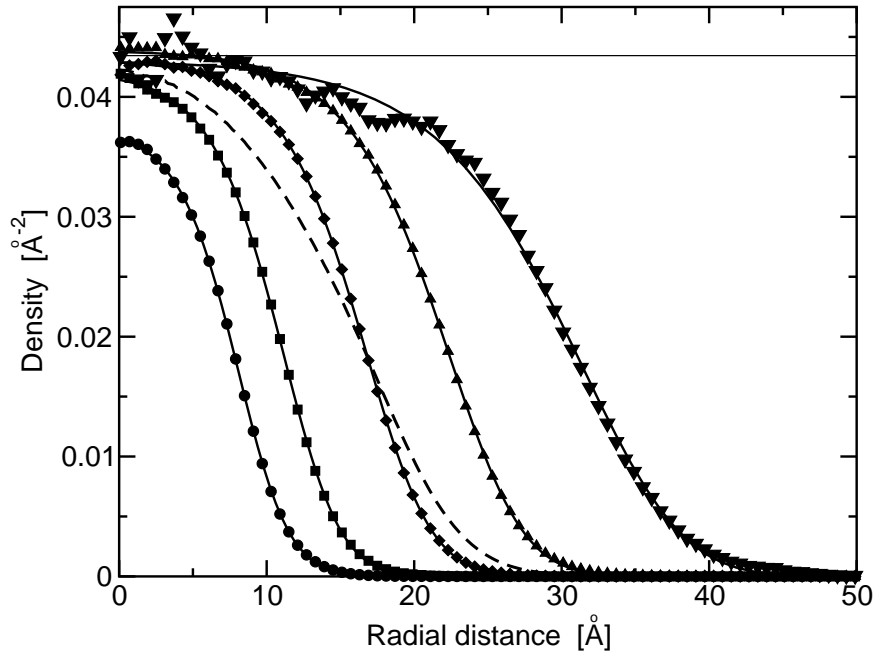}
  \end{center}%
\end{figure}

The converged pure \DMC\ estimates that we obtain in this way (with a larger
number of blocks, to reduce the statistical uncertainties that show up as
oscillations in the profiles in Fig.~\ref{fig:pure}) for the density profiles
of several puddles are plotted with symbols in Fig.~\ref{fig:profiles}.
This figure also contains an horizontal line which indicates the saturation
density of the homogeneous system, $\rho_0^{\DMC}=0.04344$ \AA$^{-2}$.
For the puddle containing 36 atoms, the \VMC\ profile obtained from a
Gaussian {\em Ansatz} [Eq.~(\ref{eq:wf})] is also shown as a dashed line for
comparison. As one can appreciate in the figure, the process of optimization
implied by the \DMC\ method changes the profile reducing its thickness, \ie,
producing a sharper surface. Indeed, by comparing this figure with
Fig.~\ref{fig:pure} one can see how the pure estimator for $M=100$ is still
influenced by the guiding function, which yields a softer boundary. This
residual influence of $\Phi_T$ vanishes completely for $M\geq 500$ for all
puddles. 

A first observation on the \dmc\ profiles is that for the smaller clusters the
central density is below $\rho_0$, while for the larger values of $N$ shown in
the figure the central density appears to be above $\rho_0$, indicating a 
so-called {\em leptodermous} behavior~\cite{str87}. One expects that,
increasing even more the number of particles, the central density will
approach $\rho_0$ from above, as in the 3D case~\cite{str87,chi92}.
It is worth noticing the oscillating behavior in the interior part of the
density profile for $N=121$. It is difficult however to decide whether these
oscillations are genuine or are simply due to a poor statistics in evaluating
the pure estimator at short central densities. Unfortunately, to discard this
last option would require an exceedingly long computing time, within the
scheme of this work.

The solid lines plotted in Fig.~\ref{fig:profiles} are fits to our \DMC\
profiles provided by a generalized Fermi function of the form:
\begin{equation}
  \rho(r)= \frac{\rho_F}{\left[ 1+\exp{\lp\frac{r-R}{c}\rp} \right]^{\nu}}
  \:.
  \label{eq:fermi}
\end{equation} 
The overall agreement with the numerical data is very good, and better than
for a usual Fermi function (\ie, fixing $\nu=1$). An exception is the droplet
with 121 atoms, where the oscillating behavior of the data cannot be
reproduced by the fitting function. However, we have already pointed out that
the physical significance of these oscillations is not clear.

The values of the parameters that best fit our data are given in
Table~\ref{tab:fermi} together with the thickness $t$ and the root mean square
(rms) radius calculated with them.
As usual, $t$ is defined as the distance over which the density falls from
90\% to 10\% of its central value; for a profile of the form of
Eq.~(\ref{eq:fermi}) this can be found analytically, and it turns out to
depend only on $c$ and $\nu$:
\begin{equation}
  t=c\:\ln\frac{0.1^{-1/\nu}-1}{0.9^{-1/\nu}-1} \:.
  \label{eq:thick}
\end{equation}

Regarding the rms radius, defined through
\begin{equation*}
  \la r^2 \ra := 
  \frac{ \int_0^\infty \mathrm{d}r\, r^3 \rho(r) }
       { \int_0^\infty \mathrm{d}r\, r   \rho(r) } \:,
\end{equation*}
we have checked that the value calculated within the \DMC\ code and
the one derived from the fit agree to better than 0.5\%, except for the
$N=121$ case, where the difference is 1\% (most probably due to the
oscillating behavior of the \dmc\ profile).
\begin{table}[t]
  \begin{center}
    \caption[Parameters of a Fermi-profile fit to the \dmc\ density profiles]
	    {\label{tab:fermi}
	      Parameters of a Fermi-profile fit to the density profiles. 
	      All lengths are in \AA\ and $\rho_F$ is in \AA$^{-2}$. 
	      The parameter $\nu$ is adimensional.
	    }
    \vspace{0.3cm}
    \newcolumntype{d}[1]{D{.}{.}{#1}} 
    \newcommand{\mul}[1]{\multicolumn{1}{c}{#1}} 
    \newcommand{\mulb}[1]{\multicolumn{1}{c|}{#1}} 
    \newcommand{\mulbb}[1]{\multicolumn{1}{|c|}{#1}} 
    \newcommand{\mulB}[1]{\multicolumn{1}{|c}{#1}} 
    \begin{tabular}{|d{0}|d{5}d{3}d{3}d{3}|d{2}d{2}|}
      \hline
      \mulbb{$N$} & \mul{$\R_F$} & \mul{$R$} & \mul{$c$} & \mulb{$\nu$} &
      \mul{$t$} & \mulb{$\langle r^2 \rangle^{1/2}$} \\
      \hline
      \hline
      8   & 0.03740 & 9.308 & 2.156 & 1.739 & 8.166 &  7.20\\
      16  & 0.04204 & 13.38 & 2.656 & 2.284 & 9.580 &  9.18\\
      36  & 0.04305 & 19.47 & 3.104 & 2.400 & 11.11 & 12.91\\
      64  & 0.04386 & 26.68 & 3.783 & 3.111 & 13.09 & 16.68\\
      121 & 0.04304 & 40.09 & 5.566 & 4.714 & 18.52 & 23.15\\
      \hline
    \end{tabular}
  \end{center}
\end{table}

For a sharp surface ($t\ll {\la r^2 \ra}^{1/2}$), one could approximate
the density profile by a step function 
$\R(r)=\R_{\rm step}\Theta(r_{\rm step}-r)$. 
The rms radius corresponding to this profile is given by
$\la r^2 \ra^{1/2} = r_{\rm step}/\sqrt2$, so that it grows with the number of
particles as $N^{1/2}$. More precisely,
\begin{equation}
  \langle r^2 \rangle^{1/2} = \sqrt{ \frac {N}{2 \pi \rho_{\rm step}}} \:.
  \label{eq:rmsrad}
\end{equation}
That is, faster than in 3D, in which case grows as $N^{1/3}$. This
behavior allows for an alternative determination of the saturation density by
performing a fit of the calculated rms radii to the above relation. The value
of $\rho_0$ extracted from the slope of a linear fit to the mean square radii
reported in Table~\ref{tab:fermi} is 0.042 \AA$^{-2}$, in reasonable agreement
(taking into account that for these droplets $t$ is not much smaller than 
$\la r^2 \ra^{1/2}$) with the determination from the calculation for the
homogeneous system. It is interesting to note that a similar treatment with
the fitted values for the parameter $R$, which intuitively gives the
`central' point of the surface, does not give a good estimate for $\rho_0$
(0.012 \AA$^{-2}$). In fact, $R$ seems to grow faster than $N^{1/2}$.

For these droplets, the thickness
is continuously increasing with $N$.  As the finite value of the thickness for
the semiinfinite system should define an asymptotic value for $t$, we expect
that for larger puddles the thickness will present a maximum and smoothly
approach this asymptotic value from above, as happens in the 3D
case~\cite{str87}. To check this numerically, one needs the profiles of much 
larger puddles, which are unaffordable in a \dmc\ calculation. However, it is
possible to find them using a Density Functional, as we will show in
Chapter~\ref{ch:heli-df}.

Finally, we also remark the asymmetric character of the density profiles 
with respect to the point at which the density falls at half its value
at the center of the droplet. Numerically, this behavior is reflected in the
value of $\nu$, which  grows with $N$, and also in the increasing difference
between the quantities $R$ and $\langle r^2 \rangle^{1/2}$.

\section{Summary and conclusions}

In this chapter we have considered strictly two-dimensional systems of liquid
$^4$He, which  are of course an idealization of a real quantum film.
They are  nevertheless interesting  because their study can enlighten 
the underlying structure of real quasi-2D systems. Of course, in the
latter case, one has to take also into account the interaction with the
substrate, which basically provides a global attractive potential.
In the ideal 2D case, the suppression of the wave function component in the 
third dimension, produces an increment of the global repulsion between atoms, 
resulting in a smaller binding energy per particle, and a decrease of the 
equilibrium density~\cite{apa}.

\index{mass formula}
We have calculated the binding energies of two-dimensional $^4$He clusters
by means of a diffusion Monte Carlo method, and we have seen that they can
be well fitted by a mass formula in powers of $z=N^{-1/2}$.
The analysis of the mass formula provides the value 
\index{line tension} $\lambda =0.121 $ K/\AA\ for the line tension, 
which significantly differs from the one obtained 
from a similar analysis of \VMC\ data and the one previously reported in the
literature~\cite{kri99}. The quadratic term of the mass formula cannot be
neglected and results in a negative value of the curvature energy, similarly
to what happens in the 3D case~\cite{panda86,chi92}. However, the studied
clusters may be too small to give physical significance to this result.

The density profiles obtained with the pure estimator have been fitted to a
generalized Fermi function, and the behavior of the rms radius and the
thickness, as well as the asymmetric character of the profile as a function 
of $N$ have been discussed. 

Due to computing limitations on \dmc\ calculations, we have restricted our
study to relatively small clusters with $N\leq 121$ atoms. However, to fully
understand the transition from finite clusters to the bulk medium, both for
the energetics and the structure of the profiles, it seems interesting to have
results for much larger $N$. We will explore these cases in the next chapter
by means of a density functional theory.


\chapter[Density Functional study of $^4$He non-homogeneous systems]
	{Density Functional study of two-dimensional $^4$He non-homogeneous
	systems}
\label{ch:heli-df}

\textsf{
  \begin{quote}
    Però jo, que sabia el cant secret de l'aigua,\\
    les lloances del foc, de la gleva i del vent,\\
    sóc endinsat en obscura presó,\\
    vaig devallar per esglaons de pedra\\
    al clos recinte de llises parets\\
    i avanço sol a l'esglai del llarg crit\\
    que deia per les voltes el meu nom.
  \end{quote}
  \begin{flushright}
     Salvador Espriu, dins {\em Final del laberint}
  \end{flushright}
}

\section{Introduction to Density Functional theory}

\index{Density Functional theory}

In this chapter we continue the study of two-dimensional $^4$He systems
initiated in Chapter~\ref{ch:heli-dmc} by analyzing the energetics and
structure of the semi-infinite medium and slabs. Also, we will take a
look at clusters larger than those studied by means of the \dmc\ technique in
order to clarify some of the points that have not been fully understood
previously, \eg, the sign and value of the curvature energy $\vareps_c$ in the
mass formula, or the behavior of the surface thickness when approaching the
bulk medium.

As it is well know, large systems are difficult to tackle with Monte Carlo
methods as these deal with 
all $d\times N$ coordinates of the system components 
($d=$ dimensionality of the system, $N=$ number of constituents of the
system). There are several alternative approximate methods, 
from mean-field calculations, where the many-body wave function is
written as a direct product of single-particle wave functions, with the
appropriate symmetry properties related to the quantum-statistical
character of the particles, to more involved techniques such as correlated
basis functions (\textsc{cbf}) or the coupled-cluster method (\textsc{ccm}). 
A very good reference to have an overview of the power and flaws of all these
techniques can be found in Ref.~\cite{ictp}. Here we will make use of the
Density Functional (\df) theory, which unites the power of computationally
simple calculations with an insight to the physics of the problem.

\subsection{The Hohenberg-Kohn theorem}
\index{Density Functional theory!Hohenberg-Kohn theorem}
The basis of the Density Functional theory relies on the famous Hohenberg-Kohn
theorem which states that
  the ground state energy of a many-body system is determined solely by its
  one-body density $\R(\rv)$%
~\cite{hk,primer-df}.
In other words, the ground-state energy can be written as a functional of the
density, $E=E[\R]$.

Unfortunately, the Hohenberg-Kohn theorem does not tell how to build this
functional; it is just an existence theorem. 
We need to resort to our understanding of the problem to build it,
which can be done in different ways.
In the context of quantum liquids, one usually follows a `phenomenological'
approach: based on the available physical information on the problem, one
defines a density functional 
\begin{equation}
  E[\R] = \int \mathrm{d}^3r\, \varepsilon[\R(\rv)] \:,
  \label{eq:dens-fun}
\end{equation}
through an energy density $\vareps[\R]$.
The corresponding ground state is then found by functional minimization, 
usually restricted by the 
\index{density normalization!in \df\ theory}normalization of the density.
This condition introduces the chemical potential into the problem as a 
Lagrange multiplier, namely
\begin{equation}
  \frac{\delta E}{\delta \R} = \mu \:.
  \label{eq:el}
\end{equation}

Several functionals have been proposed to study helium systems, either pure
$^4$He or $^3$He or mixtures thereof, see
\eg~\cite{strin2,strin3,barran93,dalfo95,szy}.
The density functional we shall use is the simplest version of the
zero-range functional intensively used in 3D calculations~\cite{strin2}. 
We have adjusted its parameters so as to reproduce some properties of the
ground state of the homogeneous two-dimensional system as obtained in \DMC\
calculations~\cite{gio96}, as well as the line tension extracted from the mass
formula in the previous chapter. 
This procedure is discussed in Section~\ref{sec:sl}, together 
with the results for the slabs.
Section~\ref{sec:dr} is devoted to the study of finite droplets, with special
emphasis for those with a large number of atoms. 
In Section~\ref{sec:pressure} a short discussion on hydrodynamic equilibrium 
in a droplet is presented,
followed in Section~\ref{sec:discussion} by a comparison of the obtained
results with those of other \qmc\ and \df\ calculations in 3D and 2D.
Finally, in Section~\ref{sec:conc} the main conclusions are summarized.

\section{Semi-infinite system and slabs}
\label{sec:sl}
Density functionals to investigate surface properties of superfluid $^4$He
were developed during the 1970's \cite{ebner}. At zero temperature and in 
the absence of currents, the order parameter of a bosonic system can be
identified simply as the square root of the one-body density $\R(\rv)$, so it
is natural to think of the energy of the system as a functional of the helium
density.
In analogy with the formalism of zero-range Skyrme interactions in nuclear
physics~\cite{vauth}, Stringari proposed a zero-range density functional for
non-homogeneous three-dimensional $^3$He systems~\cite{strin2}.
A similar form,
\index{Density Functional!for $^4$He}
\begin{equation}
  E[\R] =
   \int\,\mathrm{d}^2r
   \left\{\D\frac{|\nabla\R|^2}{4\R} +b\R^2+c\R^{2+\g}+d|\nabla\R|^2\right\},
   \label{eq:zr-func}
\end{equation}
was soon used to study $^4$He surface properties~\cite{strin3} and
clusters~\cite{str87}.

For the two-dimensional systems of our interest, we will use the same
functional form. However, the parameters $\{b,c,\g\}$ characterizing the
functional for homogeneous systems have to be recalculated by requiring that
the functional reproduces some known properties of the {\em two-dimensional}
$^4$He system, such as the ground-state energy per particle 
($e_0=-0.89706$ K), and the saturation density ($\R_0=0.04344$ \AA$^{-2}$),
which have been determined in the framework of the Diffusion Monte Carlo
method~\cite{gio96,prb03} (see also Chapter~\ref{ch:heli-dmc}). 
For the homogeneous system, the gradient terms in (\ref{eq:zr-func}) vanish in
the ground state, and one gets
\begin{align*}
  e_0 &= b\R_0+c\R_0^{1+\g}\\
  0   &= b+(1+\g)c\R_0^{\g} \:.
\end{align*}
The third equation used to fix the parameters $\{b,c,\g\}$ 
is obtained by imposing the known value of the speed of sound at saturation,
$s=92.8$ m/s~\cite{gio96}. This is related to the compressibility $\kappa$ 
of the system, and one can write it as
\begin{equation*}
  ms^2 = \inv{\kappa\rho_0} 
       = \left. \frac{\partial}{\partial\R} 
                \lp \R^2\frac{\partial e}{\partial \R} \rp 
         \right|_{\R_0}
  = 2b\R_0+(1+\g)(2+\g)c\R_0^{1+\g} \:.
\end{equation*}

Finally, the parameter $d$ is fixed by demanding that the 
\index{line tension}
line tension of the semi-infinite system equals that obtained from \DMC\ 
calculations for 2D clusters, $\lambda_{\rm DMC}=0.121$ K/\AA\ (see 
Ch.~\ref{ch:heli-dmc} of this thesis). The behavior of the energy per
particle as a function of density for the homogeneous system provided by this
functional  reproduces very well the equation of state obtained from \DMC\
calculations in a wide range of densities~\cite{gio96,luso}. 
As a first step in understanding non-homogeneous systems, we will consider 
the semi-infinite system. 

\begin{figure}[bp!]
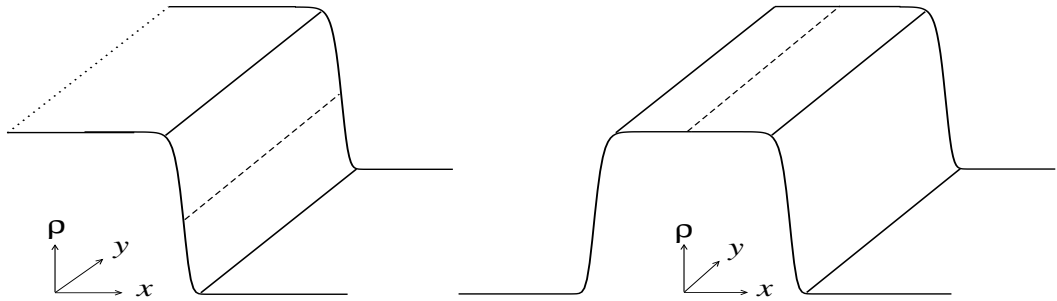

  \caption[Sketches of the semi-infinite medium and a slab]
	  {\label{fig:slab}
	    Sketches of the semi-infinite medium and a slab: the $x$ axis is 
	    perpendicular to the surfaces, the $y$ axis the symmetry axis. 
	    The dashed line indicates the projection of the $y$ axis on the
	    density profile: at $\R_0/2$ for the semi-infinite medium and at
	    $\R_c$ for the slab.
	  }
  \begin{center}
    \includegraphics*[width=6cm,height=3.9cm]{clusters/semi}%
    \includegraphics*[width=8cm,height=3.9cm]{clusters/slabs}
  \end{center}
\end{figure}

In order to determine the density profile $\R(\rv)$, in principle one should
solve the Euler-Lagrange equation (\ref{eq:el}) that results from minimizing
the energy functional~(\ref{eq:zr-func}). One should also introduce the 
Lagrange multiplier $\mu$, which ensures the 
\index{density normalization!in \df\ theory}normalization of the density 
to the desired number of particles, and that is to be identified with the 
chemical potential of the system,
\begin{equation}
  \DD\left[-\frac{2\nabla^2\R}{\R}+\frac{|\nabla\R|^2}{\R^2}\right]
    +2b\R+(2+\g)c\R^{1+\g}-2d\nabla^2\R = \mu \;.
  \label{eq:zr-EL}
\end{equation}
One should also impose the boundary conditions
$\rho(x \rightarrow -\infty)= \rho_0$ and 
$\rho'(x \rightarrow -\infty)=0$, 
where the prime denotes derivative with respect to $x$,
the coordinate perpendicular to the surface of the system
(cf. Fig.~\ref{fig:slab}).

For the particular case of a zero-range functional (\ref{eq:zr-func}), 
\index{line tension}
the line tension can be evaluated in a closed
form, \ie, without solving for the density profile:
\begin{equation}
  \lambda(d) = 2\int_0^{\R_0}\;\mathrm{d}\R\left[
    \left( \DD+\R d\right)
    \left( b\R+c\R^{1+\g}-\mu \right) \right]^{1/2} \;.
  \label{eq:zr-lambda}
\end{equation}
Then, imposing $\lambda(d)=\lambda_{\rm DMC}$, one gets an implicit equation
for $d$ which can be solved numerically, thus yielding the last parameter of the
functional. The parameters for the two-dimensional zero-range functional fixed
in this way are listed in Table~\ref{tab:func}.
\begin{table}[!hb]
  \begin{center}
    \caption[Parameters of the two-dimensional zero-range functional]
	    {\label{tab:func}
             Parameters of the two-dimensional zero-range functional
             (\ref{eq:zr-func}).
            }
    \begin{tabular}{|l@{~~~~}l|}
      \hline
      $b=-26.35$ K\AA$^2$                  & $\g=3.62$ \\
      $c=4.88\times10^5$ K\AA$^{2(1+\g)}$  & $d=359$ K\AA$^4$\\
      \hline
    \end{tabular}
  \end{center}
\end{table}

Once the functional is defined, one can study any two-dimensional $^4$He
system. The first systems we have considered are two-dimensional slabs with
varying central density $\R_c= \R(x=0)$ (see Fig.~\ref{fig:slab}).
For later reference, we note that a systematic study of 
{\em three-dimensional} $^4$He slabs with different density functionals has
been presented in Ref.~\cite{szy}.
The Euler equation for the slabs is the same as for the 
semi-infinite system [Eq.~(\ref{eq:zr-EL})].
The changes in the solution of the equation originate from the 
different geometry and boundary conditions which define the slab; 
in particular, the slab has {\em two} surfaces, while the semi-infinite 
medium only has one (located around $x=0$).
Translational symmetry implies that the density depends only on the 
coordinate perpendicular to the slab surface, which we call $x$ again. 
Then, $\nabla\R=\R'(x)$ and $\nabla^2\R=\R''(x)$.
In this way, the Euler equation~(\ref{eq:zr-EL}) can be expressed in 
a more convenient form, in which now $\R$ depends only on $x$,
\begin{equation} 
  \DD \left(- \frac{2\R''}{\R} + \frac{(\R')^2}{\R^2} \right) 
  + 2 b \R + (2+\gamma) c \R^{1+\gamma} - 2 d \R''= \mu
\label{eq:euler2}
\end{equation}
The expression in parenthesis can be rewritten as a total
derivative multiplying it by $\R'$,
\begin{equation*}
  \left( \frac{2\R''}{\R} -\frac{(\R')^2}{\R^2} \right)\R' =
  \left( \frac{2\R'\R''}{\R} -\frac{(\R')^3}{\R^2} \right) =
  \frac{d}{dx}\left(\frac{(\R')^2}{\R} \right) \:.
\end{equation*}
Thus, one can eliminate the second derivative of Eq. (\ref{eq:euler2}) by
multiplying both sides of the equation by $\R'$ and integrating with respect
to $x$, from the origin to a given value of $x$,
\begin{equation}
  \left[-\DD\frac{(\R')^2}{\R} + b\R^2 + c\R^{\gamma+2} -d(\R')^2 \right]_0^x
  = \mu \left[ \R(x) - \R(0) \right] \:.
\label{eq:euler3}
\end{equation}
Imposing the boundary conditions $\R(\infty)= \R'(\infty)=0$, and considering
that the slab is symmetric under inversion of the $x$-axis, and therefore
$\R'(0)=0$, one finds that the chemical potential as a function of the central
density of the slab reads
\begin{equation}
  \mu = b \R_c + c \R_c^{\gamma +1} .
  \label{eq:chem}
\end{equation}
It is worth to remind that this chemical potential is constant along the 
profile.

Going back to Eq. (\ref{eq:euler3}),
and using again the fact that $\R'(0)=0$, one obtains
\begin{equation}
 |\R'| = \sqrt{\frac{\R^2 (b\R + c\R^{\gamma+1} -\mu )}{ \DD + d\R} }.
\label{eq:euler4}
\end{equation}
This expression for $\R'$ is then used to calculate the number of atoms per
unit length along the $y$ axis (that is, the {\em coverage}) in a closed
expression in terms of the density
\begin{equation*}
  \frac{N}{L} = 2 \int_0^{\infty} \R(x) \mathrm{d}x 
              = 2 \int_0^{\R(0)} \R \frac{\mathrm{d}\R}{|\R'|}
	      = 2\int_0^{\R (0)} \mathrm{d}\R 
	         \sqrt{\frac{\DD + d\R}{b\R + c\R^{1+\gamma} -\mu}} \:,
\end{equation*}
where the factor 2 takes into account the symmetry of the slab under the
transformation $x\raw-x$.
This integral has a singularity when $\R \rightarrow \R_c $
that can be avoided integrating by parts. 
The final expression, free of numerical problems, reads 
\begin{multline*}
\frac{N}{L} = -\frac{4}{b} \sqrt {-\DD\mu } \\
- 4 \int_0^{\R_c} \!\!\mathrm{d}\R
\frac{ \sqrt{b\R +c \R^{1+\gamma} -\mu} 
       \left\{ \frac{d}{2} [b +c(1+\gamma) \R^{\gamma}]
              -(\DD + d \R) c \gamma (1+\gamma) \R^{\gamma-1} \right\} }
     { [b +c(1+\gamma)\R^{\gamma}]^2~(\DD + d \R )^{1/2} } .
\end{multline*}

Also useful is the energy per unit length,
\begin{equation*}
  \frac{E}{L} = 2\int_0^{\infty} \mathrm{d}x
    \left[\frac{\hbar^2}{8m}\frac{(\R')^2}{\R} +b \R^2 + c \R^{2+\gamma} + d
    (\R')^2 \right ].
\end{equation*}
Using Eq.~(\ref{eq:euler4}) and the previous definition of the coverage, one
can finally express the energy per particle $e = E/N$ in terms of the inverse
of the coverage, $\tilde x:=L/N$,
\begin{equation}
  e(\tilde x) = \mu(\tilde x) 
    + 4 \tilde x \int_0^{\R(0)} \mathrm{d} \R 
      \sqrt{\left(\frac {\hbar^2}{8m} + d \R\right) 
	         [b \R + c \R^{1+\gamma} - \mu(\tilde x)]}
\label{eq:ener1}
\end{equation}
Therefore, given a central density $\R_c \in [0, \R_0]$, 
one can calculate the chemical potential, the coverage and the energy 
per particle of the corresponding slab.

\begin{figure}[!tb]
\begin{minipage}{5.2cm}
  \caption[Energy per particle and chemical potential of slabs]
	  {\label{fig:ener-slab}
%
          \index{line tension}
	    (Top) Energy per particle (dashed line) and chemical potential 
	    (dot-dashed line) in kelvin for $^4$He slabs as a function of the
	    inverse of the coverage $\tilde x$ (in \AA). 
	    The slope of the solid line is twice the line tension
	    of the semi-infinite medium, which corresponds to the asymptotic
	    $\tilde x \raw0$ behavior of the energy per particle [see
	    Eq.~(\ref{eq:expa})].
	    (Bottom) Ratio between the central density $\R_c$ and the bulk
	    equilibrium density for $^4$He slabs as a function of the inverse
	    of the coverage.
	  }
\end{minipage}
\hfill
\begin{minipage}{7.7cm}
  \includegraphics*[width=7.7cm]{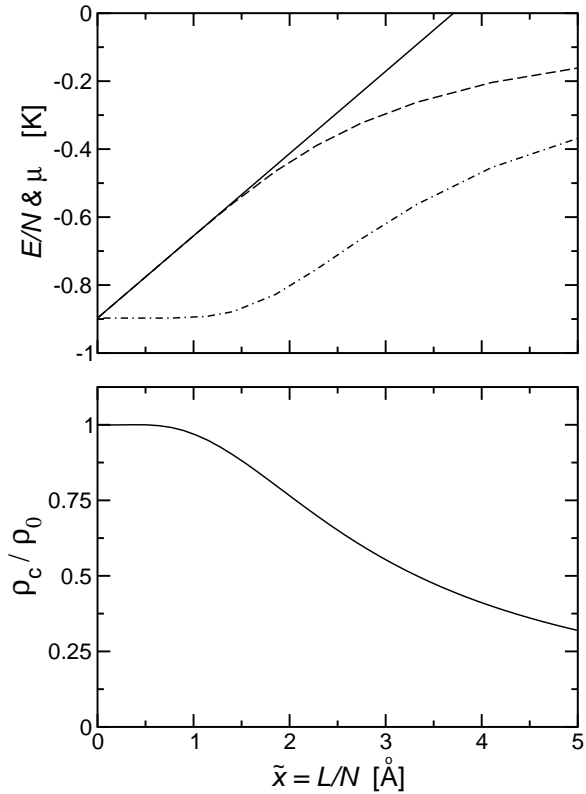}
\end{minipage}%
\end{figure}

The energy per particle and the chemical potential for $^4$He slabs are
reported in Fig.~\ref{fig:ener-slab} as a function of the inverse of the
coverage $\tilde x$.
In the limit $\tilde x \rightarrow 0$ 
one recovers the binding energy of the uniform system at $\R_0$,
which in turn coincides with the chemical potential. The energy per particle
has a very clean linear behavior at the origin as illustrated by the 
solid straight line which provides a very good description of the energy 
per particle up to $\tilde x \sim 1.5$ \AA. The slope of this line can 
\index{line tension}
be analytically derived and turns out to be twice the line tension,
which acts to reduce the binding energy per particle 
of the slab when increasing the inverse of the coverage. 
%
The derivation of this limiting behavior of the energy per particle
can be easily obtained starting from  Eq.~(\ref{eq:ener1}), which defines the
energy per particle as a function of $\tilde x$, by performing an expansion
around $\tilde x=0$. The leading terms result in
\begin{equation}
  e(\tilde x) = \mu_{\infty} + 2 \lambda \tilde x + \cdots
\label{eq:expa}
\end{equation}

For values $\tilde x\gtrsim 1.5$ \AA\ the energy per particle starts
to bend horizontally and becomes a convex function, approaching zero very
slowly.

In contrast, the chemical potential is very flat at the origin, being
determined by the central density of the slab [cf. Eq.~(\ref{eq:chem})], which 
varies only slightly for small $\tilde x$. This can be seen in the lower panel
of Fig.~\ref{fig:ener-slab}, which displays the ratio of the central density
of the slabs to the equilibrium density as a function of the inverse 
of the coverage. 
In agreement with the chemical potential, the central density
is very flat at small values of $\tilde x$. 
The flatness of the central density and the chemical potential for small 
values of $\tilde x$, indicates 
that the slab approaches very slowly the limit of the infinite system.

\begin{figure}[b!]
  \caption[Density profiles of slabs with varying central densities]
	  {\label{fig:prof-slabs}
	    Density profiles for $^4$He slabs with central densities 
	    $\R_c/\R_0$=0.05, 0.1, 0.25, 0.5, 0.75, 0.9, 0.95, where $\R_0$ 
	    is the equilibrium density of the bulk, indicated by the
	    horizontal dotted line.} 
  \begin{center}
    \includegraphics*[width=10.cm]{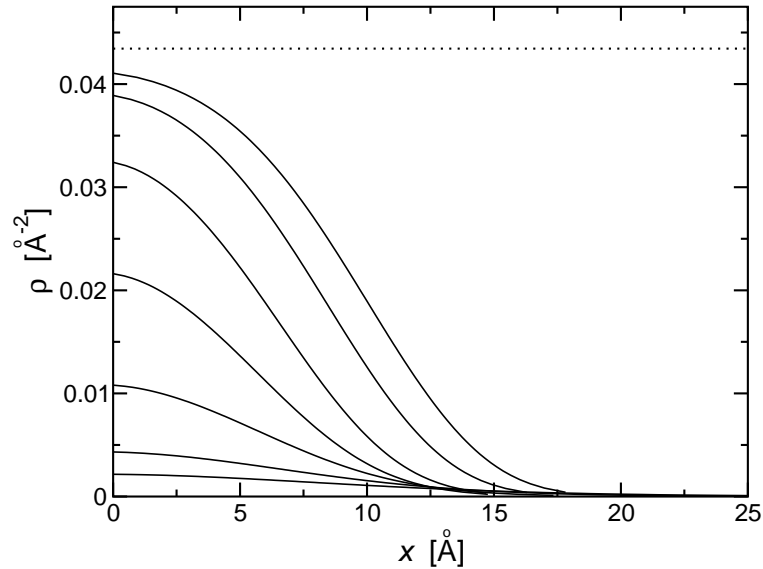}
  \end{center}
\end{figure}

Note that the central density of a slab can never go above the equilibrium
density and is always a decreasing function of $\tilde x$. More generally, 
the density profiles of the slabs can be obtained from Eq.~(\ref{eq:euler4}), 
which we now rewrite as
\begin{equation}
  x = \int_{\R_c}^{\R(x)} \mathrm{d}\R \left(\frac{d\R}{dx}\right)^{-1} =
  \int_{\R(x)}^{\R_c} \mathrm{d}\R \left[
    \frac{\DD\frac{1}{\R}+d} {b\R^2+c\R^{2+\g}-\mu\R}\right]^{1/2} \;,
  \label{eq:zr-profile}
\end{equation}
valid for $x \ge 0$. The profile for $x<0$ is found by symmetry.
Notice the presence again of a divergence when $\rho \rightarrow \rho_c$ which
can also be avoided integrating by parts.
The profiles calculated for various central densities are plotted in
Fig.~\ref{fig:prof-slabs}.

\begin{figure}[tp]
  \caption[Radius and thickness of slabs as a function of coverage]
	  {\label{fig:radius.slab}
	    Radius (left panel) and thickness (right panel) of $^4$He slabs 
	    as a function of $\tilde x$. The symbols are the calculated data 
	    while the lines are cubic splines to guide the eye.
	  }
  \begin{center}
    \includegraphics*[width=12cm]{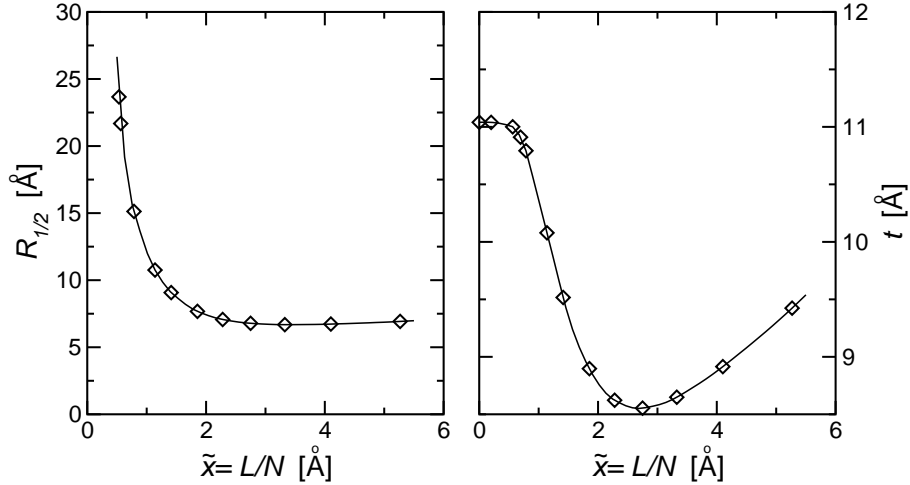}
  \end{center}
\end{figure}

The size of the slab increases with the
central density. A measure of this size is given by the radius $R_{1/2}$,
defined as the distance from the origin to the point where the density has 
decreased to half its central value. This quantity is shown on the left panel
of Fig.~\ref{fig:radius.slab} as a function of $\tilde x$.
As expected, it diverges in the limit $\rho_c \rightarrow \rho_0$ 
and is mainly a decreasing function of $\tilde x$.
However, it also presents a very shallow minimum around 
$\tilde x \sim 3.3$ \AA, after which it grows slightly. 
Physically, one can understand this last feature from the fact that 
for $\tilde x > 3$ \AA\ the density of the system is everywhere small
(cf. Figs.~\ref{fig:ener-slab}--\ref{fig:prof-slabs}): each particle will have
very few others to attach to, and the slab will become more extended.

The profiles are also characterized by the thickness $t$, 
defined as the distance between the points where the density 
has decreased from $90 \%$ to $10 \%$ of its central value. 
The dependence of the thickness on $\tilde x$ is shown on 
the right part of Fig.~\ref{fig:radius.slab}. 
Up to $\tilde x\sim 1$, the thickness is a pretty flat function, indicating
that the surface of the slab is very much the same, and is actually the radius
of the slab that grows very fast and diverges when $\tilde x \rightarrow 0$.  
The thickness is large for the slabs with larger $\tilde x$ again due
to the low density of these systems.
When the central density is increased ($\tilde x$ decreases), the thickness
decreases until it has a minimum at $\tilde x\approx 2.75$ \AA, with
$t\approx8.6$ \AA. Then, it increases again approaching a finite value at the
origin, corresponding to the semi-infinite medium, $t\approx 11.03$ \AA.
It is interesting to note the {\em plateau} that appears for $\tilde x < 1$,
which may be related to the analogous behavior for the central density
(cf. Fig.~\ref{fig:ener-slab}): the slabs are
in many aspects similar to the semi-infinite system already for 
$\tilde x \approx 1$. However, other properties (most importantly, the energy
per particle) do need to go to the limit $\tilde x \raw 0$ to get rid of its
finite-size dependencies. On the other hand, for $\tilde x > 3$ the slabs
seem to change from a strongly correlated behavior over to a more `dilute'
one.

\section{Drops and line tension}
\label{sec:dr}
\index{line tension}
As a next step, we consider 
drops of a fixed number of atoms $N$. These where already studied by \DMC\ 
techniques in Chapter~\ref{ch:heli-dmc}. However, computational limitations
allowed us to study only small values of $N$. Here, we will take advantage
of the computational feasibility of the Density Functional calculations 
and will extend the analysis to much greater $N$. In this way it will be
possible to study the asymptotic behavior of several quantities which 
characterize the drops.

Equation~(\ref{eq:zr-EL}) for the profile can be rewriten in the form of a
Schr\"odinger-like equation for $\R$:
\begin{equation}
   \widehat{\cal H}\R \equiv
   -\Dd\left[ \nabla^2\R - \frac{|\nabla\R|^2}{2\R} \right]
   +2b\R^2 + (2+\g)c\R^{2+\g} - 2d\R\nabla^2\R 
   = \mu\R
  \label{eq:zr-sch}
\end{equation}

As the number of atoms in a droplet is a well defined $N$, one would need to
be very careful in determining the central density $\rho(0)$ so that the
chemical potential adjusted exactly to $N$. However, it turns out that this
equation is more efficiently solved by means of the steepest descent method
\cite{floca}.
An initial trial $\rho_{\rm input}(r)$ is projected onto the mininum of the
functional by propagating it in imaginary time $\tau$. In practice, one
chooses a small time step $\Delta \tau$ and iterates the equation
\begin{align}
  \label{eq:propa}
  \R(r,\tau+\De \tau) &\approx \R(r,\tau) - \De \tau\, \widehat{\cal H}
                       \R(r,\tau) \\
  \R(r,0) &\equiv \R_{\rm input}(r) \notag \:,
\end{align}
\index{density normalization!in \df\ theory}
normalizing $\rho$ to the total number of atoms at each iteration. 
The time step $\Delta \tau$ that governs the rate of convergence should be
taken appropriately small in such a way that Eq.~(\ref{eq:propa}) is a valid
approximation. Convergence is reached when the local chemical potential
---defined as the value of $[\widehat{\cal H}\R] /\R$--- has a constant value
independent of position. In our calculations, typical values have been 
$\Delta \tau \sim 10^{-5}$, and some 10$^6$ iterations have been necessary to
reach convergence. This large number of iterations does not translate into
long runs, as the local nature of the functional allows for fast
calculations. Nevertheless, it is also possible that including in 
Eq.~(\ref{eq:propa}) second or higher orders in $\De\tau$, 
the computation might speed up, which we have not attempted.

\begin{figure}[t]
  \caption[Energy per particle and chemical potential of droplets]
	  {\label{fig:enerdrops}
	    Energy per particle  (empty circles) and  chemical potential
	    (full circles) of $^4$He droplets as a function of $z=N^{-1/2}$. 
	    Also shown is a quadratic fit (see text) of the results with 
	    $N \geq 512$  (solid line). The straight short-dashed line is
	    obtained when the mass formula (with terms up to $z^2$) is used 
	    to calculate the chemical potential. The empty squares are the
	    \DMC\ energies from Chapter~\ref{ch:heli-dmc}, and the dotted line
	    corresponds to the quadratic fit to these results reported there.
	  }
  \begin{center}
    \includegraphics*[width=12cm]{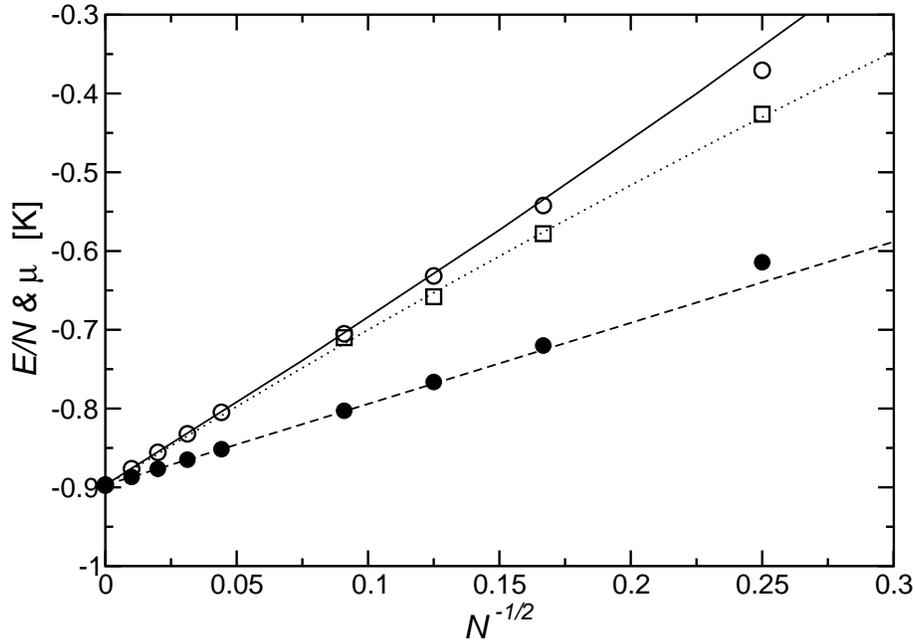}
  \end{center}
\end{figure}

The energy per particle (empty circles) and the chemical potential 
(full circles) calculated for drops with $N=16$, 36, 64, 121, 512, $1\,024$,
$2\,500$ and $10\,000$ atoms are reported in Fig.~\ref{fig:enerdrops} as a
function of $N^{-1/2}$. Also shown for comparison are the \DMC\ energies (empty
squares) and their quadratic fit to a mass formula calculated in
Chapter~\ref{ch:heli-dmc}.
The calculated energies of the droplets can again be represented very
accurately with a mass formula of the type [cf. Eq.~(\ref{eq:mass-f})]
\begin{align}
  e(N) = \varepsilon_b + \varepsilon_l z + \varepsilon_c z^2 + \cdots
  &~~~~~(z = N^{-1/2}).
  \label{eq:massfor}
\end{align}

Contrary to the \DMC\ calculations, where the largest droplet that we studied
had 121 atoms, here we have considered puddles with up to $10\,000$ atoms. In
this way we can accurately study the behavior of the energy per particle for
small values of $z$. Doing a quadratic fit to the calculated energies per
particle for $N \ge 512$ and including also the bulk binding energy for $z=0$
one gets
\begin{equation}
  y = -0.897 + 2.0587 z + 0.66466 z^2 ,
\end{equation}
which is plotted by a solid line in the figure. 
One sees that $\varepsilon_b$ accurately reproduces the bulk energy per
particle, which was used to fix the parameters of the functional; we note that
a similar agreement was found for the \dmc\ fit parameters 
(see Table~\ref{tab:coefs}).
The value of $\varepsilon_l = 2.0587$ (cf. $\eps_l=2.05$ K for the \dmc\ fit)
\index{line tension}
corresponds to a line tension $\lambda =0.121$ K/\AA, which is the same as the
value of the line tension of the semi-infinite system, used to build the
density functional. 
Note, however, that the fit, even if it has been calculated with the data for
$N \ge 512$, is rather accurate down to $N=36$. Obviously, one can not expect
a good agreement for $N=16$, which is a system where finite-size effects are
very important.

The linear behavior of the chemical potential as a function of $z$ contrasts
with the {\em plateau} that was found for the slabs. This linearity is easy
to understand using the mass formula (\ref{eq:massfor}) and the thermodynamic
definition of the chemical potential $\mu = \partial E / \partial N$, where
$E$ is the total energy of the droplet. With this prescription, the slope of
the chemical potential as a function of $z$ results to be 
$\varepsilon_l /2$. This behavior is illustrated by the short-dashed line in
the figure, which nicely follows the calculated data down to $N=36$.
A similar analysis for the three-dimensional case, would provide a behavior
of the chemical potential as a function of $z=N^{-1/3}$ dominated also by a
linear component, but with a slope at the origin given by $2 \varepsilon_s
/3$, where $\varepsilon_s$ is the surface energy associated to
three-dimensional clusters \cite{str87}.

It is interesting to remark that the coefficient of $z^2$ is positive. This
sign corresponds to the expected loose of binding energy associated to the
curvature of the contour of the cluster (cf. Fig.~\ref{fig:curvatura}), and it
is in contrast with the value of $\vareps_c$ obtained by fitting the \DMC\
results (see Table~\ref{tab:coefs}).
However, in that case the number of particles in the clusters
used to build the fit was much smaller, being $N=121$ the largest number of
particles and going down to N=8 for the smallest one. In the present fit we
have explicitly avoided the clusters with a small number of particles which
can easily distort the results 
and we have considered only the cases with $N \ge 512$.

\begin{figure}[t]
  \caption[Density profiles of droplets with varying number of atoms]
	  {\label{fig:prof-drops}
	    Density profiles for $^4$He droplets for $N$=16, 64, 121, 512,
	    $1\,024$, $2\,500$ and $10\,000$ atoms as fitted to a generalized
	    Fermi profiles [see Eq.~(\ref{eq:ferprofi}) and
	    Table~\ref{tab:fermi1}]. The dotted horizontal line indicates the
	    equilibrium density of the homogeneous system $\rho_0$.}
  \begin{center}
    \includegraphics*[width=11cm]{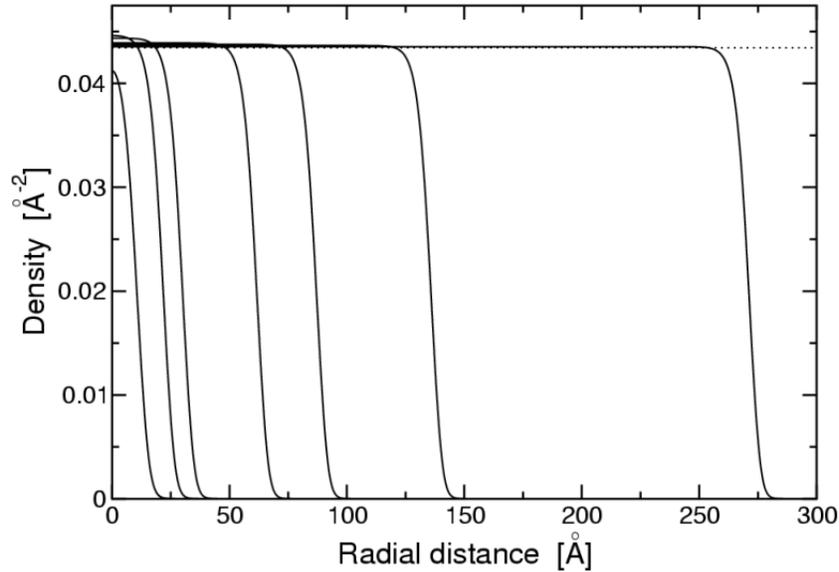}
  \end{center}%
\end{figure}

Next thing to analyze are the density profiles, which we report in
Fig.~\ref{fig:prof-drops} for different numbers of atoms. 
Contrary to the slabs, the central density of the droplets can be higher than
the saturation density, which is indicated in the figure by an horizontal
line. The profiles are well adjusted by a generalized Fermi function,
\begin{equation}
  \rho(r) = \frac {\rho_F}{\left[ 1 + e^{\frac {r-R}{c}} \right]^{\nu}} \:,
  \label{eq:ferprofi}
\end{equation}
which has an associated central density 
$\rho(0) = \rho_F /[1+e^{-R/c}]^{\nu}$. 
The parameters fitting the profiles for the different clusters are
quoted in Table \ref{tab:fermi1}, together with the thickness and
root-mean-square radius obtained thereof.
\begin{table}[ph]
  \begin{center}
    \caption[Parameters of a Fermi-profile fit to the \DF\ density profiles]
	    {\label{tab:fermi1}
	      Parameters of a generalized Fermi-profile fit
	      [Eq.~(\ref{eq:ferprofi})] to the density profiles obtained with
	      the zero-range density functional. All lengths are in \AA\ and
	      $\R_F$ is in \AA$^{-2}$. The parameter $\nu$ is adimensional.} 
    \vspace{0.3cm}
\newcolumntype{d}[1]{D{.}{.}{#1}} 
\newcommand{\mul}[1]{\multicolumn{1}{c}{#1}} 
\newcommand{\mulb}[1]{\multicolumn{1}{c|}{#1}} 
\newcommand{\mulbb}[1]{\multicolumn{1}{|c|}{#1}} 
    \begin{tabular}{|d{0}|d{4}d{4}d{5}d{5}|d{2}d{3}|}
      \hline
\mulbb{$N$} & \mul{$\R_F$} & \mul{$R$} & \mul{$c$} & \mulb{$\nu$} & \mul{$t$} 
    & \mulb{$\langle r^2\rangle^{1/2}$} \\ 
      \hline\hline
16       & 0.04321 & 13.2718 & 3.22067 & 2.13417 & 11.746 &  9.664 \\
36       & 0.04494 & 19.1852 & 3.22054 & 2.37516 & 11.548 &  12.793 \\
64       & 0.04974 & 24.6826 & 3.16845 & 2.37072 & 11.364 & 16.276 \\
121      & 0.04441 & 32.8314 & 3.12302 & 2.33636 & 11.226 & 21.721 \\
512      & 0.04392 & 64.4245 & 3.08867 & 2.32368 & 11.112 & 43.533 \\
$1\,024$ & 0.04378& 89.8497 & 3.08584 & 2.32997 & 11.097 & 61.342 \\
$2\,500$ & 0.04366& 138.628 & 3.08552 & 2.33826 & 11.090 & 95.678 \\
$10\,000$& 0.04355& 274.026 & 3.08653 & 2.34670 & 11.088 & 191.276 \\
      \hline
    \end{tabular}
  \end{center}
\end{table}

\begin{figure}[pb!]
  \caption[Central density and rms radius of various droplets]
	  {\label{fig:central-density}
	    (Left) Central density of $^4$He droplets as a function of
	    $z=N^{-1/2}$. The empty circles correspond to the results obtained
	    with the zero-range density functional, while the line stands for
	    a cubic spline fit to these data. The dotted horizontal line
	    indicates the saturation density $\rho_0$. 
	    (Right) Root-mean-square radius of $^4$He droplets as a function
	    of $N^{1/2}$. The solid line is a linear fit to the data without
	    independent term. The empty circles correspond to the results of
	    the zero-range density functional.}
  \begin{center}
    \includegraphics*[width=13cm]{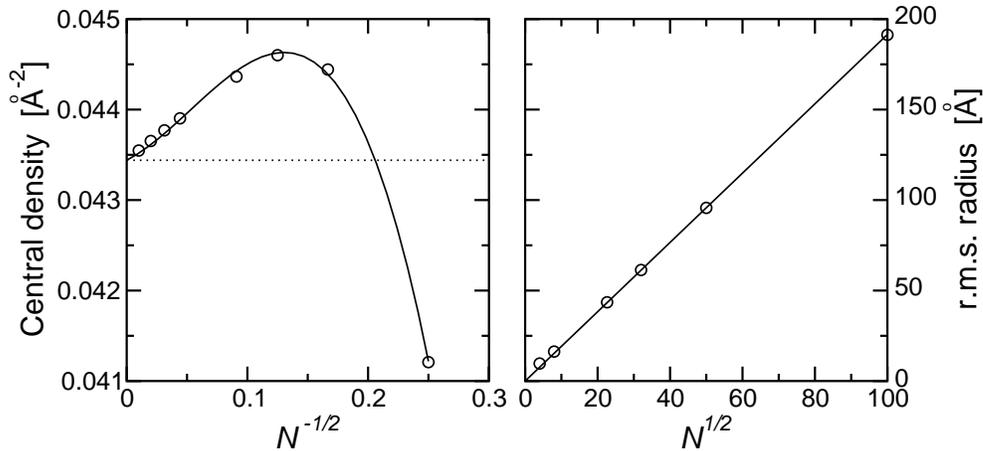}
  \end{center}%
\end{figure}

The left panel of Fig.~\ref{fig:central-density} reports the central 
density of the different droplets as a function of $z$. For large values 
of $N$, $\R(0)$ is larger than the saturation density, \ie\ the central part
of the droplet is more compressed than the bulk system, which is sometimes
referred to as a leptodermous behavior~\cite{str87}.  Of course, for $N
\rightarrow \infty$ the central density tends to $\R_0$. 
First, $\R(0)$ grows almost linearly with $z$, reaches
a maximum, for $z\approx0.13$ ($N \approx 60$), which would correspond to the
most compressed droplet and, finally, for $N \leq 25$, the central region of
the droplets becomes rapidly less compressed than the bulk system.

The root-mean-square (rms) radius corresponding to the profiles is shown in
the right panel of Fig.~\ref{fig:central-density} as a function of $N^{1/2}$. 
The expected linear behavior, associated to a constant average density
[cf. Eq.~(\ref{eq:rmsrad})],
\begin{equation}
  \langle r^2 \rangle^{1/2} = \inv{\sqrt{2\pi\overline\rho_0} } N^{1/2} \:,
\end{equation}
is rather apparent. A linear fit to the calculated values (without independent
term), from $N=16$ to $N=10\,000$, provides $\overline\rho_0 =0.0434$
\AA$^{-2}$, in very good agreement with the bulk equilibrium density used to
fix the parameters of the density functional.   

\section{Hydrodynamic equilibrium in a droplet}
\label{sec:pressure}

\index{hydrodynamic equilibrium}

We proceed now to analyze the hydrodynamic equilibrium in these helium
drops. To this end, it may be interesting to analyze the pressure in one of
these clusters.  
The study of fluid interfaces goes back to Gibbs~\cite{gibbs}. However,
a clear definition of local thermodynamic variables has not yet been
reached. It has been established that, across an interface, some variables
(such as the chemical potential) must remain constant, while for others (such
as the pressure) this is not so, and an unambiguous definition is lacking for
the latter, even though there have been several attempts in recent
years~\cite{row84,row93,row94,lovett97a,lovett97b,over-pccp,haf02}.
The usual thermodynamic expression for the pressure at zero temperature is
\begin{equation*} 
  p = -\left( \frac{\partial E}{\partial V} \right)_{N} 
    = \rho^2 \left( \frac{\partial e}{\partial \rho} \right)_{N} 
    =\rho(\mu-e)
    \:,
\end{equation*}
where we also used the definition of the chemical potential, 
$\mu=(\partial E/\partial N)_V$, $e$ stands for the energy per particle
and $\mu$ is the chemical potential, constant throughout the system.
This expression depends only on intensive variables, and is therefore 
appropriate to introduce the concept of local pressure in a finite system.

%
For our simple zero-range density functional (\ref{eq:zr-func}),
the energy per particle can be written as a function of the density thus
\begin{equation*}
  e(\R) = \DD\frac{\lp\nabla\R\rp^2}{\R^2} +b\R +c\R^{1+\g} 
    + d\frac{\lp\nabla\R\rp^2}{\R} \:,
\end{equation*}
which, combined with the equation above, will define a `pressure profile' 
$p(r)$. 
Moreover, all the contributions to $e(\R)$ as well as $\mu\R$ are quantities
computed during the imaginary-time evolution towards the ground
state. Therefore, the evaluation of $p(r)$ does not require any further
computational effort.

\begin{figure}[th]
  \caption[Pressure profile in the clusters with $N=36,10^4$ atoms]
	  {\label{fig:pressure}
	    Pressure profile in the clusters with $N=36$ (left) and
	    $N=10\,000$ (right) atoms:
	    the solid lines are the density profiles (left ordinate axes)
	    while the dashed lines are the calculated pressure profiles
	    (right ordinate axis). 
	    The thin horizontal lines indicate the saturation density 
	    $\R_0$ and the dotted lines mean zero pressure.
	  }
  \begin{center}
    \includegraphics*[width=12cm]{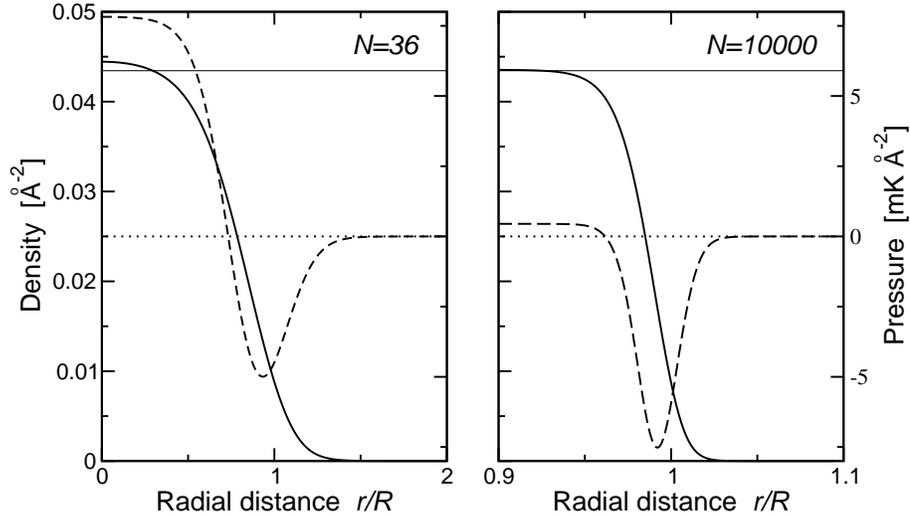}
  \end{center}%
\end{figure}

The calculated $p(r)$ for two sample droplets ($N=36,~10\,000$) is shown as a
function of a rescaled radial distance $r/R$ in
Fig.~\ref{fig:pressure}, with $R$ the fitting parameter of
Table~\ref{tab:fermi1}. 
The value $p=0$ outside the drop is due to the fact that we are working in a
vacuum at zero temperature, and therefore there is no vapor pressure exerted
on the drop.
The positive value of $p$ in the innermost region of the droplet 
is related to the density there being larger than the saturation density.
This behavior is also expected from the Young-Laplace law for a liquid-vapor
interface, that relates the pressures in the bulk of each phase with the
surface tension $\lambda$ and the mean curvature radius $R_{\rm curv}$ of the
interface~\cite{lovett97b,safran}. For the case of cylindrical
symmetry, 
\begin{equation}
  p_{\rm liq} = p_{\rm vap} + \frac{\lambda}{R_{\rm curv}} \:.
  \label{eq:lap}
\end{equation}
As mentioned above, in the zero temperature case $p_{\rm vap}=0$.
The fact that the regions where $p>0$ and $\R>\R_0$ do not coincide exactly is
due to the gradient terms of the Density Functional, that are not present for
the homogeneous case that fixes $\R_0$.
Using now Eq.~(\ref{eq:lap}) with $\lambda=\lambda_{\rm DMC}$ and the
calculated values $p(0)=7.82$ mK/\AA$^2$ ($N=36$) and $p(0)=0.447$ mK/\AA$^2$
($N=10\,000$) as the values of the pressure in the liquid phase, we find the
values for the curvature radii of the clusters with $N=36$ and $10\,000$ atoms
to be $R_{\rm curv}=15.5$ \AA\ and $R_{\rm curv}=271$ \AA, which are rather
close to the $R$-values from the Fermi-profile fit (see
Table~\ref{tab:fermi1}). This can be seen as an indication of the validity
of our definition for $p(r)$, 
specially for large drops, where the hypothesis of a sharp interface 
that resides in the Young-Laplace law is better.

Close to the surface, we find a negative value for the pressure, that is,
there is a tension located near the free surface of the puddle.
This is especially clear for the larger drop, where the region where $p<0$ is
closely restricted around the surface (note the difference in scales on the
$x$-axes of both figures), while for the $N=36$ system the separation between
a bulk and a surface regions is not easy to ascertain.
Similar results have been found in various Molecular Dynamics calculations for
liquid-vapor interfaces of different geometries (planar, cylindrical), even
if using a variety of definitions for the local pressure (see,
\eg,~\cite{row84,lovett97b}).

\begin{figure}[bt]
  \caption[Forces on an area of a two-dimensional drop]
	  {\label{fig:forces} 
	    (Left) $F(r)$ stands for the effective force that keeps the drop
	    bound.
	    (Right) The density profile (black line) and the effective force
	    $F(r)$ (red line) calculated to keep the drop with $N=10\,000$
	    atoms in equilibrium, see Eq.~(\ref{eq:forca}).}
  \begin{center}
    \includegraphics*[width=5.7cm]{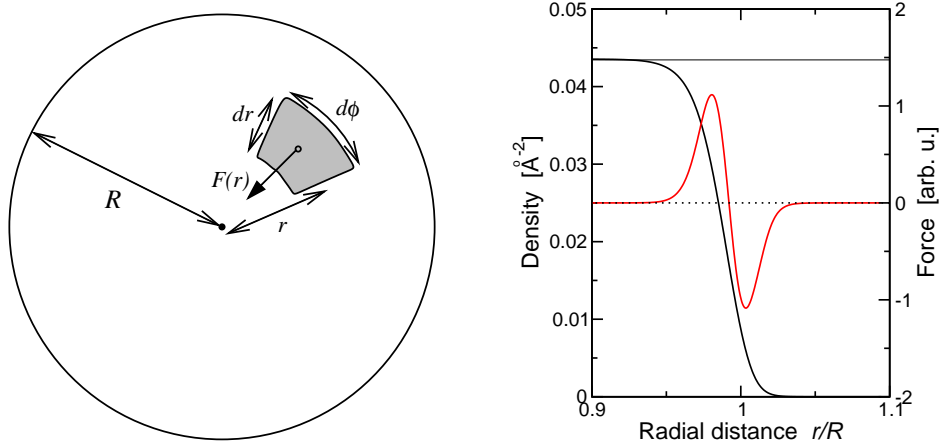}%
    \hskip1cm
    \includegraphics*[width=5.7cm]{clusters/forca}
  \end{center}%
\end{figure}

This picture of the pressure profile can be made more physically appealing 
\index{hydrodynamic equilibrium}
by interpreting it in terms of hydrodynamic equilibrium. 
To this end, let us analyze the forces exerted on an area of a drop as 
depicted on the left panel of Fig.~\ref{fig:forces}.
The area will be in equilibrium if 
there exists a force $F$ that balances 
the pressure difference on the two sides.
This force must satisfy
\begin{align}
  F(r) &= -r\,d\phi\, [ p(r+\frac{dr}{2})-p(r-\frac{dr}{2}) ] 
        = -r\,d\phi \frac{dp}{dr}\,dr
	= -dA\frac{dp}{dr} \:,
  \label{eq:forca}
\end{align}
where $dA$ is the surface of the shaded area, and the sign convention is such 
that $F>0$ pulls atoms to the center of the puddle.
\index{He-He interaction}
This force is ultimately determined by the He-He interaction.

Thus, the pressure profile can be mapped onto a central potential for 
the helium atoms.
The force calculated in this way for the drop with $N=10\,000$ atoms is
plotted on the right panel of Fig.~\ref{fig:forces}. It vanishes for positions
$r \lesssim 0.94 R$, corresponding to the region where $\R \approx \R_0$: in
this region, the system is locally equivalent to the bulk, and no net forces
are present. The force also vanishes far from the cluster: one could add a new
$^4$He atom there which would not feel the presence of the drop. In the region
close to the surface, $F(r)$ presents a maximum at 
$r_{\rm max}\approx 0.98R \approx 269$ \AA,
a node at $r_{\rm node}\approx 0.99R \approx 272$ \AA\ 
and a minimum at $r_{\rm min} \approx R \approx 275$ \AA.
The position of the node corresponds to the minimum in the curve for $p(r)$,
and indicates where an atom might be added and stay at rest. On the
contrary, if it were added at smaller or larger $r$ it would be displaced
towards $r_{\rm node}$. The linear behavior of $F(r)$ close to $r_{\rm node}$
shows that this motion would be harmonic, as expected around a minimum
of a potential.
Note that the distances between these three positions are barely equal to the
\index{He-He interaction}
characteristic length of the He-He interaction, $\sigma=2.556$ \AA.


\section{Discussion of the results}
\label{sec:discussion}

Let us finally make a comparison between the results obtained for clusters
with the present \df\ calculations and the \dmc\ ones in the previous 
chapter, and also between our results and those for three-dimensional slabs
obtained from various density functionals~\cite{szy}.

\subsection{2D and 3D slabs in Density Functional theory}
\label{ss:disc-slabs}
First of all, we compare our \df\ results for two-dimensional $^4$He
slabs (Sect.~\ref{sec:sl}) with those found in the literature for
three-dimensional slabs.
In particular, we will look at Ref.~\cite{szy} which reports a
comparison of calculations performed with a variety of density
functionals. Indeed, the functional $\vareps[\R]$ in Eq.~(\ref{eq:dens-fun}) 
is in principle rather arbitrary, and only a comparison of the obtained 
results from different choices with experimental
data (or exact calculations as provided, \eg, by the \qmc\ method) can decide
what functional dependence is better to reproduce certain properties. Thus,
for 3D helium systems a `fauna' of density functionals has developed,
starting from the simple zero-range functional originally proposed by 
Stringari~\cite{str84}, to more elaborate finite-range functionals such as
those referred to as Orsay-Paris~\cite{df-op}, Catalonia~\cite{barran93}, or
Orsay-Trento~\cite{dalfo95} in~\cite{szy}.

Of course, every functional is expected to give the same results for the
homogeneous system, and other properties used to fix its parameters. 
The differences appear when calculating features not directly incorporated in
their construction, such as surface properties of slabs or finite clusters. 
However, they tend to present some general common behaviors. 
For example, the energy per particle and chemical potential of 3D $^4$He slabs
depend on the inverse coverage in a way
quite similar to what we find
in 2D systems (compare our Fig.~\ref{fig:ener-slab} with Fig. 2 in~\cite{szy}). 

\begin{figure}[tb]
  \caption[Comparison of the 2D and 3D equations of state]
	  {\label{fig:ener-2d-3d}
	    Comparison of the 2D and 3D equations of state.
	    The thick lines are \dmc\ results, while the thin lines
	    are their extrapolations (\ref{eq:homog-df}) 
            to higher and lower densities. The triangles and circles 
            denote the spinodal and freezing points, respectively.
	    The black curves and empty symbols refer to the 3D data,
	    while the green curves and filled symbols stand for the 2D data.
	  }
  \begin{center}
    \includegraphics*[width=10cm]{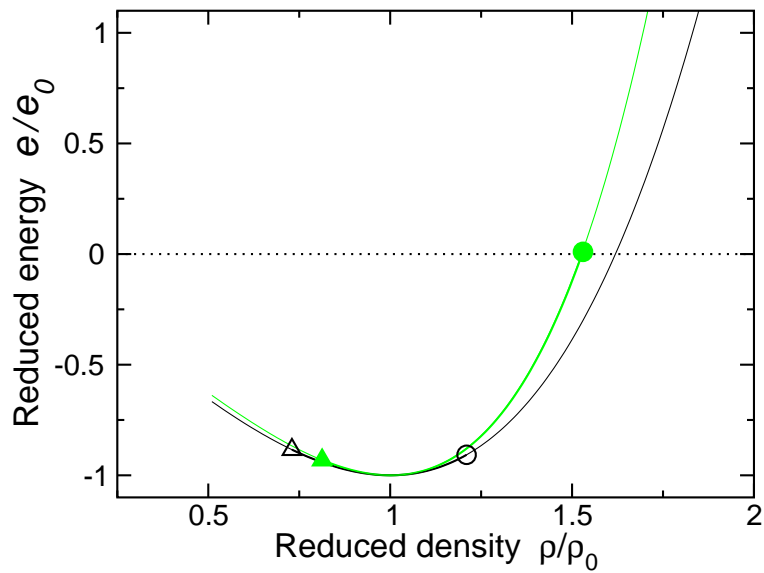}
  \end{center}%
\end{figure}

It is worth noticing, however, some intrinsic differences between the 2D and
3D homogeneous systems. To this end, let us plot 
the reduced energy per particle $e(\R)/e_0$ as a function of the
reduced density $\R/\R_0$,
where $e_0$ and $\R_0$ are the energy particle
and the density, respectively, at saturation
(see Fig.~\ref{fig:ener-2d-3d}). 
We recall that $e_0^{3D}=-7.17$ K and
$\R_0^{3D}=0.021837$ \AA$^{-3}$. 
Even though the values for $E/N$ are quite similar, the curvature of both
lines is different, as can be seen from their extrapolations to higher and
lower densities using a fit of the form
\begin{equation}
  e(\R) = b\R + c\R^{1+\g} \:.
  \label{eq:homog-df}
\end{equation}
This feature is reflected in the speed of sound at saturation density, which 
in 2D is smaller than in 3D ($s^{2D}=92.8$ m/s {\em vs.} $s^{3D}=237$ m/s). 
These different behaviors may be expected to affect also the structure of 
finite systems, where compressibility plays an important role in 
determining density profiles and other properties.
It is also remarkable the broader range of densities where 
helium remains liquid in 2D as compared to the 3D case,
cf. Fig.~\ref{fig:ener-2d-3d}.

The thickness of slabs in 3D is quite flat as a function of the
inverse of the coverage $\tilde x=L/N$ (cf. Fig. 4 in~\cite{szy}). 
This behavior contrasts with the 2D case where we observe a 
minimum in the $t(\tilde x)$ function (cf. Fig.~\ref{fig:radius.slab}). 
This discrepancy might just be due to a broader range of inverse converages
explored in our work as compared with~\cite{szy}.
In fact, our results are also quite flat in the region $\tilde x \lesssim 2$,
which relates to the larger slabs we have studied, 
$\R_c/\R_0 \gtrsim 0.7$. 
Indeed, the calculations presented in~\cite{szy} are for slabs with
$\R_c/\R_0^{3D}\gtrsim 0.6$.
However, this does not rule out an influence of dimensionality on the
behavior of the thickness, as is pointed out by the different values
for the semi-infinite system: $t_{\rm semi}^{2D}\approx 11$ \AA\
{\em vs.} $t_{\rm semi}^{3D}\approx 7$ \AA.

\subsection{2D clusters according to \DF\ and \qmc\ calculations}
\label{ss:disc-clust}

Let us now turn to finite clusters and compare the results we have obtained by
the \dmc\ (Ch.~\ref{ch:heli-dmc}) and \df\ (this chapter) methods. 
First, let us look at the energy per particle, plotted in
Fig.~\ref{fig:enerdrops}. We see that both calculations give very similar
values for $N=121$, while for $N<121$ the \df\ energies are always less bound
than the \dmc\ ones.
On the other hand, there is a nice agreement of the mass formulas for $N
\gtrsim 121$ ($z \lesssim 0.1$). This behavior is to be expected as the
parameters defining the functional ($b,c,\g,d$) have been fixed so as to
reproduce properties of the homogeneous and semi-infinite systems (see
Sect.~\ref{sec:sl}). For the same reason, the deviations for small clusters,
where surface effects are most important, are not surprising. Indeed, from 
Fig.~\ref{fig:prof-drops}, we see that the density profile of the droplets
does not present a truly `bulk' region in its interior for clusters with
less that 121 atoms. Nevertheless, the maximum relative deviation of the
energy per particle is just around 15\%.

\begin{figure}[th]
  \caption[Comparison among \vmc, \dmc\ and \df\ density profiles]
	  {\label{fig:compar-prof}
	    Density profiles of the $N=64$ cluster from the \vmc\ 
	    (solid line), \dmc\ (dashed line) and \df\ (dash-dotted line)
	    calculations. 
	    The thin horizontal line is the bulk saturation density.
	  }
  \begin{center}
    \includegraphics*[width=12cm]{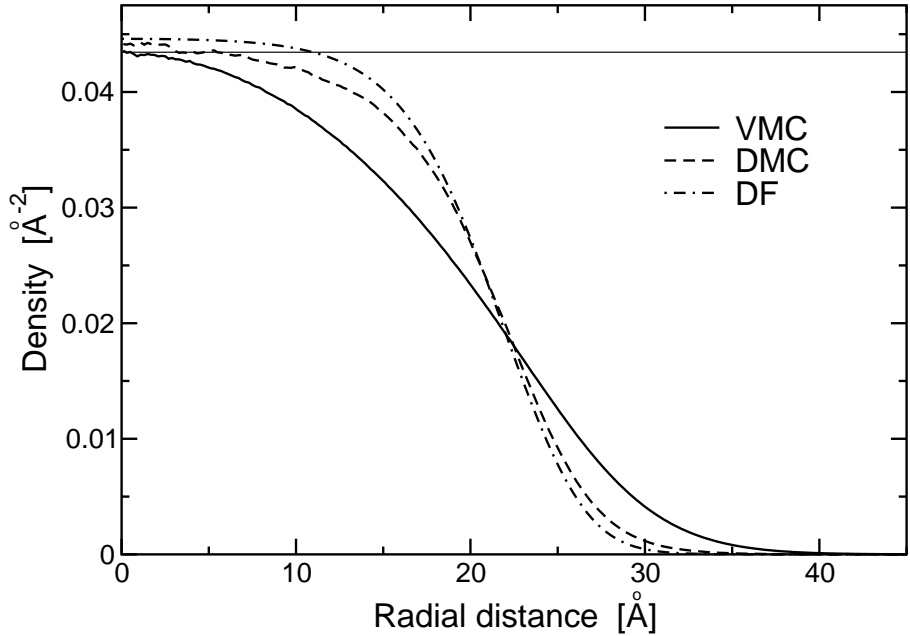}
  \end{center}%
\end{figure}

In Fig.~\ref{fig:compar-prof} we show the density profiles obtained in the
\vmc, \dmc\ and \df\ approaches for the $N=64$ drop, for which the energy 
calculated by the last two techniques is quite similar and the fluctuations in
the \dmc\ profile are small.
In this comparison one can see that the \df\ and \dmc\ profiles are
quite close to each other, with the surface location and thickness 
very similar as compared with the \vmc\ ones.
Thus, we can conclude that the description given by the zero-range Density
Functional (\ref{eq:zr-func}) is preferable over that of the simple \vmc\
calculation.
In this sense, it is worth mentioning that density profiles of 3D drops 
calculated with \dmc\ and with finite-range density functionals
show a similar agreement~\cite{bar-jltp}.
Indeed, the values for $R$ and $\la r^2 \ra^{1/2}$ of the generalized
Fermi-profile fits for the \dmc\ and \df\ data are in general in good 
agreement (cf. Tables~\ref{tab:fermi} and \ref{tab:fermi1}). 
Even though a closer look at the data shows that $R$ in the \dmc\ 
calculations is typically smaller than that from the \df\ ones, 
the behavior of the rms radius 
is very much the same, with differences at most of 6\%.

Regarding the thickness, one observes that it is quite constant for all
\df\ profiles, with a slowly decreasing behavior towards the value for the
semi-infinite system, $t_{\rm semi}=11.03$ \AA. 
This is in clear contrast with what happens with the \dmc\ data, which are
{\em growing} with $N$. 
In fact, the \df\ thickness turns out to be systematically smaller than the
\dmc\ one, \ie, the \df\ profiles present sharper surfaces than the \dmc\
ones.


\section{Summary and conclusions}
\label{sec:conc}

\index{line tension}
We have constructed a density functional suitable to study non-homogeneous
two-dimensional $^4$He systems. 
The parameters of the functional have been fixed by demanding it to reproduce
some known properties of the homogeneous and semi-infinite systems.
Then, we have considered two-dimensional slabs
and analytically shown that the line tension extracted from a mass
formula adapted to this type of geometry, is consistent with the value of the
line tension of the semi-infinite system.
We also observe that, as in three-dimensional slabs, the thickness and the
central density of the slabs approach from below the values corresponding to
the semi-infinite system, while the radius diverges.

We have also studied the energetics and structure of 2D clusters. 
In particular, clusters with a very large number of atoms have been used 
to study the behavior of the mass formula and to establish how the 
system approaches the bulk limit. 
The central density of the clusters when $N \rightarrow \infty$ approaches the
saturation density from above and, therefore, the internal regions of the
clusters are more compressed than the bulk system while the external regions
have densities which would correspond to negative pressures or even below the
spinodal point for a uniform system .
This behavior 
reflects the existence of a tension on the boundary of the drop 
that keeps it bound. 

The profiles of the clusters are very well fitted by a generalized Fermi
function. The thickness of the cluster slowly approaches the thickness of the
semi-infinite system, as it also happens in the case of the slab geometry,
but in this case from above.
This behavior contrasts with the \dmc\ data, which were however obtained for
quite small clusters, far from the bulk limit.

Finally, we have analyzed the linear behavior of the rms radius of the
droplets in terms of $N^{1/2}$, and recovered the saturation density from the
slope of this fit. We have also seen that the rms radii obtained from the \df\
calculation are in very good agreement with the \dmc\ results.

Altogether, we conclude that the proposed density functional is useful 
to perform reliable and computationally-affordable calculations 
for a variety of two-dimensional $^4$He systems and, in particular, 
for clusters where the number of particles is prohibitive for a Monte Carlo
calculation.


\fancyhead[LO]{\bfseries Conclusions and future perspectives}
\fancyhead[RE]{\bfseries Conclusions and future perspectives}
\phantomsection
\addcontentsline{toc}{part}{Conclusions and future perspectives}
\chapter*{Conclusions and future perspectives}
\label{ch:concl}

\vskip1cm
\begin{flushright}
  \textsf{%
    \begin{tabular}{r}
      Ars longa, beca brevis \\
      \rule{0pt}{0.5cm} 
      {\em Proverbium becarii}
    \end{tabular}
  }%
  ~~~~~~~~
\end{flushright}%

\vskip1.cm

The purpose of this thesis has been the study of various atomic systems where
the quantum effects are most visible, by means of a number of tools of
many-body physics. In {\bf Chapter~\ref{ch:pair-intro}} we have presented in a 
general form the \BCS\ theory of superconductivity for fermionic systems. We
have made use of the well-known Green's functions' formalism to write the gap
and number equations in a general form, capable of being used at zero and
at finite temperature, both for density-symmetric and asymmetric systems. 
We have recovered the well-known result that the zero-temperature $s$-wave gap 
for the symmetric system
in the \index{weak-coupling limit}weak-coupling limit has an exponential 
dependence on the interaction and the Fermi momentum, namely
\begin{equation*}
  \De_{\rm sym} = \frac{8}{e^2}\mu \exp\lp\frac{\pi}{2k_Fa} \rp
  \qquad (a<0) \:.
\end{equation*}
The \index{critical temperature!for pairing transition}critical temperature 
for the superfluid transition has a similar
behavior on the parameters of the system. 
One can write the corresponding result as
\begin{equation*}
  k_BT_c = \frac{e^\gamma}{\pi} \De_{\rm sym} \approx 0.567 \De_{\rm sym}
  \:.
\end{equation*}

In {\bf Chapter~\ref{ch:asym}} we have analyzed the prospect for $s$-wave
pairing in the \BCS\ framework for large density asymmetries. The main
conclusions of this chapter are:
\begin{itemize}
\item The maximum asymmetry for which \BCS\ pairing is possible is
  \begin{equation*}
    \al_{\rm max}^{BCS} \equiv \frac{\ru-\rd}{\ru+\rd} 
                        = \frac{3}{4}\frac{\De_{\rm sym}}{\mu}
    \:,
  \end{equation*}
  which in the low-density limit is a very small quantity. We have checked
  numerically this prediction by solving self-consistently the gap and number
  equations; the agreement, as shown in Fig.~\ref{fig:num_asym}, is remarkable
  for all $\al$.
\item We have also checked numerically the dependence of the value of the gap
  with temperature for the symmetric system, and seen that it follows the
  expected limiting behaviors at low temperatures $T\ll T_c$ and close to the
  critical region  $T\lesssim T_c$, cf. Fig.~\ref{fig:num_sym}.
\item Finally, we have analyzed the prospect for a \BCS\ transition in $p$-wave
  between identical fermions, mediated by another fermionic species. 
  For a density-symmetric system, the gap
  \begin{equation*}
    \frac{\De_{p\mbox{-}\rm wave}}{\mu} \propto 
    \exp\left[-13\lp\frac{\pi}{2k_Fa}\rp^2\right]
    \:,
  \end{equation*}
  is very small, but it may be optimized by appropriately adjusting the 
  density asymmetry. In practice, a sharp maximum appears for 
  $\al_{\rm opt}\approx0.478$.
\end{itemize}

In {\bf Chapter~\ref{ch:dfs}} we have studied two structures for the ground
state alternative to the usual \BCS\ one: the plane-wave \loff\ and the \dfs\
phases.
\begin{itemize}
\item 
  As it is well-known, the \loff\ phase can sustain pairing for
  larger asymmetries and has lower free energy than \bcs. We have quantified
  this effect for the case of \Lix\ at ultralow temperatures and in the
  weak-coupling regime: we have shown that the \loff\ phase if preferable to 
  the \bcs\ one for asymmetries $4\% \leq \al\leq 5.7\%$. 
\item The \dfs\ phase consists in an ellipsoidal deformation of the Fermi
  surfaces of the two pairing species. More precisely, the surface of the most
  populated species becomes prolate (\ie, cigar-like) along a
  symmetry-breaking axis randomly chosen by the system. The surface of the
  less populated species becomes oblate (\ie, pancake-like), so that the two
  Fermi surfaces have, in the end, approached each other in the plane
  perpendicular to the symmetry-breaking axis. In this case, only rotational
  symmetry is broken.
\item We have numerically evaluated the gap and free energy of the \dfs\ phase,
  and we have shown that it becomes the ground state for the \Lix\ system
  under study for asymmetries $\al\gtrsim 4\%$. 
\item We have proposed an easy scheme of detecting the
  \dfs\ phase by means of a time-of-flight determination of the momentum
  distribution for an atomic gas confined in a spherical trap.
\end{itemize}

It is our purpose to analyze in the future the excitation spectrum of 
the \dfs\ state, to identify the new Goldstone mode that shold appear due to the
breaking of rotational symmetry. 
One should also remember that more general structures for the \loff\
phase are possible, which could qualitatively modify our conclusions.

In {\bf Chapter~\ref{ch:2dbf}} we have considered the effects of the presence
of a bosonic species in a single-component fermionic system. 
\begin{itemize}
\item For the three-dimensional case, the induced pairing is optimized 
  for a bosonic density
  \begin{equation*}
    \R_B = \frac{\R_F^{2/3}}{2.88\abb} \:,
  \end{equation*}
  where $a_{BB}$ is the $s$-wave scattering length for boson-boson collisions.
\item
  For the two-dimensional case, the logarithmic dependence of the transition
  matrix on the collision energy requires a more delicate treatment, and
  the value of the optimal densities depends on the parameters determining the
  collision properties among bosons, $E_{BB}$, and between bosons
  and fermions, $E_{BF}$. We have shown that, for the case that both species
  have similar masses, $E_{BB}$ plays only a minor role, and the optimal density
  ratio is given by
  \begin{align*}
    \left.\frac{\R_B}{\R_F}\right|_{\rm opt} 
       &= -0.59\left[\ln\zeta -0.35\right] \:, \\
    \zeta(\ef) &:= \frac{m_F}{m_B}\frac{\ef}{E_{BB}} \:.
  \end{align*}
  Under these conditions, energy gaps of the order of the fermion chemical
  potential may be achieved. 
\item For the case of trapped, quasi-two-dimensional
  atomic systems, this regime seems to be achievable for mixtures with
  attractive interspecies interaction $a_{BF}<0$ such as \Rb-\K. On the other
  hand, for $a_{BF}>0$ (as in the \Lin-\Lix\ mixtures) one should also analyze 
  the stability of the system.
\end{itemize}

In {\bf Chapter~\ref{ch:spin}} we have studied the real-time evolution of a
Bose-Einstein condensate with spin degree of freedom. We have recalled that,
for the case of spin $f=1$, two possible ground-state structures are possible
\begin{itemize}
\item `Ferromagnetic': when the coupling constant in the collision channel
  with total spin $F=2$ is attractive, $c_2<0$, the spin tends to be locally
  maximized. This translates into a mixture of all spin components in each
  point.
\item `Antiferromagnetic': when $c_2>0$, the ground state is realized by
  minimizing the expectation value of the spin. As a consequence, atoms 
  with spin projection $m=\pm1$ will tend to repel $m=0$ atoms.
\end{itemize}
The dynamical evolution in the first case, which is realized for $^{87}$Rb
atoms in their electronic ground state, presents these characteristics:
\begin{itemize}
\item It shows oscillations in the populations of the various hyperfine 
  states $m=1,0,-1$. At zero temperature, and for an initial state with zero
  magnetizaton, these oscillations are around the ground-state
  configuration $(25\%, 50\%, 25\%)$. At a finite temperature $T=0.2T_c$,
  which we simulate {\em via} the inclusion of phase fluctuations,
  all components are equally populated after a fast damping process.
\item The density profiles feature small spin domains of a characteristic size
  $l_{\rm dom}\sim 10~\mu$m, that appear after $\sim$100 ms. This domain
  formation may be interpreted as the consequence of a dynamical instability
  of the spin excitation modes of the `ferromagnetic' system. At zero
  temperature, the domains formed by $m=\pm1$ atoms overlap, thus conserving 
  locally the initial magnetization. At finite temperature, the symmetry 
  between $m=1$ and $m=-1$ is broken by the fluctuating phases, and different,
  interpenetrating domains of all components appear. In this case, the
  formation of domains is much faster: it occurs for times $\sim$10 ms.
\end{itemize}
We have also analyzed a case with `antiferromagnetic' interactions for which:
\begin{itemize}
\item The population oscillations are almost coherent at zero temperature, but
  present a very small relaxing behavior at finite temperature.
\item The density profiles barely evolve at $T=0.2T_c$, while at zero
  temperature Josephson-like oscillations may be identified.
\end{itemize}
For the future, we plan studies on the role of the initial conditions and on
effects of temperature and external magnetic fields.
Also the study of condensates with higher spin values (in particular, $f=2$)
is interesting in order to interpret a number of experimental results.

\vskip0.2cm

In the second part of the thesis, we have presented two microscopic studies on
$^4$He systems in two dimensions.
In {\bf Chapter~\ref{ch:heli-dmc}} we have performed both Variational and
Diffusion Monte Carlo calculations for finite clusters. First we have built a
good variational function by including Jastrow-like correlations together with
a Gaussian term to take into account the self-binding of the drops. This
function has then been used as importance function for the \dmc\ calculation.
The main conclusions that we get from these studies are:
\begin{itemize}
\item From the results for the energy per particle as a function of the number
  of atoms in the drop, we have constructed a mass formula,
  \begin{equation*}
    \frac{E}{N} = \vareps_b + \frac{\vareps_l}{N^{1/2}} 
                + \frac{\vareps_c}{N} \:.
  \end{equation*}
  The first coefficient coincides with the calculated value for the energy per
  particle at saturation of the homogeneous system. From $\vareps_l$ we have
  extracted a value for the line energy due to the finite character of these
  systems,
  \begin{equation*}
    \lambda_{\rm DMC} = 0.121 \mbox{~K/\AA} \:,
  \end{equation*}
  which is the principal result of this chapter.

\item We have also analyzed the density profiles obtained by the forward walking
  method to evaluate pure estimators. The profiles can be very well fitted to
  a generalized Fermi function, and provide an alternative determination for
  the bulk saturation density, which roughly coincides with the value from
  \dmc\ calculations.
\end{itemize}

In {\bf Chapter~\ref{ch:heli-df}} we have used the previous results for the
homogenous system and the line tension to build a zero-range density
functional, with an energy density of the form
\begin{equation*}
  \vareps[\R] = \hm\frac{|\nabla\R|^2}{4\R} 
                +b\R^2 + c\R^{2+\gamma} +d|\nabla\R|^2 \:.
\end{equation*}
The parameters $\{b,c,\gamma\}$ have been fixed so as to reproduce the
energy per particle, density and speed of sound at saturation, while $d$ has
been determined by demanding that the line tension of the semi-infinite system
equals $\lambda_{\rm DMC}$. The conclusions that we draw from the subsequent
analysis are:
\begin{itemize}
\item The energy per particle of two-dimensional slabs
  approaches that of the bulk system slowly, in a linear way. 
  In this limit the radius of the slabs diverges as expected.
  Other physical parameters approach the corresponding bulk values much faster. 
  For instance, the chemical potential or the thickness have attained their bulk
  values for inverse coverages $\tilde x\approx 1$ \AA$^{-1}$. 
\item For clusters, the bulk limit is harder to attain, and both the energy
  and the chemical potential approach linearly the corresponding limiting
  values. An analysis of the energy per particle in the form of a mass formula
  gives a positive value for the curvature term $\vareps_c$, as one expects
  from the lack of binding energy of the atoms on the surface.
\item Some features of the density profiles of finite clusters are similar to
  their three-dimensional counterparts. For example, the central density
  presents a `leptodermous' behavior, approaching the bulk saturation
  density from above. 
\item Finally, an analysis of these profiles in terms of hydrodynamical
  equilibrium has shown that near the surface of the puddles the pressure
  presents a dip down to negative values. This has been interpreted as the
  effect of the line tension.
  Hydrodynamic equilibrium demands de presence of a force 
  (due to the He-He interaction) that close to the surface is proportional to
  the distance to it. 
\end{itemize}
Further studies to be performed on $^4$He systems are the calculation of the
natural orbits and the subsequent evaluation of the condensate fraction.

\vskip2cm
\begin{flushright}
\em
Vale.

\end{flushright}

\cleardoublepage
\renewcommand{\chaptermark}[1]{\markboth{#1}{}}
\renewcommand{\sectionmark}[1]{\markright{\thesection\ #1}}
\fancyhf{} 
\fancyhead[LE,RO]{\bfseries\thepage}
\fancyhead[LO]{\bfseries\rightmark}
\fancyhead[RE]{\bfseries\leftmark}
\appendix
\appendix
\phantomsection
\annex
\chapter{Summing up over fermionic Matsubara frequencies}
\label{app:suma}
\def\ax{\frac{\beta}{2}}

In this appendix we show how to evaluate a summation over Matsubara
frequencies. First of all, let us shortly introduce the Matsubara
technique for finite-temperature Green's functions (see~\cite{mattuck}).

At zero temperature, the solution of the Schr\"odinger equation
\begin{equation}
  i\hbar\frac{\partial}{\partial t}\psi(\xv,t)=\hat{H}\psi(\xv,t)
\end{equation}
can be reexpressed in terms of the (zero temperature) Green's function
(or \index{propagator}{\em propagator})
\begin{equation}
  G^{T=0}(\xv,t;\xvp,t') := -i\lla \Psi_0 \left| 
			           T\left[ \psio(\xv,t)\psiod(\xvp,t') 
			      \right] \right| \Psi_0 \rra 
   \:,
   \label{eq:app1}
\end{equation}
where $T$ is the time ordering operator,
\begin{equation*}
  T[\ao(t)\bo(t')] = \ao(t)\bo(t')\Theta(t-t')\pm\bo(t')\ao(t)\Theta(t'-t) \:,
\end{equation*}
and $|\Psi_0\rangle$ is the {\em exact} ground state of the sytem.
The operators in (\ref{eq:app1}) are defined int the Heisenberg picture as
\begin{equation*}
  {\cal\hat O}(t) := e^{it\hat H} {\cal\hat O} e^{-it\hat H} \:.
\end{equation*}

Analogously, for a system at $T>0$, we define a propagator as a 
{\em thermal average} of the same operator as before:
\begin{equation}
  G^T(\xv,t;\xvp,t'):= -i\lla T\left[ \psio(\xv,t)\psiod(\xvp,t') \right] \rra 
\end{equation}
where the angular brackets stand for the thermal average with the
statistical operator $\rho^T=\exp[-\beta(\hat{H}-\mu\hat{N})]$,
\begin{equation*}
  \lla{\cal O}\rra := \frac{{\rm Tr}\left[ \rho^T {\cal O}\right]}{\cal Z}\:,
  \qquad {\cal Z}={\rm Tr}\left[\rho^T\right] \:.
\end{equation*}

However, the statistical operator satisfies Bloch's equation, namely 
\begin{equation*}
  -\frac{\partial\rho}{\partial \beta} = \lp{\hat H}-\mu\hat N\rp\rho \:,
\end{equation*}
which is formaly identical to Schr\"odinger's equation with the equivalences
\begin{equation*}
  it/\hbar \leftrightarrow \beta \qquad
  \hat H \leftrightarrow \hat H -\mu\hat N\:.
\end{equation*}
This suggests to define the finite-temperature Green's function just as the
zero-temperature one, but performing these replacements everywhere, that is,
\begin{equation}
  G(\xt,\xtp) := -\lla T\left[ \psio(\xt)\psiod(\xtp)\right] \rra \:,
\end{equation}
with $\tau=it$ the so-called ``imaginary time''.
It can be shown that this \index{propagator}propagator 
(and not the $G^T$ defined above) can be
evaluated in a series expansion as it can be done with 
the zero-temperature Green's function.

We turn now to show three general properties of this $G$ that will be of
interest. Dropping for simplicity the $\xv$-dependence of the propagator,
it is easy to show that $G(\tau,\tau')=G(\tau'-\tau)$:
\begin{align*}
  G(\tau,\tau') &= 
  -\inv{\z}\tr\left[e^{-\beta K}e^{K\tau/\hbar}\psio e^{-K\tau/\hbar}
                    e^{K\tau'/\hbar}\psiod e^{-K\tau'/\hbar} \right] \\
  &=
  -\inv{\z}\tr\left[e^{K\tau/\hbar}e^{-\beta K}\psio
                    e^{K(\tau'-\tau)/\hbar}\psiod e^{-K\tau'/\hbar} 
	      \right] \\
  &=
  -\inv{\z}\tr\left[e^{-\beta K}\psio e^{K(\tau'-\tau)/\hbar}
                    \psiod e^{-K(\tau'-\tau)/\hbar} \right] 
  \equiv G(0,\tau'-\tau) \:,
\end{align*}
where we introduced the grand-canonical Hamiltonian $K=\hat H -\mu\hat N$, 
and used the cyclic property of the trace.

We can also show that, for fermion, $G(-\beta\hbar)=- G(0)$:
\begin{align*}
  G(-\beta\hbar) &= 
  -\inv{\z}\tr\left[e^{-\beta K}\psio e^{-K\beta}\psiod e^{K\beta} \right]
  =
  -\inv{\z}\tr\left[e^{-\beta K}\psiod e^{K\beta}e^{-K\beta}\psio \right] \\
  &=
  -\inv{\z}\tr\left[e^{-\beta K}\psiod \psio \right] 
  = \pm G(0) \:,
\end{align*}
where we used again the cyclic property of the trace.

With these two properties, it is easy to show that $G(\tau)$ is periodic 
with period $\beta\hbar$, and it is useful to introduce its components 
in Fourier space,
\begin{equation}
  G(\tau) = \sumn e^{-i\om_n \tau}G(\om_n) \:.
\end{equation}
Then, we have
\begin{equation*}
  G(0) = \sumn G(\om_n) \qquad
  G(-\beta\hbar) = \sumn e^{i\hbar\om_n\beta}G(\om_n) \:.
\end{equation*}
From the condition $G(-\beta\hbar)=-G(0)$ we therefore get
\begin{equation} 
  e^{i\hbar\om_n\beta}=-1 \Rightarrow \om_n=(2n+1)\frac{\pi}{\beta\hbar} 
  \:,
\end{equation}
which are called {\em fermionic Matsubara frequencies}.

Let us return to 
the evaluation of a summation over such Matsubara frequencies $\om_n$
as the one we encountered in Sect.~\ref{sub:symmetric},
evaluation of
\begin{equation}
  S=\sum_n \lp\frac{1}{i\hbar\om_n-E_k} - \frac{1}{i\hbar\om_n+E_k}\rp \;.
  \label{eq:suma}
\end{equation}
%

\begin{figure}[hbt] 
  \caption[Complex $z$-plane with the contours $C$ and $C'+\Gamma$.]
          {\label{fig:suma}
	    Complex $z$-plane with the contours $C$ and $C'+\Gamma$. 
	    The crosses indicate the poles of $\tanh(z)$ while the circles are
	    located at $\pm E_k$.}
  \begin{center}
    \includegraphics[height=0.55\textwidth]{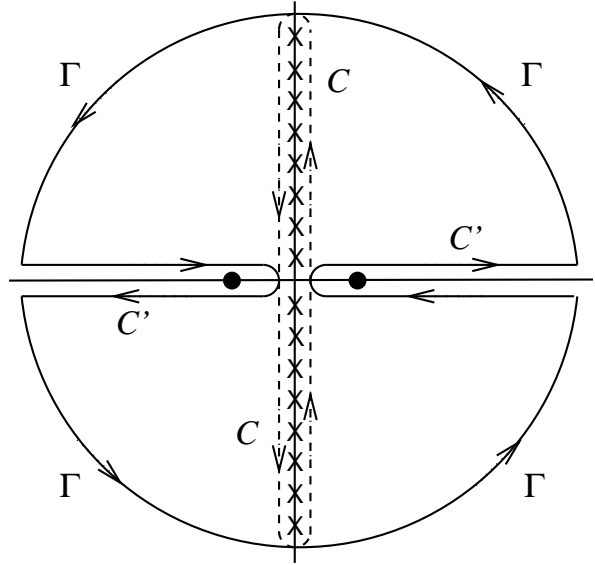}
  \end{center}
\end{figure}

The way to perform this kind of sumations is to convert them by means of
Cauchy's theorem on integrals on the complex plane~\cite{arfken}:
\begin{equation}
  \frac{1}{2\pi i}\oint_{C} dz f(z) = \sum_n \mathrm{Res}\{f(z_n)\} \;,
\end{equation}
where $f(z)$ is an analytic function 
inside the region enclosed (in the mathematically positive sense) 
by the contour $C$ (see Fig.\ref{fig:suma})
except on the points $\{z_n\}$, that are the poles of $f$;
$\mathrm{Res}\{f(z_n)\}$ is the residue of $f$ at $z_n$.

Looking at Eq.~(\ref{eq:suma}), we define a function $f$ on the complex
plane by 
\begin{equation}
  f(z) := \frac{1}{z-\ax E_k} - \frac{1}{z+\ax E_k} \;,
\end{equation}
so that now
\begin{equation}
  S=\ax\sum_n f(z_n) \;,
  \qquad z_n= \frac{\beta\hbar\om_n}{2} i = \frac{2n+1}{2}\pi i \;.
\end{equation}

It turns out that $\tanh(z)$ has poles precisely at the points $z=z_n$, 
and these poles
have unit residues. Therefore, $f(z)\tanh(z)$ will have poles at the desired
positions and with the desired residues, and can write
\begin{eqnarray}
  S = \ax \frac{1}{2\pi i}\oint_{C} dz f(z) \tanh(z) \;,
\end{eqnarray}
where the contour $C$ encloses the imaginary axis, as shown by the dashed line
in Fig.~\ref{fig:suma}, but not the poles of $f$ on the real axis.

If we deform $C$ into $C'+\Gamma$ (see again Fig.~\ref{fig:suma}),
and realize that the integral over $\Gamma$ will not contribute at
$|z|\raw\infty$ because in that limit $f(z) \propto z^{-2}$, we get
\begin{align}
  S
  &= \ax \frac{1}{2\pi i}\oint_{C'+\Gamma} dz f(z) \tanh(z) 
   = \ax \frac{1}{2\pi i}\oint_{C'} dz f(z) \tanh(z) 
  \nonumber\\
  &= \ax
  \frac{1}{2\pi i}\lp-2\pi i\rp \sum_{z=\pm\ax E_k} 
  \mathrm{Res}\{\tanh(z)\} 
  =
  -\beta\tanh\lp\beta\frac{E_k}{2}\rp \;,
\end{align}
where we evaluated the integration over $C'$ using again Cauchy's theorem
but now with a `$-$' sign because $C'$ is oriented in the mathematically
negative sense, and we used the fact that $f(z)\tanh(z)$ is analytic on the
real axis except at $z=\pm \beta E_k/2$. Thus, we obtain
\begin{equation} 
  \sum_n \lp\frac{1}{i\hbar\om_n-E_k} - \frac{1}{i\hbar\om_n+E_k}\rp 
  =
  -\beta\tanh\lp\beta\frac{E_k}{2}\rp \;,
\end{equation}
which is the result used in deriving Eqs.~(\ref{eq:gap-eq}) and
(\ref{eq:rhos}).

\chapter{Deduction of the equations of motion of spin-$f$ condensates}
\label{app:spinor}

\allowdisplaybreaks[1]

\section{Introduction}
Consider the interaction between two spin-$f$ atoms:
\begin{align}
  V(\rv_1-\rv_2) &= \delta(\rv_1-\rv_2)\sum_{F=0}^{2f}g_FP_F \\
  g_F            &= \frac{4\pi\hbar^2a_F}{M}
\end{align}
where $a_F$ is the $s$-wave scattering length in the channel of total
spin $F$, and $M$ is the mass of the atoms. 
For bosons, only even $F$'s are allowed; for fermions, only odd $F$'s. From
now on we shall consider only bosons. 
For example, for $f=1,2$, we have
\begin{align*}
  V&=g_2P_2 + g_0P_0            &(f=1)\\
  V&=g_4P_4 + g_2P_2 + g_0P_0   &(f=2)
\end{align*}

On the other hand, we also have:
\begin{align}
  \fv_1\cdot\fv_2 &= \sum_{F=0}^{2f}\lambda_FP_F \\
  \lambda_F &= \frac{F(F+1)-2f(f+1)}{2}
\end{align}
To show this, consider the addition of two spins $f$ 
to a total spin $F$:
\begin{align*}
  (\fv_1+\fv_2)^2 &= \fv_1^2 + \fv_2^2 + 2\fv_1\cdot\fv_2 
     = 2f(f+1)+2\fv_1\cdot\fv_2\\
  (\fv_1+\fv_2)^2 &= F(F+1)
\end{align*}
Therefore,
\begin{equation*}
   \fv_1\cdot\fv_2 = \frac{F(F+1)-2f(f+1)}{2} \:.
\end{equation*}
To take into account all possible channels ($F=0,...,2f$) we must sum
over $F$, projecting each contribution with the corresponding
projector $P_F$, that is
\begin{equation}
  \fv_1\cdot\fv_2 = \sum_{F=0}^{2f} \frac{F(F+1)-2f(f+1)}{2}P_F \:.
\end{equation}
For instance,
\begin{align*}
 \fv_1\cdot\fv_2 &= -2P_0 + P_2         & (f=1)\\
 \fv_1\cdot\fv_2 &= -6P_0 - 3P_2 + 4P_4 & (f=2)
\end{align*}

Finally, 
the normalization condition on the projectors provides the identity
\begin{equation}
  \sum_{F=0}^{2f} P_F=1 \:.
\end{equation}

From all these equations we can obtain:
\begin{itemize}
\item
\framebox{Spin $f=1$:}
\begin{align*}
  V               &= g_2P_2 + g_0P_0 \\
  \fv_1\cdot\fv_2 &= -2P_0 + P_2 \\
  P_0+P_2         &= 1
\end{align*}
Therefore,
\begin{equation*}
  V=\frac{2g_2+g_0}{3} +
    \frac{g_2-g_0}{3}\fv_1\cdot\fv_2 \equiv
    c_0+c_2\fv_1\cdot\fv_2  \:.
\end{equation*}
\item
\framebox{Spin $f=2$:}
\begin{align*}
  V               &= g_4P_4 + g_2P_2 + g_0P_0 \\
  \fv_1\cdot\fv_2 &= -6P_0 -3P_2 +4P_4 \\
  P_0 + P_2 + P_4 &= 1
\end{align*}
Therefore,
\begin{equation*}
  V=\frac{3g_4+4g_2}{7} +
    \frac{3g_4-10g_2+7g_0}{7}P_0 +
    \frac{g_4-g_2}{7}\fv_1\cdot\fv_2 \equiv
    c_0 + c_1P_0 + c_2\fv_1\cdot\fv_2  \:.
\end{equation*}
\end{itemize}

Thus, we can write the Hamiltonian in the form
\begin{align}
  H-\mu N &= \sum_{a=-f}^f \int \mathrm{d}^3r \left\{
    \hat\psi^+_a\left(-\hm\nabla^2\right) \hat\psi_a
    + (V_{\rm ext}-\mu)\hat\psi^+_a\hat\psi_a  \right.   \nonumber\\
    &
    +\sum_{a'=-f}^f \mig c_0\hat\psi^+_a\hat\psi^+_{a'}\hat\psi_{a'}\hat\psi_a
    +\sum_{a',b,b'=-f}^f \mig
      c_2\hat\psi^+_a\hat\psi^+_{a'}\fv_{ab}\cdot\fv_{a'b'}
         \hat\psi_{b'}\hat\psi_b
    \nonumber\\
    &
    +\sum_{a',b,b'=-f}^f \mig 
      c_1\hat\psi^+_a\hat\psi^+_{a'}(P_0)_{aa',bb'}\hat\psi_{b'}\hat\psi_b
      \mbox{~~[this term only for $f=2$]}
    \nonumber\\
    & \left.
    -\sum_{a'=-f}^f   p(z)\hat\psi^+_a(F_z)_{aa'}\hat\psi_{a'}
    +\sum_{a'=-f}^f   q\hat\psi^+_a(F^2_z)_{aa'}\hat\psi_{a'}  \:,
    \right\}
\end{align}
where $\hat\psi_a$ is the anihilation operator for an atom in state $|1,a\ra$,
Notice also that we have taken into account the effect of a magnetic field 
in the lowest two orders by
\begin{align*}
  p(z) &= g\mu_B B(z) 
     \mbox{~~~~~~~linear Zeeman effect for $\vec{B}=B\hat{z}$,}\\
  q    &= \frac{g^2\mu_B^2}{16h\nu_{hfs}}B_0^2 
     \mbox{~~~~quadratic Zeeman effect} \:.
\end{align*}
Here $g$ is the gyromagnetic factor for the atom, and $h\nu_{hfs}$ is the
hyperfine splitting characteristic of the manifold in which the atoms are.

\section{Mean-field dynamical equations for a spin-1 condensate}
We make the usual mean-field approximation for the field operators,
\begin{equation}
  \psi_a(\rv)=\langle \hat\psi_a(\rv) \rangle \:,
  \label{eq:mean-field}
\end{equation}
%
The equation of motion for the field $\psi_m$ can be derived from
the Hamiltonia by functional differentiation:
\begin{align}
  i\hbar\dt{\psi_m} &= \fm{(H-\mu N)} \notag\\
                    &= -\hM\nabla^2\psi_m +
        [V_{\rm ext}-\mu]\psi_m \nonumber \\
     &
	+ c_0\left(\sum_a\psi^*_a\psi_a\right)\psi_m
	+ c_2\left(\sum_{ab}\psi^*_a\fv_{ab}\psi_b\right)
	     \cdot\left(\fv\psi\right)_m \nonumber \\
     &
	-p(z)\left(F_z\psi\right)_m + q\left(F_z^2\psi\right)_m \:.
\end{align}

In the mean field approximation, one can rewrite the several terms 
in this way: 
\begin{itemize}
\item
\framebox{$\sum_a \psi^*_a\psi_a$:}
  \begin{equation*} 
    \sum_a \psi^*_a\psi_a \raw \sum_a n_a (\rv) = n(\rv) \:,
  \end{equation*}
which is the total density of particles at point $\rv$.
\item
\framebox{$\sum_{ab}\psi^+_a\fv_{ab}\psi_b$:}
  \begin{align*}
    \sum_{ab}\psi^+_a\fv_{ab}\psi_b 
    &\raw \sum_{ab} \sum_{i=x,y,z} \psi^*_a \lp F_i\rp_{ab}\psi_b\hat{\imath} 
    \\
    &= \sum_i \hat{\imath} \left\{
                     \lp F\rp_{11}|\psi_1|^2 +
		     \lp F\rp_{10}\psi^*_1\psi_0
		     \lp F\rp_{1-1}\psi^*_1\psi_{-1} + \cdots \right\} \\
    &= \lp\begin{array}{cc}
          [i=x] & 
	      \inv{\sqrt{2}}\left\{ \psi^*_1\psi_0 + \psi^*_0\psi_1
	                   +\psi^*_0\psi_{-1} + \psi^*_{-1}\psi_0 \right\}
			   \cr
          [i=y] & 
	      \inv{i\sqrt2}\left\{ \psi^*_1\psi_0 - \psi^*_0\psi_1
	                   +\psi^*_0\psi_{-1} - \psi^*_{-1}\psi_0 \right\}
			   \cr
          [i=z] & \left\{ n_1 - n_{-1} \right\}
	  \end{array}
	  \rp \:,
  \end{align*}
  where we have used the matrix expression of the spin operators, \ie\ the
  Pauli matrices for spin $f=1$, namely
  \begin{subequations}
  \label{eq:matrius1}
  \begin{align}
    F_x &=
    \frac{1}{\sqrt2}\left(\begin{array}{ccc}
                          0 & 1 & 0 \\
		          1 & 0 & 1 \\
		          0 & 1 & 0 \\
                        \end{array}
                  \right)
    \:, \\ 
    F_y &= 
    \frac{1}{i\sqrt2}\left(\begin{array}{ccc}
                           0 &  1 & 0 \\
		          -1 &  0 & 1 \\
		           0 & -1 & 0 \\
                         \end{array}
                  \right)
    \:, \\ 
    F_z &=
    \left(\begin{array}{ccc}
          1 & 0 &  0 \\
	  0 & 0 &  0 \\
	  0 & 0 & -1 \\
    \end{array}
    \right) \:.
  \end{align}
  \end{subequations}
\item
\framebox{$\la \fv \ra \cdot\left(\fv\psi\right)_m$:}
As above, we use the Pauli matrices to arrive at
\begin{equation*}
  \left(\sum_{ab}\psi^*_a\fv_{ab}\psi_b\right) \cdot\left(\fv\psi\right)_m
  = \lp\begin{array}{cc}
    [m=1]  & \psi_0^2\psi^*_{-1} + (n_1+n_0-n_{-1})\psi_1 \cr
    [m=0]  & (n_1+n_{-1})\psi_0 +2\psi_1\psi^*_0\psi_{-1} \cr
    [m=-1] & \psi_0^2\psi^*_{1} + (n_{-1}+n_0-n_{1})\psi_{-1}
       \end{array}
       \rp
\end{equation*}
\end{itemize}
Collecting the partial results, we get
\begin{subequations}
\label{eq:dynam}
\begin{align}
  i\hbar\parct{\psi_1} 
    &= \left[-\hM\nabla^2 + V_{\rm ext} + c_0n(\rv)\right]\psi_1 
    \notag\\
    &+c_2 \left[ \lp n_1+n_0 -n_{-1}\rp\psi_1 + \psi^2_0\psi^*_{-1} \right] 
    \notag\\
    &-p\psi_1 + q\psi_1 \\
  i\hbar\parct{\psi_0} 
    &= \left[-\hM\nabla^2 + V_{\rm ext} + c_0n(\rv)\right]\psi_0 
    \notag\\*
    &+c_2 \left[ \lp n_1+n_{-1}\rp\psi_0 + 2\psi_1\psi^*_0\psi_{-1} \right]
    \\
  i\hbar\parct{\psi_{-1}} 
    &= \left[-\hM\nabla^2 + V_{\rm ext} + c_0n(\rv)\right]\psi_{-1} 
    \notag\\
    &+c_2 \left[ \lp n_{-1}+n_0-n_1\rp\psi_{-1} + \psi^2_0\psi^*_1 \right] 
    \notag\\
    &+p\psi_{-1} + q\psi_{-1} \:.
\end{align}
\end{subequations}
We have separated in different lines the contributions: (1) equal for all
components, (2) proportional to $c_2$ and related to spin-exchange collisions,
and (3) terms due to the presence of a magnetic field. Note that, except for
the linear Zeeman term ($\propto p \propto B$), the equations for $m=1$ and
$m=-1$ are invariant under the relabelling $1\leftrightarrow-1$, indicating
our freedom to choose the orientation of the quantization axis. 
More generally, the equations are invariant under the time-reversal operator,
that transforms $1 \leftrightarrow -1$ and $B \rightarrow -B$.

These dynamical equations can be written in the shorter form
(\ref{eq:dyneqs2}) by introducing effective potentials that, in the case of a
non-vanishing magnetic field, read 
\begin{align*}
  V^\mathrm{eff}_{\pm1} &= V_\mathrm{ext} 
                           + c_0n + c_2(\pm n_1+n_0\mp n_{-1})
			   \mp p +q \\
  V^\mathrm{eff}_0 &= V_\mathrm{ext} 
                           + c_0n + c_2(n_1+n_{-1}) \:.
\end{align*}

It is also possible to derive a set of coupled continuity equations from
(\ref{eq:dynam}). To this end, we define the number density of the $m$
component as usual by
\begin{align*} 
  n_m (\rv,t) &= |\psi_m(\rv,t)|^2 \:.
\end{align*}
Applying the operator $i\hbar\partial/\partial t$ to $n_0$, we get
\begin{align*}
  i\hbar\parct{n_0} 
  &= i\hbar\parct{\psi_0^*}\psi_0 + i\hbar\parct{\psi_0}\psi_0^*\\
  &=  -\hM\left[ -\psi_0\nabla^2\psi_0^* + \psi_0^*\nabla^2\psi_0 \right]
     +2c_2\left[ -\psi_1^*\psi_0^2\psi_{-1}^* 
                 + \psi_1(\psi_0^2)^*\psi_{-1}\right] \:.
\end{align*}
Now, defining the quantum current of the $m$ component by
\begin{equation*}
  \bm{j}_m = \frac{\hbar}{2iM} (\psi_m^* {\bm \nabla} \psi_m 
                                -\psi_m {\bm \nabla} \psi_m^*) \:,
\end{equation*}
we can rewrite this result as
\begin{align*} 
  \parct{n_0} + \nablab\cdot\bm{j}_0 
  &= 2\frac{c_2}{i\hbar}\left[ -\psi_1^*\psi_0^2\psi_{-1}^* 
                               + \psi_1(\psi_0^2)^*\psi_{-1}\right] \:,
\end{align*}
which has the form of a continuity equation with a source term given by
$-(2c_2/\hbar)\,\mbox{Im}[T_0\psi_0]\in \mathbb{R}$, with
$T_0:=2\psi_1\psi_0^*\psi_{-1}$.
For the $m=\pm1$ components, a similar treatment yields
\begin{align*}
  \parct{n_{\pm1}} + \nablab\cdot\bm{j}_{\pm1} 
  &= -\frac{2c_2}{\hbar}\,\mbox{Im}[T_{\pm1}\psi_{\pm1}] \:, \\
  T_{\pm1}^* &:= \psi_0^2\psi_{\mp1}^* \:.
\end{align*}

\section{Mean-field dynamical equations for a spin-2 condensate}
The starting point is the grand-canonical Hamiltonian that, for $f=2$ reads:
\begin{align}
  H-\mu N &= \sum_{a=-2}^2 \int \mathrm{d}^3r \left\{
    \psi^\dagger_a\left(-\hM\nabla^2\right) \psi_a
    + (V_{\rm ext}-\mu)\psi^+_a\psi_a  \right.   \nonumber\\
    &
    +\sum_{a'=-2}^2 \mig c_0\psi^+_a\psi^+_{a'}\psi_{a'}\psi_a
    +\sum_{a',b,b'=-2}^2 \mig
      c_2\psi^+_a\psi^+_{a'}\fv_{ab}\cdot\fv_{a'b'}
         \psi_{b'}\psi_b
    \nonumber\\
    &
    +\sum_{a',b,b'=-2}^2 \mig 
      c_1\psi^+_a\psi^+_{a'}(P_0)_{aa',bb'}\psi_{b'}\psi_b
    \nonumber\\
    & \left.
    -\sum_{a'=-2}^2   p(z)\psi^+_a(F_z)_{aa'}\psi_{a'}
    +\sum_{a'=-2}^2   q\psi^+_a(F^2_z)_{aa'}\psi_{a'}  \:,
    \right\}
\label{ham2}
\\
  \langle\langle \hat A \rangle\rangle &:= \sum_{a,a',b,b'} 
    \xi^*_a\xi^*_{a'}A_{ab} \xi_{b'}\xi_b \:,
       \quad A_{aa',bb'} := \langle aa'|\hat A|bb' \rangle
\end{align}

Again, we derive the equations of motion of the field operators 
by a functional derivative of the Hamiltonian~(\ref{ham2}):
\begin{align}
  i\hbar\dt{\psi_m} &= \fm{(H-\mu N)} \notag\\
                    &= -\hM\nabla^2\psi_m +
        [V_{\rm ext}-\mu]\psi_m \nonumber \\
     &
	+ c_0\left(\sum_a\psi^\dagger_a\psi_a\right)\psi_m
	+ c_2\left(\sum_{ab}\psi^\dagger_a\fv_{ab}\psi_b\right)
	     \cdot\left(\fv\psi\right)_m \nonumber \\
     &  
        +c_1 \sum_{abc}\psi^\dagger_a\left(P_0\right)_{ma,bc}\psi_c\psi_b
        \nonumber\\
     &
	-p(z)\left(F_z\psi\right)_m + q\left(F_z^2\psi\right)_m \:.
\label{eq:spin2}
\end{align}
As we are working with $f=2$, all spin operators can be
represented by the following $5\times5$ Pauli matrices:
\def\a{\sqrt{\frac{3}{2}}}
\begin{subequations}
\label{eq:matrius2}
\begin{align}
  F_x &= 
  \left(\begin{array}{ccccc}
                           0 &  1 &  0 &  0 & 0 \\
		           1 &  0 & \a &  0 & 0 \\
		           0 & \a &  0 & \a & 0 \\
			   0 &  0 & \a &  0 & 1 \\
			   0 &  0 &  0 &  1 & 0 \\
                        \end{array}
                  \right) \:, \\
  F_y &=
  -i\left(\begin{array}{ccccc}
                           0 &  1 &  0 &  0 & 0 \\
		          -1 &  0 & \a &  0 & 0 \\
		           0 &-\a &  0 & \a & 0 \\
			   0 &  0 &-\a &  0 & 1 \\
			   0 &  0 &  0 & -1 & 0 \\
                        \end{array}
                  \right) \:, \\
  F_z &=
  \left(\begin{array}{ccccc}
                           2 &  0 &  0 &  0 & 0 \\
		           0 &  1 &  0 &  0 & 0 \\
		           0 &  0 &  0 &  0 & 0 \\
			   0 &  0 &  0 & -1 & 0 \\
			   0 &  0 &  0 &  0 &-2 \\
                        \end{array}
                  \right) \:.
\end{align}
\end{subequations}
With them, we can calculate most of the terms in the equations of
motion. 
For example, performing again a mean-field approximation we get
\begin{align}
  \langle F_x\rangle 
  &:= 
  \left(\psi^\dagger_2~ \psi^\dagger_1~ \psi^\dagger_0~ 
        \psi^\dagger_{-1}~ \psi^\dagger_{-2}\right)
  \left(\begin{array}{ccccc}
                           0 &  1 &  0 &  0 & 0 \\
		           1 &  0 & \a &  0 & 0 \\
		           0 & \a &  0 & \a & 0 \\
			   0 &  0 & \a &  0 & 1 \\
			   0 &  0 &  0 &  1 & 0 \\
                        \end{array}
                  \right)
  \left(\begin{array}{c}
                   \psi_2 \\
		   \psi_1 \\
		   \psi_0 \\
		   \psi_{-1} \\
		   \psi_{-2}
    \end{array}
  \right) \notag\\
  &\raw 
  \left(\psi^*_2~ \psi^*_1~ \psi^*_0~ \psi^*_{-1}~ \psi^*_{-2}\right)
  \left(\begin{array}{c}
                   \psi_1 \\
		   \psi_2+\a\psi_0 \\
		   \a\left(\psi_1+\psi_{-1}\right) \\
		   \a\psi_0+\psi_{-2} \\
		   \psi_{-1}
    \end{array}
  \right) \notag\\
  &= 
  \psi^*_2\psi_1 + \psi^*_1\left(\psi_2+\a\psi_0\right) \notag\\
  &+
     \a\psi^*_0\left(\psi_1+\psi_{-1}\right)
  +\psi^*_{-1}\left(\a\psi_0+\psi_{-2}\right)
     +\psi^*_{-2}\psi_{-1}
  \:.
\end{align}
Analogously, we get
\begin{align}
  \langle F_y\rangle 
  &= 
  -i\psi^*_2\psi_1 + i\psi^*_1\left(\psi_2-\a\psi_0\right) \notag\\ 
  &+
    i\a\psi^*_0\left(\psi_1-\psi_{-1}\right)
    +i\psi^*_{-1}\left(\a\psi_0-\psi_{-2}\right)
    +i\psi^*_{-2}\psi_{-1}
  \:, \\
  \langle F_z\rangle 
  &= 
  2|\psi_2|^2 + |\psi_1|^2 - |\psi_{-1}|^2 - 2|\psi_{-2}|^2
  \:,
\end{align}
Therefore, the term proportional to $c_2$ becomes
\begin{align*}
  \langle\fv\rangle &\cdot \fv\vec{\psi}
  = \langle F_x\rangle F_x\vec{\psi}
     + \langle F_y\rangle F_y\vec{\psi} + \langle F_z\rangle F_z\vec{\psi}
     \\
  &=
  \left(\begin{array}{cl}
       \langle F_x \rangle \psi_1 
      -i\langle F_y\rangle \psi_1
     +2\langle F_z\rangle \psi_2
     & [m=2] \\
       \langle F_x \rangle \left(\psi_2+\a\psi_0\right)
      +i\langle F_y\rangle \left(\psi_2-\a\psi_0\right)
      +\langle F_z\rangle \psi_1 
     & [m=1] \\
       \langle F_x \rangle \a\left(\psi_1+\psi_{-1}\right)
      +i\langle F_y\rangle \a\left(\psi_1-\psi_{-1}\right)
     & [m=0] \\
       \langle F_x \rangle \left(\a\psi_0+\psi_{-2}\right)
      +i\langle F_y\rangle \left(\a\psi_0-\psi_{-2}\right)
      -\langle F_z\rangle \psi_{-1}
     & [m=-1] \\
       \langle F_x \rangle \psi_{-1}
      +i\langle F_y\rangle \psi_{-1}
     -2\langle F_z\rangle \psi_{-2}
     & [m=-2]
    \end{array}
  \right)
\end{align*}

\def\b{\frac{1}{5}}
Now we must calculate the $P_0$-terms. $P_0$ projects a two-particle
state onto its $F=0$ component, $P_0=|00\rangle\langle 00|$. Therefore,
I shall use some Clebsh-Gordan coefficients to go from the uncoupled basis
of states $|fm_1fm_2\ra$ to the coupled basis $|FM\ra$:
\begin{align*}
  |bc\ra &= \sum_{FM}|FM\ra\la FM|fb,fc\ra \stackrel{f=2}=
             \sum_{FM}|FM\ra\la FM|2b,2c\ra \nonumber \\
	 &= |00\ra\la 00|2b,2c\ra 
	     + \sum_{M=-1}^{+1}|1M\ra\la1M|2b,2c\ra
	     + \sum_{M=-2}^{+2}|2M\ra\la2M|2b,2c\ra
\end{align*}
Applying the definition of $P_0$, we have
\begin{align*}
  P_0|bc\ra &= |00\ra\la00|2b,2c\ra \equiv |00\ra\la00|bc\ra \\
  \left(P_0\right)_{ma,bc} &:= \la ma|P_0|bc\ra 
  = \la 2m,2a|00\ra\la00|2b,2c\ra
\end{align*}
Note that $(P_0)_{ma,bc}=(P_0)_{am,bc}$. Thus, it is possible the merge
the two terms coming from the functional derivative into one single
term, that cancels the $\mig$ factor in front of $c_1$, as we did in
Eq.~(\ref{eq:spin2}).

The Clebsch-Gordan coefficients for two spin-2 particles that give a
total $F=0$ are:
\begin{align*}
  \la 2 m_1, 2 m_2 |00\ra = \left\{
  \begin{array}{ll}
    \frac{1}{\sqrt5}  & m_1=\pm2,m_2=\mp2 \:, m_1=m_2=0 \\
    \frac{-1}{\sqrt5} & m_1=\pm1,m_2=\mp1
  \end{array}
  \right.
\end{align*}
Therefore,
\begin{equation*}
  \left(P_0\right)_{ma,bc}=\left\{
  \begin{array}{ll}
    \frac{1}{5} & (m,a)\&(b,c)\in\{(\pm2,\mp2),(0,0)\} \\
                & \mbox{or~~~}(m,a)\&(b,c)\in\{(\pm1,\mp1)\} \\
   \frac{-1}{5} & (m,a)\in\{(\pm2,\mp2),(0,0)\}
                  \&(b,c)\in\{(\pm1,\mp1)\} \\
                & \mbox{or~~~}(m,a)\in\{(\pm1,\mp1)\}
                  \&(b,c)\in\{(\pm2,\mp2),(0,0)\}
  \end{array}
  \right.
\end{equation*}
For example, for $m=2$ the $P_0$-terms are:
\begin{align*}
  \sum_{abc}\psi^*_a\left(P_0\right)_{2a,bc}\psi_b\psi_c &=
    \sum_{abc}\psi^*_a \la 22,2a|00\ra\la00|2b,2c\ra \psi_b\psi_c
    \notag\\
  &=
    \sum_{a}\psi^*_a \la 22,2a|00\ra \sum_{bc} \la00|2b,2c\ra \psi_b\psi_c
  \notag\\
  &=
    \frac{1}{\sqrt5}\psi^*_{-2}\sum_{bc} \la00|2b,2c\ra\psi_b\psi_c 
  \notag\\
  &=
    \psi^*_{-2} \b\left[2\psi_2\psi_{-2}+\psi_0^2 - 2\psi_1\psi_{-1}\right]
\end{align*}
Note the common factor that will be the same for all $m$:
\begin{equation*}
  \Sigma'=\sum_{bc}\la00|2b,2c\ra\psi_b\psi_c=
     \frac{1}{\sqrt5}\left[2\psi_2\psi_{-2}+\psi_0^2 - 2\psi_1\psi_{-1}\right]
  \:.
\end{equation*}
The other term can be expressed as:
\begin{equation*}
  \sum_{a}\la 2m,2a|00\ra\psi_a^* = 
    (-1)^m \frac{1}{\sqrt5}\delta_{a,-m}\psi_{-m}^* \:.
\end{equation*}
Therefore, if we define
\begin{equation*}
  \Sigma=\frac{1}{\sqrt5}\Sigma' = 
     \b\left[2\psi_2\psi_{-2}+\psi_0^2 - 2\psi_1\psi_{-1}\right] \:,
\end{equation*}
we can write the equations of motion for
the different spinor components thus:
\def\bb{c_1n}
\begin{subequations}
\begin{align}
  i\hbar\dt{\psi_2} 
    &= {\cal H}_0\psi_2 \notag \\
    &+c_2n\left[\left(\la F_x\ra-i\la F_y\ra\right)\psi_1 
                    + 2\la F_z\ra\psi_2\right] \notag\\
    &+\bb\psi^*_{-2}\Sigma +2(2q-p)\psi_2 \\
  i\hbar\dt{\psi_1} 
    &= {\cal H}_0\psi_1 \notag \\
    &+
    c_2n\left[\la F_x\ra\left(\psi_2+\a\psi_0\right)
          +i\la F_y\ra\left(\psi_2-\a\psi_0\right) 
	  + \la F_z\ra\psi_1\right] \nonumber\\
    &+\bb\psi^*_{-1}\Sigma
      +(q-p)\psi_1 \\
  i\hbar\dt{\psi_0} 
    &= {\cal H}_0\psi_0 \notag \\
    &+
    c_2n\left[ \a\la F_x\ra\left(\psi_1+\psi_{-1}\right)
         +i\a\la F_y\ra\left(\psi_1-\psi_{-1}\right) \right] \notag\\
    &+\bb\psi^*_{0}\Sigma \\ 
  i\hbar\dt{ \psi_{-1} } 
    &= {\cal H}_0\psi_{-1} \notag \\
    &+
    c_2n\left[\la F_x\ra\left(\a\psi_0+\psi_{-2}\right)
          +i\la F_y\ra\left(\a\psi_0-\psi_{-2}\right) 
	  - \la F_z\ra\psi_{-1}\right] \nonumber\\
    &+\bb\psi^*_{1}\Sigma
      +(q+p)\psi_{-1} \\
  i\hbar\dt{\psi_{-2}} 
    &= {\cal H}_0\psi_{-2} \notag \\
    &+
    c_2n\left[\left(\la F_x\ra+i\la F_y\ra\right)\psi_{-1} 
                    - 2\la F_z\ra\psi_{-2}\right] \notag\\
    &+\bb\psi^*_{2}\Sigma
      +2(2q+p)\psi_{-2} \:,
\end{align}
\end{subequations}
where ${\cal H}_0=-\hM\nabla^2 + V_{\rm ext}-\mu + c_0n(\rv)$ contains the factors
common to all equations. Explicit expressions for $\la F_i\ra$
($i=x,y,z$) and $\Sigma$ have been given above.
%

\vskip 1cm
\begin{center}
\em
Forsi altro canter\`a con miglior plectio

\vskip 1cm

FINIS
\vskip 0.3cm

\rule{2cm}{0.3pt}
\end{center}



\backmatter
\bibliographystyle{jordi}
\bibliography{biblio,biblio-meva}

\cleardoublepage
\printindex

\end{document}